\documentclass[11pt]{article}

\usepackage[top=1in, bottom=1in, left=1in, right=1in]{geometry}
\usepackage{setspace}
\usepackage[T1]{fontenc}
\usepackage{times}
\usepackage{color}
\usepackage[section]{placeins} %Placeins.sty keeps floats `in their place', preventing them from floating past a "\FloatBarrier" command into another section.  To use it, declare "\usepackage{placeins}" and insert "\FloatBarrier" at places that floats should not move past, perhaps at every "\section".  
\usepackage[large, bf]{caption}
\usepackage[FIGTOPCAP]{subfigure}
\usepackage{pdfpages}
\usepackage{palatino}
\usepackage{pdflscape}
\usepackage{lscape}
\usepackage{textcomp}
\usepackage{nicefrac}
\usepackage{multirow}
\usepackage{booktabs}
\usepackage{adjustbox}	% to adjust the size of objects to fit into a page
\usepackage[hyphens]{url}
\usepackage{fancyhdr}
\usepackage[section]{placeins}

\usepackage{natbib}
\bibpunct{(}{)}{;}{a}{,}{,}

\usepackage{amsmath, amsfonts, amssymb, amsthm}

\usepackage{mathpazo} %Use Palotino fonts
\usepackage[pdftex]{hyperref}
\hypersetup{colorlinks, citecolor=blue, filecolor=blue, linkcolor=blue, urlcolor=blue}

\def\urltilda{\kern -.15em\lower .7ex\hbox{\~{}}\kern .04em}

\usepackage{titlesec}

\titlespacing*{\section}{0pt}{1.5ex plus 1ex minus .2ex}{0.8ex plus .2ex}
\titlespacing*{\subsection}{0pt}{1.2ex plus 1ex minus .2ex}{0.8ex plus .2ex}	
\usepackage[percent]{overpic}
\usepackage{dcolumn}
\newcolumntype{d}[0]{D{.}{.}{5}}

\usepackage{graphicx}
\usepackage[space]{grffile}

\title{\textbf{\LARGE{The Great Equalizer: Medicare and the Geography of Consumer Financial Strain}}\thanks{First version: April 22, 2017. This version: \today. The views expressed are those of the authors and do not necessarily reflect those of the Federal Reserve Bank of New York or the Federal Reserve System. The authors would like to thank Neale Mahoney, Matt Notowidigdo, Chima Ndumele, Mark Schlesinger, Julia Smith, Isaac Sorkin, Eric Zwick, and Trevor Gallen for helpful comments along with participants at the BU, Harvard, MIT Health Economics Seminar; Yale SOM Finance Lunch; Association for Public Policy Analysis and Management; Gerzensee European Summer Symposium in Financial Markets; NTA Annual Conference on Taxation; and Salomon Center/BPI Conference on Household Finance. Joseph Doran, Danno Lemu, Davy Perlman, and Lauren Thomas provided excellent research assistance.}}

\author{
	Paul Goldsmith-Pinkham\thanks{Yale School of Management. Email: \href{mailto:paul.goldsmith-pinkham@yale.edu}{paul.goldsmith-pinkham@yale.edu}} \and
	\and
	Maxim Pinkovskiy\thanks{Federal Reserve Bank of New York. Email: \href{mailto:maxim.pinkovskiy@ny.frb.org}{maxim.pinkovskiy@ny.frb.org}} \and
	\and
    Jacob Wallace\thanks{Yale School of Public Health. Email: \href{mailto:jacob.wallace@yale.edu}{jacob.wallace@yale.edu}} 
      }
			
\date{\today}

\begin{document}

\maketitle
\thispagestyle{empty} 
\setcounter{page}{0}

\begin{abstract}
    We use a five percent sample of Americans' credit bureau data, combined with a regression discontinuity approach, to estimate the effect of universal health insurance at age 65---when most Americans become eligible for Medicare---at the national, state, and local level. We find a 30 percent reduction in debt collections---and a two-thirds reduction in the geographic variation in collections---with limited effects on other financial outcomes. The areas that experienced larger reductions in collections debt at age 65 were concentrated in the Southern United States, and had higher shares of black residents, people with disabilities, and for-profit hospitals.
\end{abstract}

% Could consider adding the clause "even within the South, had higher shares of black residents...."
	  
%%%%%%%%%%%%%%%%%%%%%%%%%%%%%%%%%
% Text
%%%%%%%%%%%%%%%%%%%%%%%%%%%%%%%%%
	
\clearpage
\section{Introduction}

% Research question
Why does consumer financial strain vary so much across the United States \citep{keys2019}? In this paper, we examine the role that health insurance plays in shaping the geography of financial health. To do this, we use a five percent sample of Americans' credit report data, combined with a regression discontinuity (RD) approach, to estimate the effect of universal health insurance at age 65---when most Americans become eligible for Medicare---at the national, state, and commuting zone level.

% Why health insurance?

% How we use the place-based estimates
We use our location-specific estimates of Medicare's effects for three purposes. First, we show that Medicare reduces geographic variation in debt collections by two-thirds at age 65. Second, we show that the gains in financial health due to Medicare are greatest in the South, where a higher share of the near-elderly (i.e., 55-64 year olds) are uninsured and the financial health improvements \emph{per newly-insured individual} are largest. We show that the commuting zones (CZs) experiencing the largest gains in financial health at age 65 had larger shares of black residents, people with disabilities, and for-profit hospitals. Third, using shrinkage estimators, we construct forecasts of the causal effect of expanding coverage to the near-elderly in each of the 741 CZs in the United States, which we use to evaluate existing policies and potential future expansions.

% 1.1 | Quantify much Medicare reduces geographic disparities in consumer financial strain at age 65.
To quantify how Medicare reduces geographic disparities in consumer financial strain, we construct counterfactual estimates of consumer financial outcomes at age 65, with and without Medicare, for each locality. With these location-specific counterfactuals, we construct an estimator of the nationwide, cross-locality reduction in the \emph{variance} of consumer financial outcomes. Using this approach, we show that Medicare reduces the geographic variation in collections by two-thirds at age 65, highlighting an understudied aspect of the Medicare program---that it largely eliminates geographic disparities in access to insurance (which fall by 93.5\% at age 65), and substantially reduces geographic disparities in collections-related financial strain. However, we do not find evidence that Medicare reduces geographic variation in the other financial health outcomes we study (e.g., credit score, bankruptcy, delinquent debt), though our confidence intervals are quite wide.

% 2.  | Characteristics of CZs that experience the largest improvements in financial health at 65.

% 2.1 | Changes in insurance and changes in collections at 65 --- consider changing to topic sentence like: We begin by showing that the reductions in financial strain are mainly driven by reductions in uninsurance, rather than changes in the composition of health insurance.
Second, we explore why the effect of Medicare on collections debt, where we find a large reduction in geographic variation, differs so much across localities. 
 %We begin by asking whether the reductions in consumer financial strain are driven by reductions in the uninsurance rate, or changes in the composition of health insurance.%\footnote{For individuals with no insurance prior to Medicare, turning age 65 provides a big increase in risk protection. For individuals with insurance at age 65, the transition to Medicare can lead to a change in premiums, benefits, their provider network, and the set of supply incentives providers face when treating them (\citet{clemens2017shadow}).} 
%Given that Medicare is a national policy, people living in all locations are eligible to benefit. However, it is reasonable to expect that the impact may vary depending on individual (or local) characteristics. Since we cannot directly observe health insurance status in our credit report data, we focus on the relationship between insurance coverage and financial health at the locality-level. 
We find that reductions in collections debt are higher in areas that experienced larger increases in the insurance rate at age 65, suggesting that the gains in financial health are primarily driven by reducing the number of uninsured, rather than changes in the composition of coverage.\footnote{Our state-level results imply a reduction in collection balances of \$584 per newly-insured individual at 65, which falls within the range of estimates from prior work on the effects of Medicaid coverage \citep{finkelstein2012oregon,hu2018effect}.} Motivated by this finding, we then examine how CZ-level reductions in collections differ between areas on a per capita and per newly-insured basis, with the latter done by scaling the change in financial health estimates by the estimated effect of Medicare on insurance rates. %This allows us to separately examine an extensive margin coverage effect---locations with large changes in the number of uninsured---and the intensive margin---locations with large effects per newly-insured individual. 
While our analysis suggests that a \emph{primary} mechanism through which Medicare affects financial strain is by reducing the uninsurance rate, it is unlikely that the exclusion restriction holds.\footnote{For individuals with no insurance prior to Medicare, turning 65 provides a big increase in risk protection. However, for individuals with insurance at age 65, the transition to Medicare changes premiums, benefits, provider networks, and the set of incentives their providers face (\citet{clemens2017shadow}). Hence, the treatment we study is a combination of the effect of Medicare for those who were previously uninsured and those with other forms of coverage at age 64.} As a result, we view this as an informative scaling exercise rather than an estimate of the causal effect of health insurance coverage on debt collections. To understand why the effects of Medicare differ across areas, we examine the demographic and healthcare market characteristics associated with the 741 causal estimates of CZ-level reductions in per capita collections debt at age 65. We find that the effect of Medicare on collections debt is larger in areas with larger shares of black residents, people with disabilities, and for-profit hospitals.

% 3.1 | Introduction and methods
Third, we construct forecasts of the causal effect of Medicare on financial health in each CZ. This gives us a local approximation to the effect of lowering the Medicare eligibility age, a popular policy proposal. To reduce noise, we follow \cite{chetty2018impactsb} and construct forecasts using a shrinkage estimator that combines our unbiased RD estimates and a predicted effect for each CZ based on its demographic and healthcare market characteristics. Maps of the forecast reductions in per capita and per newly-insured debt collections are strikingly similar, with the largest values in both concentrated in the South. For example, a coverage expansion to the near-elderly is forecast to reduce collection balances by 53 dollars per capita in Raleigh, NC (one of the largest forecast reductions). In contrast, the same treatment in San Francisco, CA, is only forecast to reduce collection balances by 8 dollars per capita. This is not simply because there are a greater share of uninsured in Raleigh; in fact, the near-elderly uninsurance rates in the two places are similar.\footnote{The near-elderly uninsurance rates in Raleigh, NC and San Francisco, CA during the period 2014-2017 were 6.5 and 5.9 percent, respectively.} The difference in the forecasts arises primarily due to large differences in the forecast reductions in collections balances \textit{per newly-insured} individual in the two locations. In Raleigh, NC, the forecast reduction in collection balances was 956 dollars per newly-insured individual, 785\% higher than the analogous forecast in San Francisco, CA.

% 3.3 | How the ACA has changed the forecast
Lastly, we examine how CZ-level forecasts changed due to the passage of the Affordable Care Act (ACA) in 2010, federal health reform legislation that substantially expanded coverage \citep[e.g.,][]{frean2017premium}. Forecasts of the causal effect of expanding coverage on collections are smaller after the ACA's implementation in 2014 and have become more geographically concentrated in the South. This is because forecasted effects decreased by only 30\% in the South (states like Mississippi, Texas, and Georgia) after the ACA's implementation, while they decreased by 50\% in the rest of the country. Using a Kitagawa-Oaxaca-Blinder style decomposition, we show that this was due to the uniformity of ACA coverage gains across regions for the near-elderly---despite higher rates of uninsurance in the South---and, within the South, the fact that near-elderly uninsurance rates remain highest in areas where the financial health gains of expanding coverage per newly-insured are largest. These findings highlight a potential limitation of policies, such as the ACA, that delegate states considerable latitude in policy implementation, and a relative advantage of programs, such as Medicare, that are federally-administered---specifically, that the former may exacerbate geographic disparities while the latter tend to reduce them.

%one-quarter of this gap is due to smaller gains in health insurance in the South and one-third of the gap is due improved targeting (on debt collections) in the rest of the country, and zero change in targeting in the South. The remaining share is due to a growth in per newly-insured effects in the South.

%First, the near-elderly uninsured have become increasingly concentrated in the South. This is a result of the geographically diffuse coverage gains associated with the ACA. Second, the forecast reductions in collections debt per newly-insured individual are largest in the South (before and after the ACA), compounding the growing disparity in coverage.

% Contribution | Health insurance and financial risk protection
This paper makes three primary contributions. First, we contribute to a small existing literature that examines the financial risk protection provided by Medicare to elderly Americans \citep{finkelstein2008did,engelhardt2011medicare,barcellos2015effects,dobkin2018economic,doi:10.1086/706623,Batty2020} and a broader literature on the risk protection provided by health insurance \citep[e.g., ][]{gross2011health,finkelstein2012oregon,mazumder2016effects,brevoort2020credit,hu2018effect} and the economic consequences of health shocks \citep[e.g., ][]{cochrane1991simple,charles2003longitudinal,poterba2017asset,dobkin2018economic,meyer2019disability}. We contribute to this literature in two ways. First, we examine the effect of Medicare at age 65 on a broad set of financial health outcomes from administrative credit report data. These results expand the outcomes of \cite{doi:10.1086/706623} beyond just debt collections, and highlight that the financial health benefits of Medicare are concentrated in debt collections, with limited effects for other consumer credit outcomes. Second, we exploit our location-specific estimates of Medicare's causal effect to explore the effects of Medicare on the extensive margin (i.e., changes in insurance rates) and intensive margin (i.e., changes in financial health per newly-insured individual), and document considerable heterogeneity across geography in both.

% Contribution | Geographic variation in financial health

Second, our work contributes to a growing literature on the geography of health and economic opportunity. Prior work has documented the important role that geography plays in economic opportunity \citep{chetty2014land}, healthcare utilization and spending \citep{finkelstein2016sources,cooper2018price}, and mortality \citep{finkelstein2019place}. Recent work has also documented geographic concentration in financial strain \citep{keys2019}, but causal evidence on the effects of geography on consumer financial strain is limited \citep{miller2018neighborhoods,keys2019}. %We contribute to this literature by using nearly universal eligibility for Medicare to examine the role health insurance plays in shaping the geography of financial health. 
We contribute a set of methods for estimating location-specific effects of Medicare and the extent to which they reduce geographic variation in financial health outcomes. Our findings suggest that differential access to health insurance is a key driver of the geographic variation in collections debt for the near-elderly, but  less important in explaining differences across areas in other financial health outcomes.

Third, we use our locality-level estimates to investigate the incidence of Medicare. At \$750 billion in annual spending (and growing), Medicare's incidence as one of the largest public programs is of first-order policy importance \citep{ms2006jpube,bl2006jpube}. We show that the gains in financial health due to Medicare are greatest in the South (and particularly the Deep South) where, in addition to there being a greater number of the uninsured, the financial health improvements per newly-insured person are the greatest.

\section{Study data}\label{background_data}
In this section we briefly describe the data we use to construct area-level financial, demographic, health insurance, and healthcare market characteristics. Appendix \ref{apx:data} provides additional detail.

\subsection{Financial outcomes data}\label{background_ccp_data}
The main dataset used in our analysis is the Federal Reserve Bank of New York's Equifax Consumer Credit Panel (CCP). The CCP is a five percent random sample of all individuals in the U.S. with credit reports.\footnote{\citet{lee2010introduction} show that the CCP is reasonably representative of the U.S. population.} The data include a comprehensive set of consumer credit outcomes quarterly from 1999 to 2017, including credit scores (originating from Equifax Risk Score 3.0), unsecured credit lines, auto loans, and mortgages. The data also include year of birth and zip code. No other demographic information is available. A major virtue of the CCP is its large sample size, which allows us to estimate the effect of Medicare separately for all 50 states and 741 commuting zones in the country. Our datasets include ages 55-64 (``the near-elderly'') and 65-75 (``the elderly'').

\subsection{Demographic and health insurance data}
For demographic and health insurance information, we draw on the American Community Survey \citep{ipums2019}. All analyses use samples constructed from the Public Use Microdata Area (PUMA) and state datasets, linked to the CZ- and state-level.\footnote{Our cross-walk from PUMA to CZ uses David Dorn's crosswalks: \url{https://www.ddorn.net/data.htm}.} Demographic and health insurance variables from the ACS allow us to test for covariate smoothness to validate our RD design and examine the correlates of geographic heterogeneity in our treatment effects.

\subsection{Additional area-level characteristics data}
We constructed additional characteristics at the PUMA-level using data from the Healthcare Cost Report Information System (HCRIS) and the Dartmouth Atlas. From the HCRIS data, we construct PUMA-level measures of the share of hospital patient days at for-profit hospitals, teaching hospitals, and public hospitals in addition to other healthcare market characteristics. From the Dartmouth Atlas data, we measure the PUMA-level risk-adjusted Medicare spending per enrollee \citep{dartmouth2019}.
\section{Empirical strategy: Regression discontinuity design}\label{sec:methods}

% In this section, we describe our empirical methodology for estimating the effect of Medicare. We then report the results of covariate smoothness tests, and discuss our approach to forecasting the effects of a coverage expansion to the near-elderly. 

% Describing empirical strategy
\subsection{Econometric model}\label{subsec:empirical_framework_rd}

To estimate the causal impact of Medicare, we use an RD design that takes advantage of the sharp change in eligibility at age 65. We compare individuals just above and below the age 65 eligibility threshold under the assumption that individuals around the discontinuity are similar on observable and unobservable characteristics. Under this assumption, any discontinuities in financial health around age 65 can be attributed to the change in coverage as individuals age onto Medicare.
%\footnote{Since there are individuals at age 64 with and without coverage, and with different benefit designs and plan structures, the treatment we're studying is a weighted average of the effect of the transition to Medicare for those who were previously uninsured and those with different forms of coverage at age 64. Later, we exploit geographic variation in the distribution of the uninsurance rate to explore how our effects vary based on the impact of Medicare on uninsurance.} 
%This research design has been used to study the impact of Medicare on healthcare use, spending, mortality, and a limited set of financial health measures (e.g. \cite{card2008impact, card2009does, wallace2016traditional, barcellos2015effects, doi:10.1086/706623}).

We estimate our regression discontinuity analyses both at the national level, and separately for each state and commuting zone,\footnote{Commuting zones are groups of counties representing local labor markets \citep{david2013growth,dorn2019work}} using equations of the following form:
\begin{equation} 
\label{eq:rd_locations}
y_{i,l,t}(\text{age}) = \gamma \times 1(\text{age}>65) + f\left(\text{age}\right)\times 1(\text{age}\leq 65) + g\left(\text{age}\right)\times 1(\text{age}>65) + \epsilon_{i,l,t}(\text{age}).
\end{equation}
where $y_{i,l,t}(\text{age})$ is an outcome for individuals $i$ in location $l$ in time period $t$ of a given age. The functions $f\left(\text{age}\right)$ and $g\left(\text{age}\right)$ are the age profile of $y_{i,l,t}$ for those below the age of 65 and those above the age of 65, respectively, and control for the direct effect of age on outcomes. We denote the national-level effect of Medicare at age 65 as $\gamma$. We denote a set of $\gamma_{l}$, one for each location (either state or commuting zone) in our sample, as the location-level effect of Medicare at age 65, where $f(age)$ and $g(age)$ are allowed to vary by location.
%The $\epsilon_{i,l,t}$ term captures other omitted factors that may drive differences in financial health across locations, time, and age. 
%Due to smaller sample sizes, each locality-level estimate is noisier than the overall national effects. %Intuitively, our estimates of $\gamma_{l}$ have more inherent noise and variation than the true underlying estimates of $\gamma_{l}$ due to estimation error (in part due to smaller sample sizes). 
%To address this, we use a standard empirical Bayes approach to shrink each estimate towards the overall average of the locality-level effects. When we use this approach, we report statistical significance based on the confidence intervals for the raw estimates, but use  shrinkage when presenting the estimates for the different localities, and the comparisons across characteristics (for additional details, see Appendix \ref{apx:methods}).

%We estimate the effects using data collapsed to location-age or location-age-year cells, but weighting by the underlying population counts.

%%% Estimation difficulties
The estimation of Equation \ref{eq:rd_locations} is complicated by two features of our data; First, our running variable, age, is measured discretely by year in our sample; Second, because we only observe birth year in the CCP data, and the data is quarterly, we measure age with noise.  %\footnote{To see this, consider an individual born in 1954. In 2019, they can be either age 64 or 65. As a result, our data will combine both Medicare-treated and non-treated individuals at age 65.} 
As a result, both the estimation and inference of $f(\cdot)$, $g(\cdot)$, and $\gamma$ are more challenging, as error in measuring age 65 forces us to omit age 65 in our estimation procedure and use a ``donut'' RD, and the discreteness of the age variable requires further extrapolation.\footnote{The donut RD is a common solution to this problem in the literature \citep[e.g.,][]{barreca2011saving}.} To account for both issues, we follow the "honest" confidence intervals approach outlined in \cite{kolesar2018inference}, and \cite{armstrong2018optimal, armstrong2018simple}. Briefly, this approach bounds the second derivative of the true $f(\cdot)$ and $g(\cdot)$ functions near age 65, and uses this bound to estimate the maximum potential bias due to extrapolation. %This bias is used to bias-adjust the confidence intervals. 
%This method requires an additional tuning parameter, $K$, which imposes an upper bound on the absolute value of the second derivative of the conditional expectation function. Intuitively, this method places a bound on how quickly the functions $f(\cdot)$ and $g(\cdot)$ can change.\footnote{To choose our value of $K$ for our main estimates, we follow an approach similar to the approach advocated in \cite{imbens2019optimized}. We take a large window to the left of the RD cutoff and fit a quadratic function of age to the data. We take the coefficient on the quadratic term (the second derivative), take the absolute value, and multiply it by four. We take this as our estimate of $K$.} Similar to robustness exercises with bandwidths in previous RD methods, we present additional robustness tests which vary the value of $K$. 
In our estimation, we report our point estimate and bias-adjusted 95\% confidence intervals. See Appendix \ref{apx:methods} for additional details.

%% Advantage of RD
Like all RD approaches, our design allows us to easily visualize the treatment effect $\gamma$ using graphical methods. However, more uniquely in our setting, we also use our estimates to consider how Medicare changes the cross-location distribution of counterfactual outcomes at age 65. Denote the average counterfactual at age 65 for location $l$ as $y_{l}(65-)$ and $y_{l}(65+)$, and define the causal effect of Medicare on the variance of outcomes across locations as $\phi = \frac{Var\left(y_{l}(65-)\right) - Var\left(y_{l}(65+)\right)}{Var\left(y_{l}(65-)\right)},$ where the variance is taken across locations. This measure captures the change in geographic \emph{variance} of our outcomes due to Medicare, rather than just the average level. We estimate standard errors and bias-adjusted confidence intervals for $\phi$ using the delta method following \cite{armstrong2018simple}, and report an estimated drop in variance for each of our outcomes along with bias-adjusted 95\% confidence intervals. See Appendix \ref{apx:methods} for additional details. 

\subsection{Forecasting the causal effects of Medicare by location}\label{subsec:empirical_forecasts}

This section describes our approach to measuring the area-level factors associated with reductions in consumer financial strain at age 65 and constructing forecasts of the causal effects of Medicare by location. 

We begin by estimating bivariate regressions between our locality-level causal effects of Medicare and location characteristics:
\begin{equation}\label{eq:bivariate}
    \hat{\gamma}_{l} = \alpha + X_{l}\omega + v_{l}.
\end{equation}
where $X_{l}$ is a scalar containing a single area-level characteristic (e.g. the share of black residents) and $\hat{\gamma}_{l}$ is our RD estimate for location $l$. We separately estimate $\omega$ for each characteristic, weighting the regression by each location's near-elderly population.

Given that many of the area-level characteristics we study are highly correlated, we re-estimate Equation \ref{eq:bivariate} with the full set of area-level covariates:
\begin{equation}\label{eq:lasso}
    \hat{\gamma}_{l} = \alpha + \vec{X}_{l}\vec{\omega} + v_{l}.
\end{equation}
where $\vec{X}_{l}$ is the full set of area-level covariates. We estimate Equation \ref{eq:lasso} in two ways. First, we estimate the model using ordinary least squares (OLS) to recover the marginal association of each area-level characteristic with our locality-level causal estimates. Second, since the dimension of $X_{l}$ is large (and many of the covariates are highly correlated), we use Lasso to perform covariate selection on $\vec{X}_{l}$, and then re-estimate the model using OLS \citep{belloni2013least}. This Lasso procedure lets us trade off between including multiple characteristics and constructing predictions of $\hat{\gamma}_{l}$ with lower mean squared error.
%Briefly, the Lasso estimation procedure penalizes covariates and shrinks terms in the estimated $\omega_{l}$ towards zero, in order to minimize mean squared error. As a result, the estimation procedure will select a subset of the covariates in $\vec{X}_{l}$, to have non-zero parameters, and set the remaining parameters to zero.\footnote{We implement this using a ten-fold cross-validation over the penalization parameter, implemented using R \texttt{glmnet} package.}

We next construct forecasts of the causal effect of expanding Medicare (i.e., a reduction in the Medicare eligibility age) in each commuting zone. While our RD estimates of $\gamma_{l}$ are unbiased forecasts, many are estimated with substantial estimation error. To reduce noise, we build on \cite{chetty2018impactsb} and construct forecasts using a shrinkage estimator that combines our unbiased RD estimates and a predicted effect for each commuting zone using the covariates selected from our Lasso procedure (see Appendix \ref{apx:methods} for additional details).

We repeat our approach for a scaled estimate $\beta_{l} = \gamma_{l} \big/ \gamma_{l}^{h}$,  where $\gamma_{l}$ is a location-specific estimate of the effects of Medicare on financial health outcomes and $\gamma_{l}^{h}$ is a location-specific estimate of the effects of Medicare on the insurance rate. This provides a measure of the effect of Medicare on financial health outcomes \emph{per newly-insured individual} at each location. This helps account for the mechanical effect that areas with high near-elderly uninsurance rates will likely see large changes in financial health alongside increases in coverage. However, while we estimate $\beta_{l}$ using fuzzy RD, it does \emph{not} estimate the causal effect of insurance on financial health, as the characteristics of health insurance coverage also change for the previously-insured as they enter Medicare \citep{card2008impact}.

\section{Results}
% In this section, we report the effect of Medicare eligibility on health insurance and various financial health outcomes at the national, state, and commuting zone (CZ) levels. 
% We initially focus on the state level when presenting our primary estimates of the geographic heterogeneity in the effect of Medicare. We then turn to the CZ-level to examine the characteristics of areas that experienced the largest improvements in financial health at age 65, forecast the causal effects of health insurance, and examine how the ACA has changed those forecasts.

% Subsection on national results
\subsection{Medicare, health insurance, and financial health, nationally and across states}
\begin{figure}
\centering
  \caption{Changes in health insurance, financial health, and covariates at age 65}
  \label{fig:main_ageRD}
  \makebox[\textwidth][c]{
\begin{tabular}{cc}
  \textit{Panel A:} Share with any coverage &   \textit{Panel B:} Debt in collections \\
\includegraphics[width=3.25in]{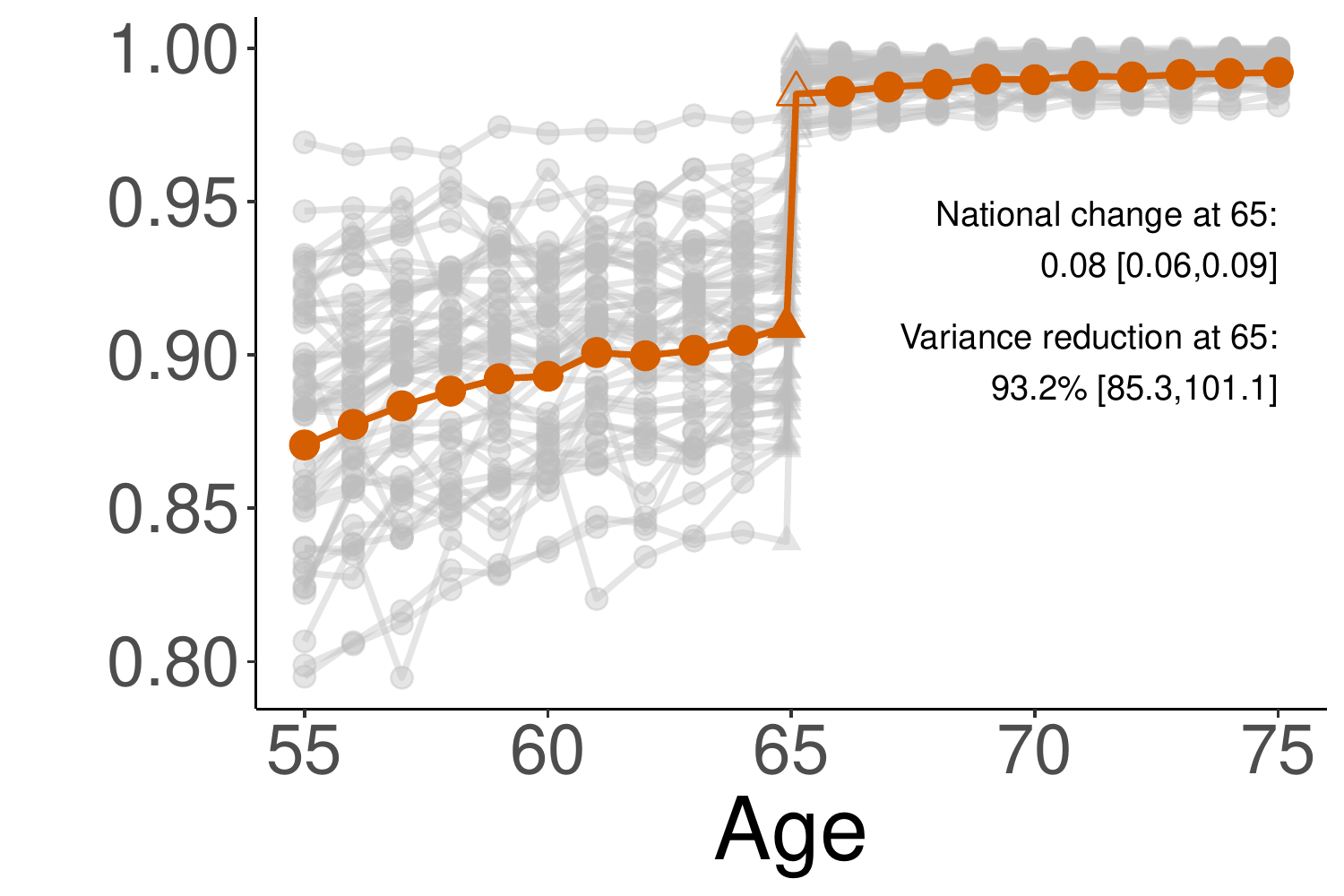} & \includegraphics[width=3.25in]{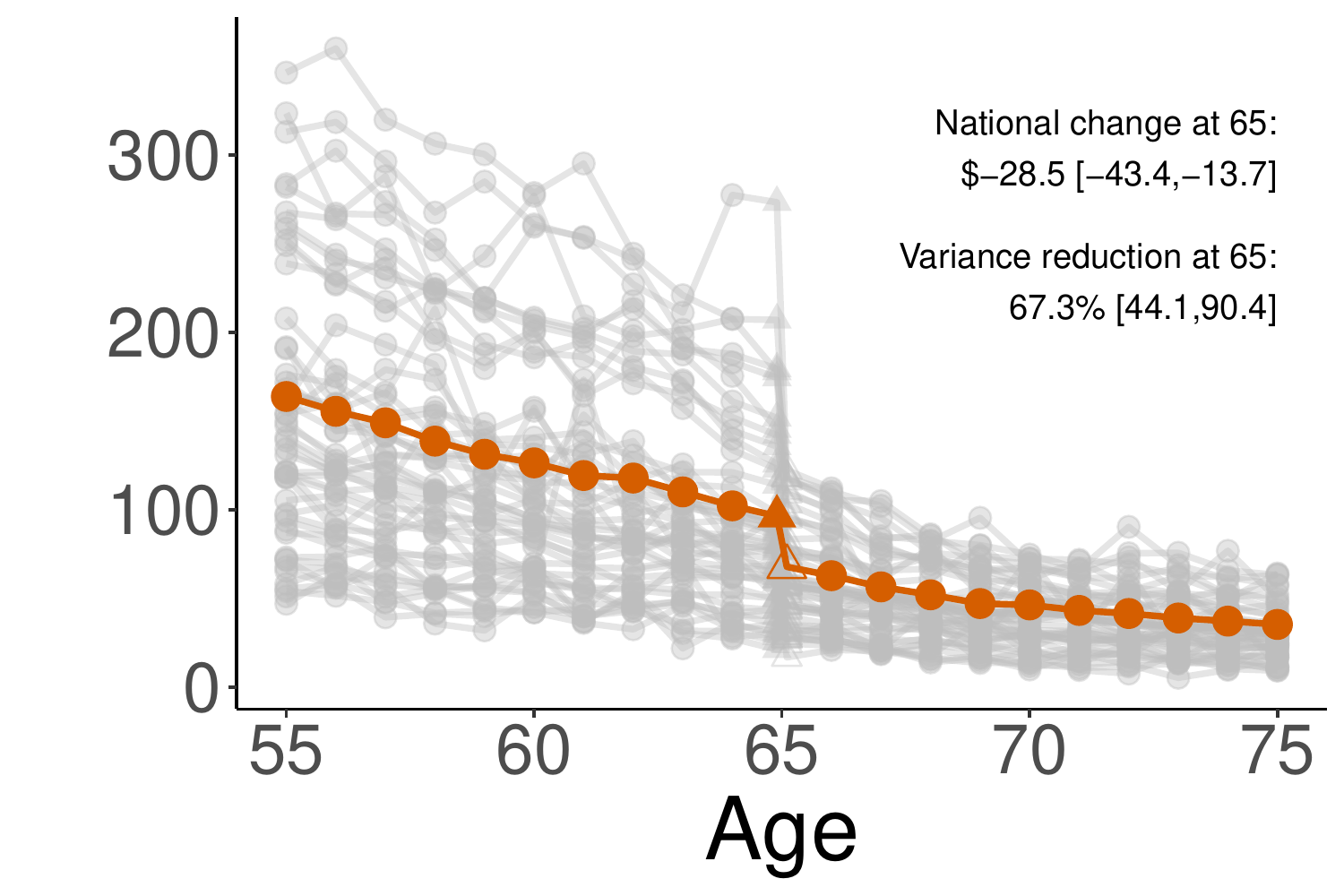} \\ 
\textit{Panel C:} Credit Score & \textit{Panel D:} Bankruptcy (p.p.) \\
\includegraphics[width=3.25in]{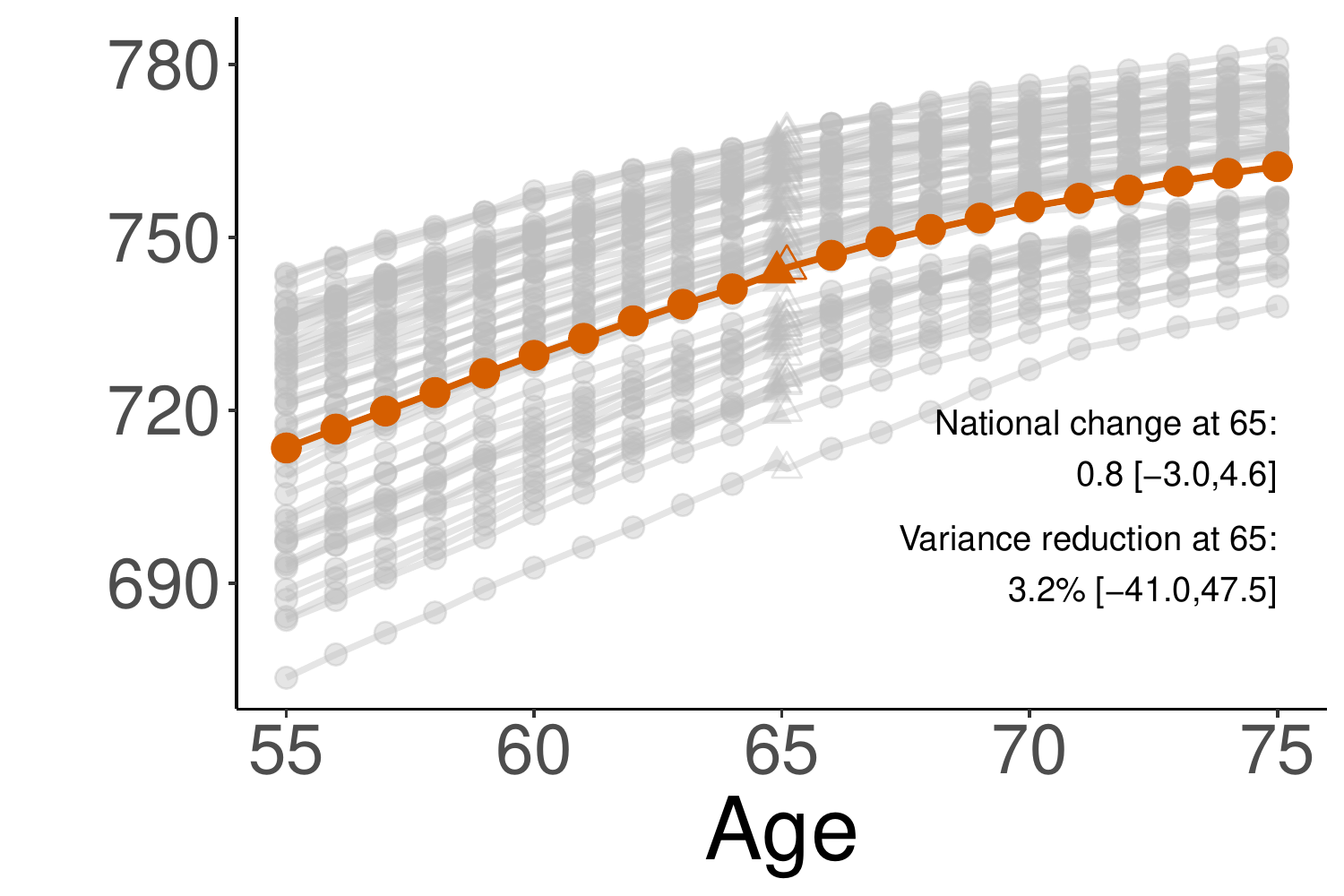} & \includegraphics[width=3.25in]{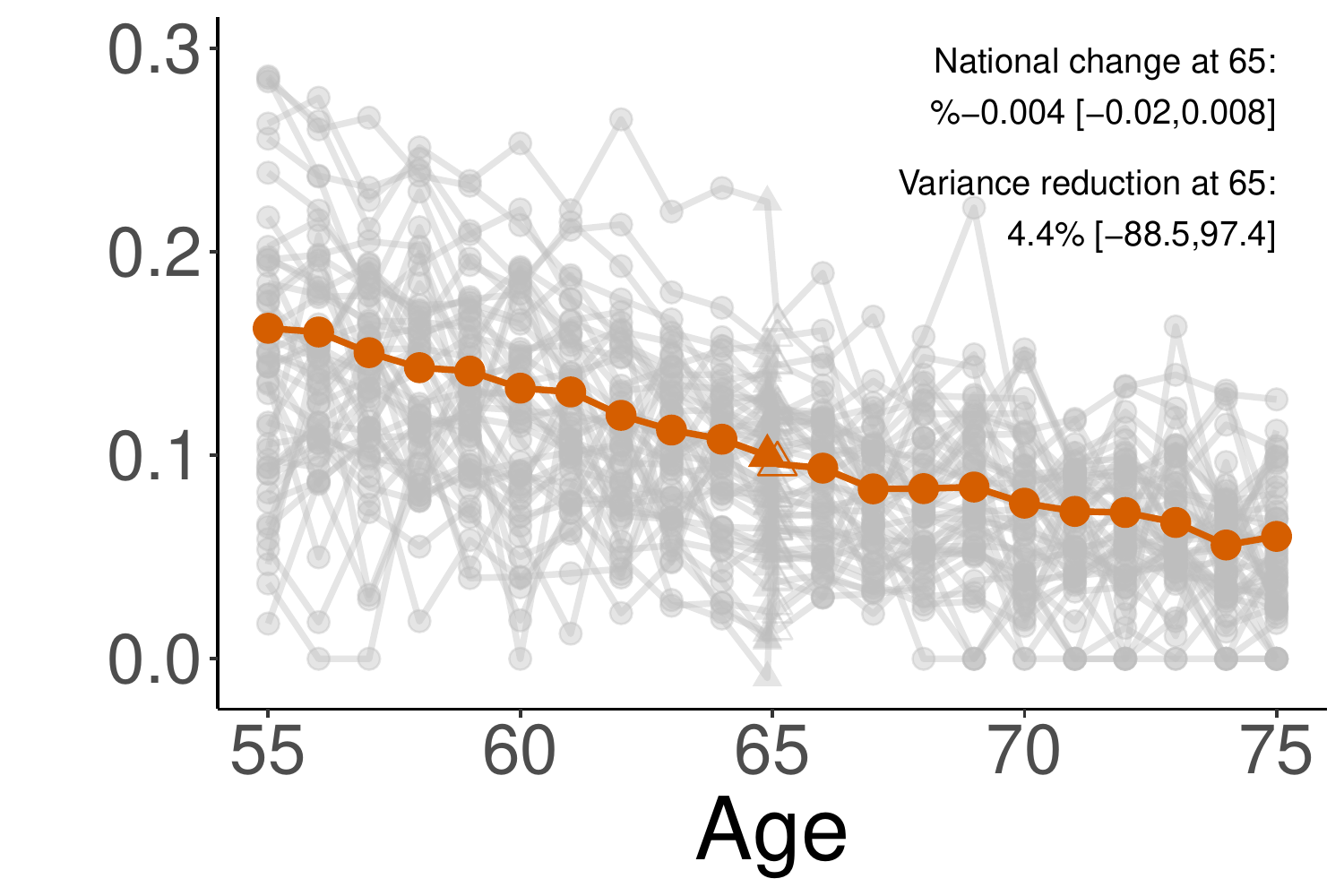} \\
\textit{Panel E:} Share employed &   \textit{Panel F:} Income \\
\includegraphics[width=3.25in]{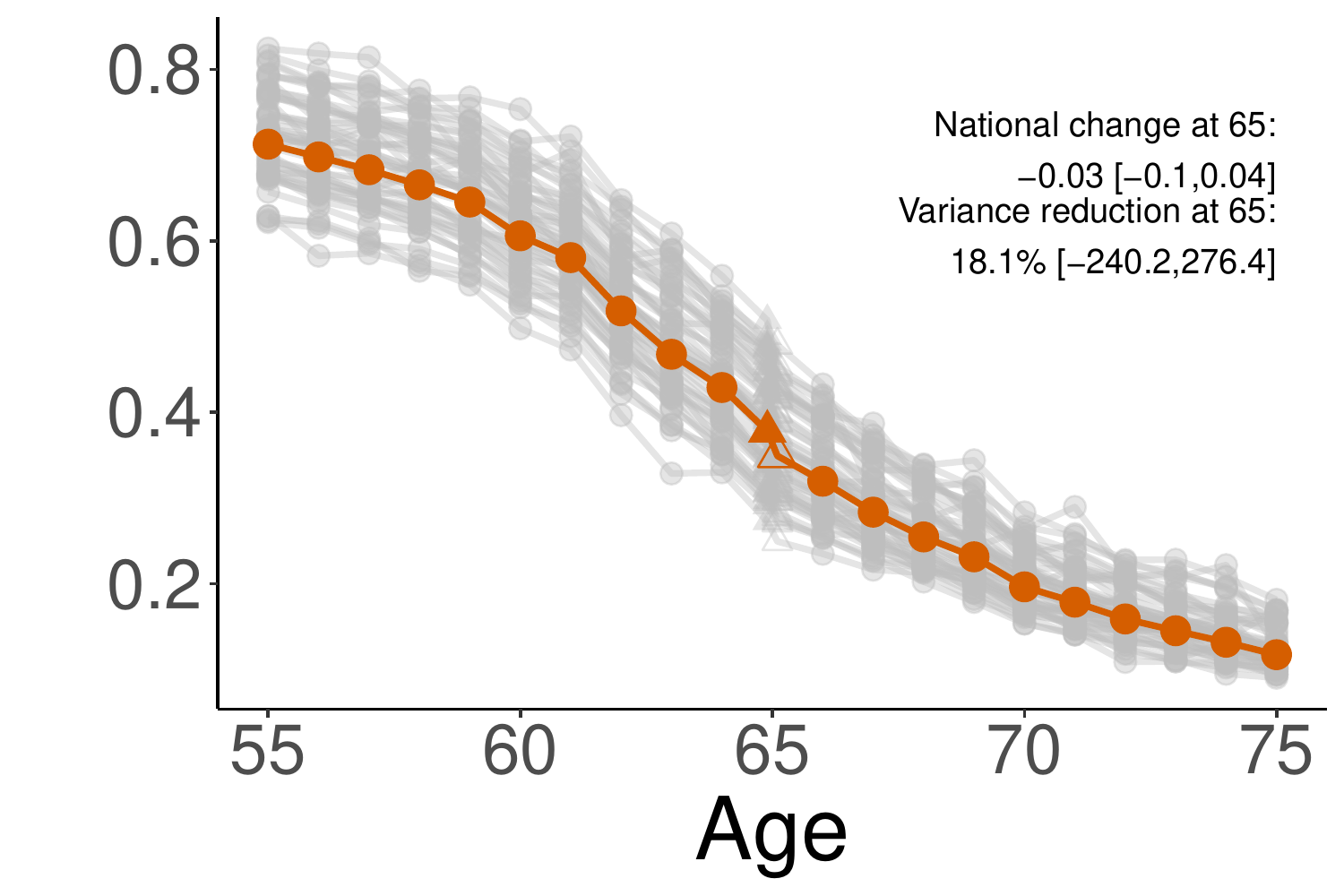} & \includegraphics[width=3.25in]{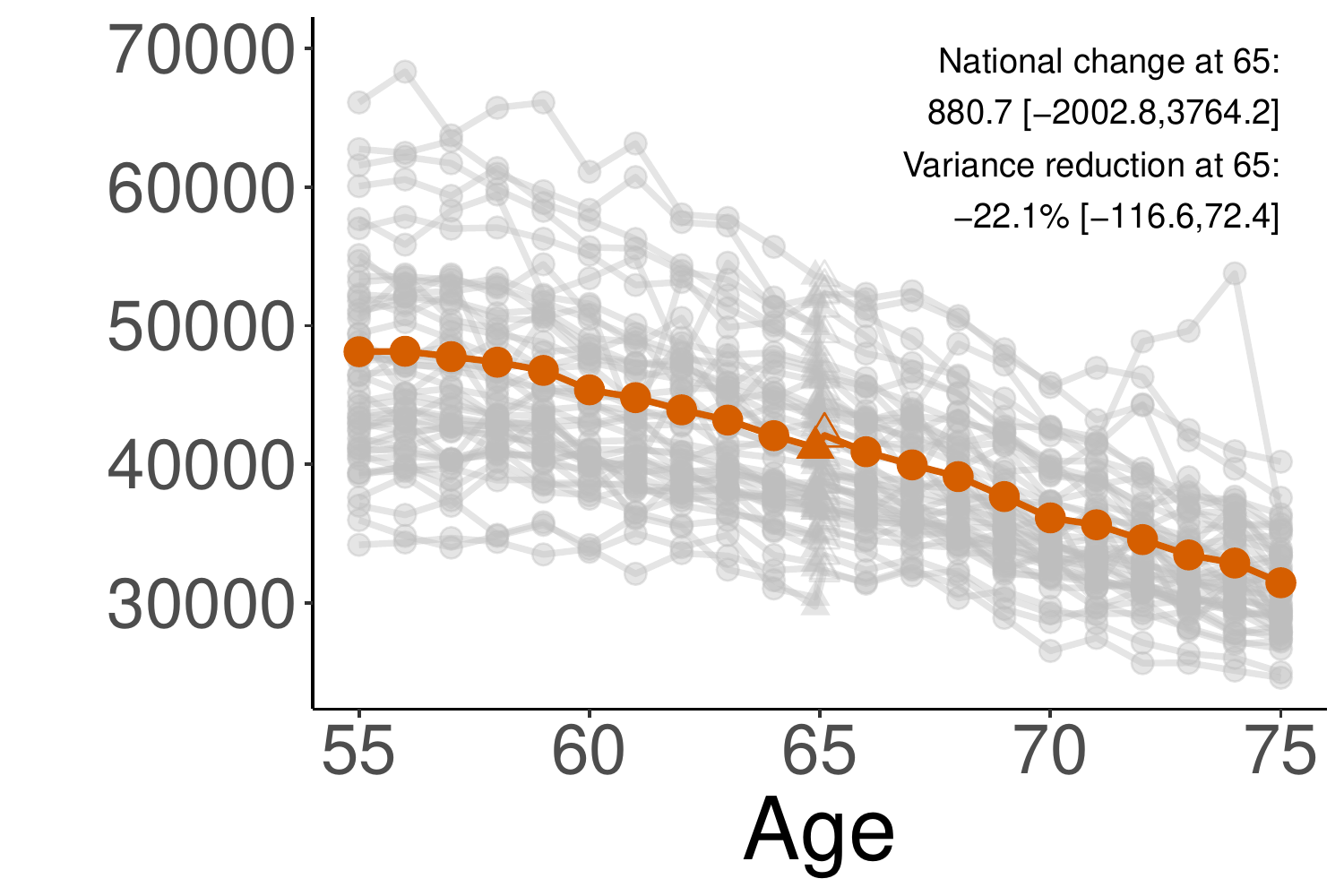} 
\end{tabular}}
\begin{minipage} {0.95\textwidth} \setstretch{.9} \medskip
  \scriptsize{\textbf{Note:} This figure plots the average outcomes by age at both the national level (in red) and across the fifty U.S. states (in grey). The horizontal axis denotes age in years. The points plotted as circles reflect empirical averages from the data, while the triangles reflect imputed counterfactual values at age 65 from a local linear regression from the left and right (following \cite{kolesar2018inference} and \cite{armstrong2018simple}) with a bandwidth of 5. The imputed value without Medicare at age 65 is plotted at age 64.9, while the imputed value with Medicare at age 65 is plotted at age 65.1. Panel A plots the share of individuals with any form of health insurance. Panel B plots the average debts in collections over the past 12 months. Panel C plots the average Equifax Risk Score 3.0. Panel D plots the share of individuals with new bankruptcies (p.p.). Panel E plots the share of individuals employed. Panel F plots the average total income. The sample includes individuals who were age 55-75 between 2008 and 2017. See Section \ref{background_data} for additional details on the outcomes and sample. Source: American Community Survey, 2008-2017 and New York Fed Consumer Credit Panel/Equifax, 2008-2017.}
  \end{minipage}
\end{figure}

% Medicare and health insurance
Figure \ref{fig:main_ageRD} presents the effect of Medicare at age 65 at the national level and for each state. In solid red circles, we plot the average national outcome for each age. At age 65, we plot two points: in the solid red triangle, we plot $y(65-)$, the national average at age 65 without Medicare and in the hollow red triangle we plot $y(65+)$, the national average at age 65 with Medicare. In gray, we repeat the same exercise for each of the states in our sample. For each outcome, we report the estimated  national effect (with the bias-adjusted 95\% confidence interval) and the estimated percent reduction in variance across states (with the bias-adjusted 95\% confidence interval).

We plot the share of the population with health insurance coverage in Panel A of Figure \ref{fig:main_ageRD}. As has been documented previously \citep[e.g.,][]{card2008impact}, the effect of Medicare eligibility on the share of individuals with any form of coverage at age 65 is large---rising by 7.9 percentage points, relative to an overall insurance rate of almost 91 percent prior to Medicare. %\footnote{Appendix \ref{apx:results} discusses estimated changes in other forms of coverage at 65.}
We estimate a sharp reduction in geographical variation in health insurance rates of 93.2\% (95\% CI: 85.3 to 101.1) at age 65 due to Medicare eligibility.\footnote{Due to the asymptotic nature of these confidence intervals, the 95\%  CI includes values larger than the maximum possible value, 100\%. %PGP: note sure what else to put here
} This suggests that Medicare, as expected, eliminates almost all  variation across states in health insurance rates.

In Panel B of Figure \ref{fig:main_ageRD}, we also estimate a large national reduction in collections debt at age 65, with a sharp drop of 28.5 dollars (95\% CI: -48.3 to -13.7).\footnote{We find a decline in especially large collections balances (Figure \ref{fig:collections_dist_RD}), consistent with evidence that Medicare curbs the upper tail of medical spending \citep{finkelstein2008did, doi:10.1086/706623}.} We also estimate a corresponding reduction of 67.3\% (95\% CI: 44.1\%-90.4\%) in the overall cross-state variance of collections debt at age 65, consistent with the drop in variance for health insurance, implying that Medicare reduced the differences in collections debt across states by two-thirds. 

%\footnote{For each of the financial health outcomes, the age profiles are consistent with improvement in the measure as individuals get older. This is consistent with the literature on the lifecycle of financial strain \citep{brown2016graying}.}

In contrast, for our other financial health measures, we find statistically insignificant effects on credit score (Panel C) and bankruptcy (Panel D). The estimated variance reduction and national effects are small with large confidence bounds, suggesting that Medicare had limited effects on cross-state variance, despite large baseline differences across states. We also examine a variety of other financial health outcomes, including delinquent debt and foreclosure, and find small, though noisy, effects, with no corresponding reduction in national variance (Figure \ref{fig:appendix_main_ageRD}). In Appendix \ref{apx:methods} we demonstrate that these results (and lack thereof) are robust to alternative specifications (Figures \ref{fig:bound_scaling_robustness_apx}, \ref{fig:age_bandwidth_robustness_apx}, \ref{fig:bound_scaling_var_robustness_apx}, \ref{fig:age_bandwidth_var_robustness_apx}, and Table \ref{tab:main_agerd_appendix_othermodels_apx}).

% Robustness and covariate smoothness (could be expanded but we're tight on space)
In Panels E and F of Figure \ref{fig:main_ageRD}, we test our key identifying assumption that non-Medicare characteristics that affect outcomes do not jump discontinuously at age 65. For example, given that many individuals tend to retire in their early to mid-60s, we test whether this coincides with the age of eligibility for Medicare. We do not find evidence of discontinuities in non-Medicare characteristics at the national or local level at age 65.  For both figures, we cannot reject the null that the effect size is zero at the national level, and that there is no change in the variance across states.

In Appendix \ref{apx:methods} we examine potential discontinuities in additional covariates and present our estimates from state- and CZ-level covariate smoothness tests (Figures \ref{fig:covariates_ageRD_apx}, \ref{fig:cov_smoothness_state_apx}, and \ref{fig:cov_smoothness_cz_apx}). We find little evidence of discontinuities in the average values of covariates at age 65. Intuitively, while the early to mid-60s are a time of transition for many individuals, the precise age of 65 is no longer a focal point for retirement decisions, which smooths out the timing of other lifestyle changes in a way that does not confound them with Medicare eligibility.

%% CZ-level results, focused on collections debt
\subsection{Medicare and the geography of financial strain}
% First sentence: Pivot to debt collections b/c geography effect is similar and important because X Y Z
In this section, we seek to understand why the effect of Medicare on collections debt, where we find a large reduction in geographic variation, varies so much across localities. 

\begin{figure}
  \centering
  \caption{Counterfactual levels of collections debt by commuting zone at age 65 with and without Medicare}
  \label{fig:collections_prepost_cz_map_shrink}
  \makebox[\linewidth][c]{
  \begin{tabular}{cc}
  \textit{Panel A:} Without Medicare &\textit{Panel B:} With Medicare \\ 
\includegraphics[width=3.5in,trim=4 4 4 4,clip]{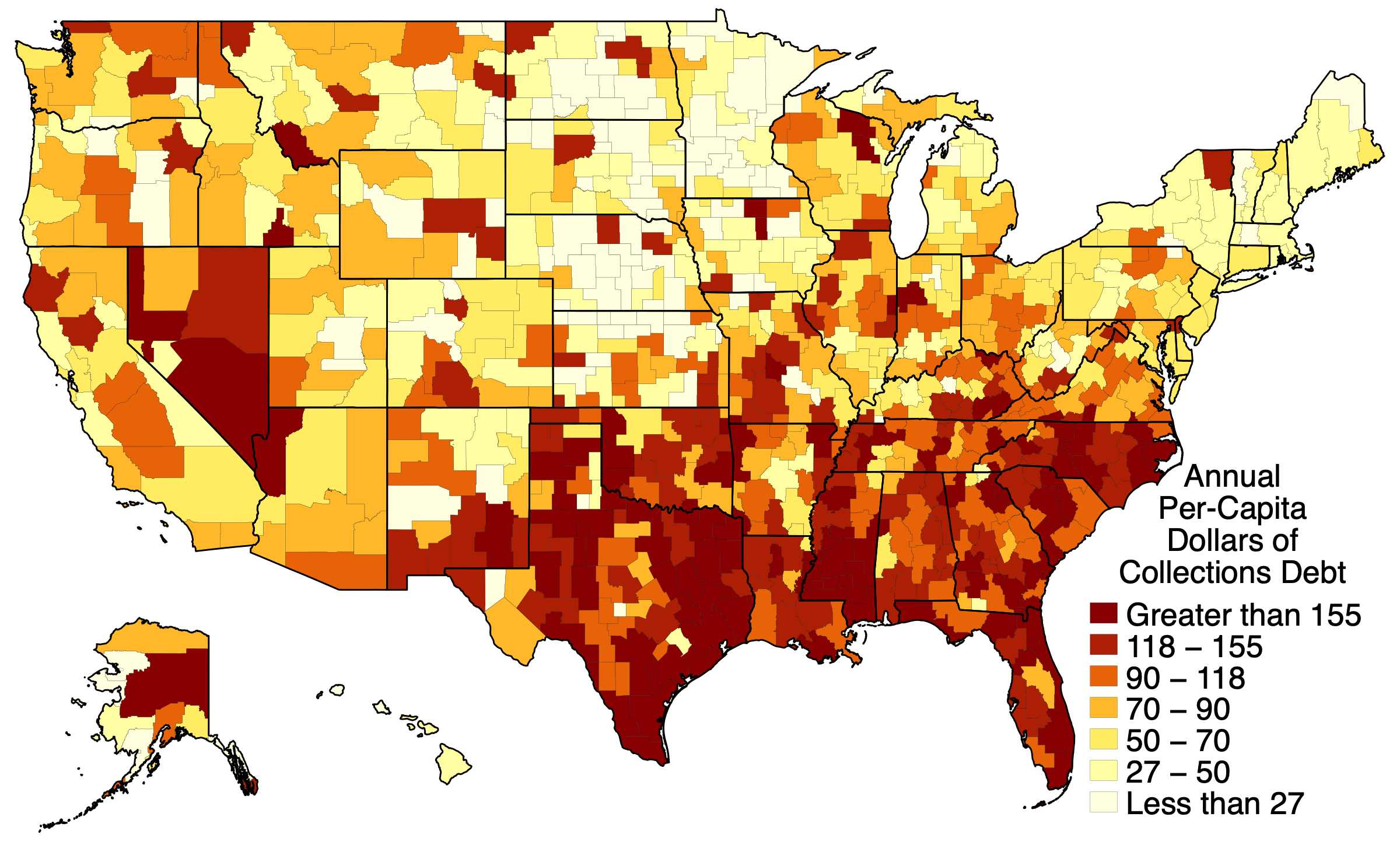} & \includegraphics[width=3.5in,trim=4 4 4 4,clip]{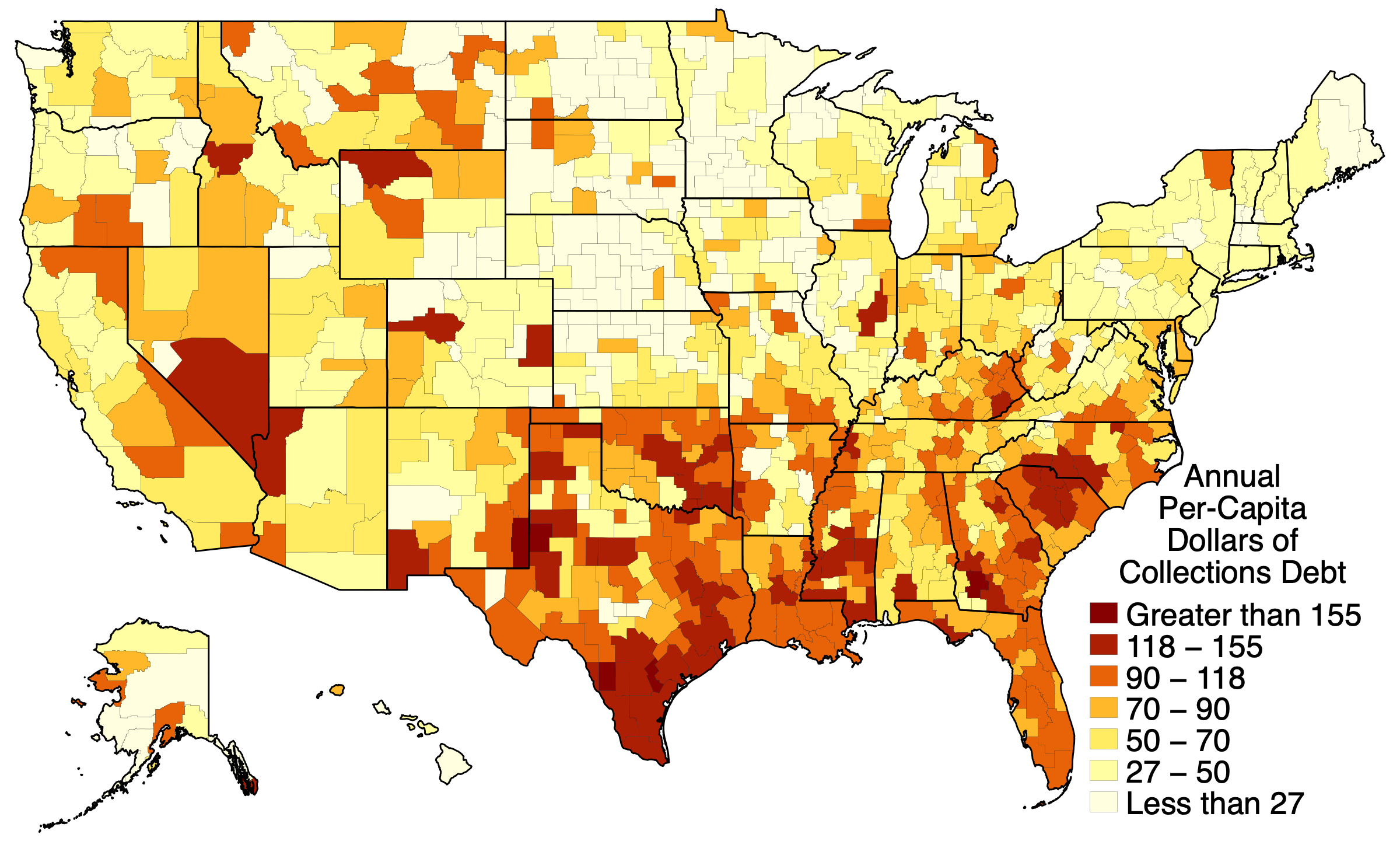}
\end{tabular}
 }
  \begin{minipage} {0.9\textwidth} \setstretch{.9} \medskip
    \footnotesize{\textbf{Note:} This figure plots our counterfactual estimates of the flow of newly reported collections debt (within the past year) per capita at age 65, with and without Medicare. The CZ-level variance reduction at age 65---the difference in CZ-level variance between the two panels---was 63.8\% (95\% CI: 30.7\%-97.0\%). The counterfactuals are based on local linear regressions, done separately by commuting zone, using the methods from \cite{kolesar2018inference}. These estimates are then shrunk using empirical Bayes, described in Appendix \ref{apx:methods}.  Darker regions correspond to higher counterfactual collections debt per capita. Source: Consumer credit outcomes are based on 137,340,577 person-year observations from the New York Fed Consumer Credit Panel/Equifax, 2008-2017.}
  \end{minipage}
\end{figure}

Prior to individuals gaining Medicare eligibility at age 65, we observe large differences in collections debt across areas, with particularly high levels in the South. This feature of the US landscape of financial health is apparent in Figure \ref{fig:collections_prepost_cz_map_shrink}, where we map the commuting zone estimates of counterfactual collections debt flows at age 65, with and without Medicare.\footnote{For this map, due to smaller sample sizes, we use an empirical Bayes approach to shrink each locality-level estimate towards the overall average of the effects (see Appendix \ref{apx:methods}).} It is clear from Panel A that, absent Medicare, collections debt for the near-elderly varies widely across the country, with low levels in the Midwest and Northeast, and with high levels of debt collection concentrated in the South. At age 65, we observe a large reduction in collections, concentrated in the South. Panel B shows that much of the geographic variation in collections debt disappears at age 65, with lower (though still elevated) levels of collections debt in states like Mississippi, Texas, and Nevada.\footnote{At the CZ-level, this reduction equates to a drop in the across-CZ variance in collections debt of 70\% (very similar to our state-level estimate of 67.3\%).}

% Area-level correlates | change in collections vs. changes in insurance
Why are collections so concentrated before Medicare and much less so after? One clear candidate is geographic differences in the uninsurance rate for the near-elderly. Prior work documents a link between health insurance coverage and collections debt \citep[see e.g.,][]{finkelstein2012oregon}, and we find similar associations between area-level health insurance rates and financial health outcomes among the near-elderly (Figure \ref{fig:uninsurance_vs_finhealth_state_apx}). We compare the estimated increase in health insurance due to Medicare  to the drop in debt collections at the state-level and CZ-level in Figure \ref{fig:drop_pctui_binscatter}. This ``extensive margin'' effect of Medicare on coverage explains a surprisingly large share of the variation ($R^{2}$ = 0.38),  with small estimated reductions in collections for states with small estimated changes in the insurance rate at age 65. 

%Appendix Figure \ref{fig:drop_pctui_binscatter} examines the correlation between the causal effect of Medicare on health insurance rates (i.e., the ``extensive margin'') and debt collections at the state- and CZ-level. The increase in health insurance coverage explains a surprisingly large share of the debt collections effect at the state-level (R$^2$ of 0.38), with small estimated reductions in collections for states with small estimated changes in the insurance rate at age 65. Both of these facts suggest that the effect of Medicare eligibility on debt collections may be driven by individuals who gain coverage, rather than those whose primary source of coverage is changing. 

These facts suggest that the effect of Medicare eligibility on debt collections may be driven by individuals who gain coverage, rather than those whose primary source of coverage is changing. Motivated by this finding, we construct a scaled version of our CZ-level estimates, $\beta_{l}$, that measures the reduction in collections debt \textit{per newly-insured}, and examine it alongside our \textit{per capita} estimates going forward. This allows us to compare how the effects of Medicare differ across locations with different baseline levels of uninsurance among the near-elderly.

% Area-level correlates | change in collections vs. other characteristics
We now examine what other area-level factors are associated with the reductions in collections debt at age 65. We present evidence that commuting zones with larger shares of black residents, people with disabilities, and for-profit hospitals experience the largest gains in financial health at age 65, across a variety of estimation approaches.

\begin{figure}
  \centering
  \caption{Commuting zone characteristics correlated with the reduction in collections debt at age 65}
  \label{fig:correlates_drop_collections_cz}
 \makebox[\textwidth][c]{ \begin{tabular}{c}
  \textit{Panel A:} Demographic characteristics \\[.25cm]
\includegraphics[width=6in]{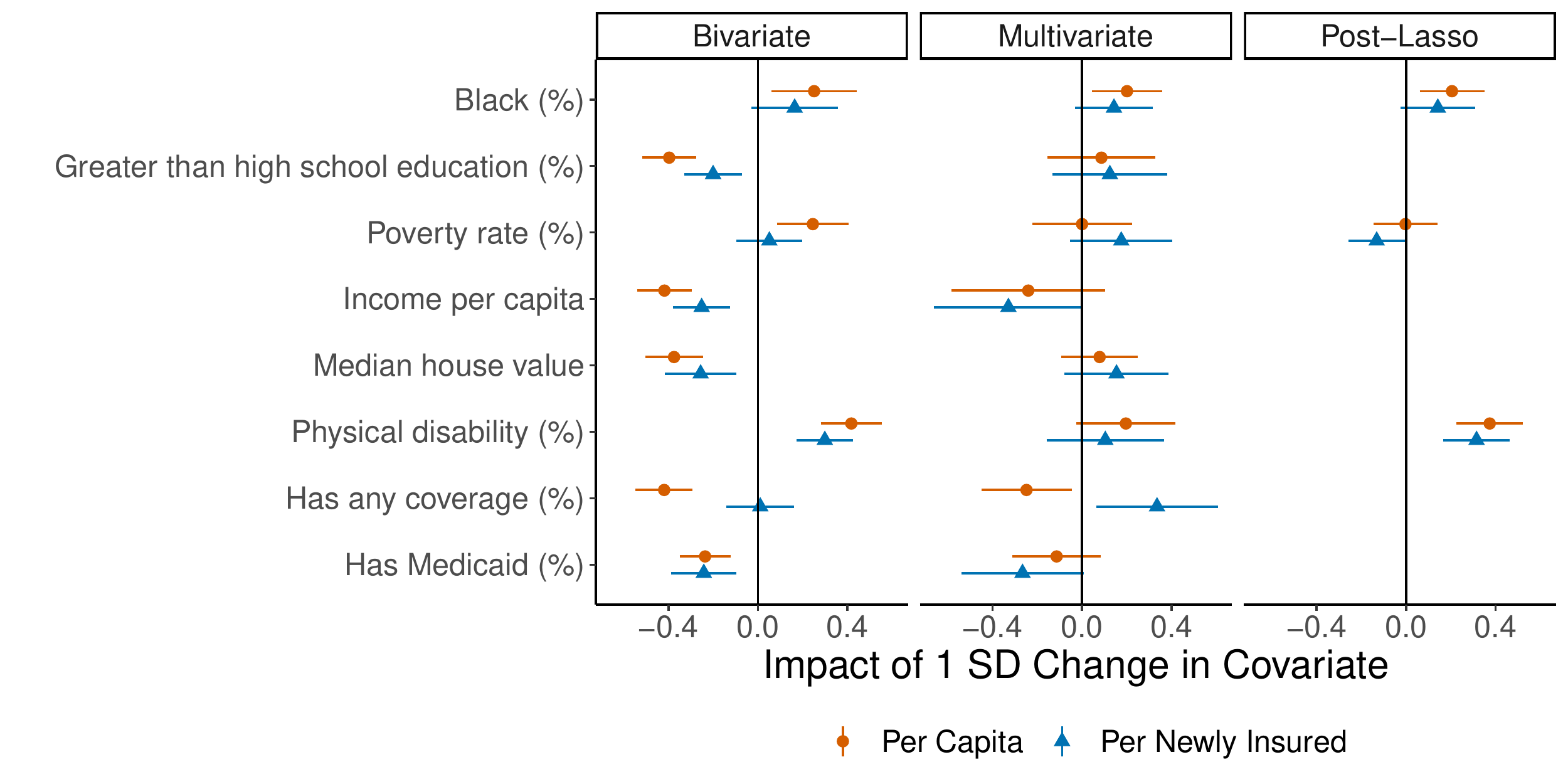} \\
\\
  \textit{Panel B:} Healthcare market characteristics \\[.25cm]
\includegraphics[width=6in]{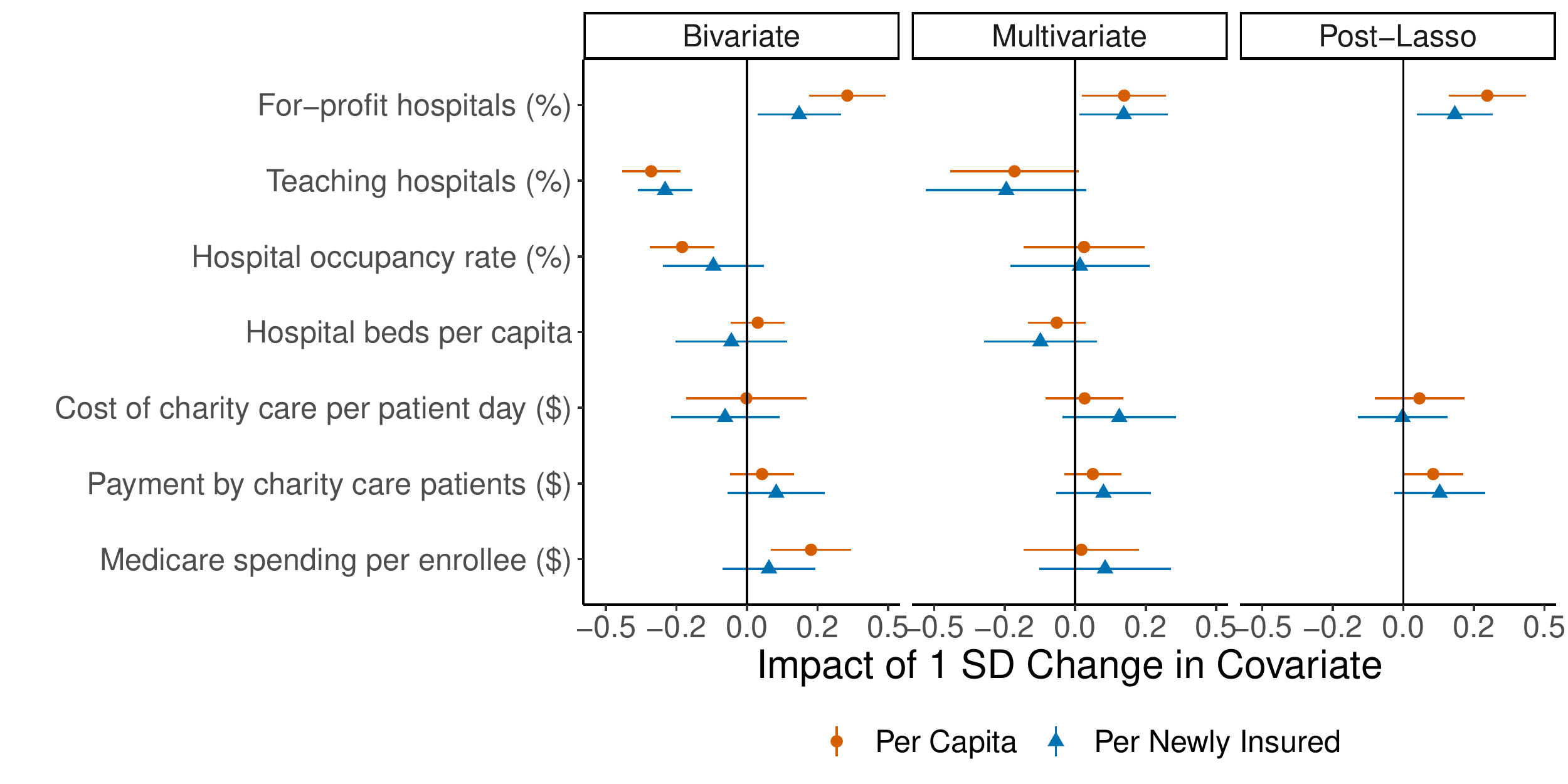} \\
\end{tabular}}
 
  \begin{minipage} {\textwidth} \setstretch{.9} \medskip
    \footnotesize{\textbf{Note:} This figure plots bivariate OLS estimates (left panel), multivariate OLS estimates (center panel), and post-Lasso multivariate estimates (right panel) of CZ-level estimated reductions in collections debt per capita on a set of CZ-level characteristics. We standardize all the variables so the coefficients reflect the strength of the association between a one standard deviation change in the covariate and the estimated reduction in collections debt at age 65. The horizontal bars are 95\% confidence intervals. The multivariate OLS regression results and post-Lasso multivariate regression results are both run on the full set of characteristics in Panels A and B. For post-Lasso, we first estimate a Lasso regression on the full set of characteristics and then report the results of multivariate OLS run on the characteristics chosen by the Lasso regression. Tabular versions of these results are in Table \ref{tab:tau_beta_correlates_apx}. Source: Consumer credit outcomes are based on 137,340,577 person-year observations from the New York Fed Consumer Credit Panel/Equifax, 2008-2017. CZ-level uninsurance rates are from the American Community Survey, 2008-2017. Healthcare market characteristics are from the Healthcare Cost Report Information System (HCRIS) and the Dartmouth Atlas. For additional details on the data see Section \ref{background_data}.}
  \end{minipage}
\end{figure}

Figure \ref{fig:correlates_drop_collections_cz} presents correlations between our area-level characteristics and estimated effects of Medicare. The leftmost panel presents the coefficients from separate bivariate OLS regressions of our regression discontinuity estimates of CZ-level reductions in collections debt \textit{per capita} (in red circles) and \textit{per newly-insured} (in blue triangles) on CZ-level demographic and healthcare market characteristics, with bars representing the 95\% confidence intervals. Since many of the area-level characteristics are highly correlated, the center and right panels plot multivariate and post-Lasso analyses describing the partial correlations between the characteristics and our locality-level causal estimates. We separate the demographic and healthcare market characteristics into two panels for presentation purposes, but all covariates are jointly combined in estimation for the multivariate and post-Lasso models. To facilitate comparison across the area-level correlates and the \emph{per capita} and \emph{per newly-insured} measures, we standardize all the area-level correlates, and then divide the coefficients by the respective national \textit{per capita} or \textit{per newly-insured} estimates. Hence, plotted coefficients in Figure \ref{fig:correlates_drop_collections_cz} correspond to the effect of a one standard deviation change in the covariate on a percentage change in the reduction in debt collections at age 65. Hence, a coefficient of 0.1 implies a 10\% increase in the effect of Medicare in reducing debt collections from a one standard deviation change in the covariate. 

% Discussion of the demographic characteristics
Panel A of Figure \ref{fig:correlates_drop_collections_cz} shows that the share of high school graduates, income per capita, and median house values in a CZ were all associated with smaller reductions in \textit{per capita} collections debt at age 65. Unsurprisingly, the share of residents with health insurance (or Medicaid), was also associated with a smaller reduction in per capita collections debt at age 65. The share of black residents, the poverty rate, and the share of people with disabilities, on the other hand, were associated with larger reductions in collections at age 65. In multivariate and post-Lasso analyses, only the near-elderly health insurance rate and the high school graduation rate were associated with smaller reductions in per capita collections, while the share of black residents and people with disabilities in a CZ were consistently associated with larger reductions in per capita collections.\footnote{While the Lasso procedure did not select the near-elderly health insurance rate, we note a very high correlation between that measure and the high school graduation rate ($\rho = 0.45$).}  Including Census region or division fixed effects---and using only the within-area, across-CZ variation that remains---does not qualitatively change our results (Figure \ref{fig:correlates_drop_collections_cz_fe}). Once we restrict to only using variation across CZs, but within states, however, the association between the share of black residents and per capita reductions in collections is severely attenuated.

% Healthcare market characteristics
Panel B of Figure \ref{fig:correlates_drop_collections_cz} presents the healthcare market characteristics correlated with our estimates. In the bivariate OLS model, a higher share of for-profit hospitals and higher risk-adjusted spending per Medicare beneficiary were both associated with larger reductions in collections debt at age 65, while a higher share of teaching hospitals and higher hospital occupancy rates were associated with smaller reductions in collections at age 65. In multivariate OLS and post-Lasso analyses, only the CZ-level share of for-profit hospitals was associated with our causal CZ-level effects, with a one standard deviation increase in the share of for-profit hospitals associated with a 40\% larger reduction in \textit{per capita} collections at 65. Unlike not-for-profit hospitals, for-profits are not required to provide charity (discounted or free) care and evidence suggests that they offer less charity care than not-for-profit hospitals \citep[e.g.,][]{horwitz2005making,schlesinger2006nonprofits,valdovinos2015california}.\footnote{In addition, hospitals in markets with a higher share of for-profits respond to competition by reducing charity care and trying to avoid the uninsured \citep{frank1990market}.} Figure \ref{fig:correlates_drop_collections_cz_fe} demonstrates that the relationship between for-profit hospital share and CZ-level reductions in per capita collections debt at age 65 is robust to the inclusion of fixed effects for census regions or divisions, but not states.

Figure \ref{fig:correlates_drop_collections_cz} also examines the demographic and healthcare market characteristics associated with reductions in collections debt \textit{per newly-insured}. This exercise accounts for the change in the uninsurance rate at age 65, to identify CZs that experienced larger or smaller reductions in debt collections not mechanically driven by Medicare's ``extensive margin'' effect on coverage. The share of people with disabilities and for-profit hospitals in a CZ is consistently associated with larger reductions in collections debt per newly-insured. In addition, CZs with a larger share of black residents---where we see large reductions in \textit{per capita} collections debt---also appear to experience larger reductions in collections \textit{per newly-insured}. The other area-level characteristics were not consistently associated with the estimated reductions in collections debt per newly-insured.

\subsection{Forecasts of the causal effects of Medicare on financial strain}\label{subsec:results_forecast}
Given that the effect of Medicare varies substantially across localities, where would the effects of a broad expansion of coverage to the near-elderly (i.e., by lowering the Medicare eligibility age) be the largest? %Our CZ-level RD estimates would provide an unbiased estimate of this effect, but due to noise we ... 
In what follows, we discuss our forecasts of CZ-level causal effects, how those forecasts have changed post-ACA, and their implications for future potential coverage expansions.

\begin{figure}[htpb!]
  \centering
  \caption{Forecasts of causal reductions in collections debt from expanding health insurance to the near-elderly by commuting zone}
\label{fig:forecast_drop_collections_cz}
\makebox[\linewidth][c]{
\begin{tabular}{cc}
    \textit{Panel A:} Per capita, 2008-2017 &     \textit{Panel B:} Per newly insured, 2008-2017 \\
    \includegraphics[width=3.5in,trim=4 4 4 4,clip]{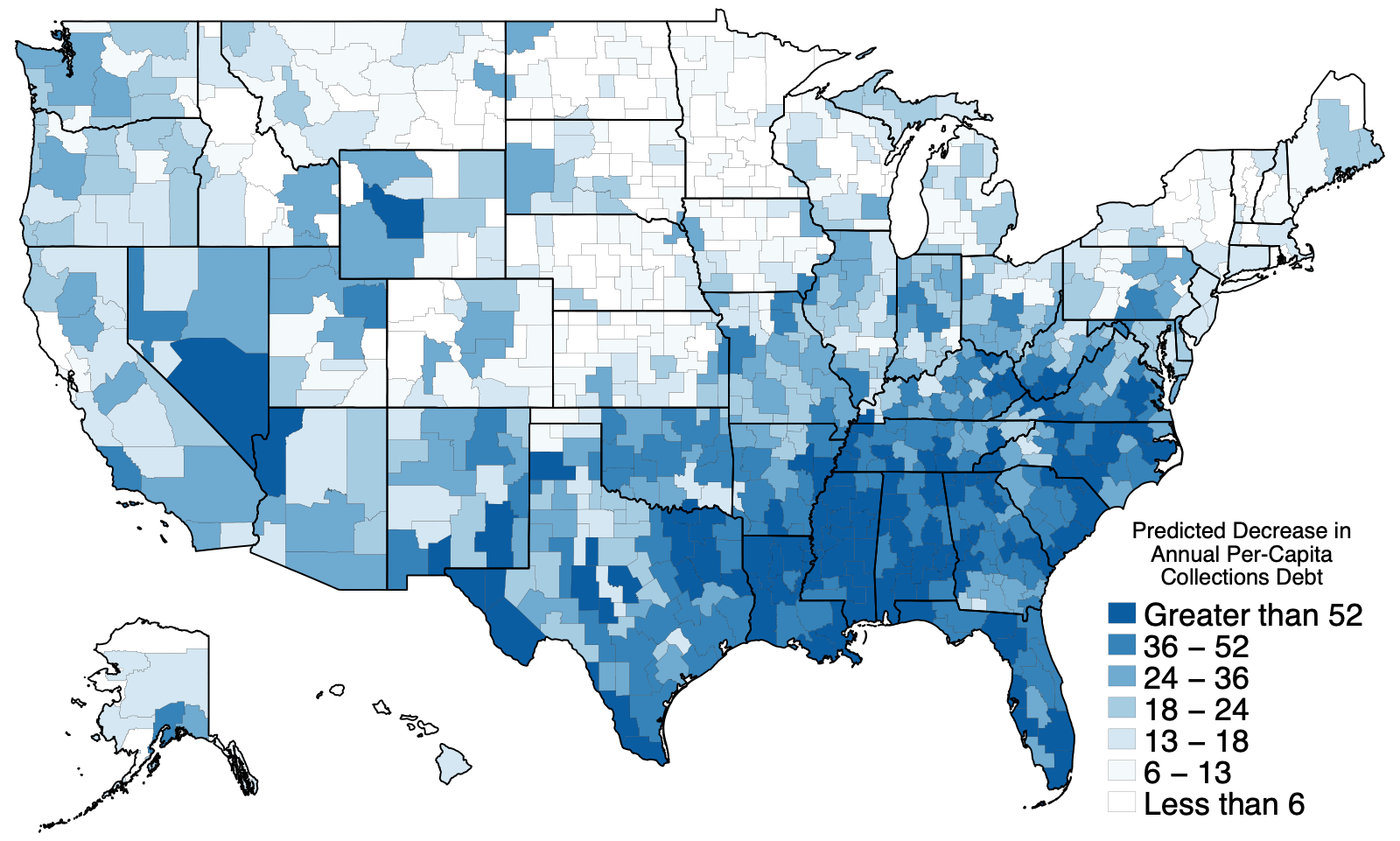} & \includegraphics[width=3.5in,trim=4 4 4 4,clip]{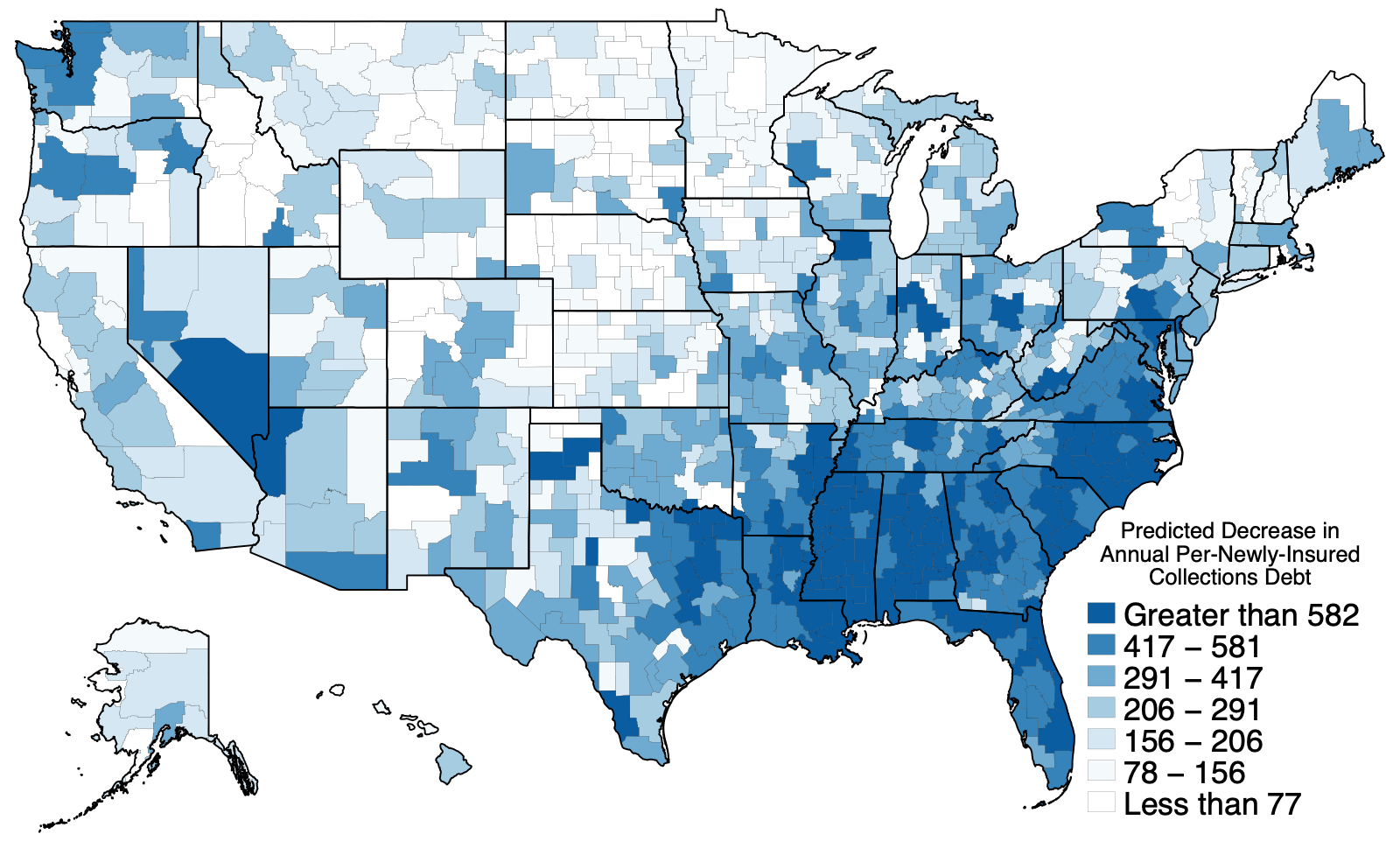}
\end{tabular}}
\begin{minipage} {0.9\textwidth} \setstretch{.9} \medskip
  \footnotesize{\textbf{Note:} This figure plots mean square error (MSE)-minimizing forecasts of the reductions in collections debt per capita (Panel A) and the reduction in collections debt per newly-insured (Panel B). We construct the MSE-minimizing forecasting by first running a Lasso regression to predict the CZ-level reductions in collections debt per capita (or per newly-insured) as discussed in Section \ref{sec:methods}. This generates a prediction for each CZ, which we call $\hat{\gamma_{l}}$. Following \cite{chetty2018impactsb} we then combine the $\hat{\gamma_{l}}$ estimates with our estimates of $\gamma_{l}$ to construct the MSE-minimizing forecast for each commuting zone, $\gamma_{l}^{f}$. Source: Consumer credit outcomes are based on 137,340,577 person-year observations from the New York Fed Consumer Credit Panel/Equifax, 2008-2017. CZ-level uninsurance rates are from the American Community Survey, 2008-2017. Healthcare market characteristics are from the Healthcare Cost Report Information System (HCRIS) and the Dartmouth Atlas. For additional details on the data see Section \ref{background_data}.}
\end{minipage}
\end{figure}

In Panel A of Figure \ref{fig:forecast_drop_collections_cz}, we map the \textit{per capita} mean-square error minimizing forecast causal effects, $\hat{\gamma}_{l}^{f}$, across CZs for the near-elderly, with darker colors depicting areas predicted to experience larger reductions in collections-related strain associated with an expansion of (nearly) universal health insurance to the near-elderly. The largest forecast reductions are concentrated in the South, ranging from \$20-\$50 in most CZs. The opposite is true in the Midwest, where forecast reductions in consumer financial strain are small across all CZs.\footnote{Table \ref{tab:czone_estimates_table} lists the forecasts for the 50 commuting zones with the largest near-elderly populations (accounting for 53.2\% of the near-elderly US population during our sample period).} In Panel B, we map the forecasts per newly-insured at age 65, $\hat{\beta}_{l}^{f}$. Despite large geographic differences in the near-elderly uninsurance rate, the maps are strikingly similar, with the largest forecast reductions in collections debt per newly-insured also concentrated in the South. This is consistent with the Lasso procedure selecting similar area-level characteristics when predicting changes in per capita and per newly-insured debt collections at age 65.\footnote{We plot the relationship between the two forecasts across CZs in Figure \ref{fig:tau_vs_beta_scatter} and find an R$^2$ of 0.8213.}

\begin{figure}
  \centering
  \caption{Forecasts of causal reductions in collections debt at the commuting zone level before and after the Affordable Care Act (ACA)}
  \label{fig:collections_prepost_aca}
  \makebox[\linewidth][c]{
  \begin{tabular}{cc}
  Panel A:  Pre and Post-ACA Per Capita Forecasts & Panel B: Decomposing Changes in Forecasts\\
\includegraphics[width=3in]{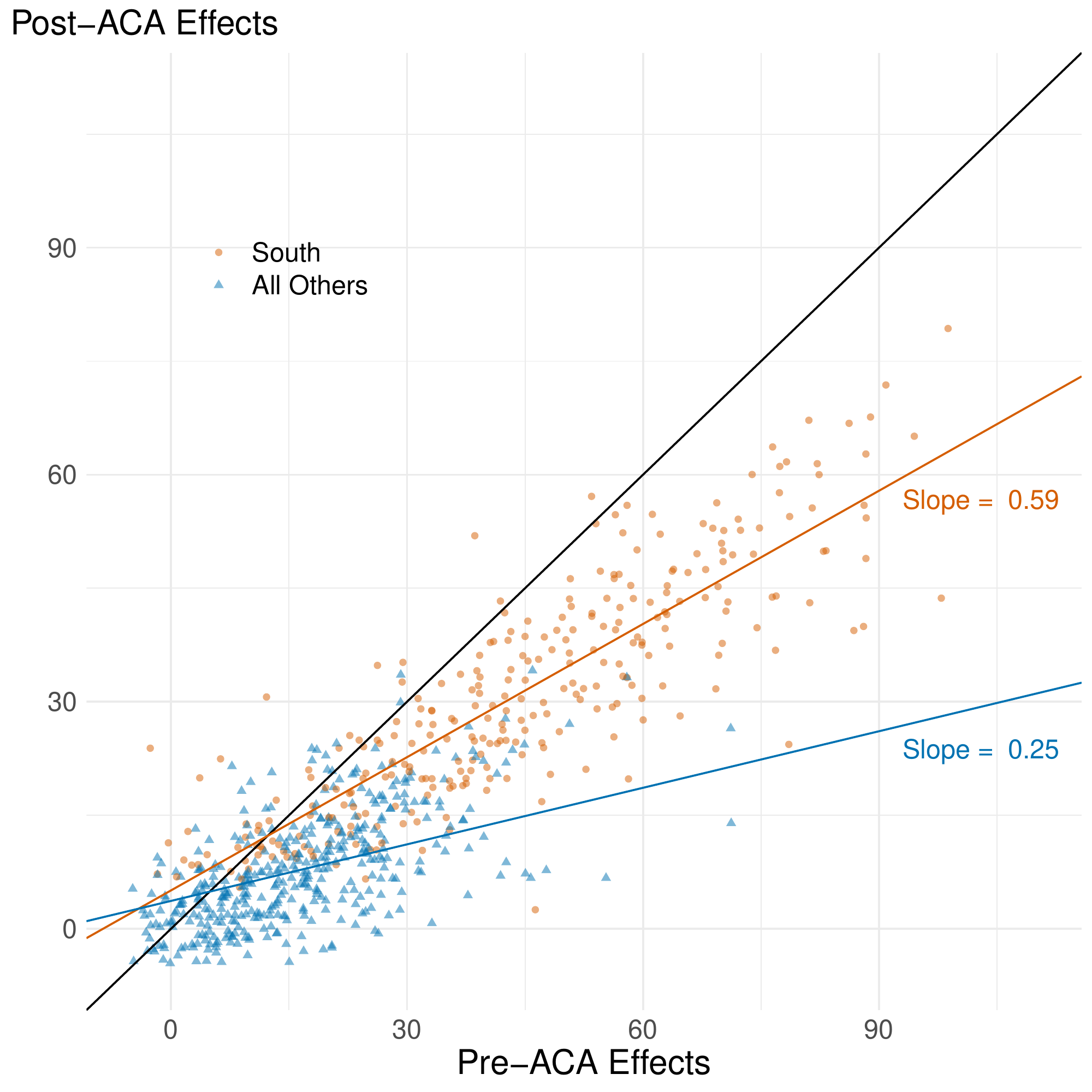} & \includegraphics[width=3in]{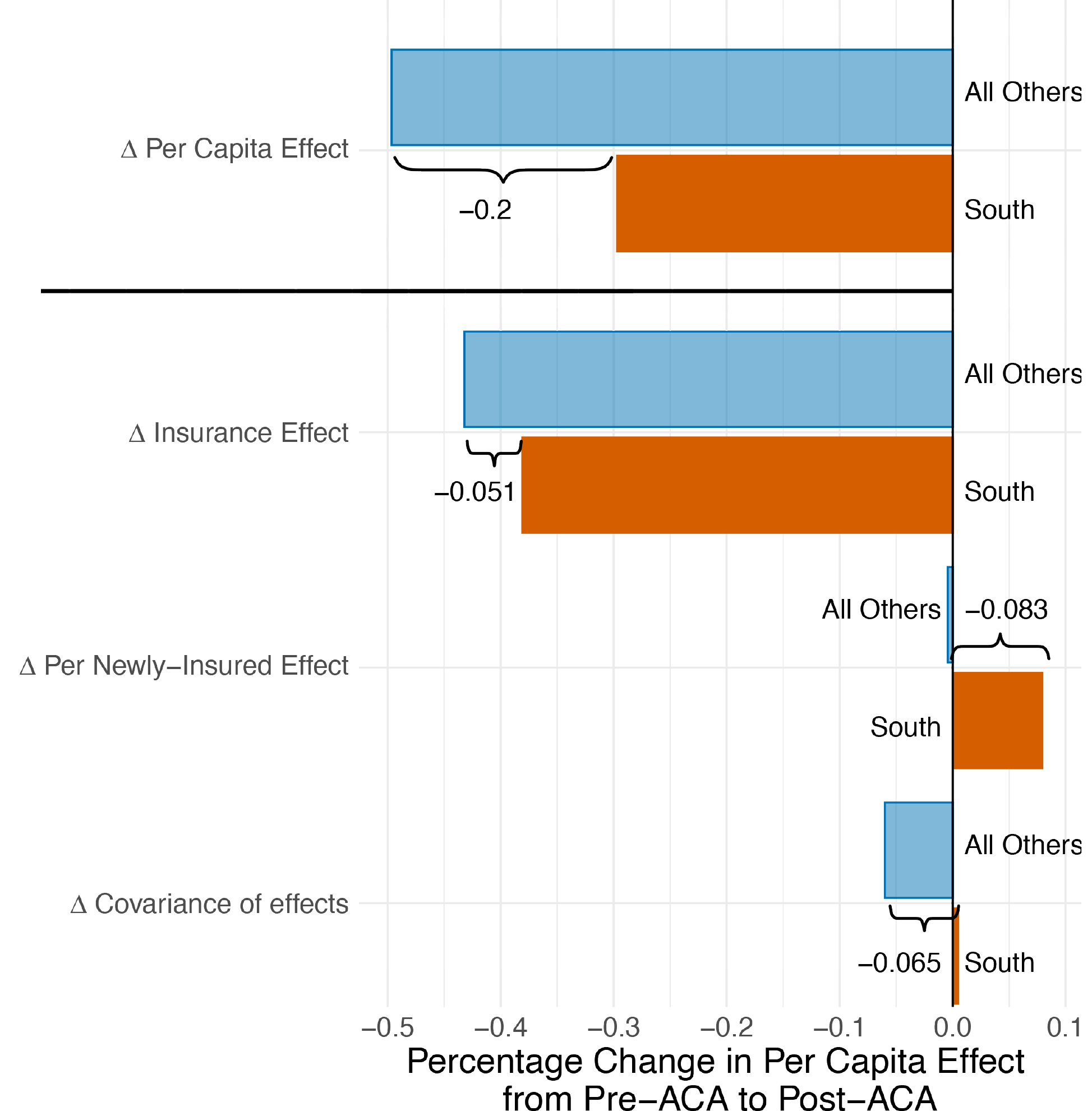} 
\end{tabular}}
 
  \begin{minipage} {0.9\textwidth} \setstretch{.9} \medskip
    \scriptsize{\textbf{Note:} Panel A of this figure plots mean square error (MSE)-minimizing forecasts of the CZ-level reductions in collections debt per capita in the pre-ACA period on the x-axis against the analogous post-ACA period forecast on the y-axis. We separately plot commuting zones in the South in orange circles and all other commuting zones with blue triangles. Fitted lines are constructed using bivariate OLS. We construct the MSE-minimizing forecasting by first running a Lasso regression to predict the CZ-level reductions in collections debt per capita separate for each period as discussed in Section \ref{sec:methods}. This generates a prediction for each CZ in each period, which we call $\hat{\gamma_{l}}$. Following \cite{chetty2018impactsb} we then combine the $\hat{\gamma_{l}}$ estimates with our estimates of $\gamma_{l}$ to construct the mean square error-minimizing forecast for each commuting zone in each period, $\gamma_{l}^{f}$. 
    Panel B of this figure plots the average percentage decline in per capita forecasted effects for the South and non-South regions, and the decomposition described in the text and Appendix \ref{apx:methods}. The top bars present the average percentage decline in per capita forecasted effects for the South and non-South regions. The second set of bars present the decline due to the change in insurance coverage rates, holding fixed the per newly-insured effect. The third set of bars present the decline due to the change in the per newly-insured effect, holding fixed the impact on insurance coverage rates. The last set of bars is due to the change in covariance between the per newly-insured effect and the effect on insurance. In curly braces for each set of bars we report the difference between the non-South and South averages.
    Source: Consumer credit outcomes are based on 137,340,577 person-year observations from the New York Fed Consumer Credit Panel/Equifax, 2008-2017. CZ-level uninsurance rates are from the American Community Survey, 2008-2017. For additional details on the data see Section \ref{background_data}.}
  \end{minipage}
\end{figure}

%% ACA section
We next examine how these forecasts have changed due to the ACA, federal health reform legislation that substantially expanded coverage \citep[][]{frean2017premium}. Panel A of Figure \ref{fig:collections_prepost_aca} presents the forecasts using the sample before and after the implementation of the ACA. On the x-axis, we plot the pre-ACA per capita forecast reductions in collections, and on the y-axis, the post-ACA per capita forecast reductions. The forecast reductions post-ACA are generally smaller than pre-ACA, which results in the majority of commuting zones below the 45-degree line. Rather than having a uniform effect across CZs on collections forecasts, which would appear as a vertical shift in the cloud of points, the ACA led to a ``rotation'' in the forecasts; CZs with larger pre-ACA forecasts experienced larger changes in forecast pre- to post-ACA. This is consistent with the increase in health insurance coverage due to the ACA \citep{frean2017premium}. However, the degree of rotation varied significantly by geography.
%In the South, the slope of 0.59 implies that a CZ with a forecast reduction in collections of \$100 pre-ACA has, on average, a forecast reduction in collections of \$59 post-ACA (a one-third reduction). The slope of 0.25, for the rest of the country, implies that a CZ with a forecast reduction in collections of \$10 pre-ACA has, on average, a forecast reduction of only \$2.50 post-ACA (a three-quarters reduction). 
The forecasts fell less in the South than elsewhere. As a result, the effect of Medicare on collections-related financial strain have become much more geographically-concentrated in the South, and particularly the ``Deep South'' region comprised of Louisiana, Alabama, Mississippi, Georgia, South Carolina, and parts of Texas and Florida (Figure \ref{fig:ca_forecast_tau_aca}).

Why did the forecast reductions in collections fall less in the South than elsewhere? Panel B of Figure  \ref{fig:collections_prepost_aca} documents that average CZ-level forecasts decreased by only 30\% in the South after the ACA's implementation, but 50\% elsewhere. We decompose the differential change in forecasts (i.e., the 30\% vs. 50\%) using a Kitagawa-Oaxaca-Blinder style decomposition \citep{kitagawa1955components,oaxaca1973male, blinder1973wage}. Intuitively, this decomposes the change in the forecast reductions in collections debt per capita into changes in the extensive margin ``any health insurance'' effect, changes in the intensive margin ``reduction in collections per newly-insured'' effect, and a residual term that reflects the changing covariance between the two (see Appendix \ref{apx:methods} for additional details). We find that the differential change in forecasts between regions was driven by all three pieces. First, uniform gains in ACA coverage across regions for the near-elderly (Figure \ref{fig:diffindisc_insurance})---despite higher pre-existing rates of uninsurance in the South---led to smaller reductions, percentage-wise, in uninsurance rates in the South as compared to elsewhere.\footnote{This left the uninsured near-elderly population more concentrated in the South after the implementation of the ACA (Figure \ref{fig:insurance_pre_cz_map_shrink_aca}).} This accounted for one-quarter of the differential change in forecasts between regions (0.051/0.2). Second, the forecast reductions in collections debt per newly-insured \textit{increased} in the South (where they were larger to begin with) after the ACA, but remained unchanged elsewhere (Figure \ref{fig:cz_forecast_beta_aca}). This accounted for two-fifths of the differential change in forecasts (0.083/0.2). Lastly, in the non-south CZs, the covariance between the per newly-insured effect and the effect on the share of individuals covered \emph{decreased} post-ACA, whereas it was unchanged in the South. This suggests that the ACA expansions in the South were not as well targeted (on this dimension) as those elsewhere in the country. The poorer targeting explained the remaining third (0.065/0.2) of the differential change in forecasts between regions.

%This is due to the uniform gains across regions in coverage for the near-elderly due to the ACA---despite higher pre-existing rates of uninsurance in the South---and, within the South, the fact that uninsurance rates remain high in areas where the financial health gains per newly-insured are largest.

% Consider the following exercise:

% \begin{align}
%     \gamma_{l} &= \beta_{l}\gamma_{l}^{h}\\
%     E(\gamma_{l}) &= E( \beta_{l}\gamma_{l}^{h})\\
%     &= E( \beta_{l})E(\gamma_{l}^{h}) + Cov(\beta_{l}, \gamma_{l}^{h})\\
%     \frac{E(\gamma_{l}^{South,Post}- \gamma_{l}^{South,Pre})}{E(\gamma_{l}^{South,Pre})} &= \frac{ A + B + C}{E(\gamma_{l}^{South,Pre})}\\
%     A & = \left(E( \beta_{l}^{Post, South}) -  E( \beta_{l}^{Pre, South})\right)E(\gamma_{l}^{h,Post, South})\\
%     B &= E( \beta_{l}^{Pre, South})\left(E(\gamma_{l}^{h,Post, South}) - E(\gamma_{l}^{h,Pre, South})\right)\\
%     C & =  Cov(\beta_{l}^{Post, South}, \gamma_{l}^{h,Post, South}) -
%  Cov(\beta_{l}^{Pre, South}, \gamma_{l}^{h,Pre, South}) 
% \end{align}

\section{Conclusion} \label{section_conclusion}
This paper examines the relationship between health insurance and financial health by studying financial outcomes for individuals as they age onto Medicare at 65. We find a 30 percent reduction in debt collections---and a two-thirds reduction in the geographic variation in collections---at age 65, with limited effects on other financial outcomes. Areas that experienced larger gains in financial health at age 65 had higher shares of black residents, people with disabilities, and for-profit hospitals. 

Our data suggest that the financial health benefits of potential future coverage expansions to the near-elderly have become more geographically concentrated in the South after the passage of the ACA. This is due to the uniform gains across regions in coverage for the near-elderly due to the ACA---despite higher pre-existing rates of uninsurance in the South---and, within the South, the fact that uninsurance rates remain high in areas where the financial health gains per newly-insured are largest. These findings highlight a potential limitation of policies, such as the ACA, that delegate states considerable latitude in policy implementation, and a relative advantage of programs, such as Medicare, that are federally-administered---specifically, that the former may exacerbate geographic disparities while the latter tend to reduce them.
	
\newpage
\singlespacing
\bibliographystyle{aea}
\bibliography{library}

@article{armstrong2018optimal,
  title={Optimal inference in a class of regression models},
  author={Armstrong, Timothy B and Koles{\'a}r, Michal},
  journal={Econometrica},
  volume={86},
  number={2},
  pages={655--683},
  year={2018},
  publisher={Wiley Online Library}
}

@article{frandsen2012quantile,
  title={Quantile treatment effects in the regression discontinuity design},
  author={Frandsen, Brigham R and Fr{\"o}lich, Markus and Melly, Blaise},
  journal={Journal of Econometrics},
  volume={168},
  number={2},
  pages={382--395},
  year={2012},
  publisher={Elsevier}
}

@techreport{armstrong2018simple,
  title={Simple and honest confidence intervals in nonparametric regression},
  author={Armstrong, Timothy and Koles{\'a}r, Michal},
  year={2018},
  institution={Cowles Foundation Discussion Paper}
}

@techreport{miller2018neighborhoods,
  title={Do Neighborhoods Affect Credit Market Decisions of Low-Income Borrowers? Evidence from the Moving to Opportunity Experiment},
  author={Miller, Sarah and Soo, Cindy K},
  year={2018},
  institution={National Bureau of Economic Research}
}

@techreport{keys2019,
  title={What Determines Consumer Financial Distress? Place-and Person-Based Factors},
  author={Keys, Benjamin J and Mahoney, Neale and Yang, Hanbin},
  year={2020},
  institution={National Bureau of Economic Research}
}

@techreport{lee2010introduction,
  title={An introduction to the frbny consumer credit panel},
  author={Lee, Donghoon and Van der Klaauw, Wilbert},
  institution={FRB of New York Staff Report},
  number={479},
  year={2010}
}

@article{lee2008regression,
  title={Regression discontinuity inference with specification error},
  author={Lee, David S and Card, David},
  journal={Journal of Econometrics},
  volume={142},
  number={2},
  pages={655--674},
  year={2008},
  publisher={Elsevier}
}

@article{ms2006jpube,
  title={The incidence of Medicare},
  author={McClellan, Mark and Jonathan Skinner},
  journal={Journal of Public Economics},
  volume={90},
  number={1-2},
  pages={257-276},
  year={2006},
  publisher={Elsevier}
}

@article{bl2006jpube,
  title={Does Medicare benefit the poor?},
  author={Bhattacharya, Jay and Darius Lakdawalla},
 journal={Journal of Public Economics},
  volume={90},
  number={1-2},
  pages={277-292},
  year={2006},
  publisher={Elsevier}
}

@article{barcellos2015effects,
  title={The effects of Medicare on medical expenditure risk and financial strain},
  author={Barcellos, Silvia Helena and Jacobson, Mireille},
  journal={American Economic Journal: Economic Policy},
  volume={7},
  number={4},
  pages={41--70},
  year={2015},
  publisher={American Economic Association}
}

@article{morris1983parametric,
  title={Parametric empirical Bayes inference: theory and applications},
  author={Morris, Carl N},
  journal={Journal of the American statistical Association},
  volume={78},
  number={381},
  pages={47--55},
  year={1983},
  publisher={Taylor \& Francis Group}
}

@article{clemens2017shadow,
  title={In the shadow of a giant: Medicare’s influence on private physician payments},
  author={Clemens, Jeffrey and Gottlieb, Joshua D},
  journal={Journal of Political Economy},
  volume={125},
  number={1},
  pages={000--000},
  year={2017},
  publisher={University of Chicago Press Chicago, IL}
}

@article{calonico2015rdrobust,
  title={rdrobust: An r package for robust nonparametric inference in regression-discontinuity designs},
  author={Calonico, Sebastian and Cattaneo, Matias D and Titiunik, Rocio},
  journal={R Journal},
  volume={7},
  number={1},
  pages={38--51},
  year={2015},
  publisher={Citeseer}
}

@article{gross2011health,
  title={Health insurance and the consumer bankruptcy decision: Evidence from expansions of Medicaid},
  author={Gross, Tal and Notowidigdo, Matthew J},
  journal={Journal of Public Economics},
  volume={95},
  number={7},
  pages={767--778},
  year={2011},
  publisher={Elsevier}
}

@article{mazumder2016effects,
  title={The effects of the Massachusetts health reform on financial distress},
  author={Mazumder, Bhashkar and Miller, Sarah},
  journal={American Economic Journal: Economic Policy},
  volume={8},
  number={3},
  pages={284--313},
  year={2016},
}

@article{finkelstein2012oregon,
  title={The Oregon health insurance experiment: evidence from the first year},
  author={Finkelstein, Amy and Taubman, Sarah and Wright, Bill and Bernstein, Mira and Gruber, Jonathan and Newhouse, Joseph P and Allen, Heidi and Baicker, Katherine and Oregon Health Study Group},
  journal={The Quarterly journal of economics},
  volume={127},
  number={3},
  pages={1057--1106},
  year={2012},
  publisher={MIT Press}
}

@article{finkelstein2008did,
  title={What did Medicare do? The initial impact of Medicare on mortality and out of pocket medical spending},
  author={Finkelstein, Amy and McKnight, Robin},
  journal={Journal of public economics},
  volume={92},
  number={7},
  pages={1644--1668},
  year={2008},
  publisher={Elsevier}
}

@article{card2008impact,
  title={The impact of nearly universal insurance coverage on health care utilization: evidence from Medicare},
  author={Card, David and Dobkin, Carlos and Maestas, Nicole},
  journal={American Economic Review},
  volume={98},
  number={5},
  pages={2242--58},
  year={2008}
}

@article{engelhardt2011medicare,
  title={Medicare Part D and the financial protection of the elderly},
  author={Engelhardt, Gary V and Gruber, Jonathan},
  journal={American Economic Journal: Economic Policy},
  volume={3},
  number={4},
  pages={77--102},
  year={2011}
}

@article{imbens2019optimized,
  title={Optimized regression discontinuity designs},
  author={Imbens, Guido and Wager, Stefan},
  journal={Review of Economics and Statistics},
  volume={101},
  number={2},
  pages={264--278},
  year={2019},
  publisher={MIT Press}
}

@article{kolesar2018inference,
  title={Inference in regression discontinuity designs with a discrete running variable},
  author={Koles{\'a}r, Michal and Rothe, Christoph},
  journal={American Economic Review},
  volume={108},
  number={8},
  pages={2277--2304},
  year={2018}
}

@techreport{duggan2019impact,
  title={The Impact of the Affordable Care Act: Evidence from California's Hospital Sector},
  author={Duggan, Mark and Gupta, Atul and Jackson, Emilie},
  year={2019},
  institution={National Bureau of Economic Research}
}

@article{hu2018effect,
  title={The effect of the affordable care act Medicaid expansions on financial wellbeing},
  author={Hu, Luojia and Kaestner, Robert and Mazumder, Bhashkar and Miller, Sarah and Wong, Ashley},
  journal={Journal of public economics},
  volume={163},
  pages={99--112},
  year={2018},
  publisher={Elsevier}
}

@article{brevoort2020credit,
  title={The credit consequences of unpaid medical bills},
  author={Brevoort, Kenneth and Grodzicki, Daniel and Hackmann, Martin B},
  journal={Journal of Public Economics},
  volume={187},
  pages={104203},
  year={2020},
  publisher={Elsevier}
}

@article{frean2017premium,
  title={Premium subsidies, the mandate, and Medicaid expansion: Coverage effects of the Affordable Care Act},
  author={Frean, Molly and Gruber, Jonathan and Sommers, Benjamin D},
  journal={Journal of Health Economics},
  volume={53},
  pages={72--86},
  year={2017},
  publisher={Elsevier}
}

@article{chetty2018impactsb,
  title={The impacts of neighborhoods on intergenerational mobility II: County-level estimates},
  author={Chetty, Raj and Hendren, Nathaniel},
  journal={The Quarterly Journal of Economics},
  volume={133},
  number={3},
  pages={1163--1228},
  year={2018},
  publisher={Oxford University Press}
}

@techreport{finkelstein2019place,
  title={Place-based drivers of mortality: Evidence from migration},
  author={Finkelstein, Amy and Gentzkow, Matthew and Williams, Heidi L},
  year={2019},
  institution={National Bureau of Economic Research}
}

@article{doi:10.1086/706623,
author = {Caswell, Kyle J. and Goddeeris, John H.},
title = {Does Medicare Reduce Medical Debt?},
journal = {American Journal of Health Economics},
volume = {0},
number = {ja},
pages = {null},
year = {2019},
doi = {10.1086/706623},

URL = { 
        https://doi.org/10.1086/706623
    
},
eprint = { 
        https://doi.org/10.1086/706623
    
}

}

@article{belloni2013least,
  title={Least squares after model selection in high-dimensional sparse models},
  author={Belloni, Alexandre and Chernozhukov, Victor},
  journal={Bernoulli},
  volume={19},
  number={2},
  pages={521--547},
  year={2013},
  publisher={Bernoulli Society for Mathematical Statistics and Probability}
}

@article{david2013growth,
  title={The growth of low-skill service jobs and the polarization of the US labor market},
  author={David, H and Dorn, David},
  journal={American Economic Review},
  volume={103},
  number={5},
  pages={1553--97},
  year={2013}
}

@article{dorn2019work,
  title={When Work Disappears: Manufacturing Decline and the Falling Marriage Market Value of Young Men},
  author={Dorn, David and Hanson, Gordon and others},
  journal={American Economic Review: Insights},
  volume={1},
  number={2},
  pages={161--78},
  year={2019}
}

@article{barreca2011saving,
  title={Saving babies? Revisiting the effect of very low birth weight classification},
  author={Barreca, Alan I and Guldi, Melanie and Lindo, Jason M and Waddell, Glen R},
  journal={The Quarterly Journal of Economics},
  volume={126},
  number={4},
  pages={2117--2123},
  year={2011},
  publisher={Oxford University Press}
}

@misc{ipums2019,
author = {Ruggles, Steven and Flood, Sarah and Goeken, Ronald and Grover, Josiah and Meyer, Erin and Pacas, Jose and Sobek, Matthew},
editor = {Minneapolis, MN: IPUMS},
title = {IPUMS USA: Version 9.0 [dataset]},
year = {2019},
url = {https://doi.org/10.18128/D010.V9.0},
}

@article{oaxaca1973male,
  title={Male-female wage differentials in urban labor markets},
  author={Oaxaca, Ronald},
  journal={International economic review},
  pages={693--709},
  year={1973},
  publisher={JSTOR}
}

@article{blinder1973wage,
  title={Wage discrimination: reduced form and structural estimates},
  author={Blinder, Alan S},
  journal={Journal of Human resources},
  pages={436--455},
  year={1973},
  publisher={JSTOR}
}

@article{cattaneo2019binscatter,
  title={On binscatter},
  author={Cattaneo, Matias D and Crump, Richard K and Farrell, Max H and Feng, Yingjie},
  journal={arXiv preprint arXiv:1902.09608},
  year={2019}
}

@misc{dartmouth2019,
editor = {Dartmouth Institute},
title = {Dartmouth Atlas of Health Care},
author = {{Dartmouth Institute}},
year = {2019},
url = {http://www.dartmouthatlas.org/tools/downloads.aspx (accessed November 2019).},
}

@article{finkelstein2016sources,
  title={Sources of geographic variation in health care: Evidence from patient migration},
  author={Finkelstein, Amy and Gentzkow, Matthew and Williams, Heidi},
  journal={The quarterly journal of economics},
  volume={131},
  number={4},
  pages={1681--1726},
  year={2016},
  publisher={MIT Press}
}

@article{valdovinos2015california,
  title={In California, not-for-profit hospitals spent more operating expenses on charity care than for-profit hospitals spent},
  author={Valdovinos, Erica and Le, Sidney and Hsia, Renee Y},
  journal={Health Affairs},
  volume={34},
  number={8},
  pages={1296--1303},
  year={2015}
}

@article{frank1990market,
  title={Market forces and the public good: competition among hospitals and provision of indigent care.},
  author={Frank, RG and Salkever, DS and Mitchell, J},
  journal={Advances in health economics and health services research},
  volume={11},
  pages={159},
  year={1990}
}

@article{schlesinger2006nonprofits,
  title={How Nonprofits Matter In American Medicine, And What To Do About It: Reports of the demise of nonprofits in US health care are premature.},
  author={Schlesinger, Mark and Gray, Bradford H},
  journal={Health affairs},
  volume={25},
  number={Suppl1},
  pages={W287--W303},
  year={2006},
  publisher={Project HOPE-The People-to-People Health Foundation, Inc.}
}

@article{horwitz2005making,
  title={Making profits and providing care: comparing nonprofit, for-profit, and government hospitals},
  author={Horwitz, Jill R},
  journal={Health affairs},
  volume={24},
  number={3},
  pages={790--801},
  year={2005},
  publisher={Project HOPE-The People-to-People Health Foundation, Inc.}
}

@article{chetty2014land,
  title={Where is the land of opportunity? The geography of intergenerational mobility in the United States},
  author={Chetty, Raj and Hendren, Nathaniel and Kline, Patrick and Saez, Emmanuel},
  journal={The Quarterly Journal of Economics},
  volume={129},
  number={4},
  pages={1553--1623},
  year={2014},
  publisher={Oxford University Press}
}

@article{cooper2018price,
  title={The price ain’t right? Hospital prices and health spending on the privately insured},
  author={Cooper, Zack and Craig, Stuart V and Gaynor, Martin and Van Reenen, John},
  journal={The Quarterly Journal of Economics},
  volume={134},
  number={1},
  pages={51--107},
  year={2018},
  publisher={Oxford University Press}
}

@article{Batty2020,
author = {Batty, Michael and Gibbs, Christa and Ippolito, Benedic},
journal = {AEI Economics Working Paper},
number = {June},
title = {{Health insurance , medical debt , and financial}},
year = {2020}
}

@article{dobkin2018economic,
  title={The economic consequences of hospital admissions},
  author={Dobkin, Carlos and Finkelstein, Amy and Kluender, Raymond and Notowidigdo, Matthew J},
  journal={American Economic Review},
  volume={108},
  number={2},
  pages={308--52},
  year={2018}
}

@article{kitagawa1955components,
  title={Components of a difference between two rates},
  author={Kitagawa, Evelyn M},
  journal={Journal of the american statistical association},
  volume={50},
  number={272},
  pages={1168--1194},
  year={1955},
  publisher={Taylor \& Francis}
}

@article{cochrane1991simple,
  title={A simple test of consumption insurance},
  author={Cochrane, John H},
  journal={Journal of political economy},
  volume={99},
  number={5},
  pages={957--976},
  year={1991},
  publisher={The University of Chicago Press}
}

@article{meyer2019disability,
  title={Disability, earnings, income and consumption},
  author={Meyer, Bruce D and Mok, Wallace KC},
  journal={Journal of Public Economics},
  volume={171},
  pages={51--69},
  year={2019},
  publisher={Elsevier}
}

@article{charles2003longitudinal,
  title={The longitudinal structure of earnings losses among work-limited disabled workers},
  author={Charles, Kerwin Kofi},
  journal={Journal of human Resources},
  volume={38},
  number={3},
  pages={618--646},
  year={2003},
  publisher={University of Wisconsin Press}
}

@article{poterba2017asset,
  title={The asset cost of poor health},
  author={Poterba, James M and Venti, Steven F and Wise, David A},
  journal={The Journal of the Economics of Ageing},
  volume={9},
  pages={172--184},
  year={2017},
  publisher={Elsevier}
}

\onehalfspacing

\makeatletter
\setlength{\@fptop}{5pt}
\makeatother

\clearpage  

\appendix

\begin{center}
{\Large For Online Publication \\ Appendix for:\\\textbf{The Great Equalizer: Medicare and the Geography of Consumer Financial Strain}}\\
\end{center}
\singlespacing
% \pagestyle{fancy}
% % \lhead{Online Appendix}
% % \rhead{}

\addtocounter{section}{0}
\renewcommand{\tablename}{Appendix Table}
\renewcommand{\figurename}{Appendix Figure}

\renewcommand{\thetable}{A\arabic{table}}
\setcounter{table}{0}
\renewcommand{\thefigure}{A\arabic{figure}}
\setcounter{figure}{0}
\setcounter{page}{1}
\setcounter{footnote}{0}

%\section{Background}\label{apx:background}

% Some details on how to think about Medicare transition
%Since there are individuals at age 64 with and without coverage, and with different benefit designs and plan structures, the treatment we're studying is a weighted average of the effect of the transition to Medicare for those who were previously uninsured and those with different forms of coverage at age 64. Later, we exploit geographic variation in the distribution of the uninsurance rate to explore how our effects vary based on the impact of Medicare on uninsurance. 

\section{Study data}\label{apx:data}

\subsection{Financial outcomes data}\label{background_ccp_data_apx}
The main dataset used in our analysis is the Federal Reserve Bank of New York's Equifax Consumer Credit Panel (CCP). The CCP is a five percent random sample of all individuals in the U.S. with credit reports. The CCP data is a representative sample of all individuals with a credit report but it does not include the roughly 11 percent of the U.S. population without credit reports. As a result, the CCP data is more representative for high-income individuals than for low-income individuals, and it is more representative for older than younger people. \citet{lee2010introduction} show that the CCP is reasonably representative of the U.S. population with the possible exception of very young adults, suggesting that sample representativeness should not be a concern in our application. 

The data include a comprehensive set of consumer credit outcomes from quarterly from the first quarter of 1999 to the fourth quarter of 2017, including information on credit scores (originating from Equifax Risk Score 3.0), unsecured credit lines, auto loans, and mortgages. The data also include year of birth and precise geographic location at the census block level. No other demographic information is available at the individual level. 

A major virtue of the CCP is its large sample size, which allows us to measure financial outcomes at granular geographic levels with precision. This is key to our RD estimation strategy across geographies, where we estimate the effect of Medicare separately for 50 states and 741 commuting zones in the country. For our analyses, we aggregate the data to locality-by-age-by-year cells and weight by the underlying population in each cell. Since we only observe birth year, and the data is quarterly, age is measured with noise. For example, all individuals with birth year 1940 are measured as age 65 in the first quarter of 2005, while in reality some of these individuals will turn 65 later in the year. We address this using a ``donut'' RD procedure which we discuss in more detail in Section \ref{sec:methods}. 

The financial health variables that we focus on from the CCP are the size of accounts sent to collection agencies (usually, these are accounts that have been delinquent for over 90 days), the size of accounts that are delinquent, the Equifax Risk Score, as well as additional financial health outcomes (e.g., bankruptcy). Refer to Appendix Table \ref{tab:ccp_descriptions} for the definitions of each of the financial health outcomes we use from the CCP.

We examine the impact of Medicare on the distributions of three of our outcome measures: amount of debt in collections, total amount of debt in delinquency, and amount of credit card debt in delinquency. For all three of these cases, we would expect that large out-of-pocket expenses would cause increases in the right tail of the distribution. To examine this, we calculate the share of the population in a county-year-age bin that has amounts in the following bins: 1-500, 501-1,000, 1,001-2,500, 2,501-5,000, 5,001-10,000, and greater than 10,000 dollars. The residual category is any person with 0 dollars. We use the share within a given bin as the outcome variable in our main specification, so that our estimate is the change in the relative share of individuals within each bin due to Medicare eligibility.\footnote{An alternative approach would be to directly estimate quantile treatment effects using regression discontinuity, such as those proposed in \cite{frandsen2012quantile}. However, we are not able to easily account for the discrete running variable in our estimation process using quantile treatment effects. As a result, we focus on our share-based approach.}

\subsection{Demographic and health insurance data}
For demographic and health insurance information, we draw on the American Community Survey \citep{ipums2019}. All analyses use samples constructed from the PUMA and state datasets, linked to the Commuting Zone (CZ)- and state-level. Our cross-walk from PUMA to Commuting Zones uses David Dorn's crosswalks (available here: \url{https://www.ddorn.net/data.htm}).

\textit{Demographic data}. We construct demographic variables from the ACS at the PUMA-by-age-by-year level and then crosswalk to the CZ- and state-level to test for covariate smoothness in validating our RD design and to examine the correlates of geographic heterogeneity in our treatment effects. From the ACS, we measure the homeownership rate, marital status, race, educational attainment, employment status, usual hours worked per week, total personal income, social security income, poverty status, and disability rate.

\textit{Health insurance data}. The ACS also allows us to construct health insurance variables from the ACS at the PUMA-by-age-by-year level to test for changes in health insurance at age 65 and to examine the correlates of geographic heterogeneity in our treatment effects. From the ACS, we measure the share of individuals in each cell with any health insurance coverage.%, private health insurance coverage, health insurance through employer/union, health insurance purchased directly, health insurance through Medicaid, health insurance through VA, health insurance through Indian Health Services. From this data, we also constructed an indicator for whether an individual had multiple sources of coverage and used that to measure the share of individuals in each cell with more than one source of coverage.

\subsection{Additional area-level characteristics data}
In addition to the CCP and ACS data, we constructed additional characteristics at the PUMA-level. These characteristics are drawn from several places, including the Healthcare Cost Report Information System (HCRIS) and the Dartmouth Atlas. From the HCRIS data, we construct PUMA-level measures of the share of hospital patient days at for-profit hospitals, teaching hospitals, and public hospitals. We also measure the average hospital occupancy rate at the PUMA-level and the hospital beds per capita. In addition, for the period 2010-2017 we measure PUMA-level reported charity care costs per patient day and payments recovered by hospitals from charity care patients by patient day. From the Dartmouth Atlas data, we measure the PUMA-level risk-adjusted Medicare spending per enrollee \citep{dartmouth2019}.

\section{Additional Methods and Robustness}\label{apx:methods}

\subsection{Honest RD and Shrinkage Estimators}

\paragraph{Robustness using Honest RD}
As discussed in Section \ref{sec:methods}, we account for discreteness and measurement error in our running variable, age, by following the "honest" confidence intervals approach outlined in \cite{kolesar2018inference}, and \cite{armstrong2018optimal, armstrong2018simple}. This method requires an additional tuning parameter, $K$, which imposes an upper bound on the absolute value of the second derivative of the conditional expectation function. Intuitively, this method places a bound on how quickly the functions $f(\cdot)$ and $g(\cdot)$ can change. To choose our value of $K$ for our main estimates, we follow an approach similar to the approach advocated in \cite{imbens2019optimized}. We take a large window to the left of the RD cutoff and fit a quadratic function of age to the data. We take the coefficient on the quadratic term (the second derivative), take the absolute value, and multiply it by four. We take this as our estimate of $K$. Similar to robustness exercises with bandwidths in previous RD methods, we present additional robustness tests which vary the value of $K$ by changing the number that we scale this second derivative by in Appendix Figures \ref{fig:bound_scaling_robustness_apx} and \ref{fig:bound_scaling_var_robustness_apx}. %In all cases for our estimation, we report four numbers: our point estimate, the estimated standard error, and bias-adjusted confidence intervals.

%We present an example of the shrinkage estimates compared to the raw estimates in Appendix Figures \ref{fig:shrinkage_insurance_comparison} and \ref{fig:shrinkage_collections_comparison}. These shrinkage estimates are distinct from our forecasts in Section \ref{subsec:results_forecast}.

\paragraph{Inference for variance reduction}
Our estimate of cross-state and cross-CZ variance reduction due to the Medicare is a non-linear functional of the different estimates of a local non-parametric estimate. Specifically, we are interested in $T(f,g) = \phi(f,g)$, where 
$$\phi(f,g) = 1 - \frac{\left(g(0)'g(0) - (L^{-1}g(0)'\iota)^2)\right) }{\left(f(0)'f(0) - (L^{-1}f(0)'\iota)^2)\right) } = 1- \frac{L^{-1}\left(\sum_{l} g_{l}(0)^{2} - \left(L^{-1}\sum_{m}  g_{m}(0)\right)^{2}\right)}{L^{-1}\left(\sum_{l} f_{l}(0)^{2} - \left(L^{-1}\sum_{m}  f_{m}(0)\right)^{2}\right)},$$ 

where $f$ and $g$  are the vector of functions $f_{l}$ and $g_l$ that are estimated using local linear regression and $\iota$ is an $L \times 1$ vector of ones.\footnote{Note that we additionally population-weight our estimates. We omit this notation here for simplicity's sake.} 

To construct confidence intervals for this estimate that correctly account for the discreteness of our outcome variable, we apply the delta method following Appendix B.1.1 of \cite{armstrong2018simple}. Let the numerator and denominator of $\phi$ be $A$ and $B$, respectively. Then, $dA/dg = 2g(0)' - 2(L^{-1}g(0)'\iota) L^{-1} \iota'$ and  $dB/df = 2f(0)' - 2(L^{-1}f(0)'\iota) L^{-1} \iota'$. The cross-derivatives are zero. Hence,  
$d\phi(f,g)/df = -d(A B^{-1})/df = -(dA/df )B^{-1}$ and $d\phi(f,g)/dg = -d(A B^{-1})/dg = -A (dB^{-1})/dg = A (B^{-2})dB/dg$. 

Thus, our bias term will be $B = \sum_{l} |\phi'_{l} B_{l}|$, where $B_{l}$ is the bias determined from the underlying estimation.

We next consider the  covariance matrix $\Sigma$ of our stacked $f$ and $g$. Since we estimate each $f_l$ and $g_{l}$ separately, $\Sigma$ is simply a diagonal matrix of the $S_{l}^{2}$ estimates for each $f_{l}(0)$ and $g_{l}(0)$ estimate. Hence, our variance estimate is $S^2 = \phi'(f,g) ' \Sigma \phi'(f,g)$. 

Finally, to calculate the confidence intervals around our estimate $\hat{T}(f,g)$, we follow \cite{armstrong2018simple} and calculate the 95\% confidence intervals around our estimate of the ratio as $$\hat{T}(f,g) \pm \text{cv}_{0.95}(t) \cdot \hat{se},$$ where $t =B/S$ is our bias-sd ratio and $\hat{se} = \sqrt{S^2}$. We note that $\text{cv}_{0.95}(t)$ is the quantile of the folded normal distribution with mean equal to $t$ (see the note in Table 1 of \cite{armstrong2018simple}). 

\paragraph{Shrinkage Estimators}
Due to smaller sample sizes, the locality-level estimates are noisier than estimates of the overall national effects (or counterfactuals). Hence, our estimates of $\gamma_{l}$, $y_{l}(65-)$, and $y_{l}(65+)$ have more inherent noise and variation than the true underlying estimates due to estimation error (in part due to smaller sample sizes). Here we provide additional details on the shrinkage estimator we use to address this.

Formally, using our estimates of state-level discontinuities as an example, we calculate the shrinkage estimator by assuming that the $\gamma_{s} \sim \mathcal{N}(\gamma_{0}, \sigma^{2})$. We estimate these two parameters directly. Then, using the standard errors estimated for each $\gamma_{s}$, $\hat{\sigma}_s$, and following the standard James-Stein estimator approach \citep{morris1983parametric}, we construct $B = \frac{\hat{\sigma}^{2}}{\hat{\sigma}_{s}^{2} + \hat{\sigma}^{2}}$, and our shrinkage estimator is $\tilde{\gamma}_{s} = B \hat{\gamma}_{s} + (1-B) \gamma_{0}$. The CZ-level counterfactuals in Figure \ref{fig:collections_prepost_cz_map_shrink}, for example, are shrunk using this method.

\subsection{Robustness checks}
\label{apx:main_agerd_robustness}
\begin{figure}[htpb!]
  \centering
  \caption{Robustness of Age RD Estimates to Bound Scaling Factor}
  \label{fig:bound_scaling_robustness_apx}
\begin{tabular}{ccc}
  \textit{Panel A:} Share with coverage &    \textit{Panel B:} Total Collections (\$) &  \textit{Panel C:} Risk Score \\
 \includegraphics[width=4.5cm]{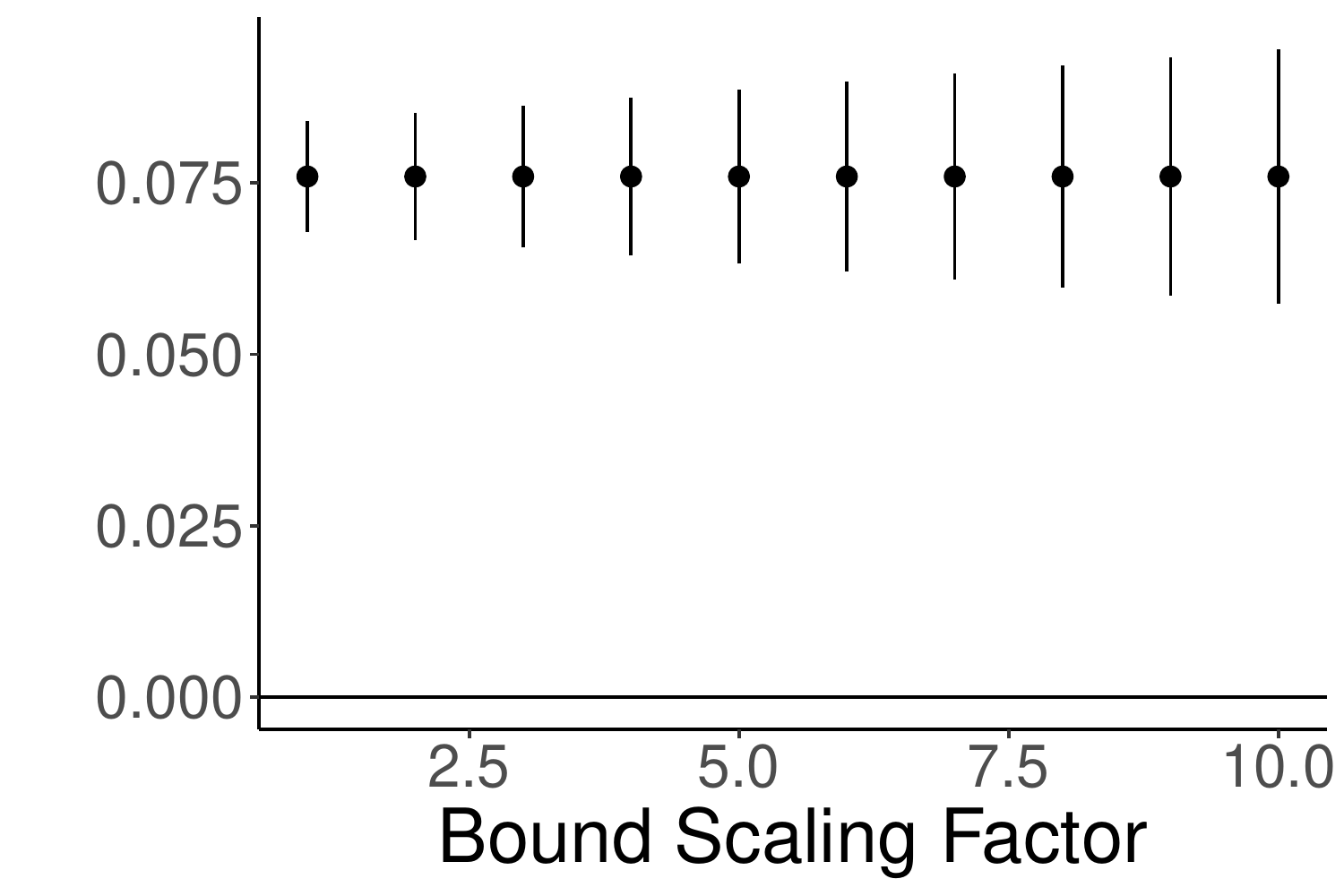} &   \includegraphics[width=4.5cm]{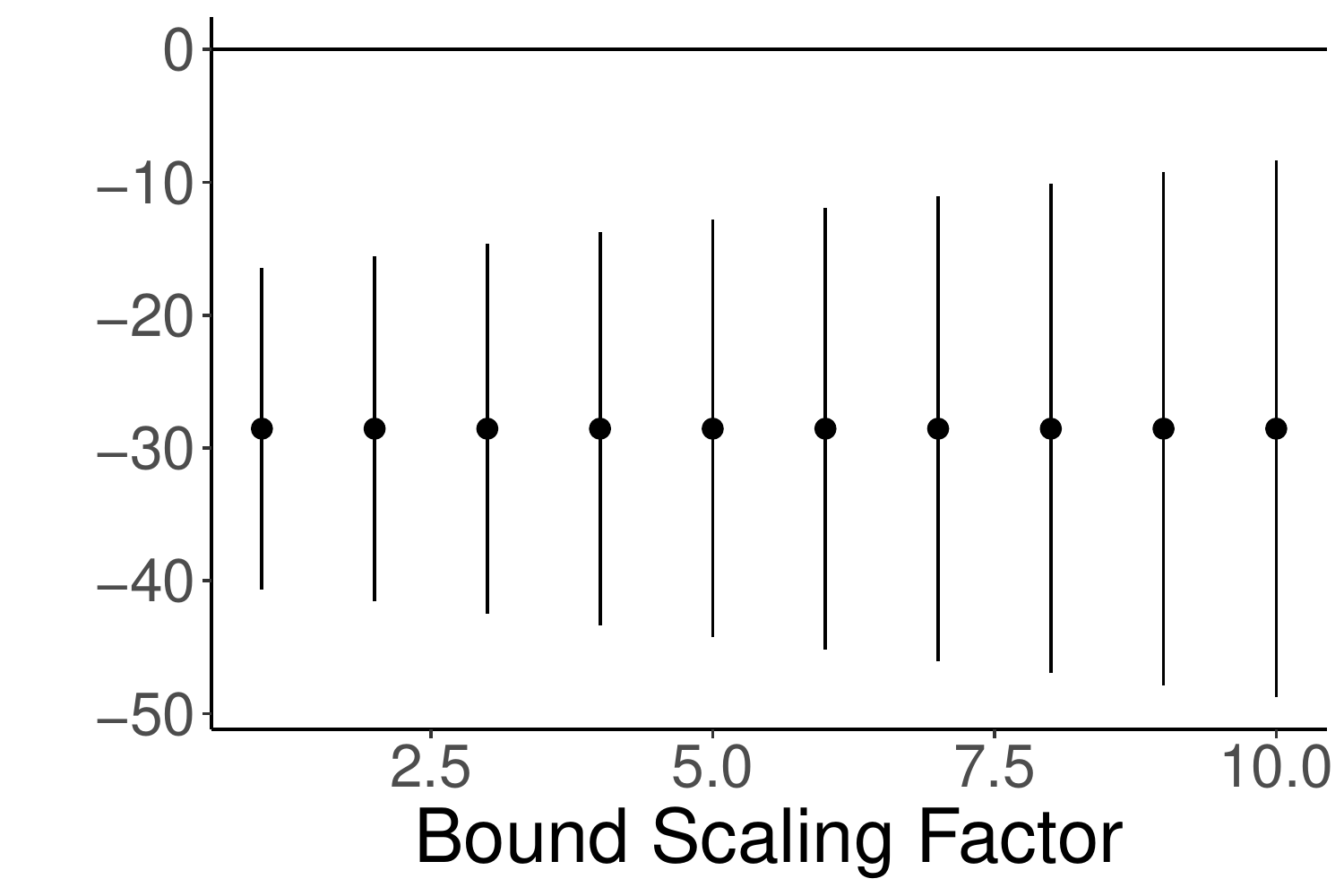} &   \includegraphics[width=4.5cm]{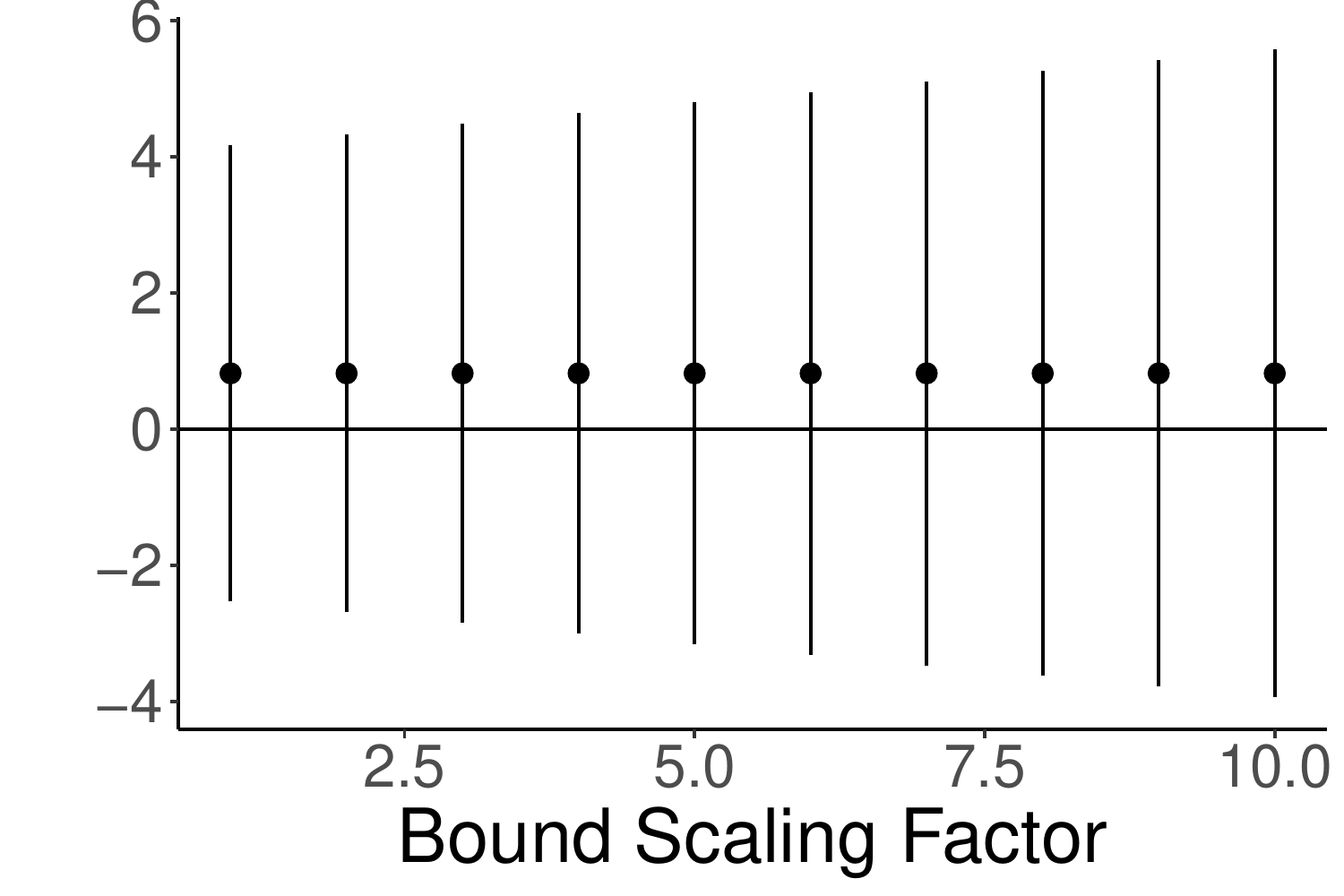} \\ [.5cm]
\textit{Panel D:} Bankruptcy &  \textit{Panel E:} Total debt past due &  \textit{Panel F:} Mortgage debt past due \\
 \includegraphics[width=4.5cm]{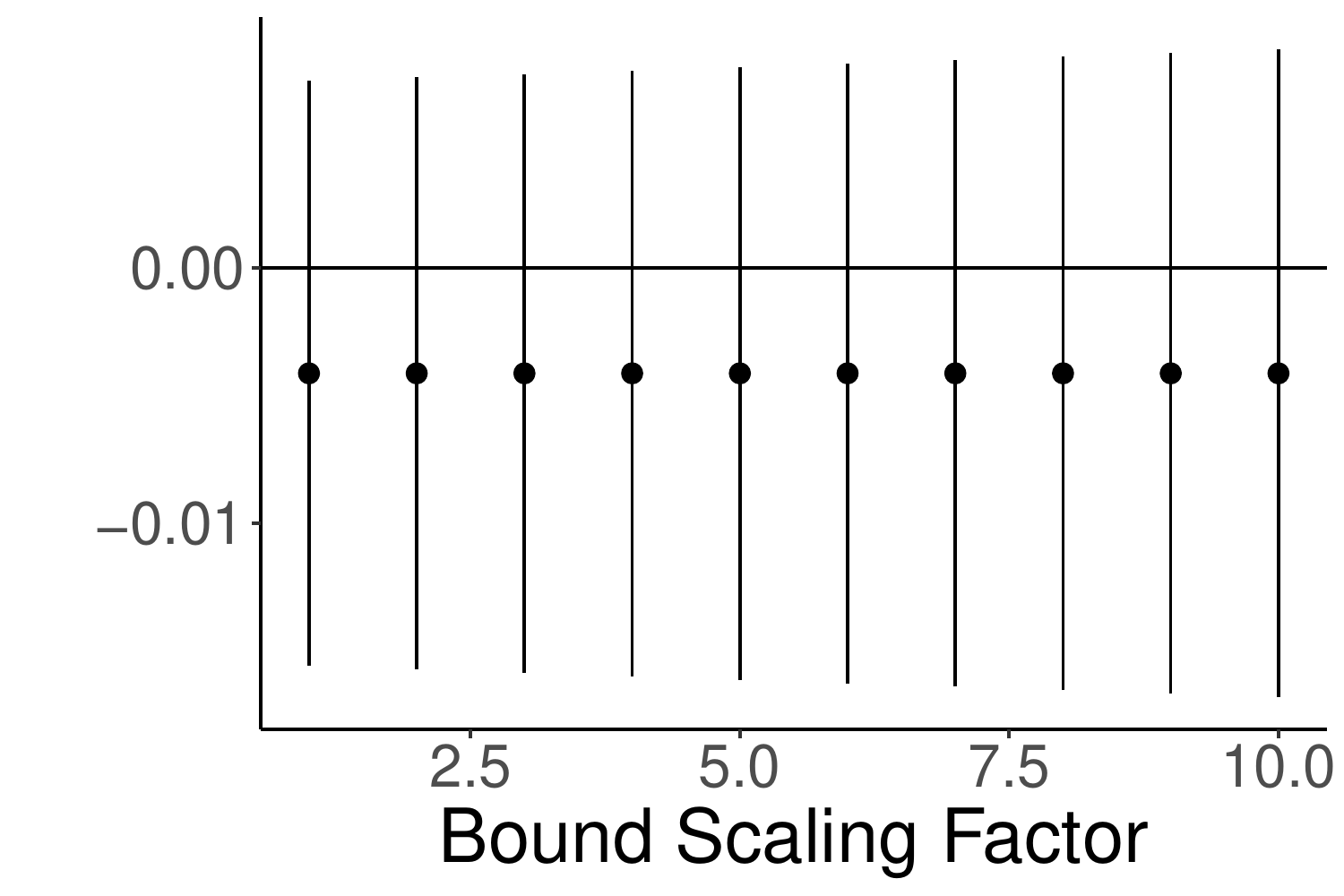} &   \includegraphics[width=4.5cm]{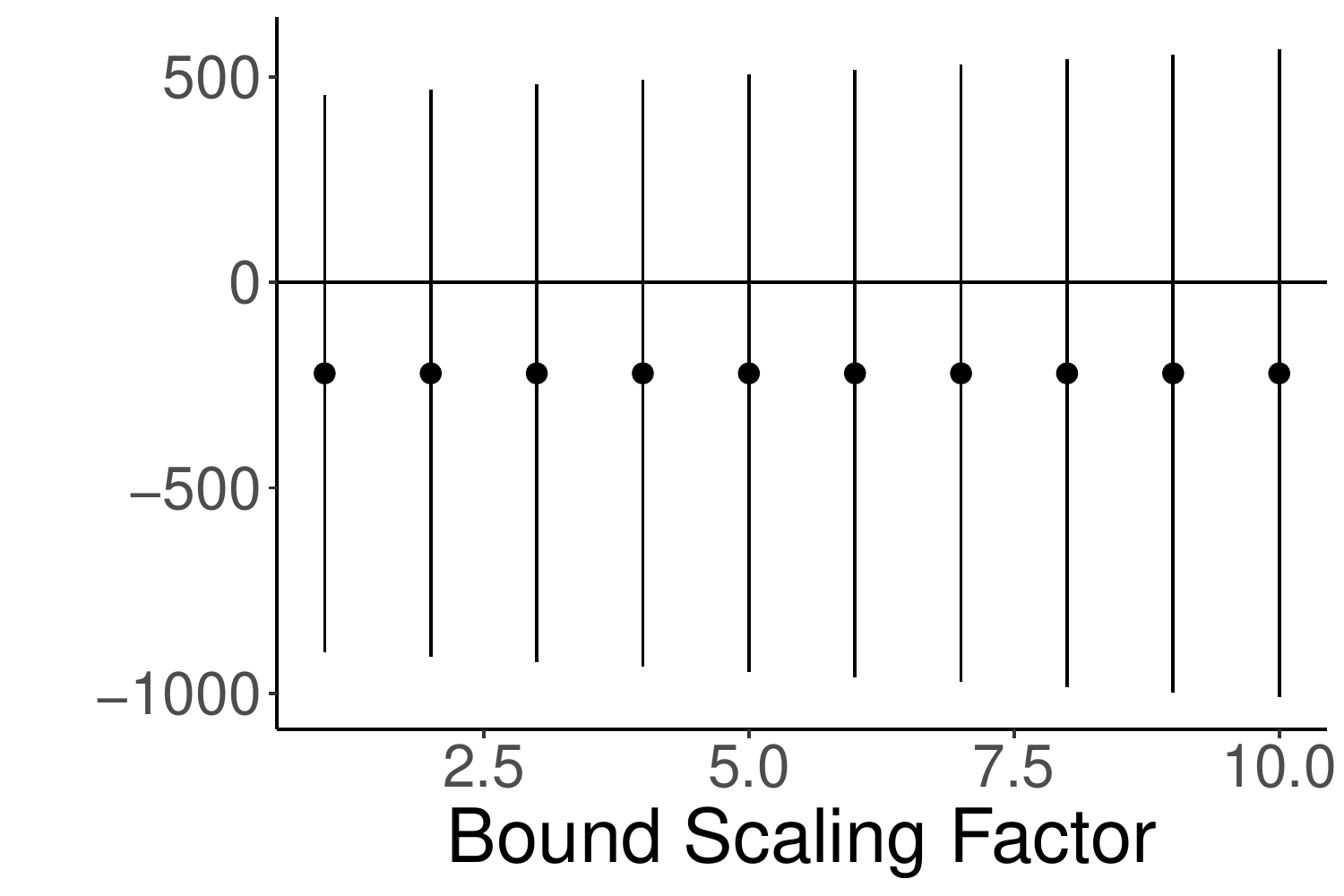} &   \includegraphics[width=4.5cm]{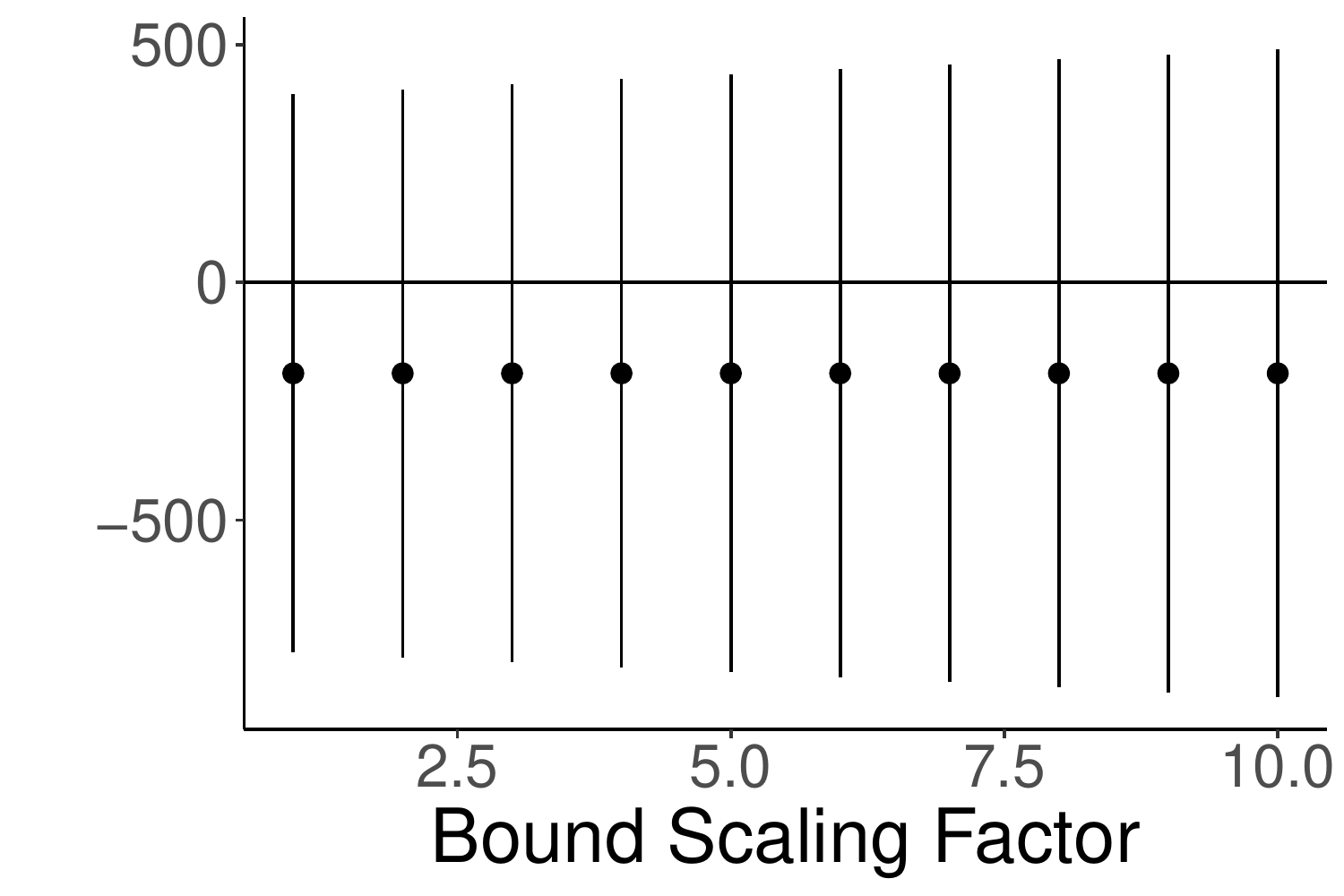} \\ [.5cm]
 \textit{Panel G:} CC debt past due &  \textit{Panel H:} Share debt past due &  \textit{Panel I:} Foreclosure \\
 \includegraphics[width=5cm]{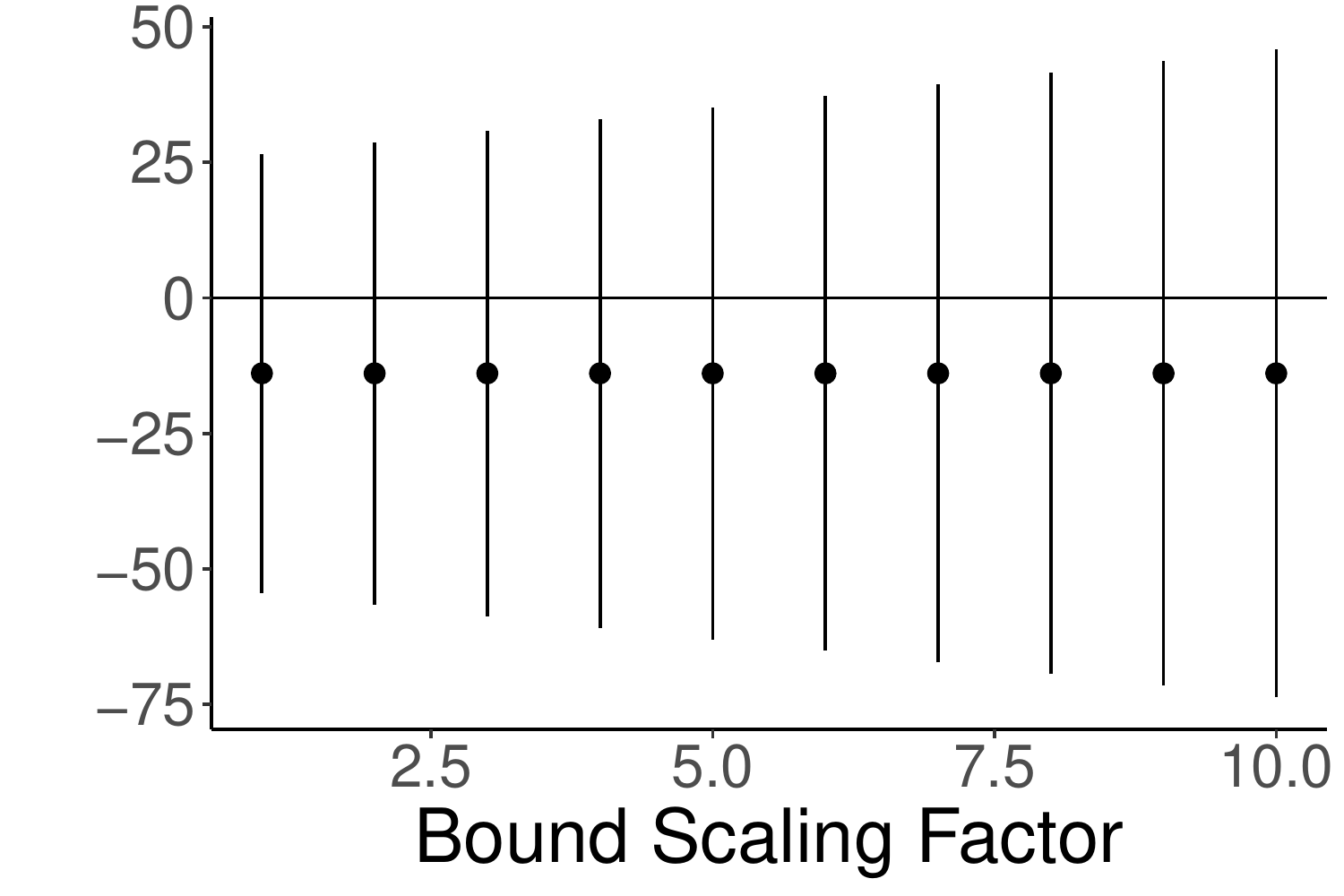} &   \includegraphics[width=5cm]{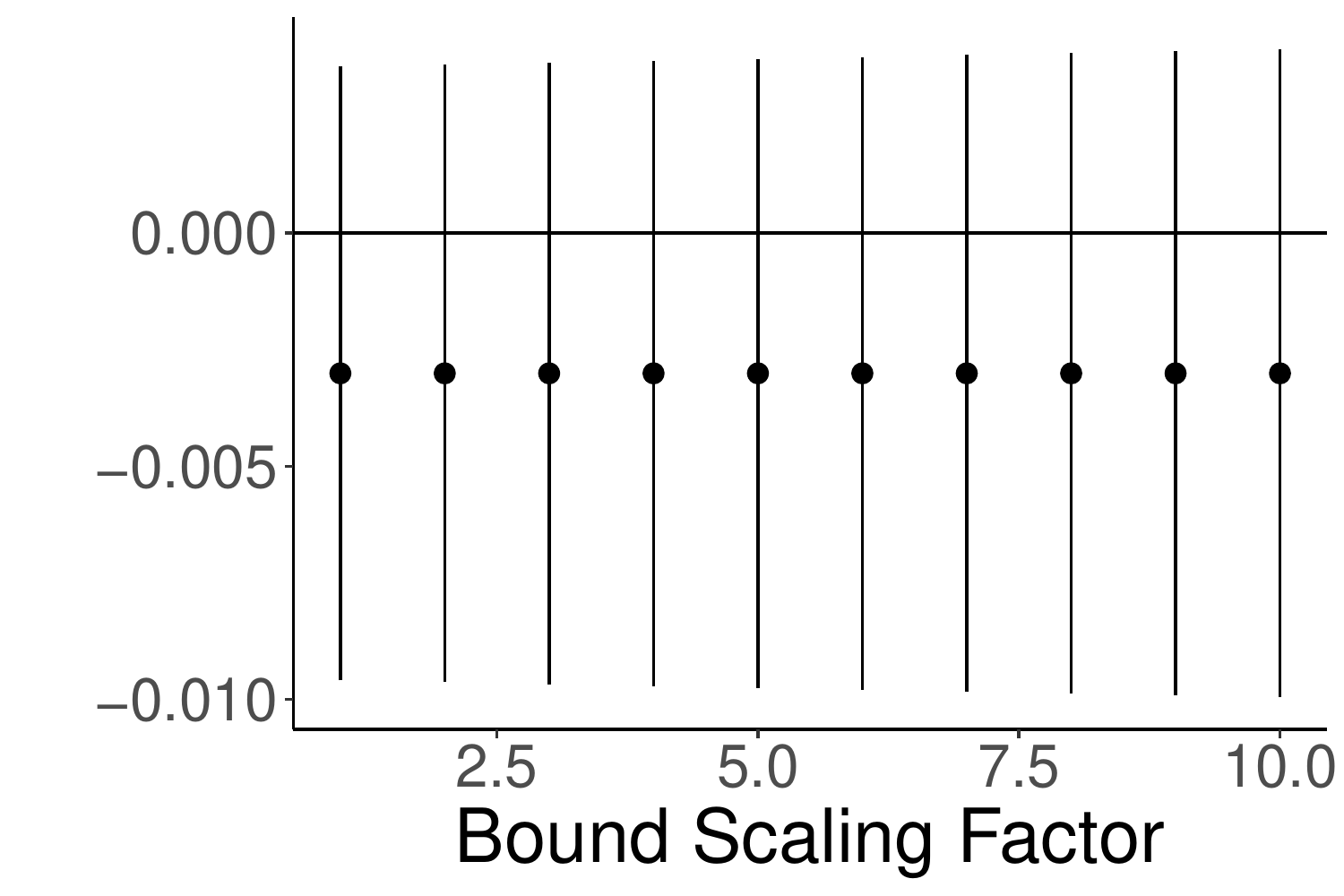} &   \includegraphics[width=5cm]{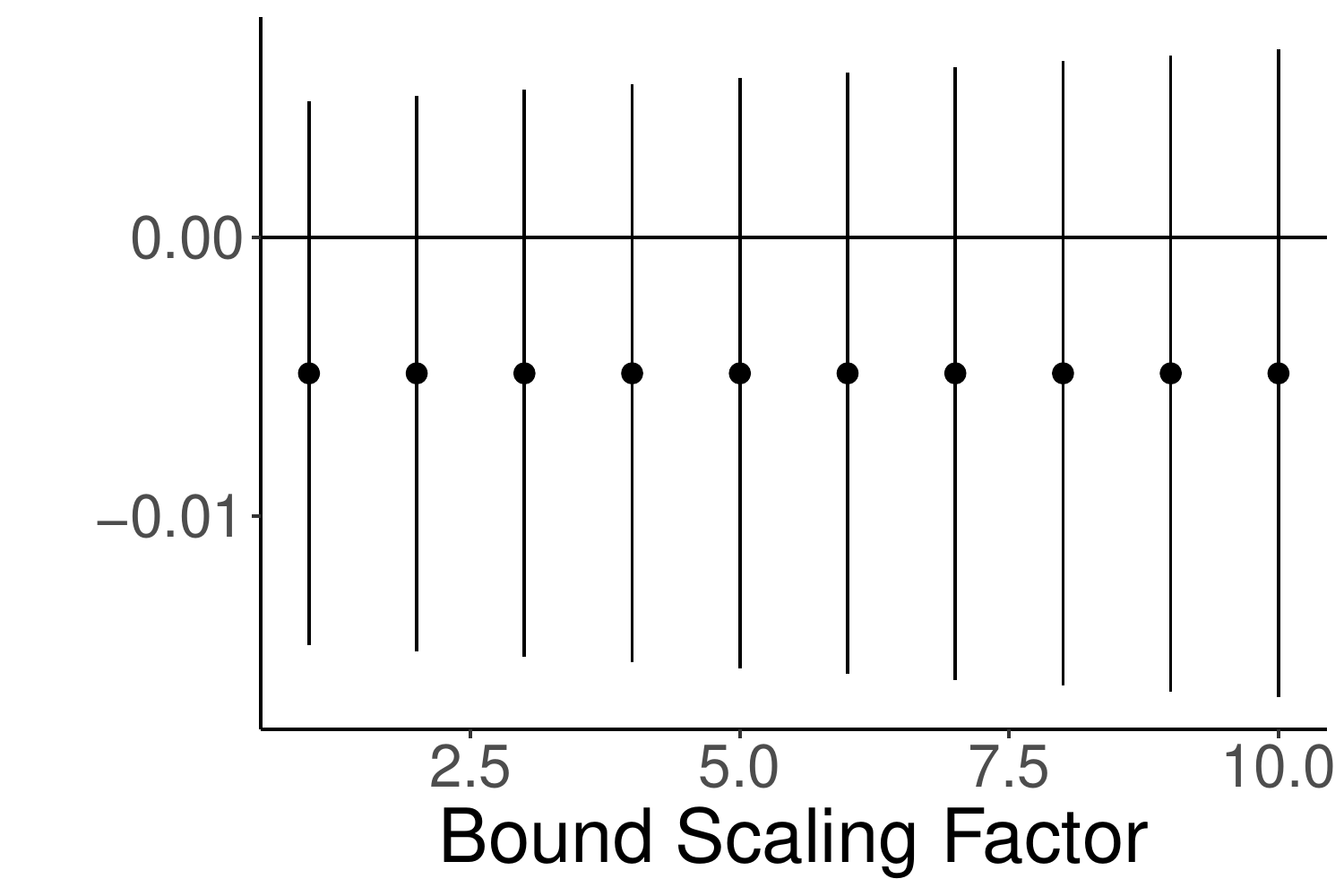} 
\end{tabular}
	\begin{minipage} {0.9\textwidth} \setstretch{.9} \medskip
          \footnotesize{\textbf{Note:} This figure plots the robustness of our regression discontinuity estimates to the choice of the bound scaling factor used in the \cite{kolesar2018inference} estimation procedure. Panel A plots the robustness of the share of the population with any coverage estimates. Panel B plots the robustness of the average debt in collections in dollars RD estimates. Panel C plots the robustness of the risk score RD estimates based on the Equifax Riskscore 3.0. Panel D plots the robustness of the bankruptcy RD estimates. Panel E plots the robustness of the average debt past due RD estimates. Panel F plots the robustness of the average mortgage debt past due RD estimates. Panel G plots the robustness of the average credit card debt past due RD estimates. Panel H plots the robustness of the share of debt past due RD estimates. Panel I plots the robustness of the foreclosure RD estimates. The sample includes individuals who were age 55-75 between 2008 and 2017. See Section \ref{background_data} for additional details on the outcomes and sample. Source: The financial health outcomes are based on 137,340,577 person-year observations from the New York Fed Consumer Credit Panel / Equifax, 2008-2017.}
	\end{minipage}
\end{figure}

\begin{figure}[htpb!]
  \centering
  \caption{Robustness of Age RD Estimates to Bandwidth Selection}
  \label{fig:age_bandwidth_robustness_apx}
\begin{tabular}{ccc}
\textit{Panel A:} Share with coverage &    \textit{Panel B:} Total Collections (\$) &  \textit{Panel C:} Risk Score \\
 \includegraphics[width=4.5cm]{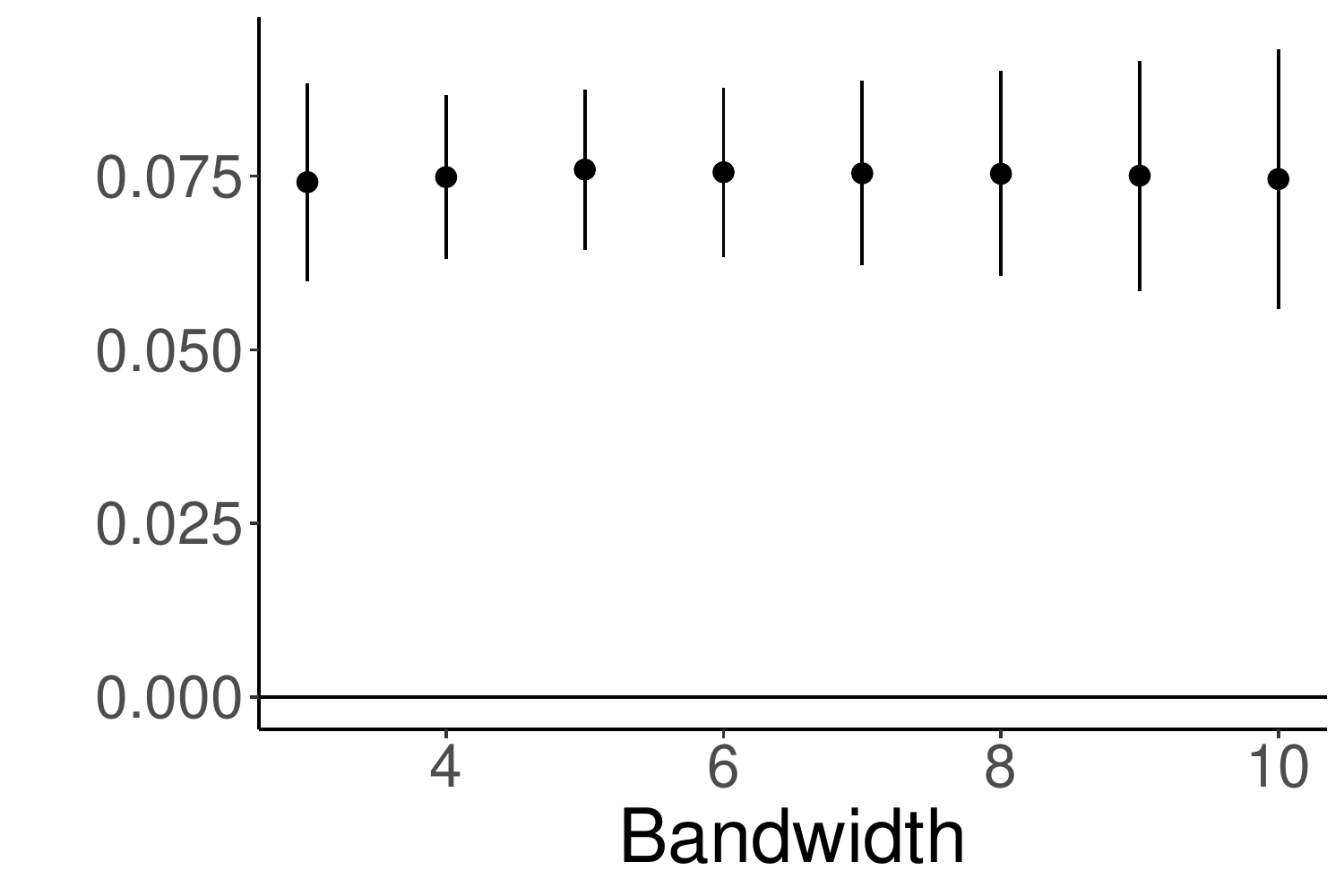} &   \includegraphics[width=4.5cm]{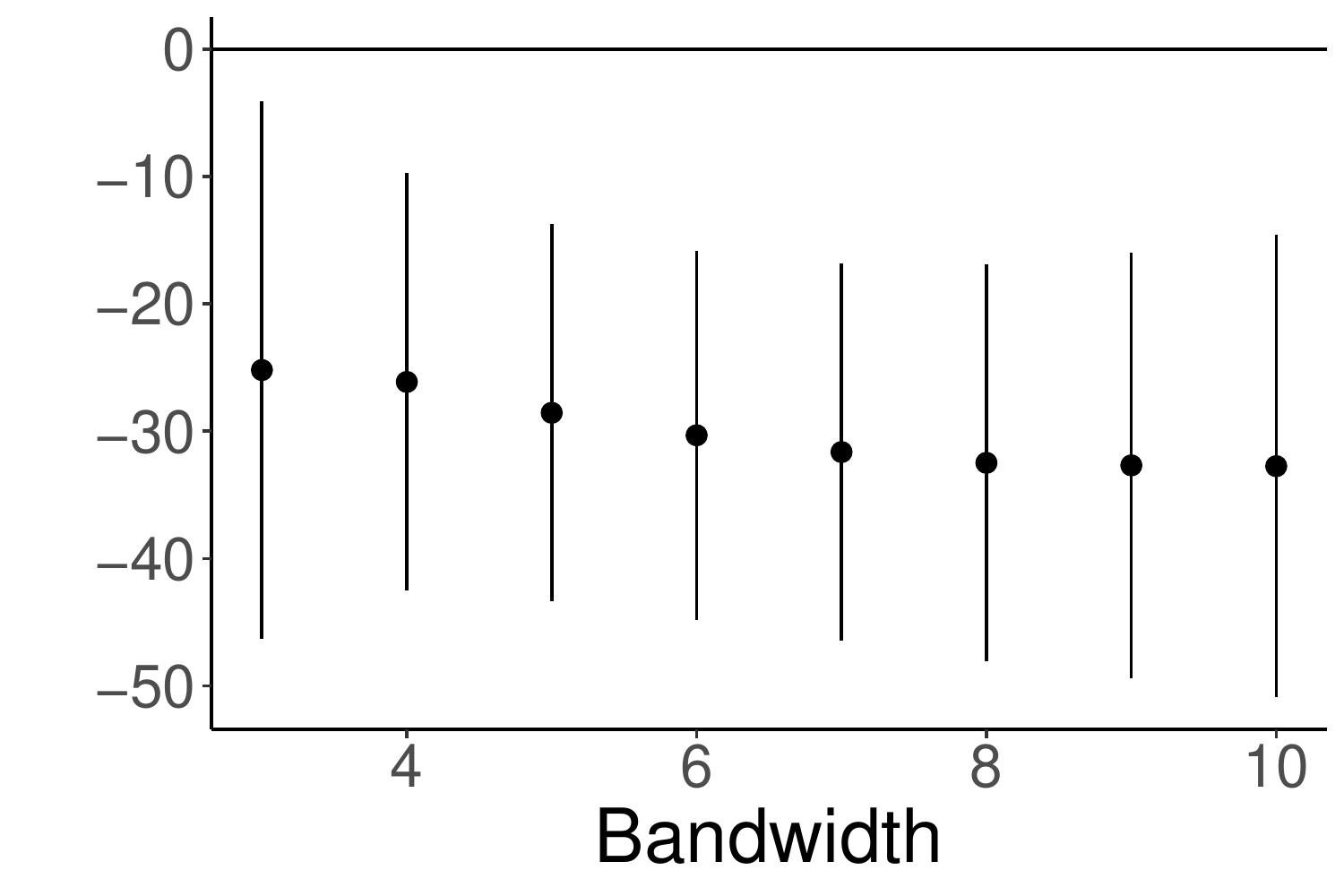} &   \includegraphics[width=4.5cm]{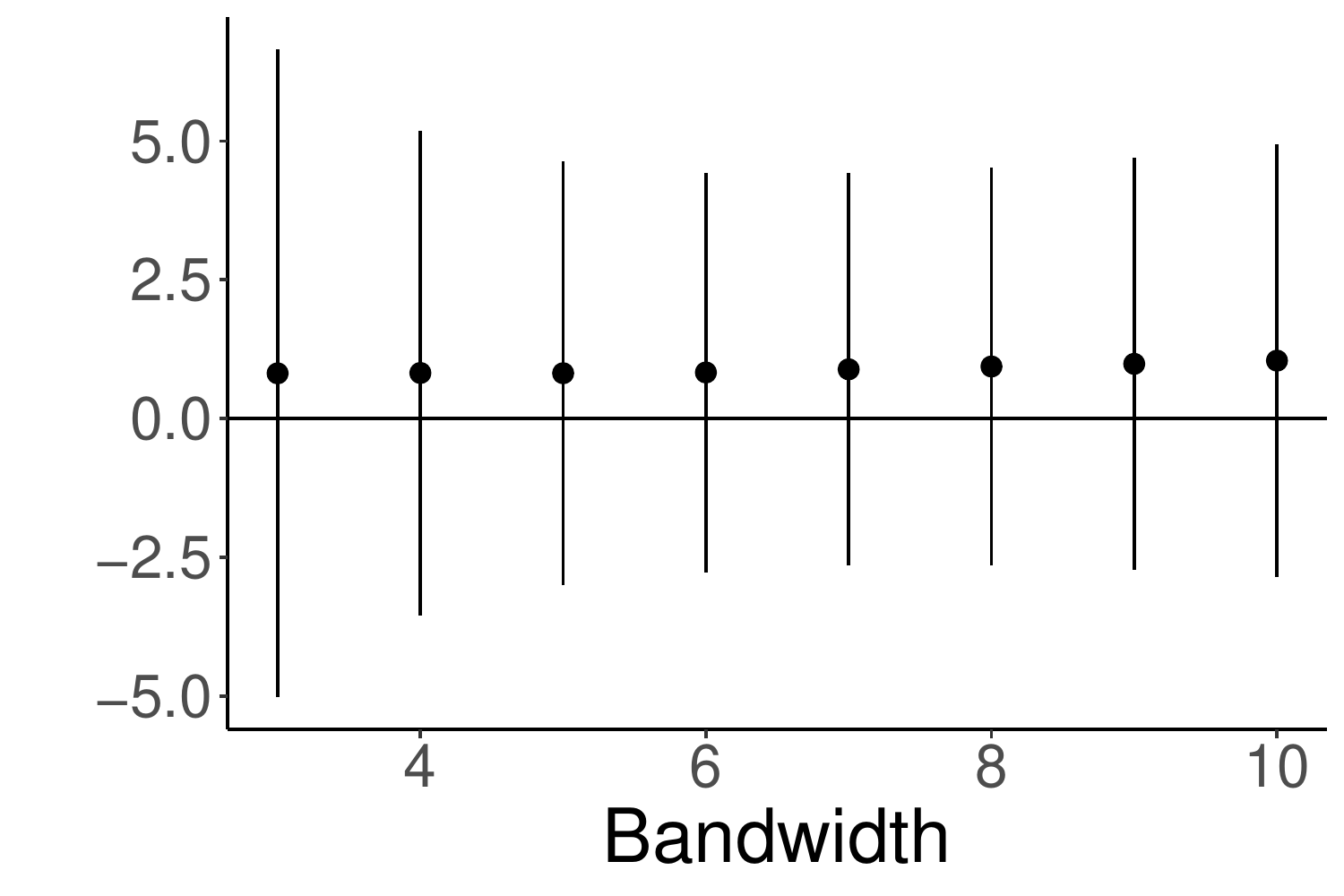} \\ [.5cm]
\textit{Panel D:} Bankruptcy &  \textit{Panel E:} Total debt past due &  \textit{Panel F:} Mortgage debt past due \\
 \includegraphics[width=4.5cm]{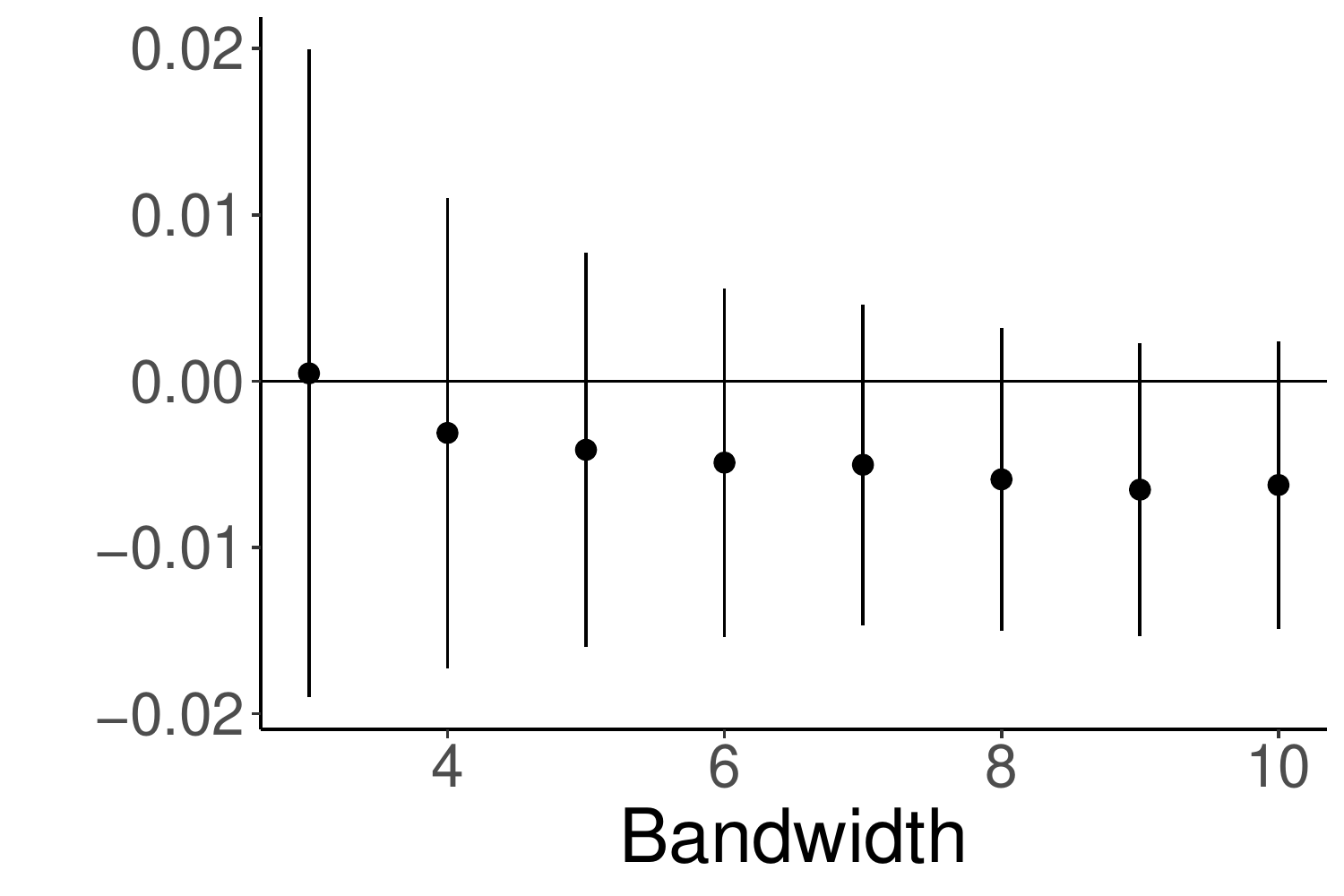} &   \includegraphics[width=4.5cm]{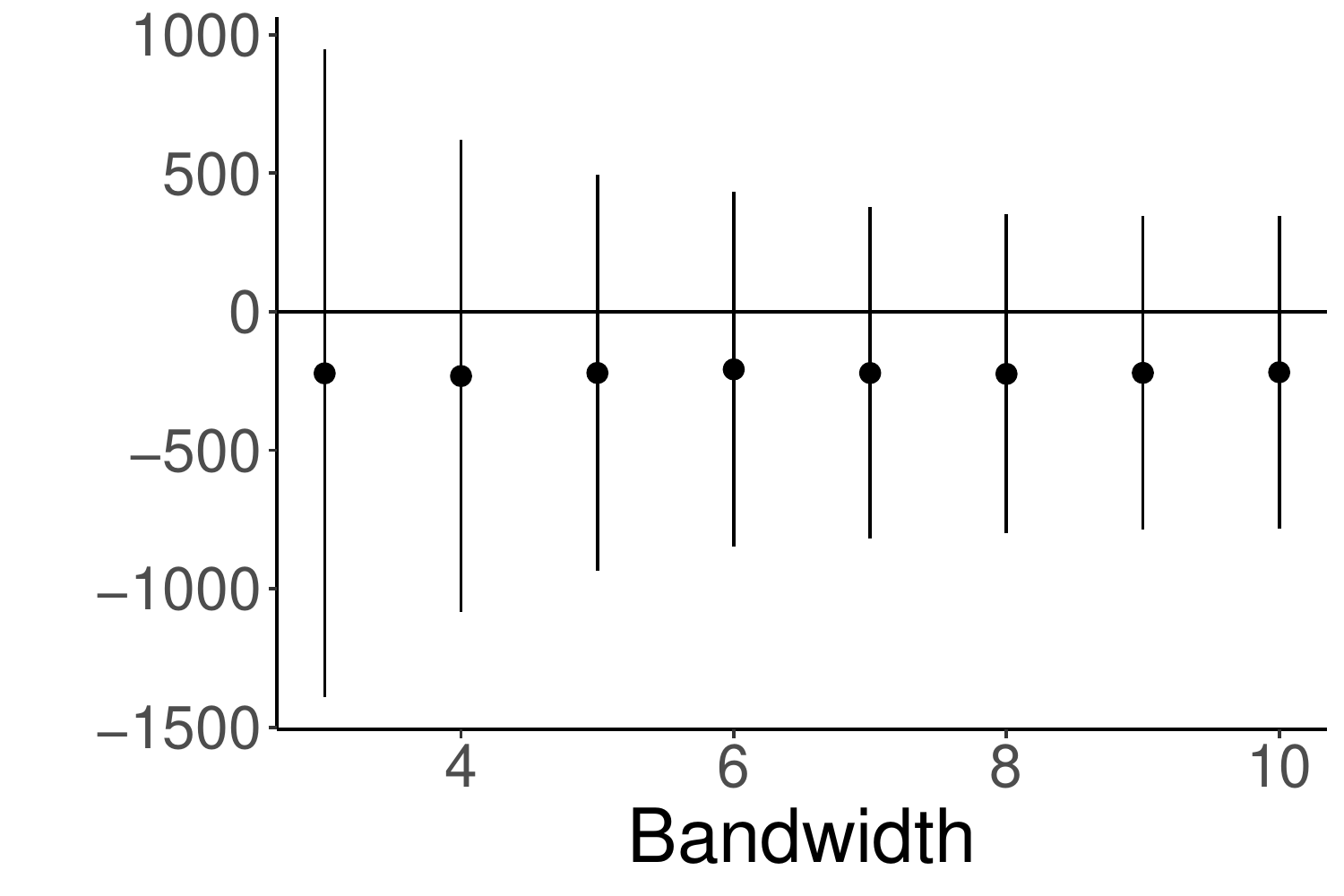} &   \includegraphics[width=4.5cm]{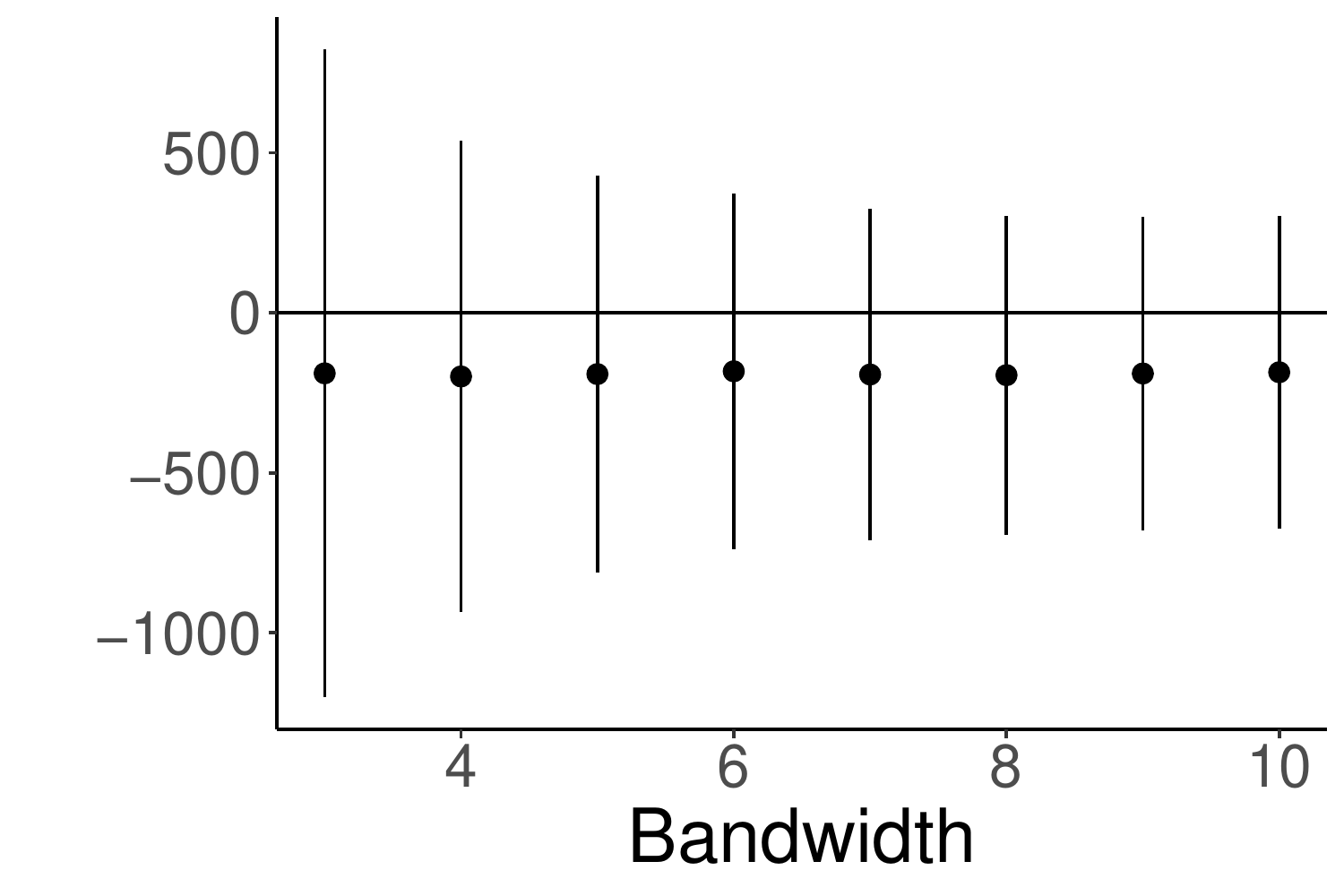} \\ [.5cm]
 \textit{Panel G:} CC debt past due &  \textit{Panel H:} Share debt past due &  \textit{Panel I:} Foreclosure \\
 \includegraphics[width=5cm]{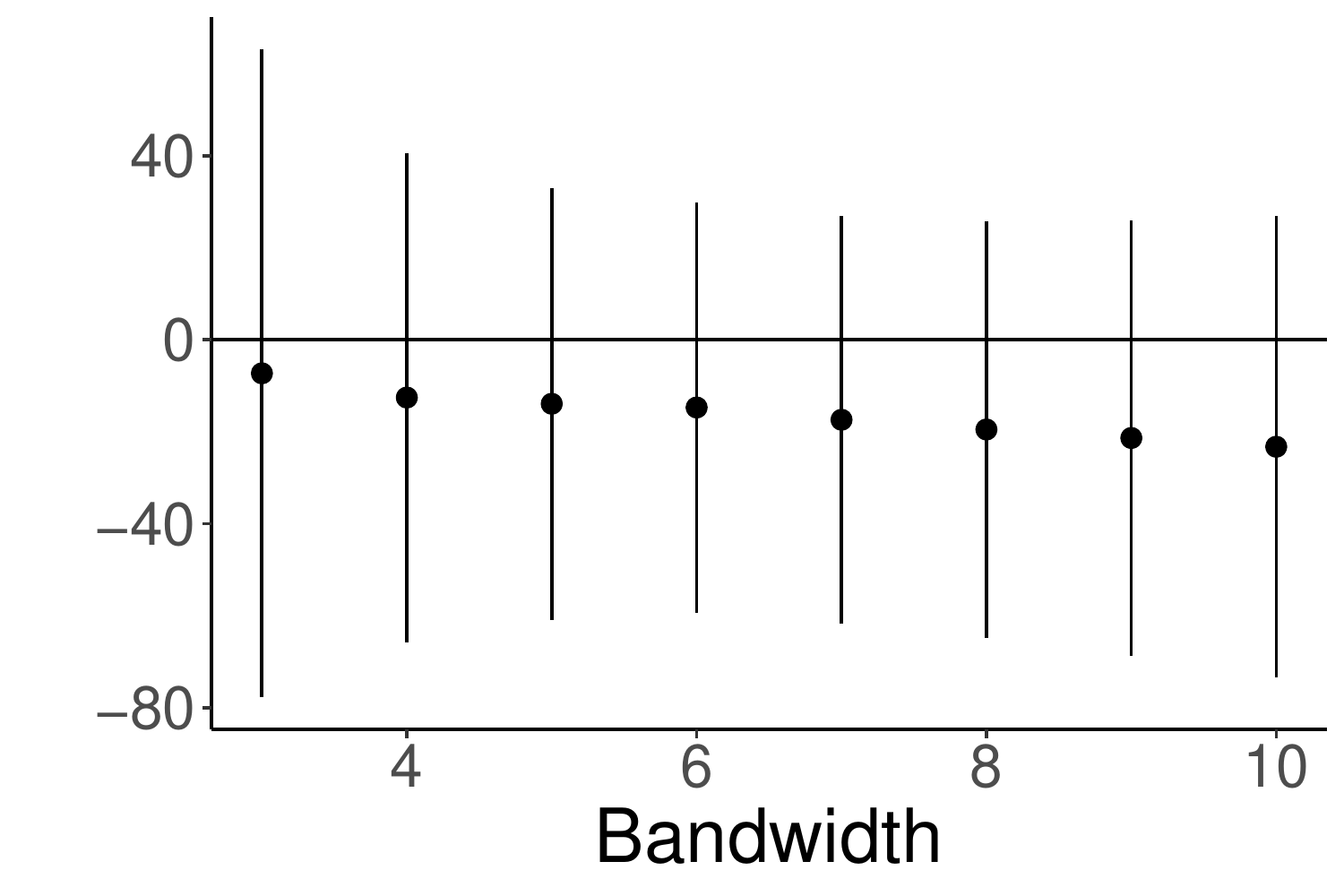} &   \includegraphics[width=5cm]{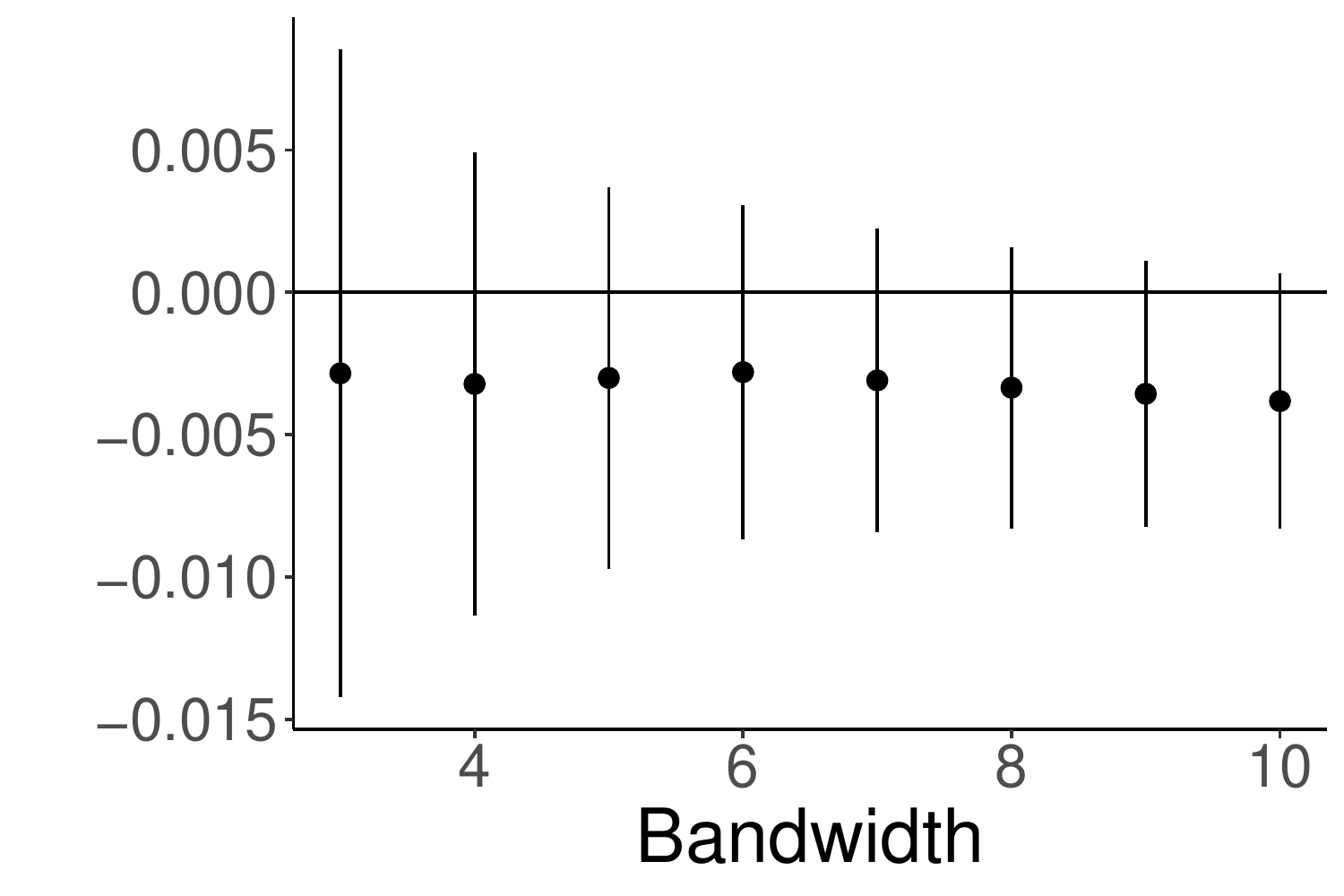} &   \includegraphics[width=5cm]{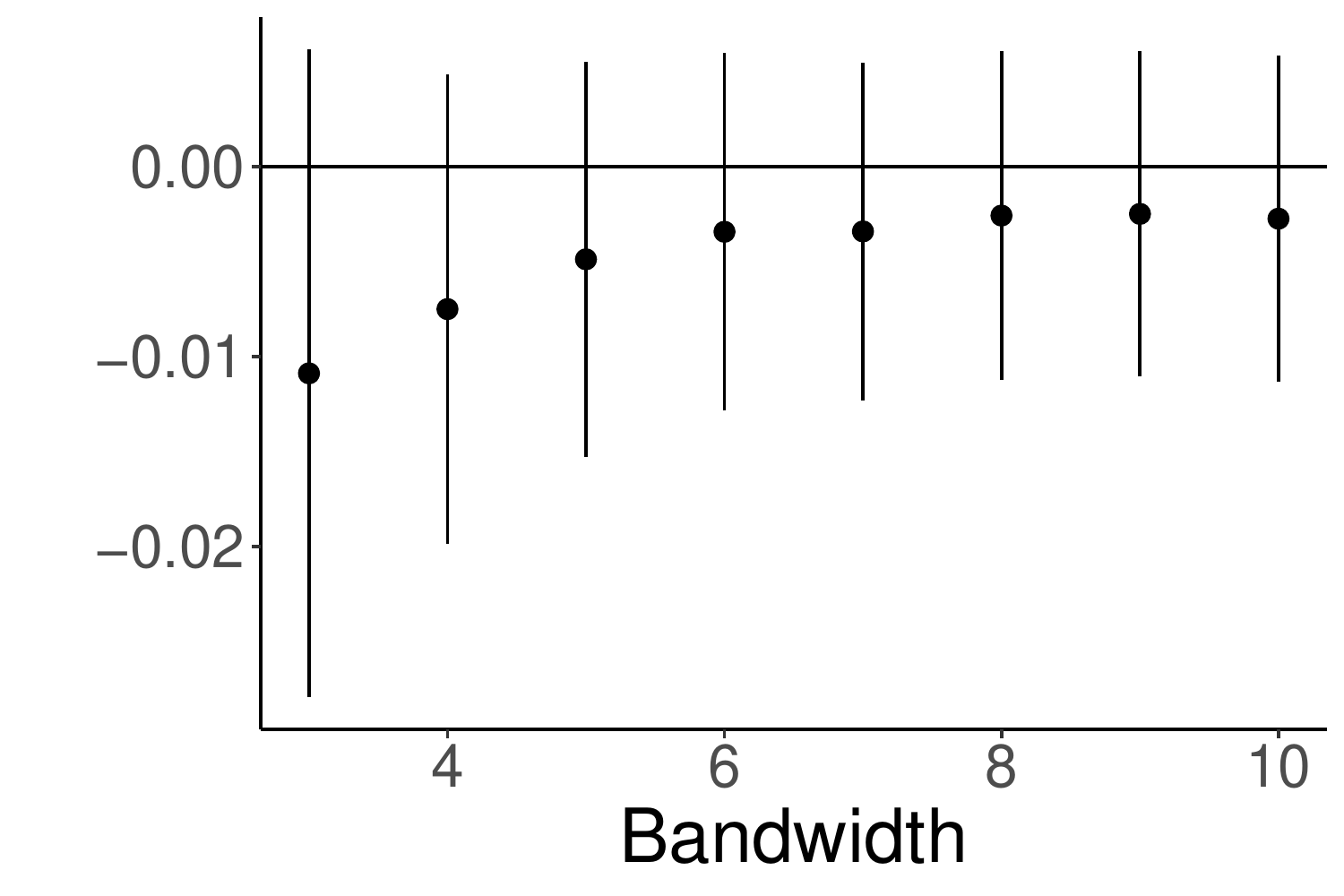} 
\end{tabular}
	\begin{minipage} {0.9\textwidth} \setstretch{.9} \medskip
          \footnotesize{\textbf{Note:} This figure plots the robustness of our regression discontinuity estimates to the bandwidth selection used in the \cite{kolesar2018inference} estimation procedure. Panel A plots the robustness of the share of the population with any coverage estimates. Panel B plots the robustness of the average debt in collections in dollars RD estimates. Panel C plots the robustness of the risk score RD estimates based on the Equifax Riskscore 3.0. Panel D plots the robustness of the bankruptcy RD estimates. Panel E plots the robustness of the average debt past due RD estimates. Panel F plots the robustness of the average mortgage debt past due RD estimates. Panel G plots the robustness of the average credit card debt past due RD estimates. Panel H plots the robustness of the share of debt past due RD estimates. Panel I plots the robustness of the foreclosure RD estimates. The sample includes individuals who were age 55-75 between 2008 and 2017. The regressions include 26,120,830 person-year-quarter observations for 2,977,952 unique individuals. See Section \ref{background_data} for additional details on the outcomes and sample. Source: New York Fed Consumer Credit Panel / Equifax.}
	\end{minipage}
\end{figure}

\begin{figure}[htpb!]
  \centering
  \caption{Robustness of Variance Reduction Estimates to Bound Scaling Factor}
  \label{fig:bound_scaling_var_robustness_apx}
\begin{tabular}{ccc}
 \textit{Panel A:} Share with coverage &    \textit{Panel B:} Total Collections &  \textit{Panel C:} Risk Score \\
 \includegraphics[width=4.5cm]{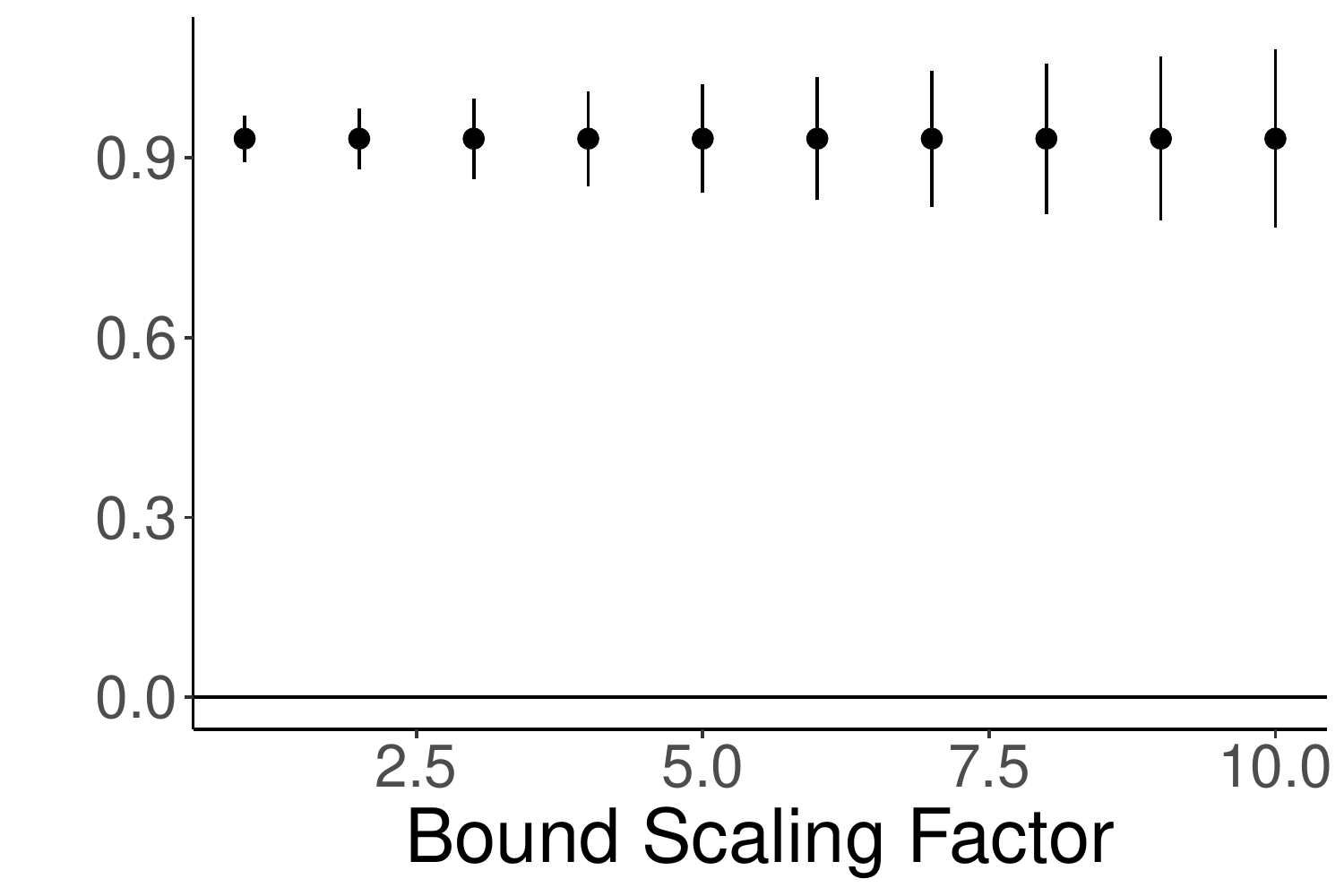} &   \includegraphics[width=4.5cm]{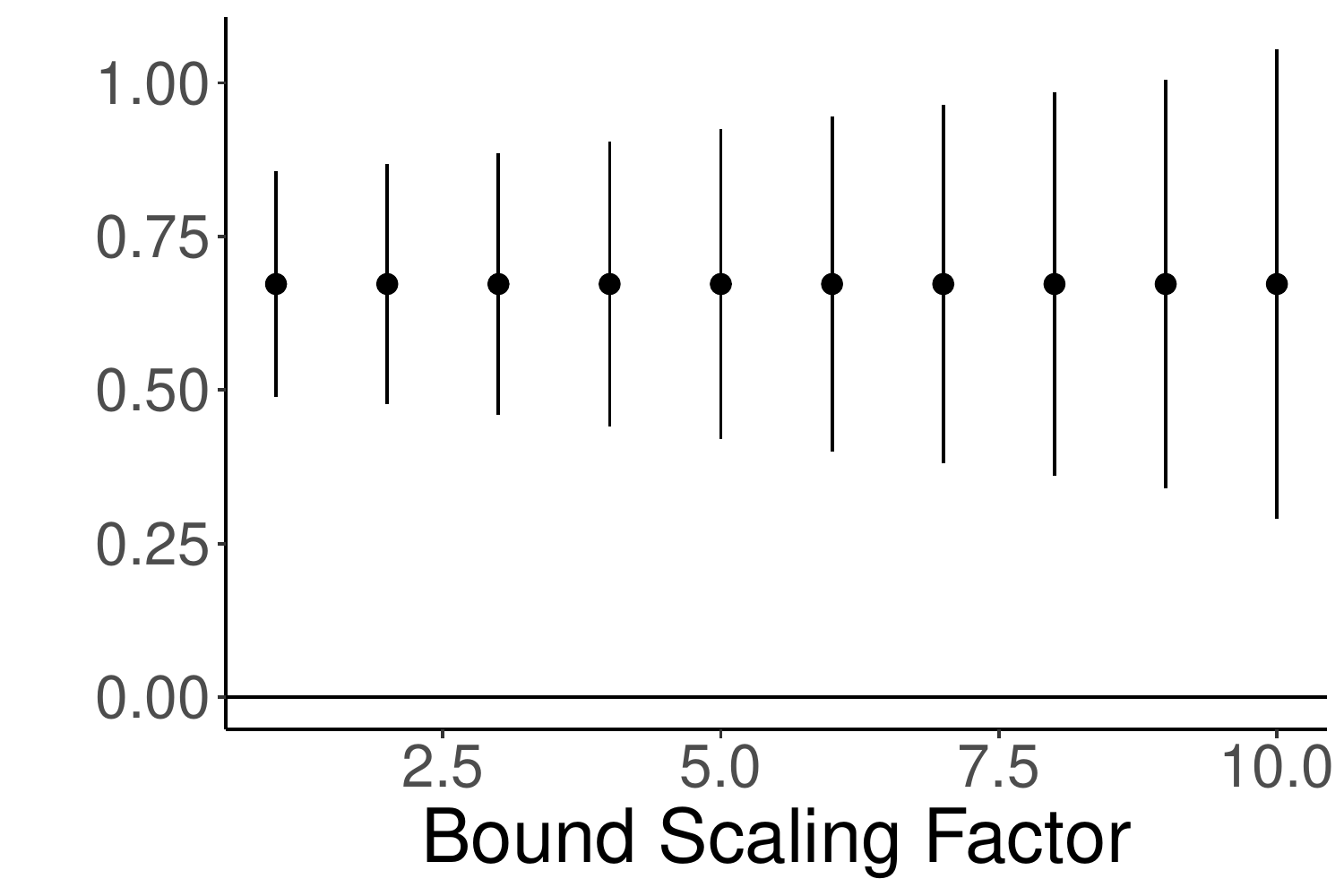} &   \includegraphics[width=4.5cm]{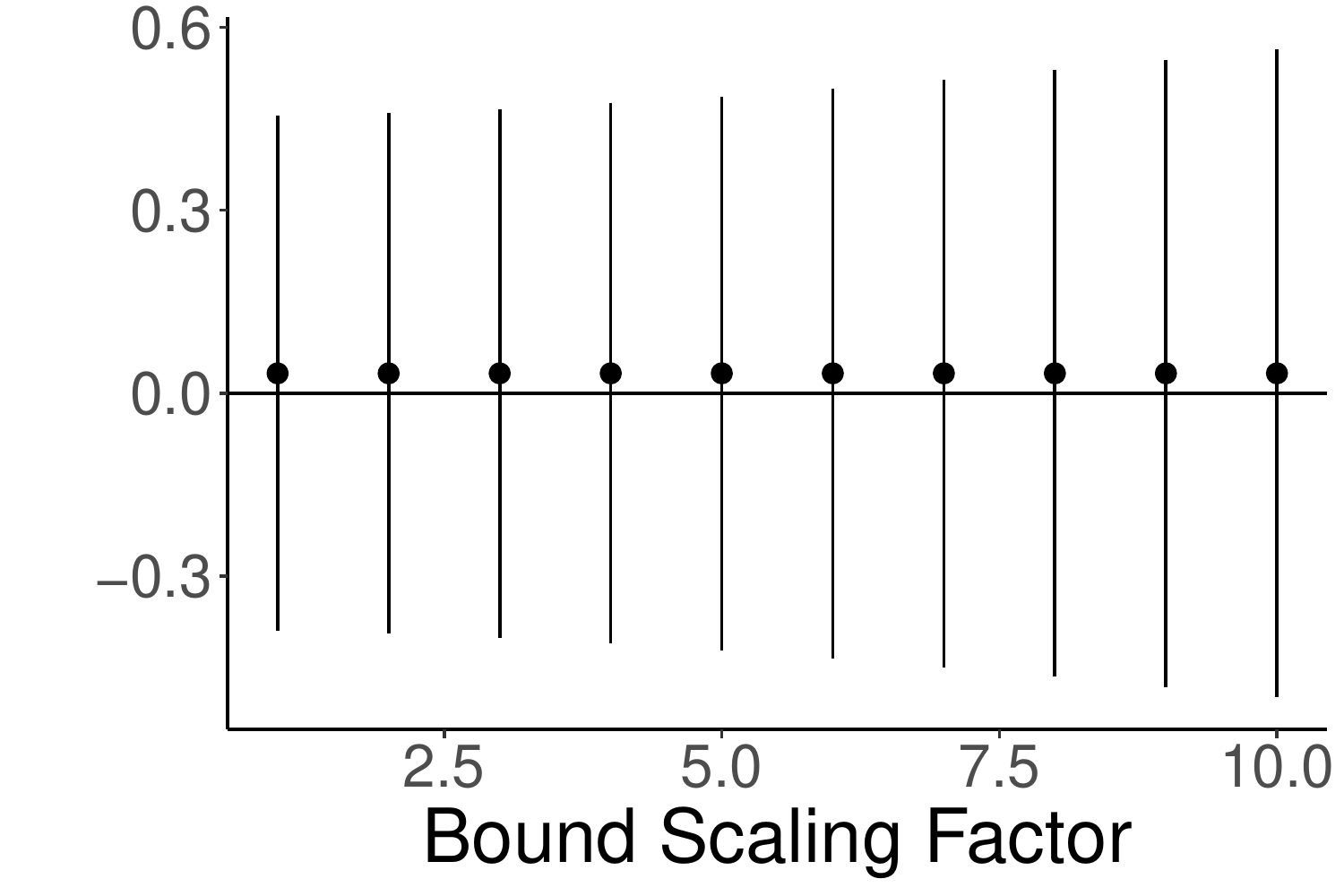} \\ [.5cm]
\textit{Panel D:} Bankruptcy &  \textit{Panel E:} Total debt past due &  \textit{Panel F:} Mortgage debt past due \\
 \includegraphics[width=4.5cm]{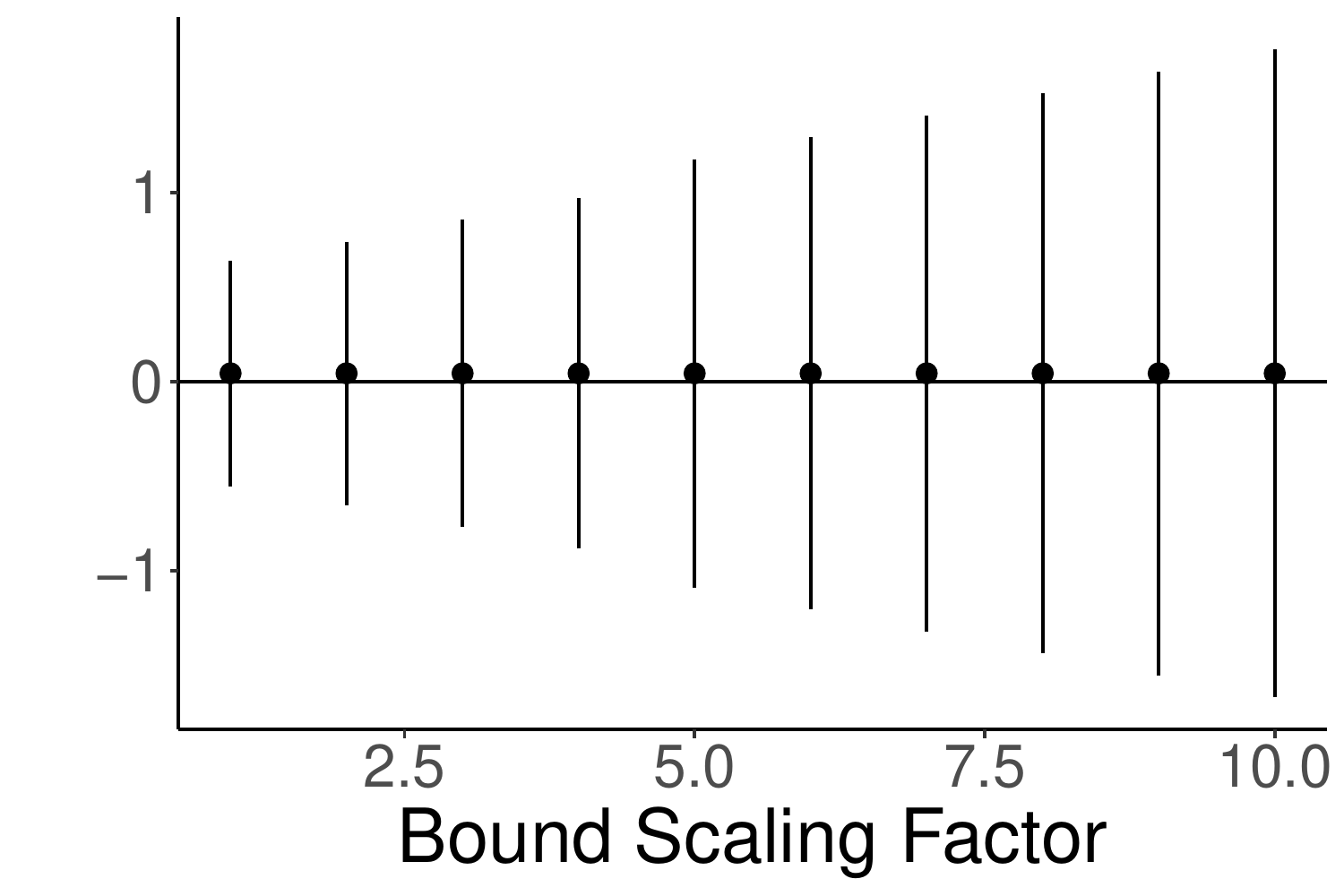} &   \includegraphics[width=4.5cm]{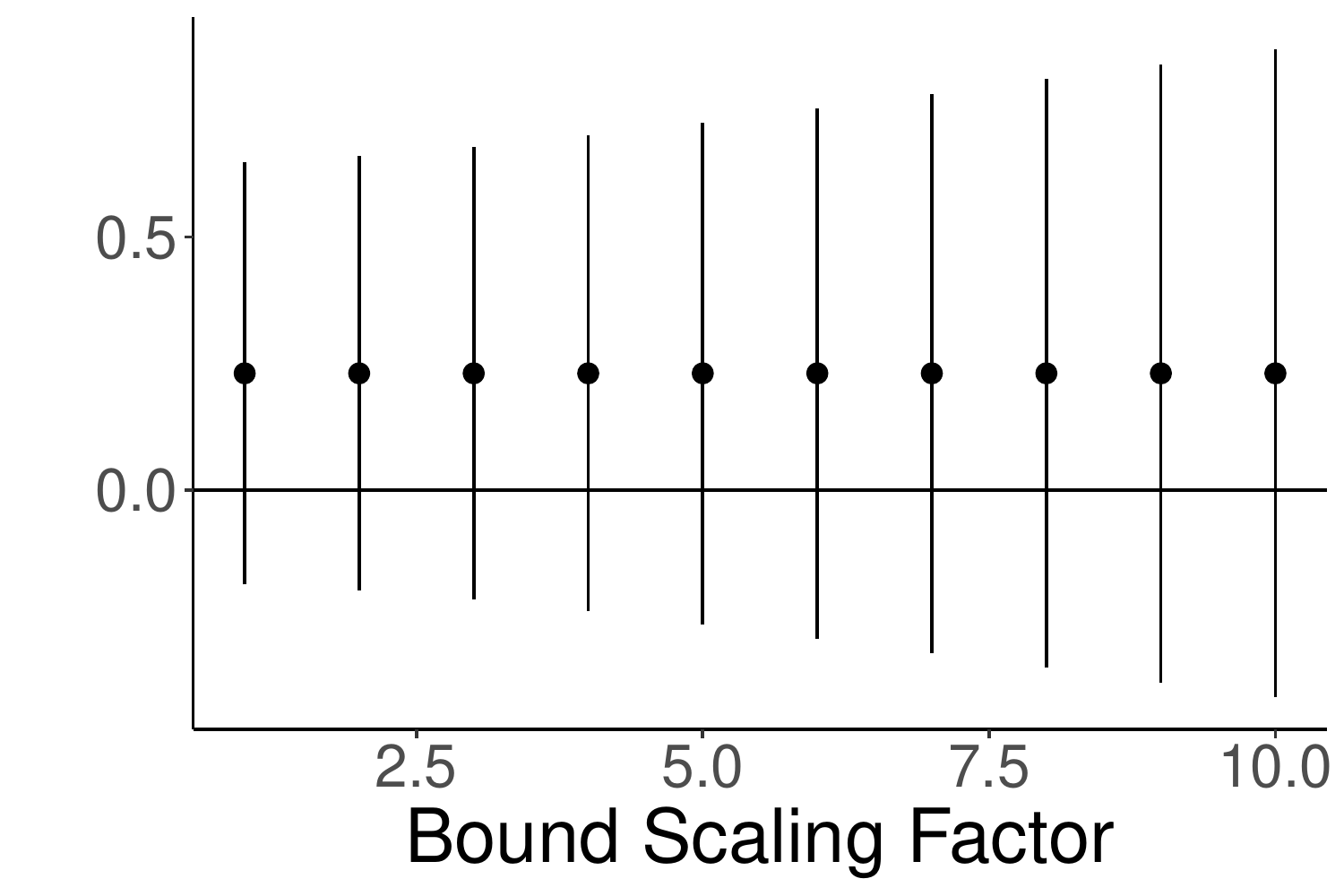} &   \includegraphics[width=4.5cm]{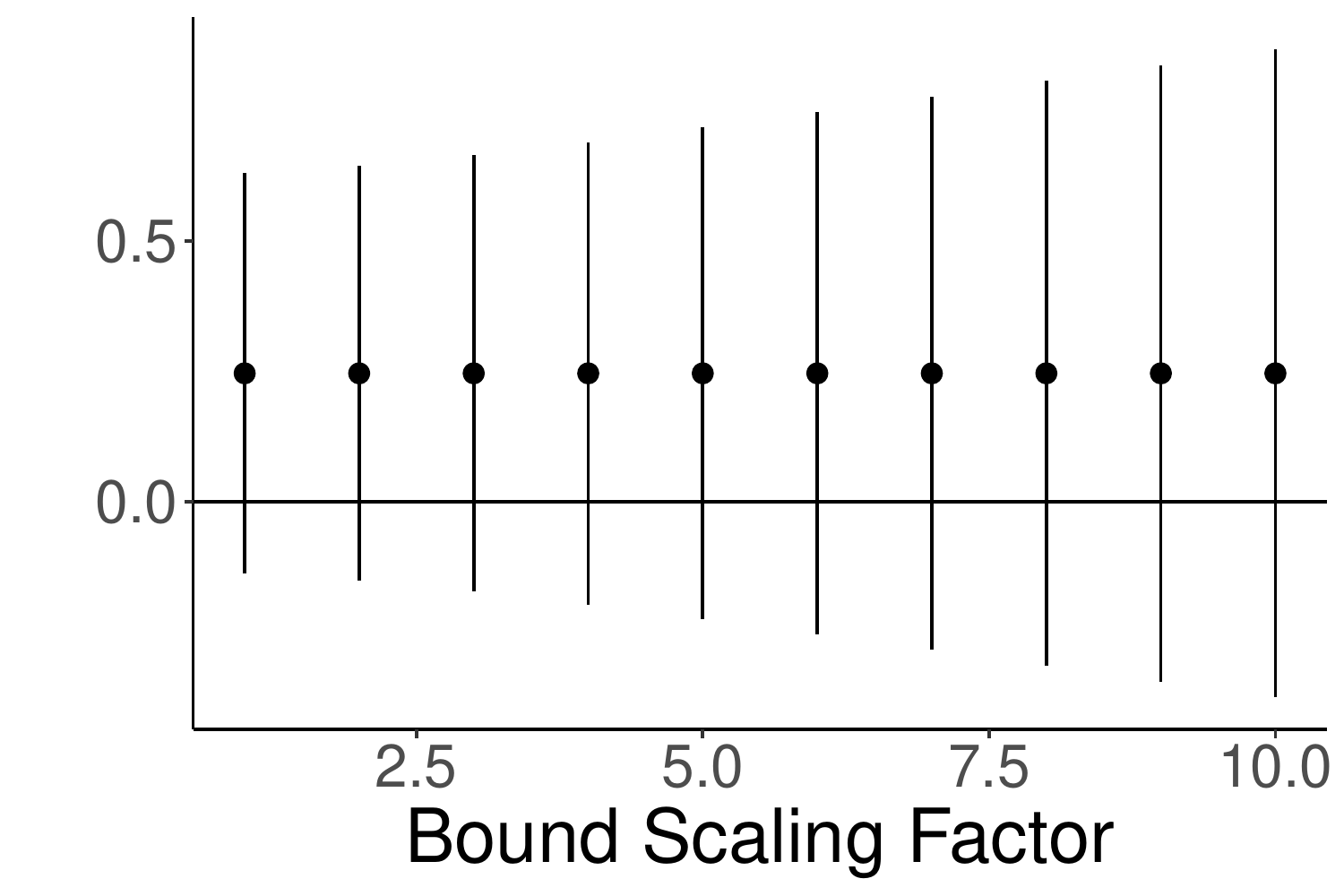} \\ [.5cm]
 \textit{Panel G:} CC debt past due &  \textit{Panel H:} Share debt past due &  \textit{Panel I:} Foreclosure \\
 \includegraphics[width=5cm]{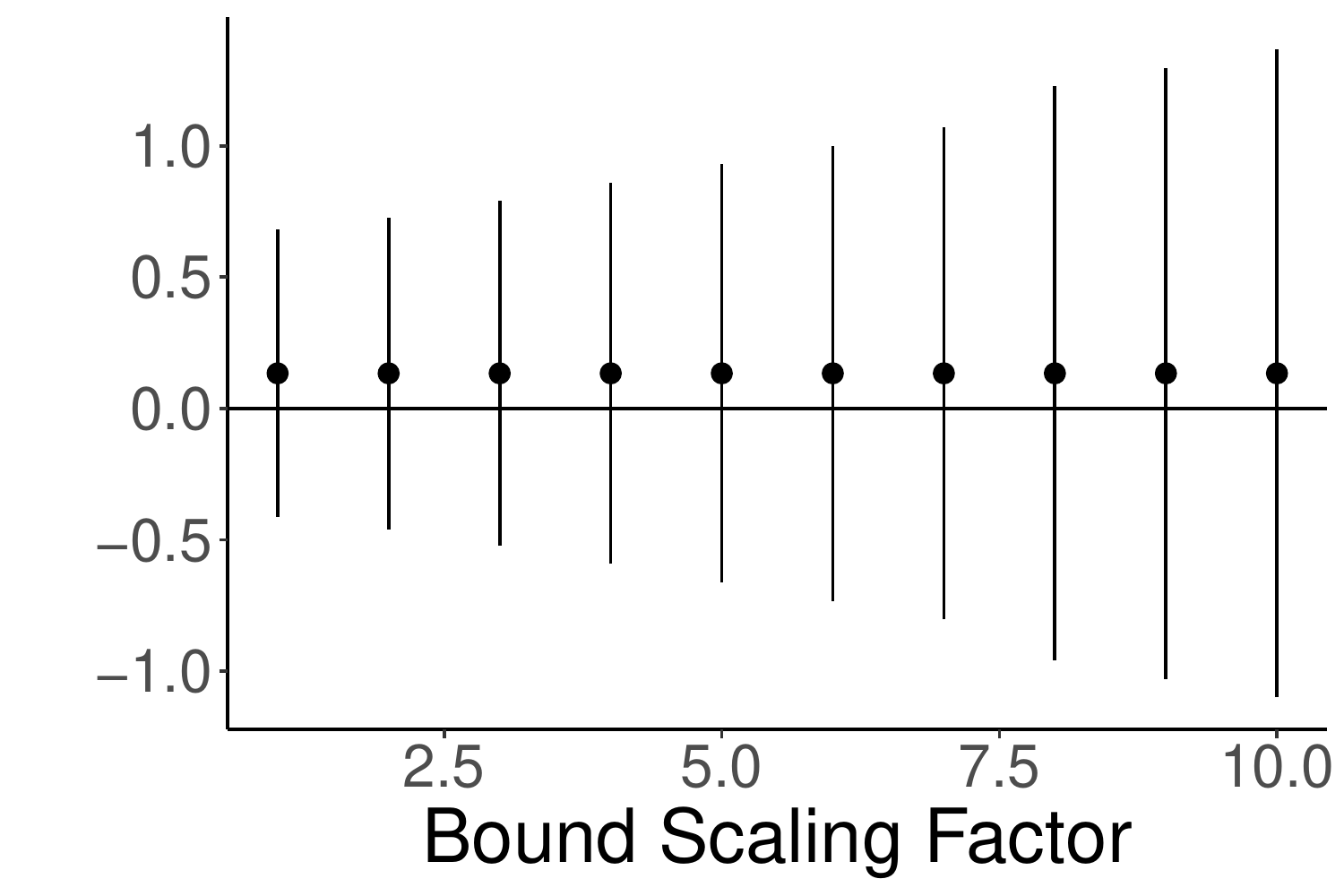} &   \includegraphics[width=5cm]{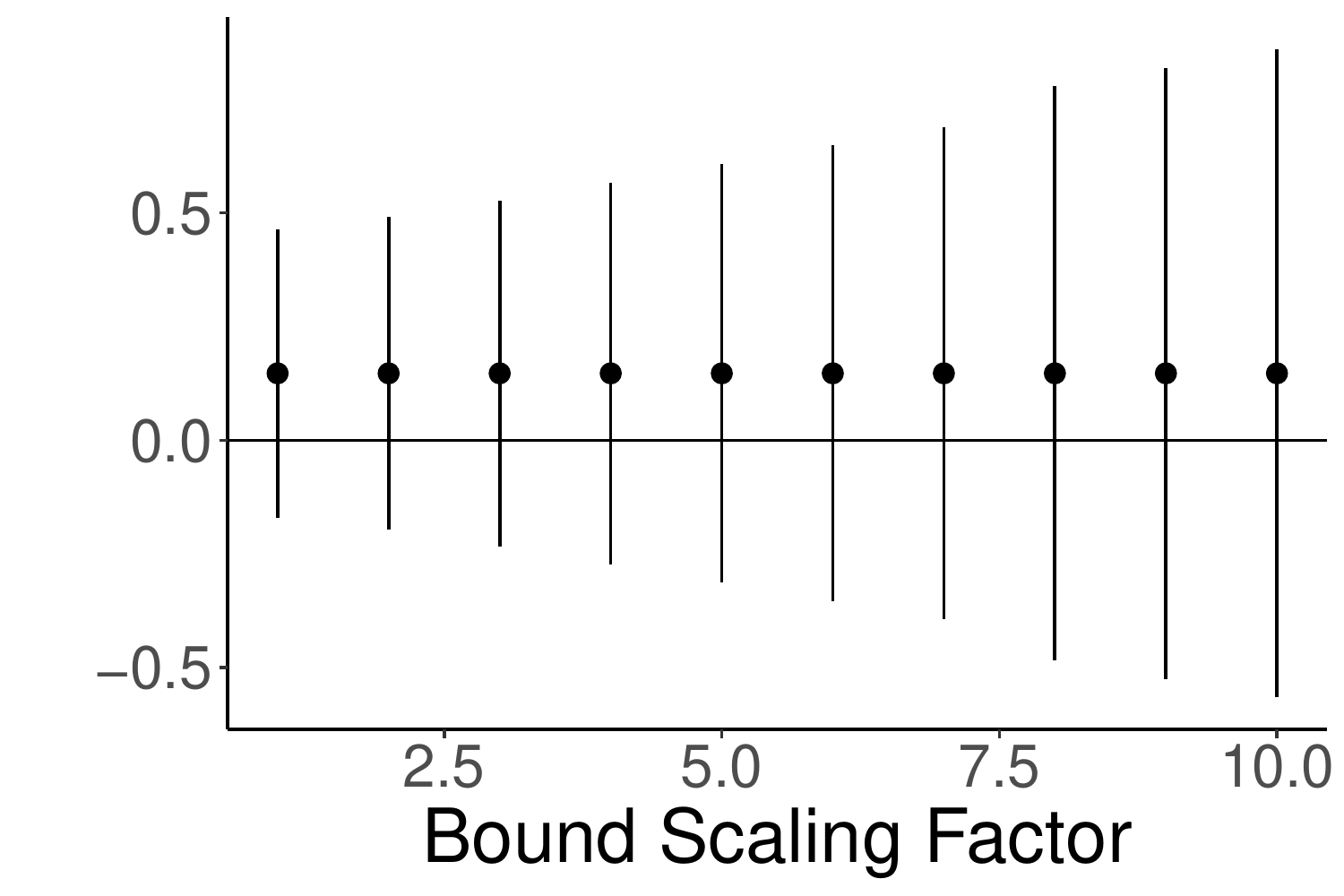} &   \includegraphics[width=5cm]{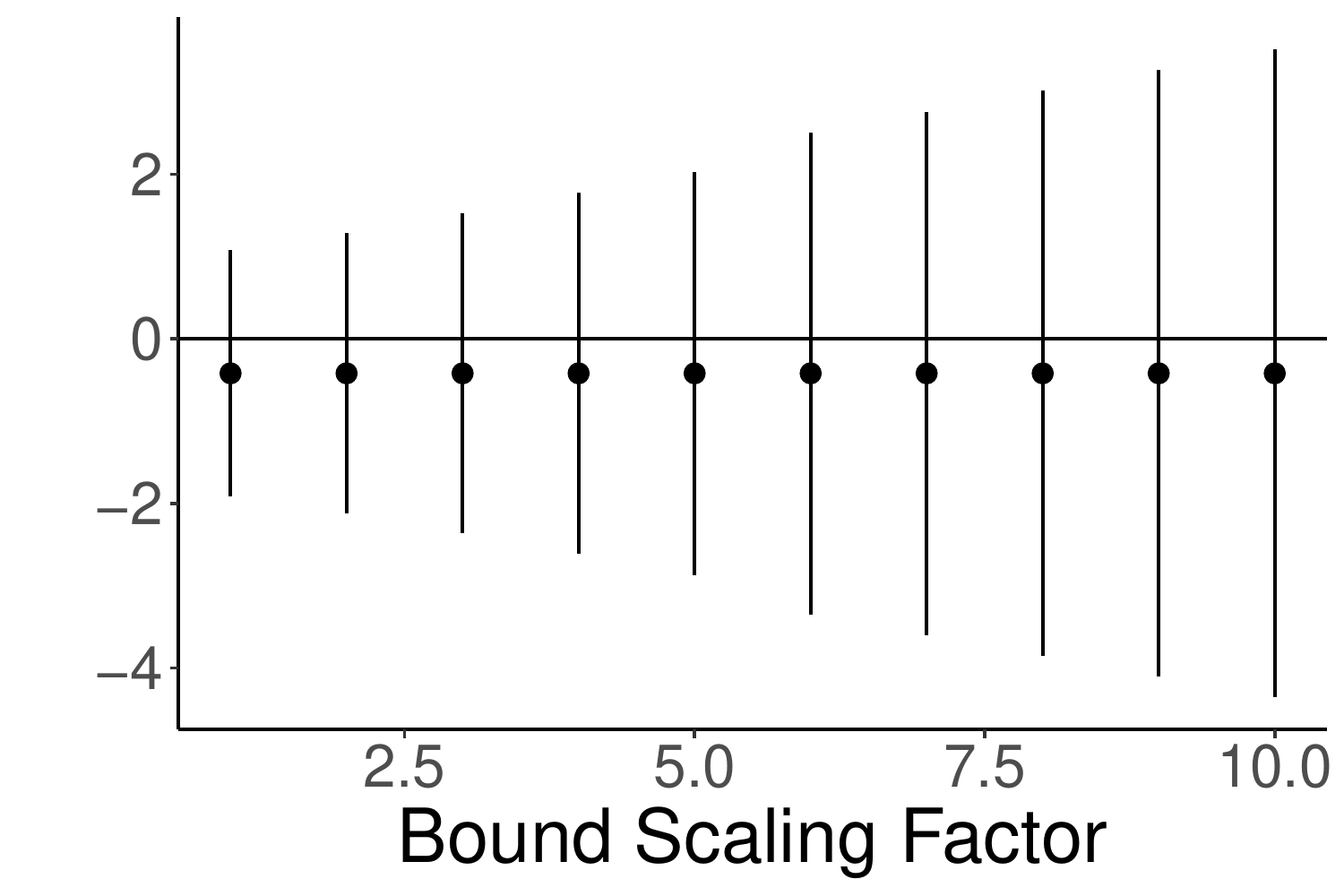}
\end{tabular}
	\begin{minipage} {0.9\textwidth} \setstretch{.9} \medskip
          \footnotesize{\textbf{Note:} This figure plots the robustness of our regression discontinuity estimates to the choice of the bound scaling factor used in the \cite{kolesar2018inference} estimation procedure. Panel A plots the robustness of the share of the population with any coverage estimates. Panel B plots the robustness of the average debt in collections in dollars RD estimates. Panel C plots the robustness of the risk score RD estimates based on the Equifax Riskscore 3.0. Panel D plots the robustness of the bankruptcy RD estimates. Panel E plots the robustness of the average debt past due RD estimates. Panel F plots the robustness of the average mortgage debt past due RD estimates. Panel G plots the robustness of the average credit card debt past due RD estimates. Panel H plots the robustness of the share of debt past due RD estimates. Panel I plots the robustness of the foreclosure RD estimates. The sample includes individuals who were age 55-75 between 2008 and 2017. See Section \ref{background_data} for additional details on the outcomes and sample. Source: The financial health outcomes are based on 137,340,577 person-year observations from the New York Fed Consumer Credit Panel / Equifax, 2008-2017.}
	\end{minipage}
\end{figure}

\begin{figure}[htpb!]
  \centering
  \caption{Robustness of Variance Reduction Estimate to Bandwidth Selection}
  \label{fig:age_bandwidth_var_robustness_apx}
\begin{tabular}{ccc}
  \textit{Panel A:} Share with coverage &    \textit{Panel B:} Total Collections &  \textit{Panel C:} Risk Score \\
 \includegraphics[width=4.5cm]{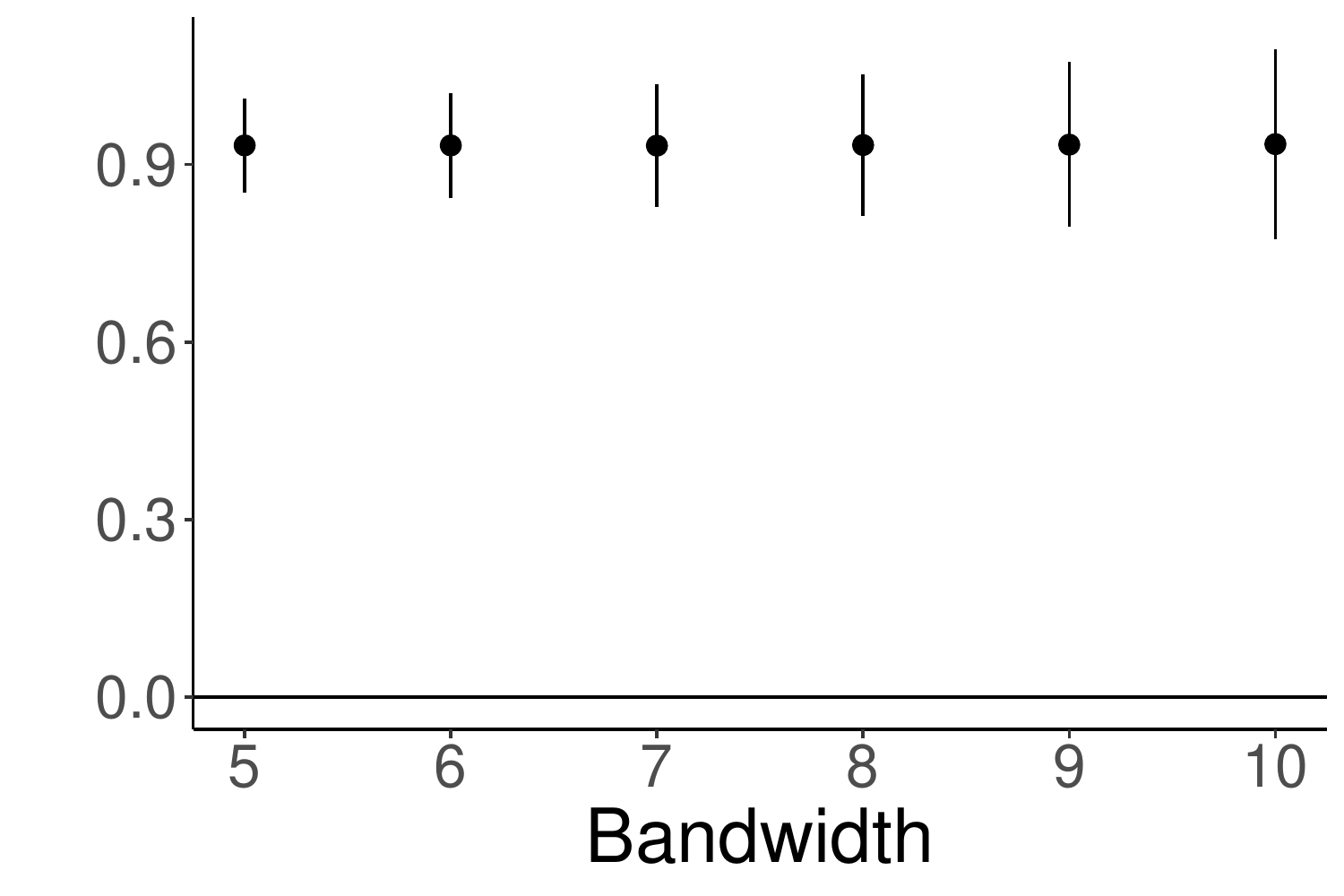} &   \includegraphics[width=4.5cm]{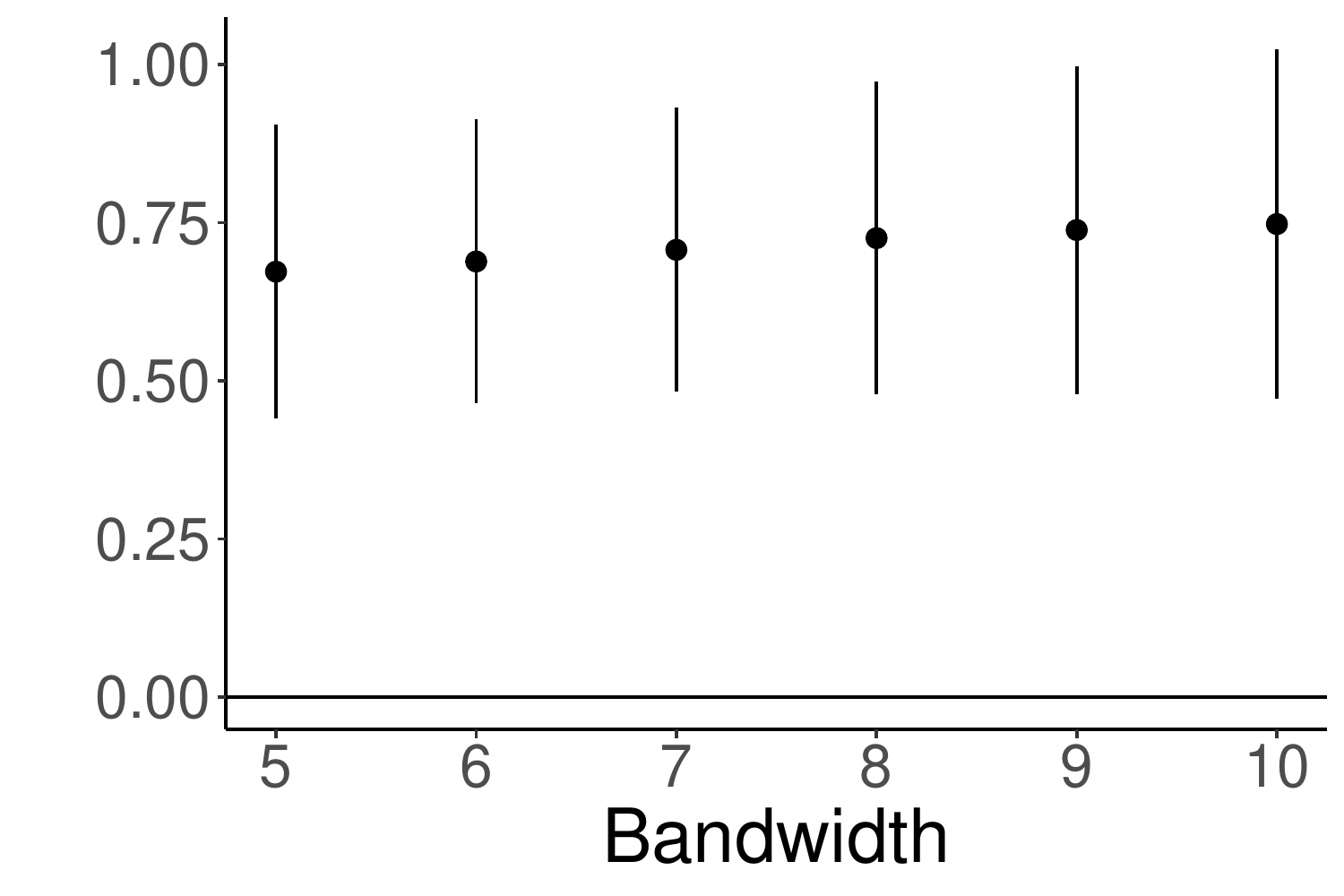} &   \includegraphics[width=4.5cm]{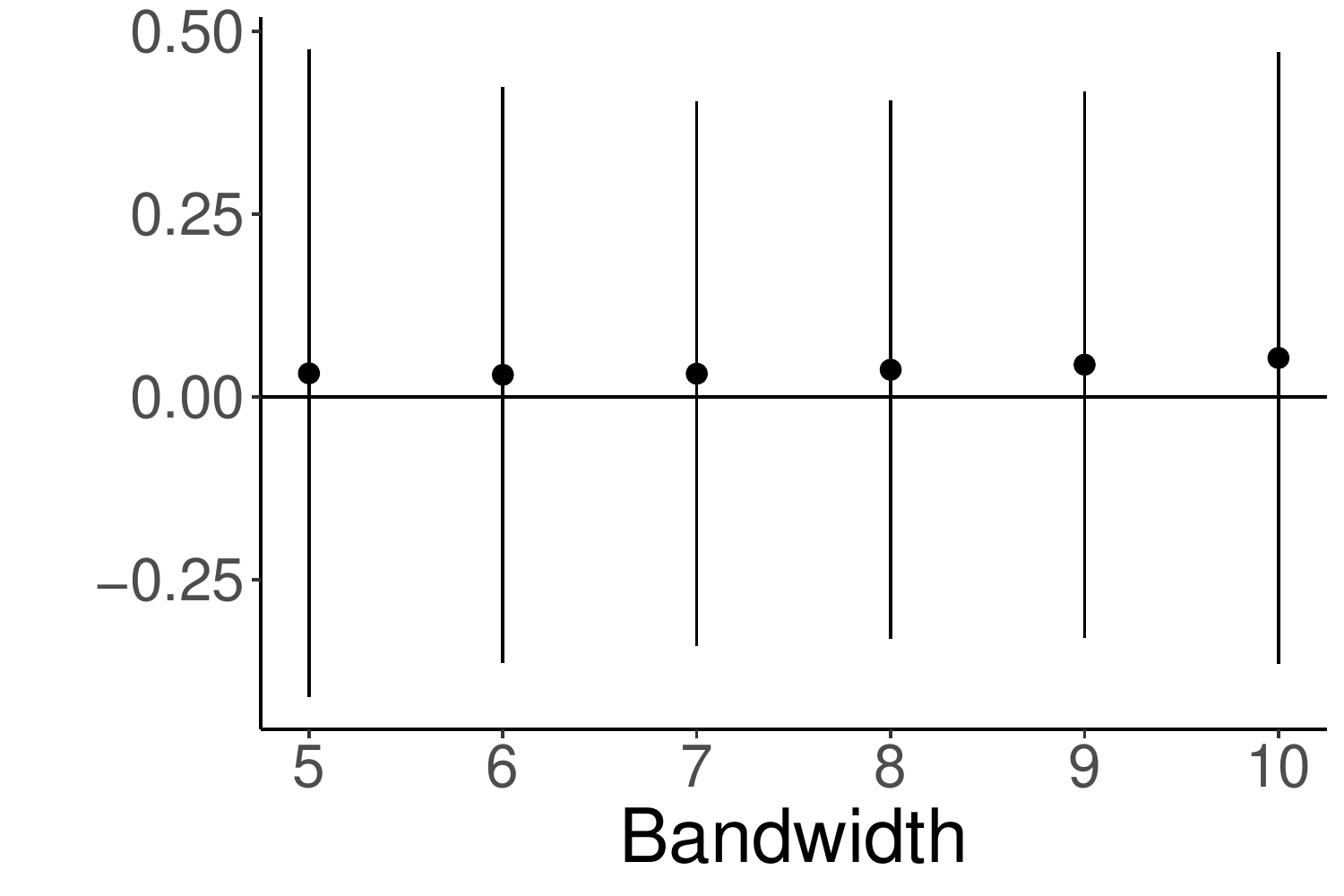} \\ [.5cm]
\textit{Panel D:} Bankruptcy &  \textit{Panel E:} Total debt past due &  \textit{Panel F:} Mortgage debt past due \\
 \includegraphics[width=4.5cm]{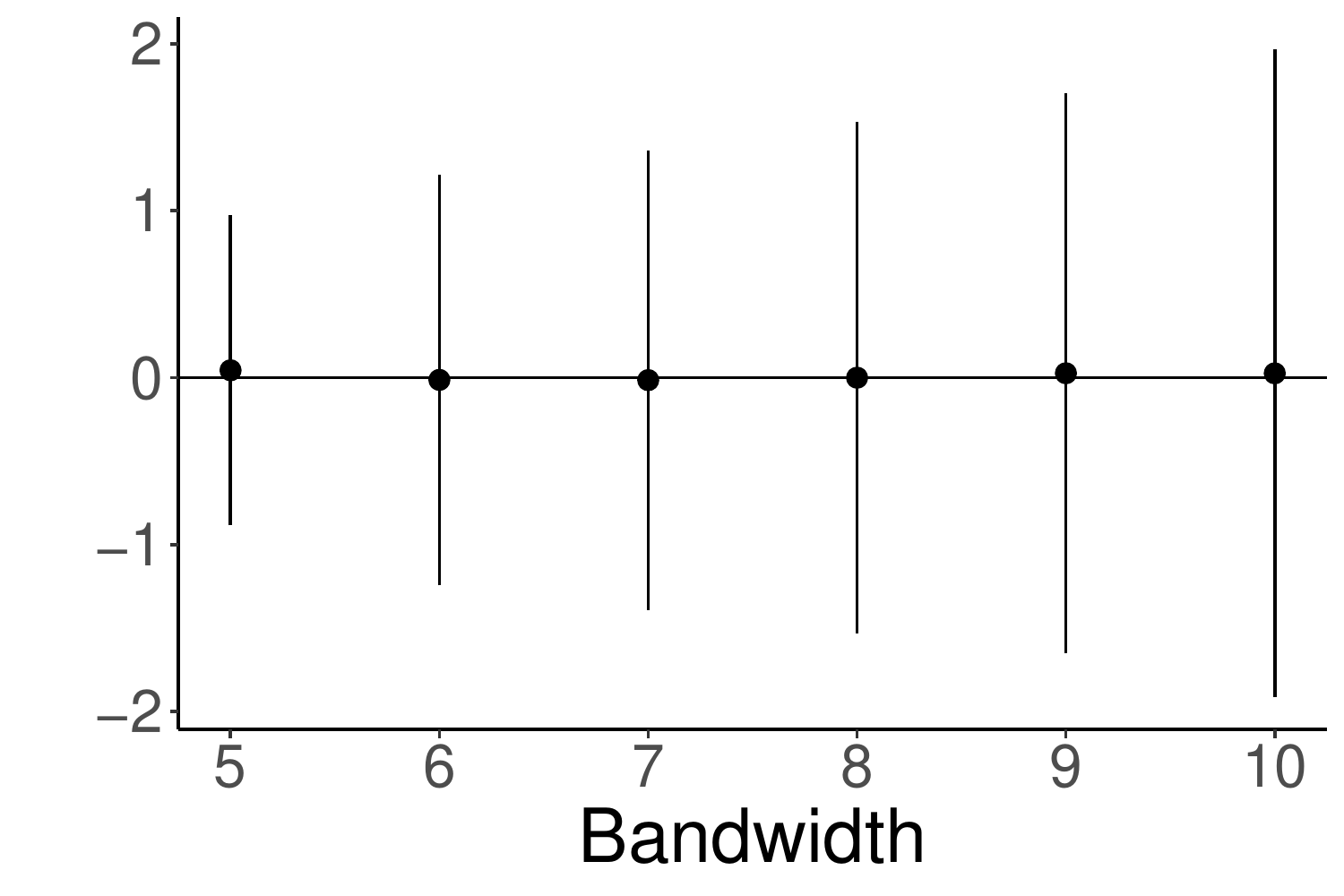} &   \includegraphics[width=4.5cm]{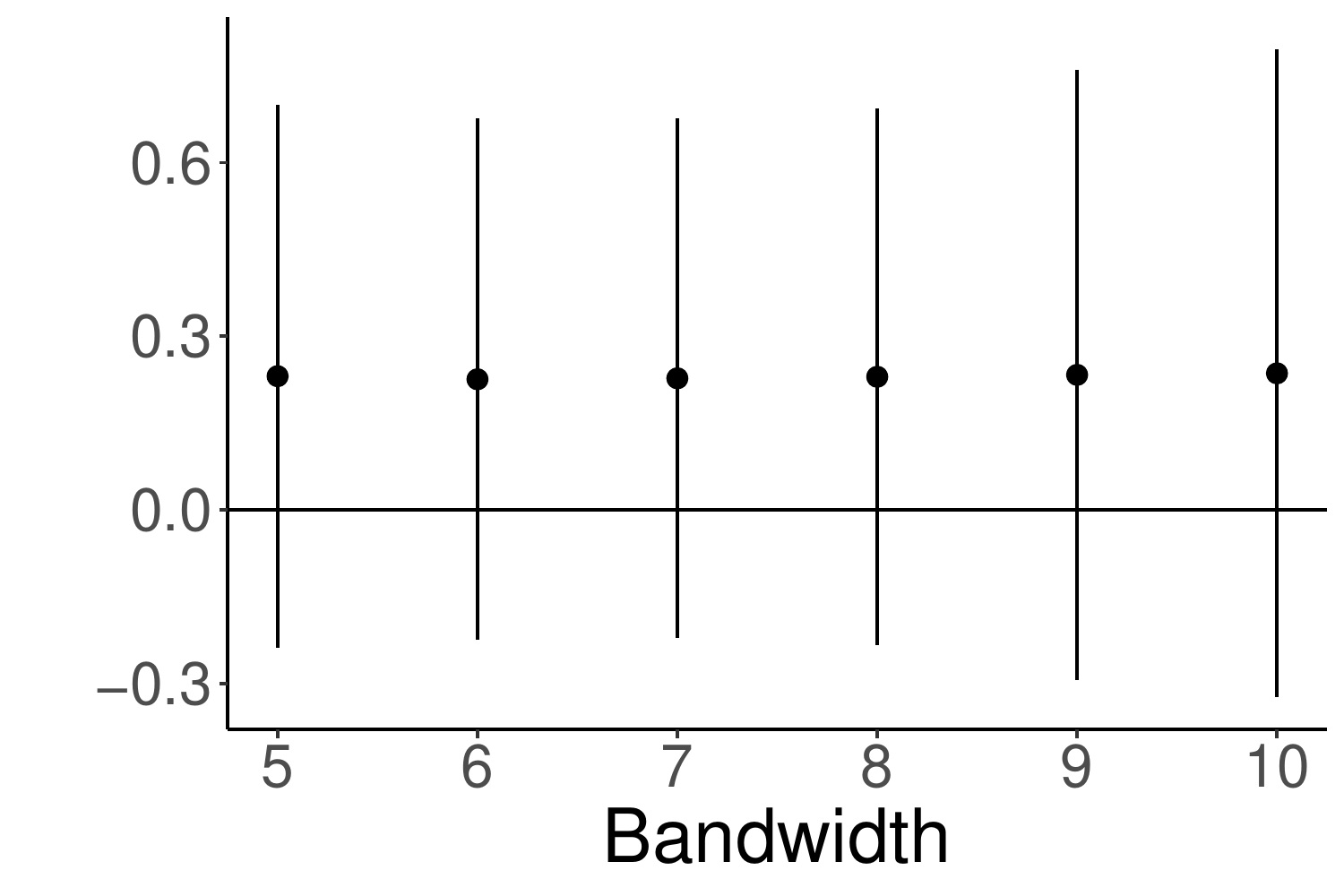} &   \includegraphics[width=4.5cm]{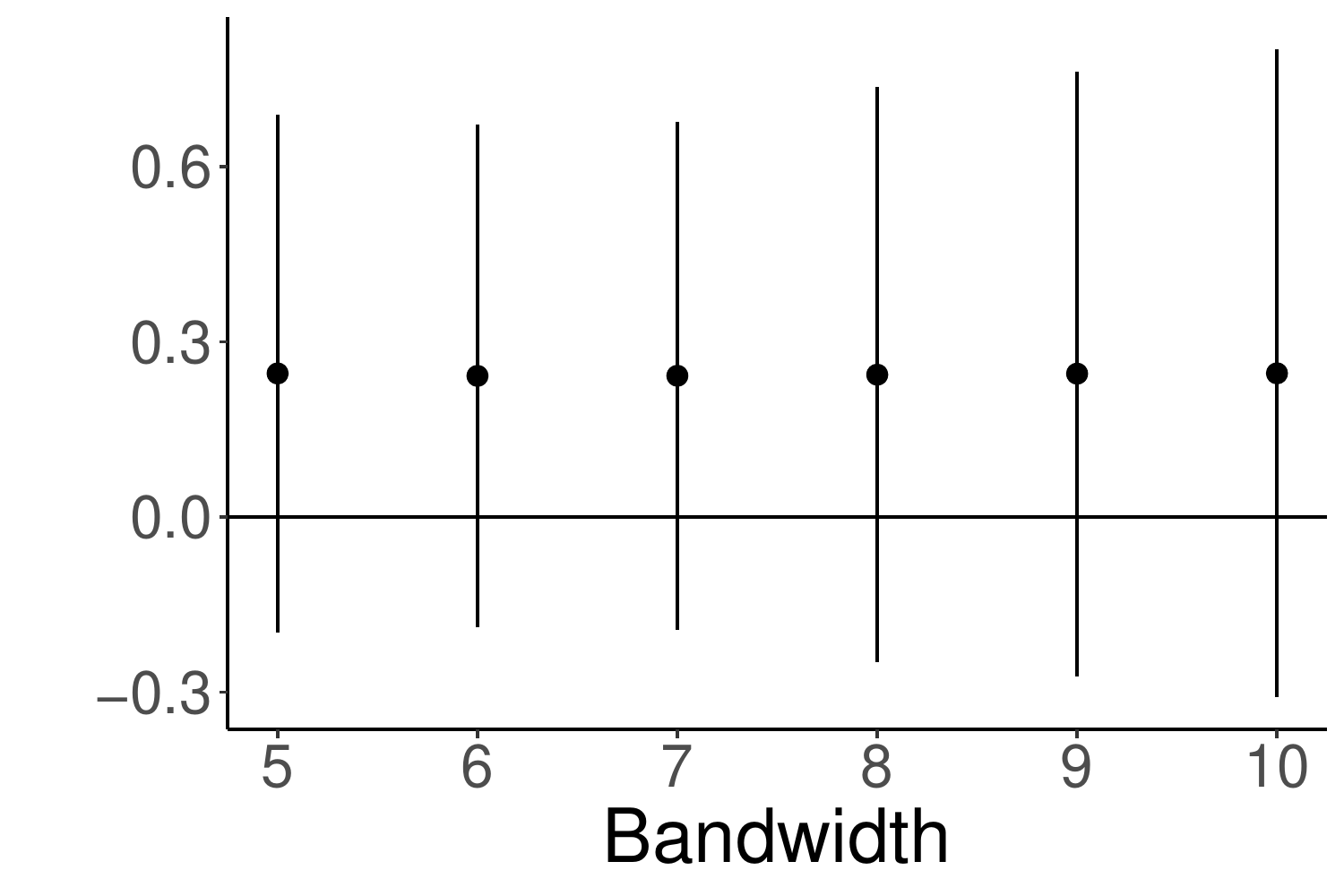} \\ [.5cm]
 \textit{Panel G:} CC debt past due &  \textit{Panel H:} Share debt past due &  \textit{Panel I:} Foreclosure \\
 \includegraphics[width=5cm]{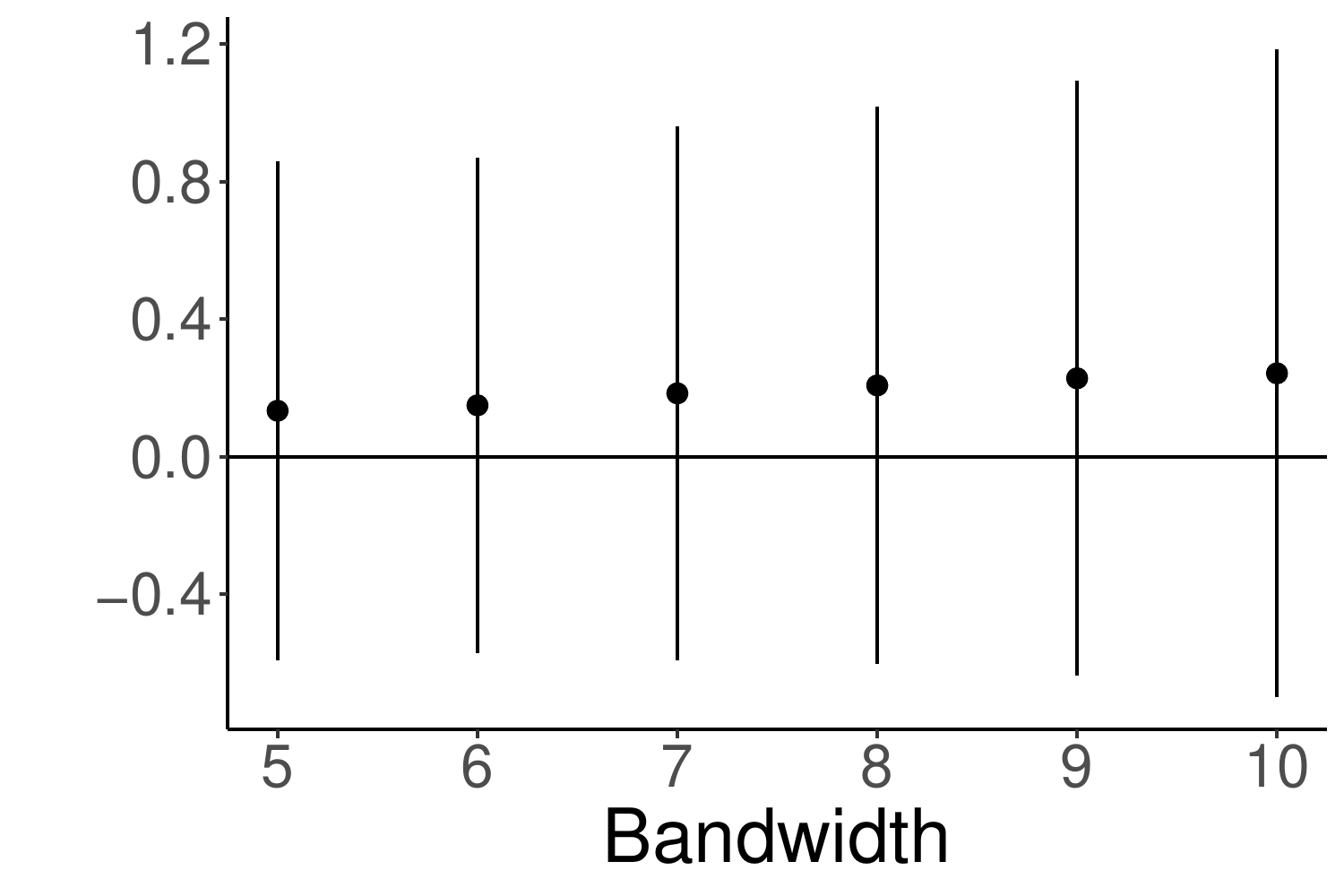} &   \includegraphics[width=5cm]{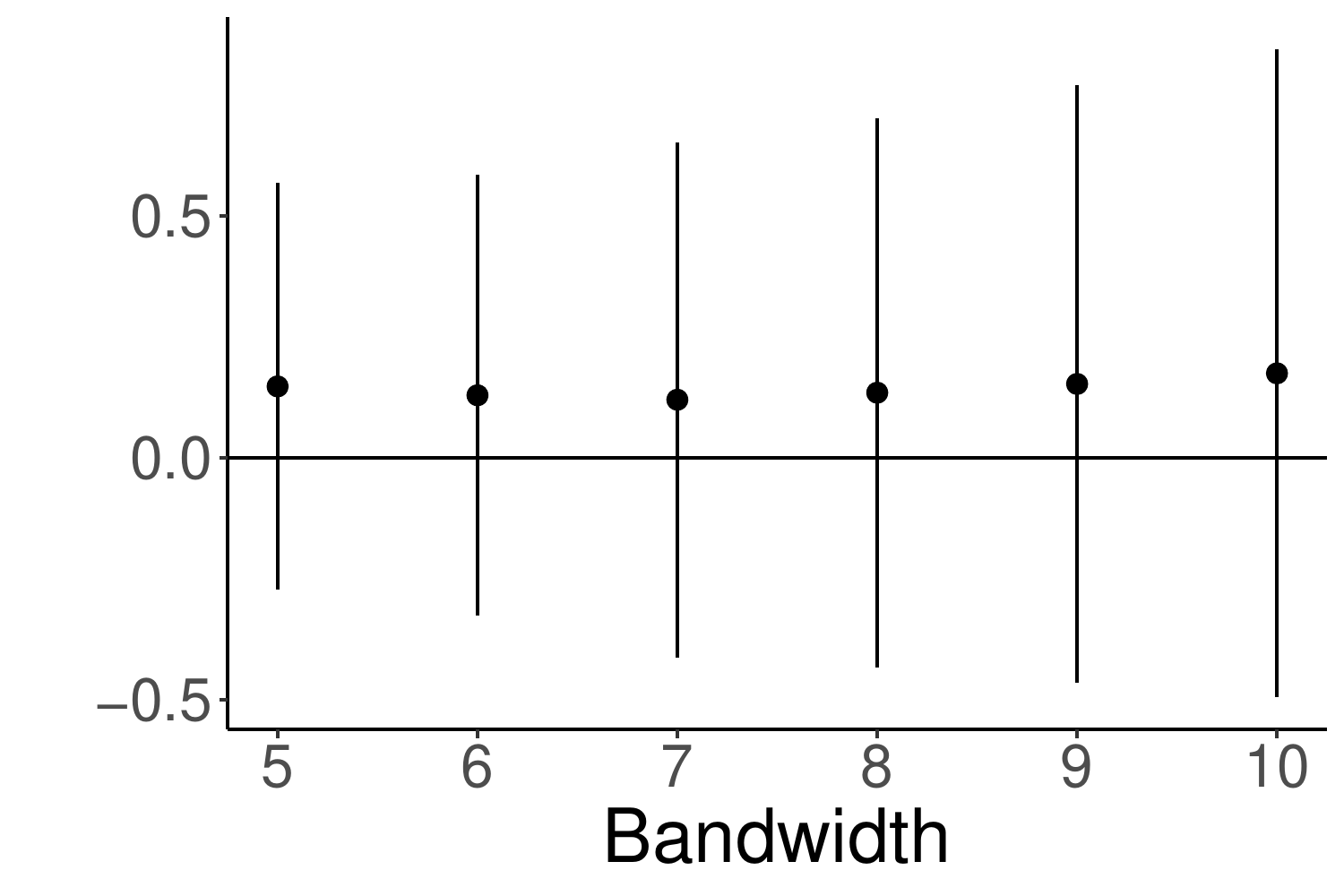} &   \includegraphics[width=5cm]{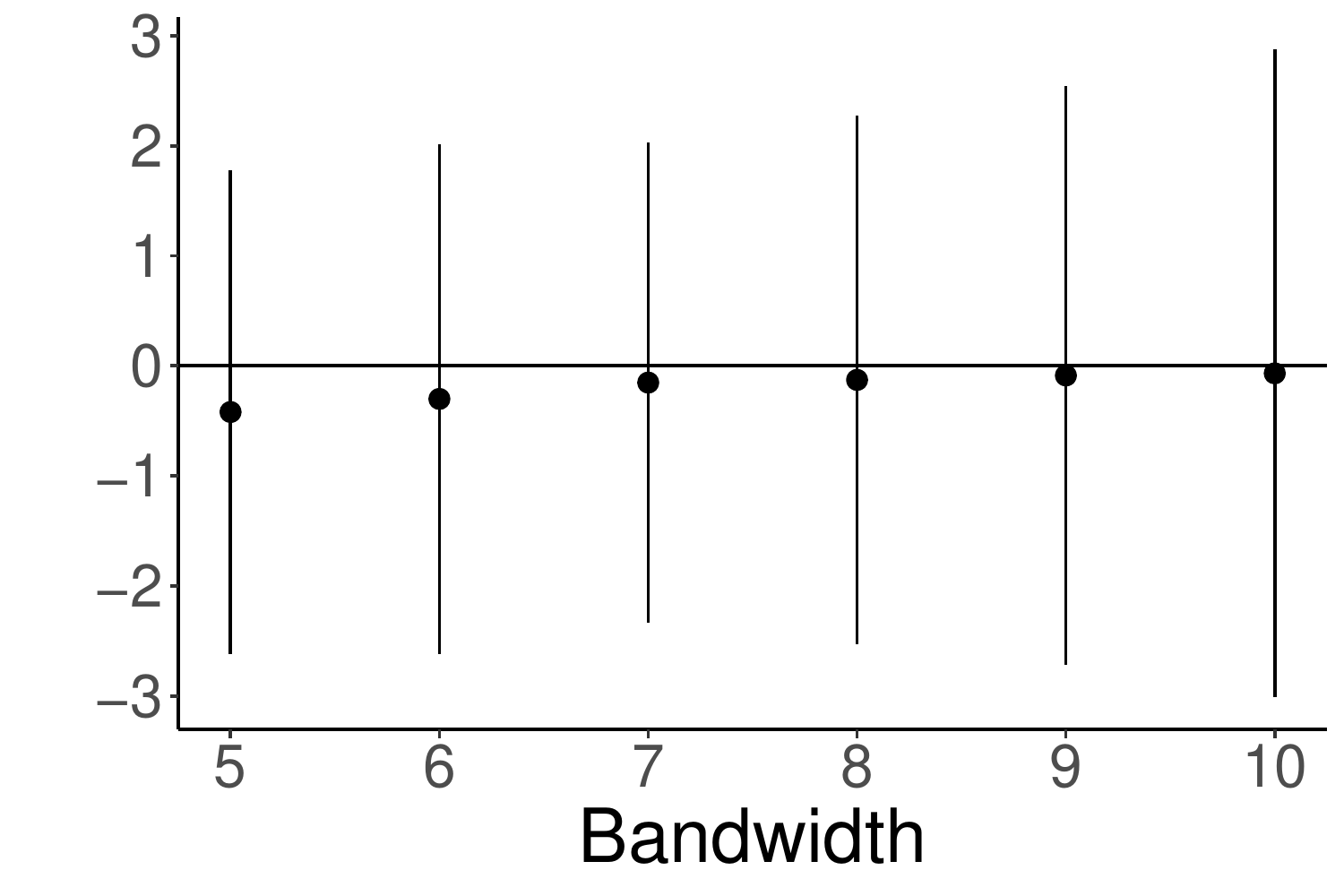} 
\end{tabular}
	\begin{minipage} {0.9\textwidth} \setstretch{.9} \medskip
          \footnotesize{\textbf{Note:} This figure plots the robustness of our regression discontinuity estimates to the bandwidth selection used in the \cite{kolesar2018inference} estimation procedure. Panel A plots the robustness of the share of the population with any coverage estimates. Panel B plots the robustness of the average debt in collections in dollars RD estimates. Panel C plots the robustness of the risk score RD estimates based on the Equifax Riskscore 3.0. Panel D plots the robustness of the bankruptcy RD estimates. Panel E plots the robustness of the average debt past due RD estimates. Panel F plots the robustness of the average mortgage debt past due RD estimates. Panel G plots the robustness of the average credit card debt past due RD estimates. Panel H plots the robustness of the share of debt past due RD estimates. Panel I plots the robustness of the foreclosure RD estimates. The sample includes individuals who were age 55-75 between 2008 and 2017. The regressions include 26,120,830 person-year-quarter observations for 2,977,952 unique individuals. See Section \ref{background_data} for additional details on the outcomes and sample. Source: New York Fed Consumer Credit Panel / Equifax.}
	\end{minipage}
\end{figure}

In this section, we discuss our approach to assessing the robustness of our results to alternative specifications and bandwidths. As discussed in Section \ref{apx:methods}, our empirical methodology requires two tuning parameters: the bandwidth (standard regression discontinuity applications) and our upper bound on the magnitude of the second derivative. In Figures \ref{fig:bound_scaling_robustness_apx} and \ref{fig:age_bandwidth_robustness_apx}, we present sensitivity tests for our main RD estimates for various outcome measures to the choice of bandwidth and our upper bound. Our results are qualitatively unchanged across our choice of bandwidth and upper bound. In Figures \ref{fig:bound_scaling_var_robustness_apx} and \ref{fig:age_bandwidth_var_robustness_apx} we demonstrate that our estimated reductions in the variance of health insurance and collections debt are robust to alternative bound scaling factors and bandwidths.

\begin{figure}[htpb!]
  \centering
  \caption{Additional outcomes for changes in financial health at age 65}
  \label{fig:appendix_main_ageRD}
  \makebox[\textwidth][c]{
\begin{tabular}{cc}
  \textit{Panel A:} Total debt past due  &   \textit{Panel B:} Mortgage debt past due \\
\includegraphics[width=3.25in]{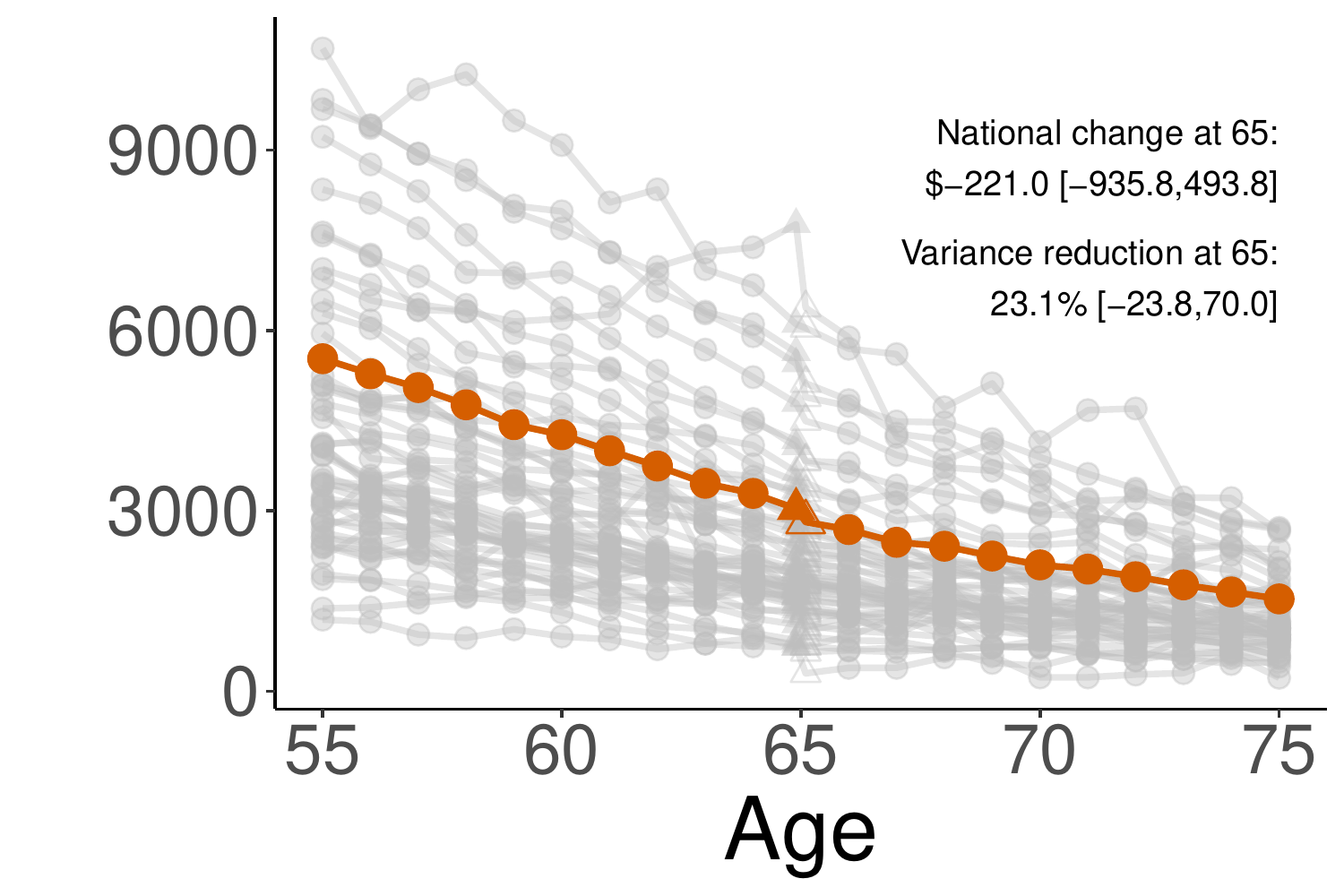} & \includegraphics[width=3.25in]{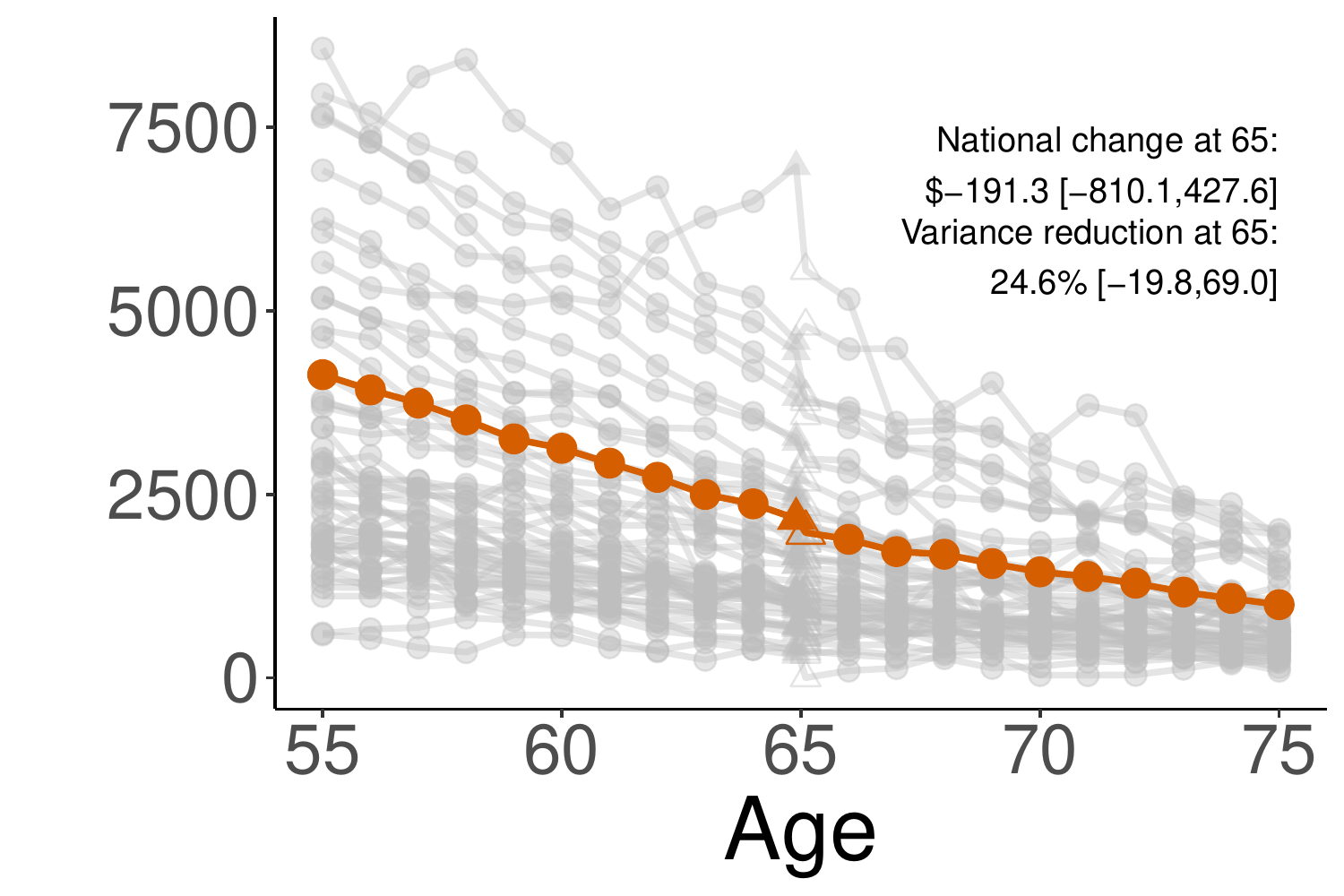}\\
  \textit{Panel C:} Credit card debt past due &  \textit{Panel D:} Foreclosure  \\
\includegraphics[width=3.25in]{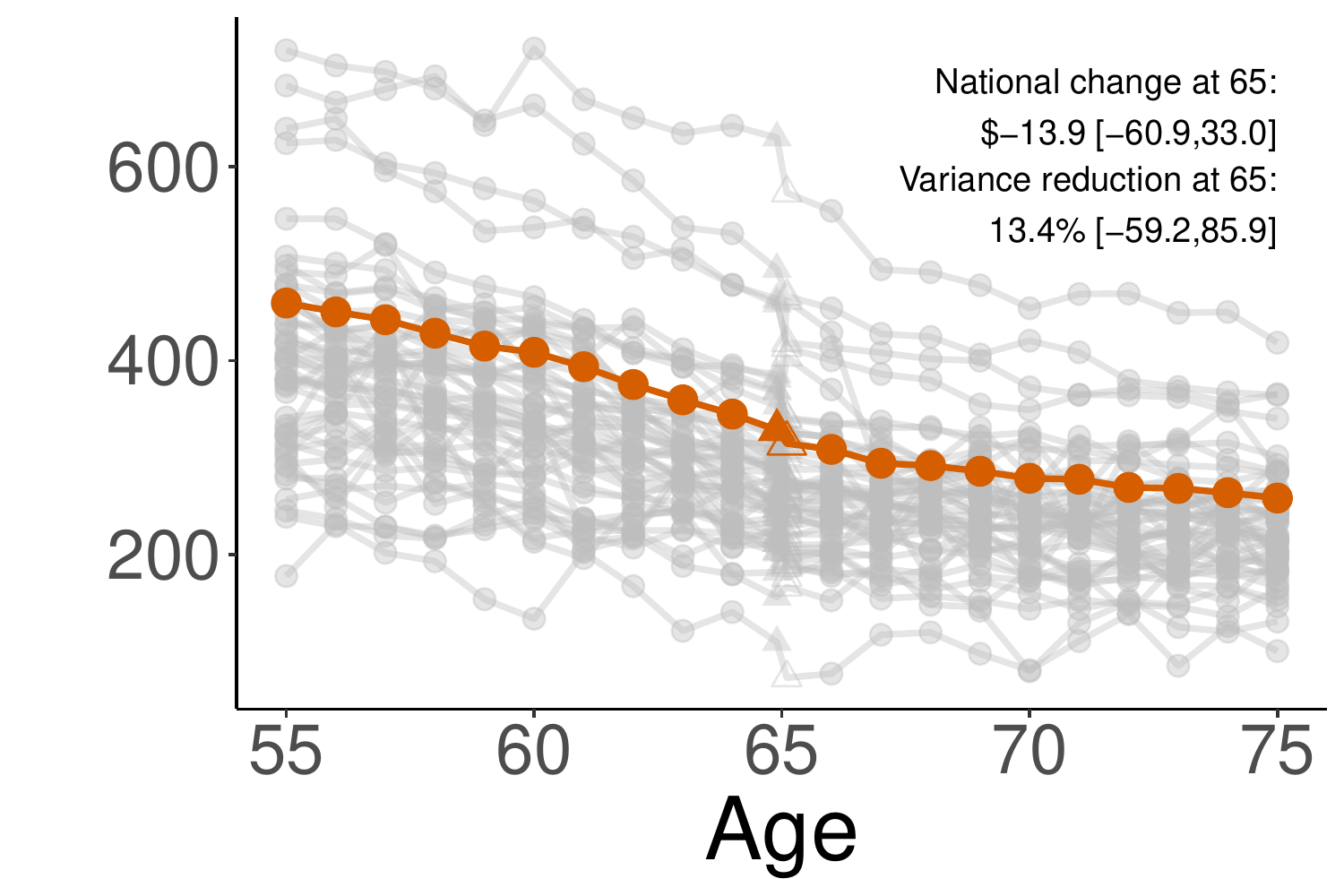} & \includegraphics[width=3.25in]{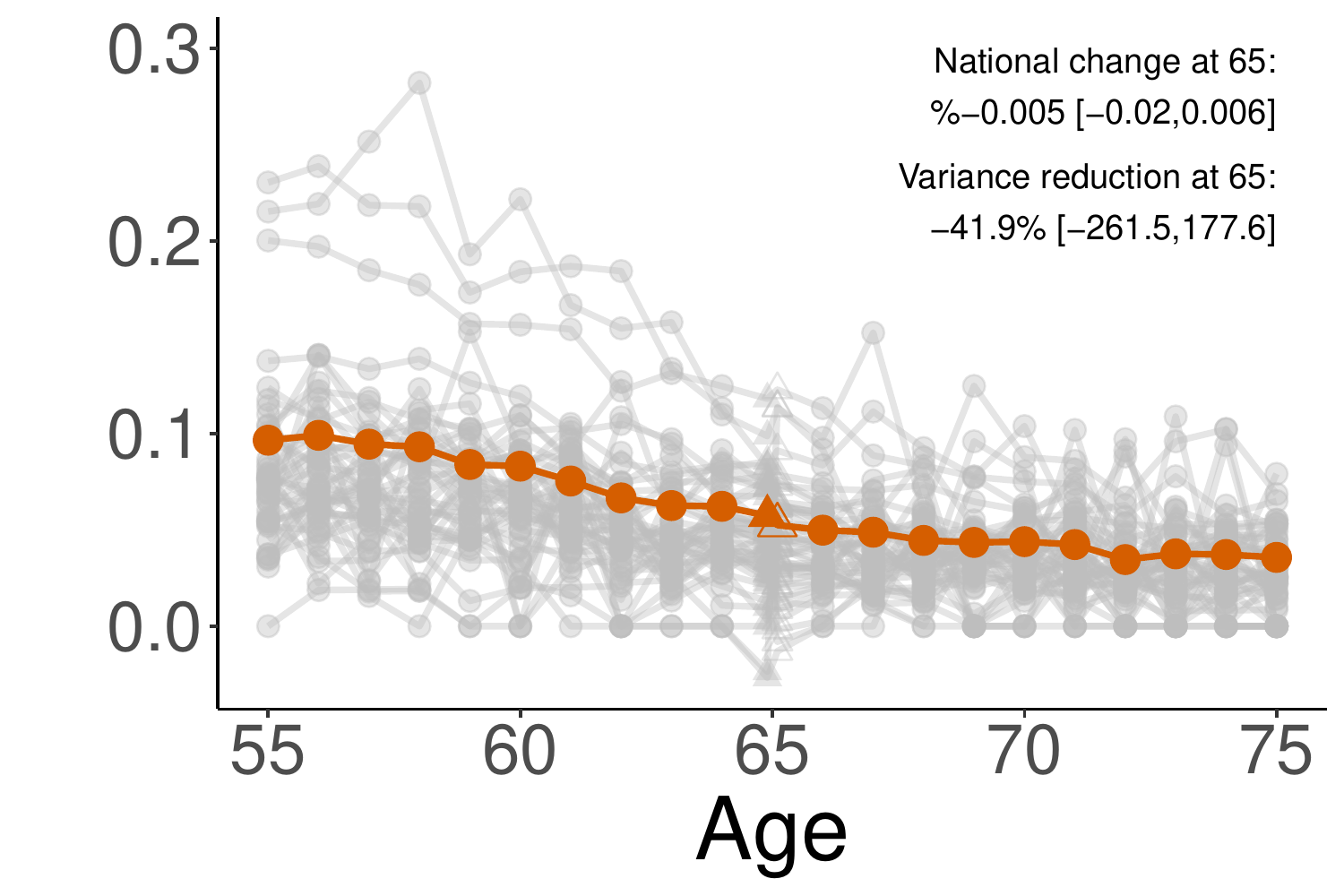}\\
  \textit{Panel E:} Share of credit card debt past due &  \textit{Panel F:} Share of mortgage debt past due  \\
\includegraphics[width=3.25in]{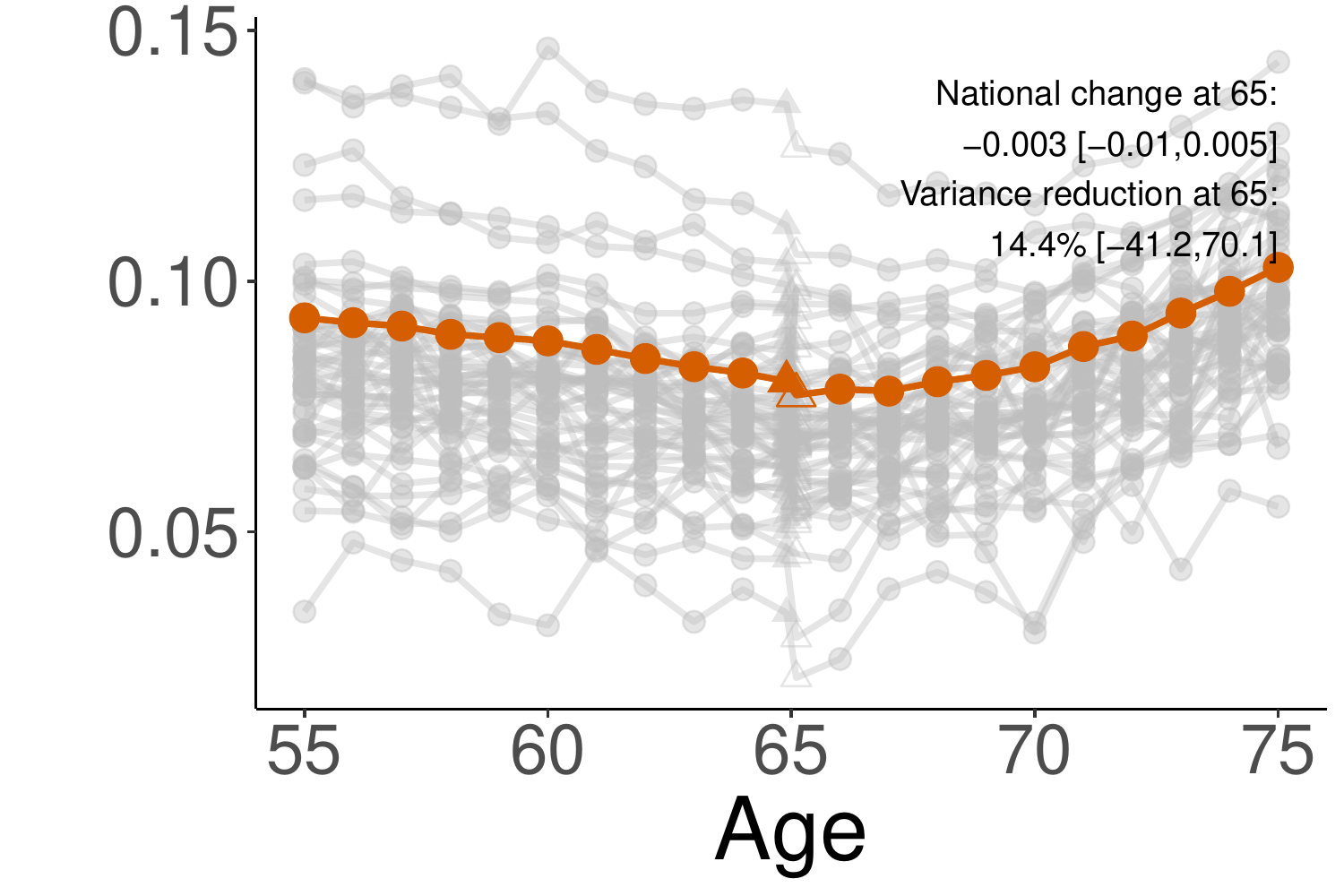} & \includegraphics[width=3.25in]{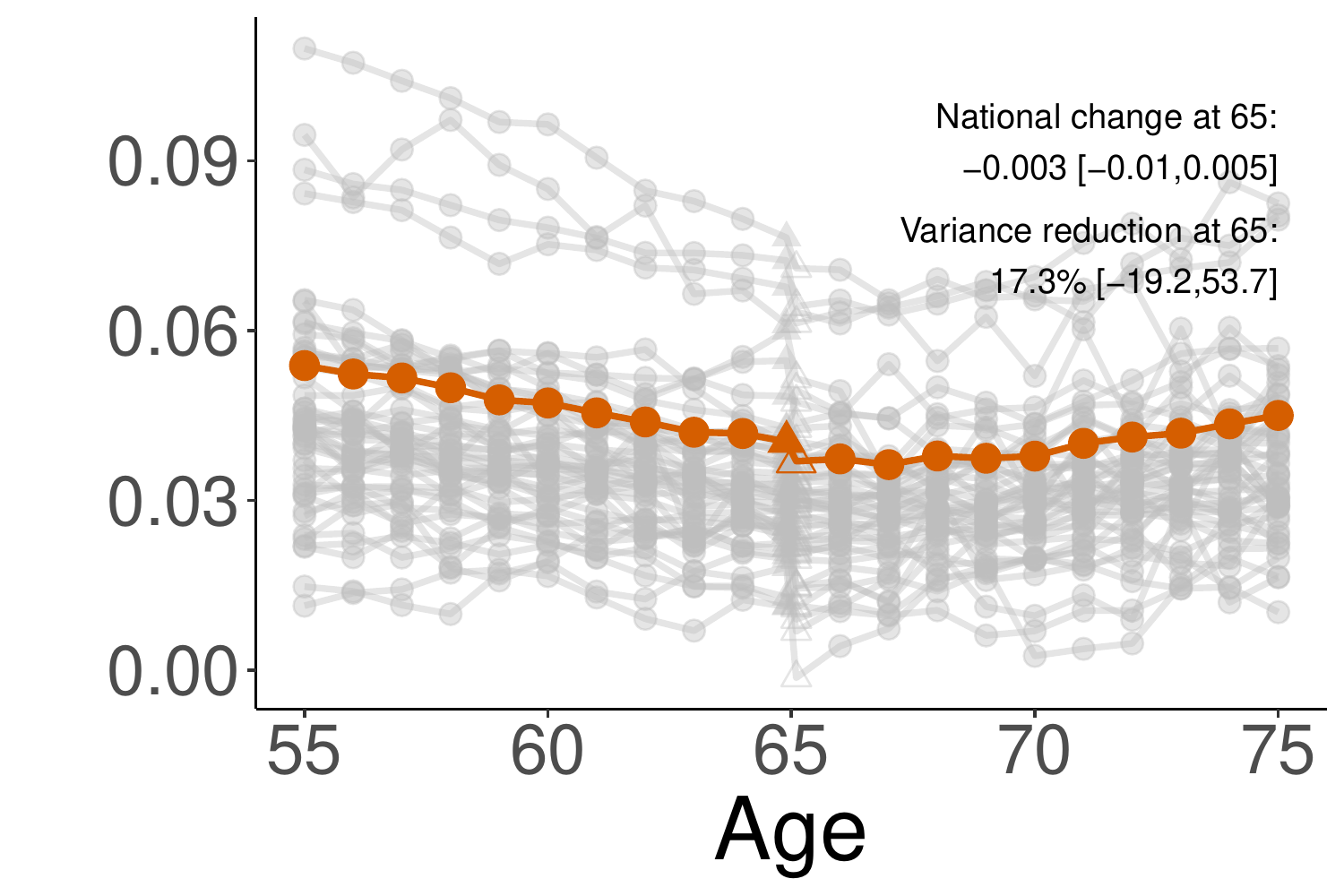}\\
\end{tabular}}
\begin{minipage} {0.9\textwidth} \setstretch{.9} \medskip
  \scriptsize{\textbf{Note:} This figure plots consumer credit outcomes by age. The horizontal axis denotes age in years. A local linear regression is fit on each side of the Medicare eligibility threshold using methods from \cite{kolesar2018inference}. We include hollow points that are the predicted counterfactual outcomes with and without Medicare at 65. The blue hollow dot is the predicted outcome without Medicare at age 65 and the red hollow dot is the predicted consumer credit outcome with Medicare at age 65. Panel A plots average amount of debt that is more than 30 days past due by age. Panel B plots the average amount of mortgage debt that is more than 30 days past due by age. Panel C plots the average amount of credit card debt that is more than 30 days past due by age. Panel D plots the share of individuals experiencing a foreclosure by age. Panel E reports the share of credit card that is more than 30 days past due. Panel F reports the share of mortgage debt that is more than 30 days past due. The share debt past due outcomes are calculated as the average individual debt past due, divided by the average total debt of all individuals of the same age living in that state. We divide by this average, rather than individuals' own debt levels, to avoid the divide-by-zero problem. The sample includes individuals who were age 55-75 between 2008 and 2017. See Section \ref{background_data} for additional details on the outcomes and sample. Source: The financial health outcomes are based on 137,340,577 person-year observations from the New York Fed Consumer Credit Panel / Equifax, 2008-2017.}
	\end{minipage}
\end{figure}

In Appendix Table \ref{tab:main_agerd_appendix_othermodels_apx}, we repeat our estimates for our consumer credit outcomes, using alternative RD methodologies. We do so in four ways. First, in Column 1, we replicate our main estimates from Figures \ref{fig:main_ageRD} and \ref{fig:appendix_main_ageRD}. Second, we consider three parametric models, fitting a linear, quadratic and cubic model in age on either side of the discontinuity and estimating the jump at age 65. For inference, we use heteroskedasticity-robust standard errors. We report these estimates in Columns 2, 3 and 4. Third, we estimate the same models, but cluster on the running variable of age, as suggested in \cite{lee2008regression} (and subsequently shown to have coverage issues in \cite{kolesar2018inference}). We report these models in Columns 4, 5 and 6. Finally, we use local linear estimation using the \texttt{RDRobust} package from \cite{calonico2015rdrobust}. We report this estimate in Column 8.

Our estimates are similar across various estimation methodologies. However, many outcomes do appear statistically significant when we cluster on the running variable, unlike in our main specification. This is likely due to incorrect coverage, as highlighted in \cite{kolesar2018inference}. When using heteroskedasticity-robust standard errors, there are also additional significant estimates, but fewer, and they are not consistent across various parametric forms. Our estimates are quite similar, qualitatively, to using the \cite{calonico2015rdrobust} method, but our preferred estimate's point estimate is larger in magnitude and the confidence interval is smaller for our collection estimates.

\subsection{Forecasting the causal effects of Medicare by location}

% Brief description of what is discussed in this section.
This section provides additional details about how we forecast the causal effects of Medicare by location. We are interested in the forecastable components of both $\gamma_{l}$ and $\beta_{l}$, where $\beta_{l}$ is our fuzzy-RD estimates of CZ-level reductions in collections debt \textit{per newly-insured} (see Section \ref{sec:methods} for details). We are interested in the best \emph{predictions} of $\gamma_{l}$ and $\beta_{l}$. 

Ideally, each forecast would be the unbiased causal estimate for the location from our RD design. However, in many locations, the near-elderly population is small and the estimates are noisy. To reduce noise, we follow \cite{chetty2018impactsb} and  construct  forecasts using a shrinkage estimator that combines  our unbiased RD estimates and the predicted effect for each commuting zone based on its demographic and healthcare market characteristics. Since the dimension of our set of predictors, $\vec{X}_{l}$, is large (and many of the covariates are highly correlated), we use our Lasso predictions from Section \ref{subsec:empirical_forecasts} in order to minimize over-fitting.

% Brief background on Lasso
We denote our predictions of $\gamma_{l}$ and $\beta_{l}$ estimated using our Lasso model as $\hat{\gamma_{l}}$ and $\hat{\beta_{l}}$, respectively. Briefly, the Lasso estimation procedure penalizes covariates and shrinks terms in the estimated $\omega_{l}$ towards zero, in order to minimize mean squared error. As a result, the estimation procedure will select a subset of the covariates in $\vec{X}_{l}$, to have non-zero parameters, and set the remaining parameters to zero. We implement this using a ten-fold cross-validation over the penalization parameter, implemented using R \texttt{glmnet} package.

% Additional details on the forecast
To forecast the causal effects of Medicare by location, we then combine the Lasso estimates together with our RD estimates of $\gamma_{l}$ order to construct the mean square error-minimizing forecast for each location, defined as $\hat{\gamma}_{l}^{f}$.  This MSE-minimizing forecast is constructed using the following formula \citep{chetty2018impactsb}:\footnote{See Appendix D of \cite{chetty2018impactsb} for the explicit derivation of this approach. Our approach deviates from \cite{chetty2018impactsb} in that we use the Lasso predicted estimate, rather than the estimated mean value of residents (as \cite{chetty2018impactsb} do). This extension is discussed in their Appendix D. Additionally, since our estimates are not mean zero by construction, we demean our estimates for the purposes of the shrinkage, and then add the overall mean back in. Our approach is otherwise identical.} 

\begin{equation}
    \hat{\gamma}_{l}^{f} = \bigg(\frac{\chi^{2}}{\chi^{2} + s_{l}^{2}}(\gamma_{l}-\overline{\gamma}_{l}) + \frac{s_{l}^{2}}{\chi^{2} + s^{2}_{l}}\tau(\hat{\gamma}_{l} - \overline{\hat{\gamma}}_{l})\bigg) + \overline{\gamma}_{l},
\end{equation}
where 
$\overline{\gamma}_{l}$ is the average RD prediction across locations, $\overline{\hat{\gamma}}_{l}$ is the average Lasso prediction across locations, $\tau = Cov(\hat{\gamma}_{l},\gamma_{l}) \big/ Var(\hat{\gamma}_{l})$ is the coefficient of a regression of $\gamma_{l}$ on $\hat{\gamma}_{l}$, $\chi^{2}$ is the residual place effect variation after subtracting off the variance due to estimation of $\gamma_{l}$, and $s_{l}^{2}$ the squared standard error of the $\gamma_{l}$. For the purposes of the shrinkage, we demean our estimates and then add the overall mean back, such that the shrinkage is around the variation around the overall mean.  We estimate $\tau$ using linear regression of the demeaned values, and calculate $\chi^{2}$ as
\begin{equation*}
    \chi^{2} = Var(\gamma_{l} - \tau(\hat{\gamma}_{l} - \overline{\hat{\gamma}}_{l})) - E(s^{2}_{l}),
\end{equation*}
where $E(s^{2}_{l})$ is the average sampling variance across locations. In all calculations, we weight by the precision of the fixed effect estimates ($1/s^{2}_{l}$) to maximize efficiency. 

Note that this approach will shrink our estimates towards the predicted $\hat{\gamma}_{l}$ when the original estimate is noisy and the shrinkage will only occur if the lasso prediction has predictive power for $\gamma_{l}$. If this prediction has limited value, then $\tau$ will be zero, and the shrinkage will shrink towards the overall mean. By a similar argument, as $s^{2}$ goes to zero, the forecasted estimate will be exactly $\gamma_{l}$. We follow the same procedure to construct forecasts for $\beta_{l}$, defined as $\hat{\beta}_{l}^{f}$. 

To calculate the prediction errors of the forecasts for Table \ref{tab:czone_estimates_table}, we follow \cite{chetty2018impactsb}, where the root mean-squared error of the prediction is:
\begin{equation*}
    \sqrt{e^{2}_{l}} = \sqrt{\frac{1}{\frac{1}{s^{2}_{l}} + \frac{1}{\chi^{2}}}}.
\end{equation*}

Note that as the variance for our unbiased estimate ($s^{2}_{l}$) grows, $\chi^{2}$ places an upper bound on the size of the root MSE. In contrast, if the sampling error gets very small, the forecast will place all the weight on the unbiased estimate, and send the root MSE to zero.

\subsection{Estimating the effects of Medicare eligibility before and after implementation of the ACA}

We briefly describe the methods used to document differences in the effect of Medicare eligibility before and after the ACA. We also document our approach to quantifying the changes in health insurance and financial health for the near-elderly from the ACA.

To examine the relationship between Medicare eligibility, health insurance,
and financial health before and the after the implementation of the ACA we re-estimate our primary specification separately, pre- and post-ACA: 
\begin{equation} 
\label{eq:rd_locations_aca}
y_{i,l,t}(\text{age}) = \gamma_{T} \times 1(\text{age}>65) + f_{T}\left(\text{age}\right)\times 1(\text{age}\leq 65) + g_{T}\left(\text{age}\right)\times 1(\text{age}>65) + \epsilon_{i,l,t}(\text{age}).
\end{equation}
where $T$ indexes the pre-ACA (2008-2013) and post-ACA (2014-2017)
periods. The coefficients of interest are $\gamma_{T}$ which
measure the change in health insurance and financial strain at age 65 before
($T=0$) and after ($T=1$) the implementation of the ACA. As before, the above specification allows for flexible age trends on both sides of the discontinuity, with standard errors constructed using
methods outlined in \cite{kolesar2018inference} and discussed previously.

To quantify the changes in health insurance and financial health for the near-elderly from the ACA, we follow \cite{duggan2019impact} and
estimate a regression discontinuity difference-in-differences
(``difference-in-discontinuity'') research design. Intuitively, this design exploits the fact that for 65-year-olds (and older), the expansion of the ACA was limited (compared to the near-elderly). 
%By considering how the effect at the discontinuity changed post-ACA, this allows us to back out the causal effect of the ACA on the non-elderly. This requires two assumptions: first, we need to identify the relevant counterfactual discontinuity for the post-ACA period. We do this following \cite{duggan2019impact}, and use the 2010-2013 period as the baseline comparison. Second, this appraoch assumes that the only time-varying difference in the age 65 discontinuity that occurred between 2010-2013 and 2014-2017 is due to the ACA. %This approach is partially testable in that we can use 2008 and 2009 as a placebo period. 
To implement this approach, we construct $\Delta y_{i,l,t}(\text{age})) = y_{i,l,t}(\text{age})) - \bar{y}_{i,l,t,2010-2013}(age)$, where $\bar{y}_{i,l,t,2010-2013}(age)$ is the average outcome in a given location-age from 2010-2013. We then re-estimate our regression discontinuity approach using this modified outcome variable: 
\begin{equation}   \label{eq:diff_in_disc}
\Delta y_{i,l,t}(\text{age}) = \tilde{\gamma} \times 1(\text{age}>65) + \tilde{f}\left(\text{age}\right)\times 1(\text{age}\leq 65) + \tilde{g}\left(\text{age}\right)\times 1(\text{age}>65) + \tilde{\epsilon}_{i,l,t}.
\end{equation}
with the standard errors constructed using methods outlined in \cite{kolesar2018inference} and discussed previously.

\subsection{Decomposing the change in forecast reductions in per capita collections before and after implementation of the ACA}
\label{sec:decomp_apx}
In this section, we describe how we use our estimates to provide insight into why forecast reductions in collections have become increasingly concentrated in the Deep South. We are interested in understanding why the changes in the average CZ-level forecast from pre- to post-ACA differed between the South and other regions of the country.

To formally decompose this, we define the relative percentage change before and after the ACA , $\eta = \left(E(\gamma_{l}^{Post}) - E(\gamma_{l}^{Pre})\right) \big/ E(\gamma_{l}^{Pre})$, for both regions, South and All Others. This change can be written as three parts: the change in the effect of Medicare on health insurance rates, the change in the effect of Medicare on collections debt per newly-insured, and the change in the covariance between the two. Formally,
\begin{align*}
\eta &= \frac{\eta_{1} + \eta_{2} + \eta_{3}}{E(\gamma_{l}^{Pre})}\\
    \eta_{1} &= \left(E( \beta_{l}^{Post}) -  E(\beta_{l}^{Pre})\right)E(\gamma_{l}^{h,Post})\\
    \eta_{2} &= E( \beta_{l}^{Pre})\left(E(\gamma_{l}^{h,Post}) - E(\gamma_{l}^{h,Pre})\right)\\
     \eta_{3} &= Cov(\beta_{l}^{Post}, \gamma_{l}^{h,Post}) - Cov(\beta_{l}^{Pre}, \gamma_{l}^{h,Pre})
\end{align*}
This derivation follows from the fact that $E(\gamma_{l}) = E(\gamma_{l}^{h}\beta_{l}) = E(\gamma_{l}^{h})E(\beta_{l}) + Cov(\gamma_{l}^{h}, \beta_{l})$, where $\gamma_{l}$ is the reduction in collections debt per capita at age 65, $\gamma_{l}^{h}$ is the change in the insurance rate at age 65, and $\beta_{l}$ is the  reductions in debt collections \textit{per newly-insured} at age 65. We then rearrange terms to derive the above expression. An important note is that since we are focusing on the forecasts, rather than the underlying parameters, there are small differences because we use the shrinkage estimates. E.g. $\gamma_{l} = \gamma_{l}^{h}\beta_{l}$, but we use $\hat{\gamma}_{l}$. As a result, it is useful to rewrite $\hat{\gamma}_{l} = \gamma_{l} + \epsilon_{\gamma, l}$, and note that $\hat{\gamma}_{l} = \gamma^{h}_{l}\beta_{l} + \epsilon_{\gamma, l}$. We can use these approximations to redefine our approximation in terms of the forecasted estimates, which will leave us with additional error terms. In our results, these terms are captured in our last decomposition piece, $\eta_{3}$. 

For transparency, we present the underlying quantities in Appendix Table \ref{tab:decomp_components_apx} and plot the three components of the decomposition (the $\eta$s) in Panel B of Figure \ref{fig:collections_prepost_aca}.

\makeatletter
\setlength{\@fptop}{5pt}
\makeatother

%%%%%%%%%%%%%%%%%%%%%%%%%%%%%%%%%%%%%%%%%
% APPENDIX MATERIAL REFERENCED IN PAPER %
%%%%%%%%%%%%%%%%%%%%%%%%%%%%%%%%%%%%%%%%%

%%%%%%%%%%%%%%%%%%%%%%%%%%%%%%%%%%%%%%%%%%%%%%%%%%%%%%
% Collections distribution outcomes with main age RD %
%%%%%%%%%%%%%%%%%%%%%%%%%%%%%%%%%%%%%%%%%%%%%%%%%%%%%%
\begin{figure}[htpb!]
  \centering
  \caption{Changes in the distribution of collections debt at age 65}
  \label{fig:collections_dist_RD}
\begin{tabular}{cc}
  \textit{Panel A:} Share \$1-500  &   \textit{Panel B:} Share \$500-1,000 \\
    \includegraphics[width=7.5cm]{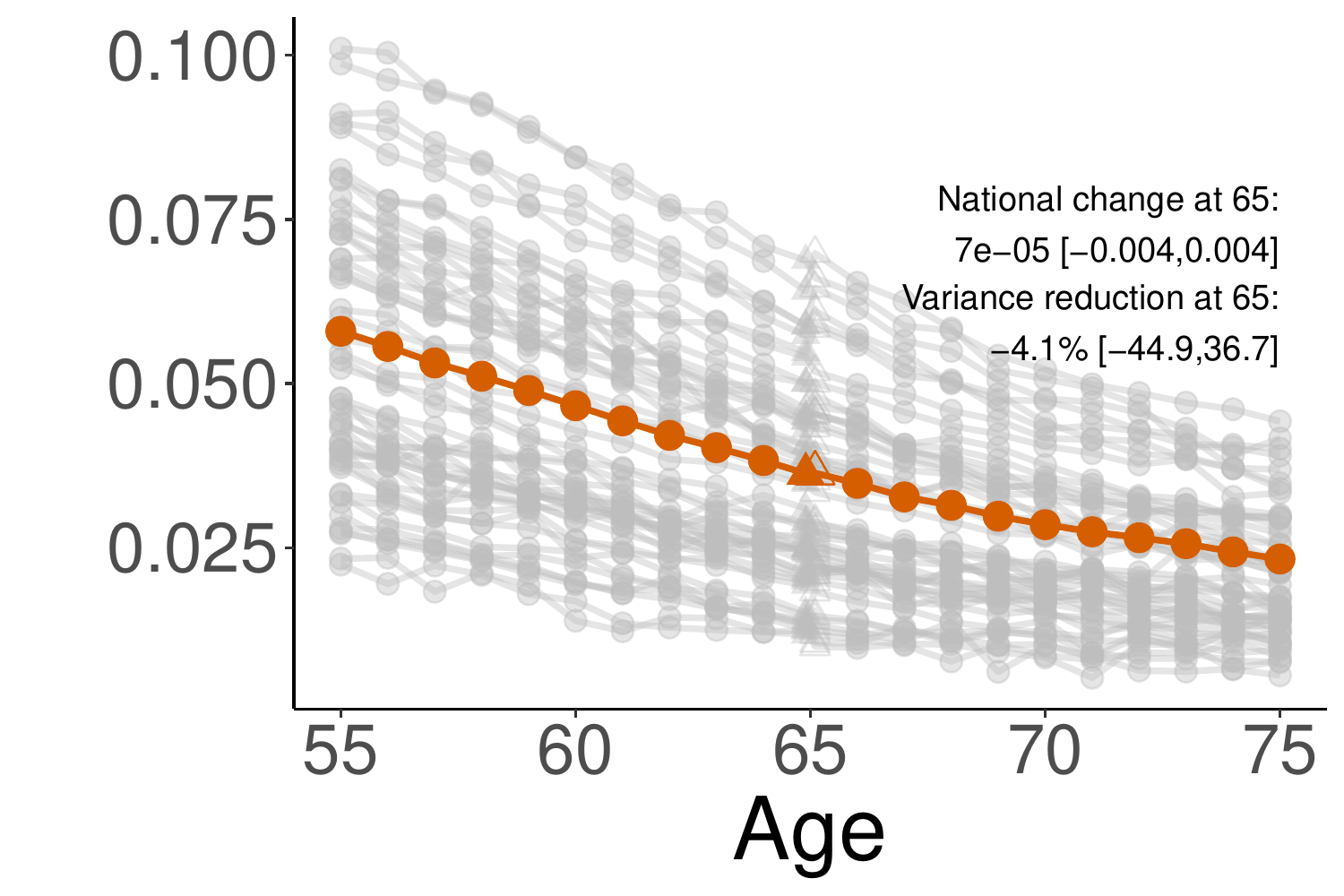} & \includegraphics[width=7.5cm]{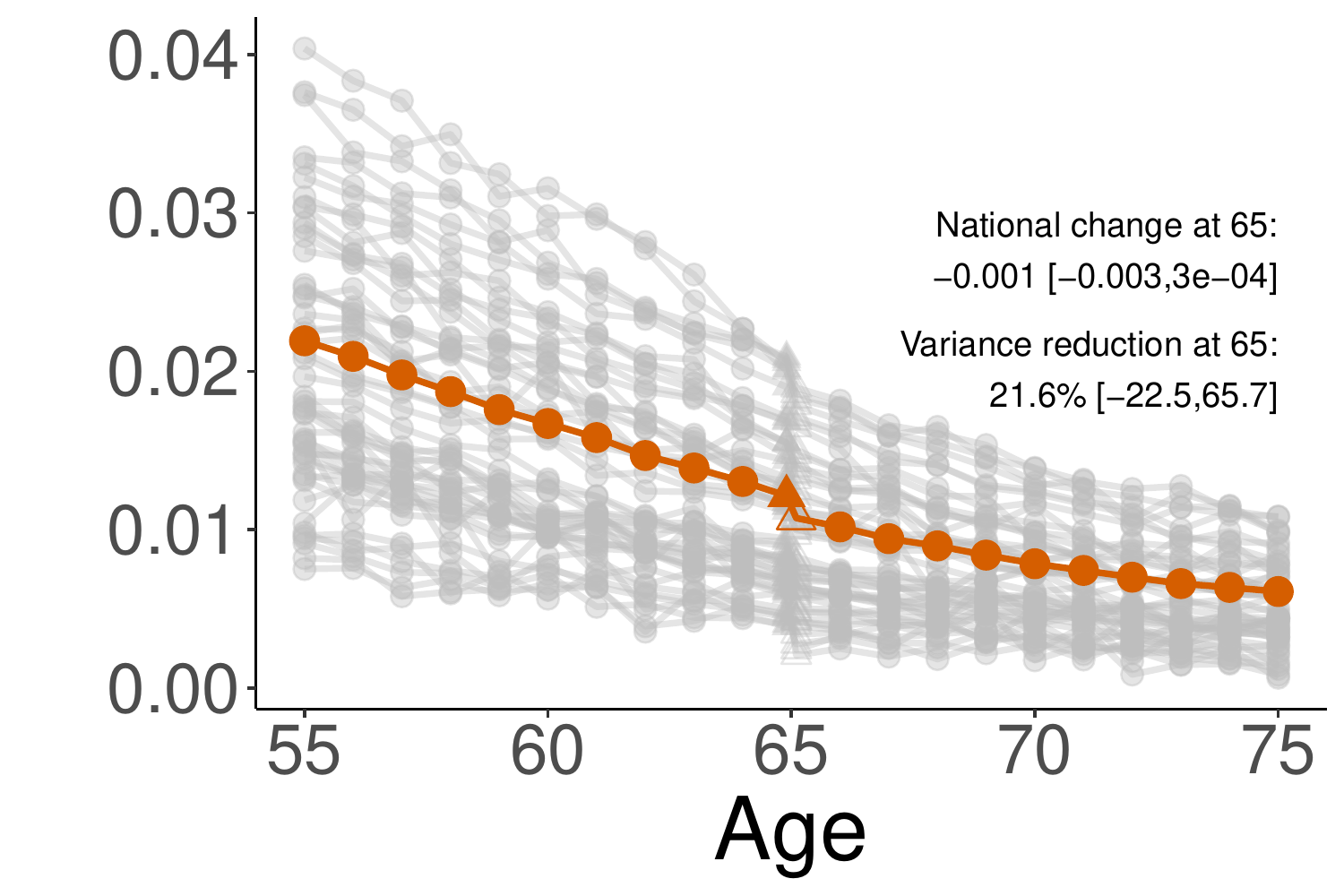}\\
  \textit{Panel C:} Share \$1,000-2,500 &  \textit{Panel D:} Share \$2,501-5,000  \\
    \includegraphics[width=7.5cm]{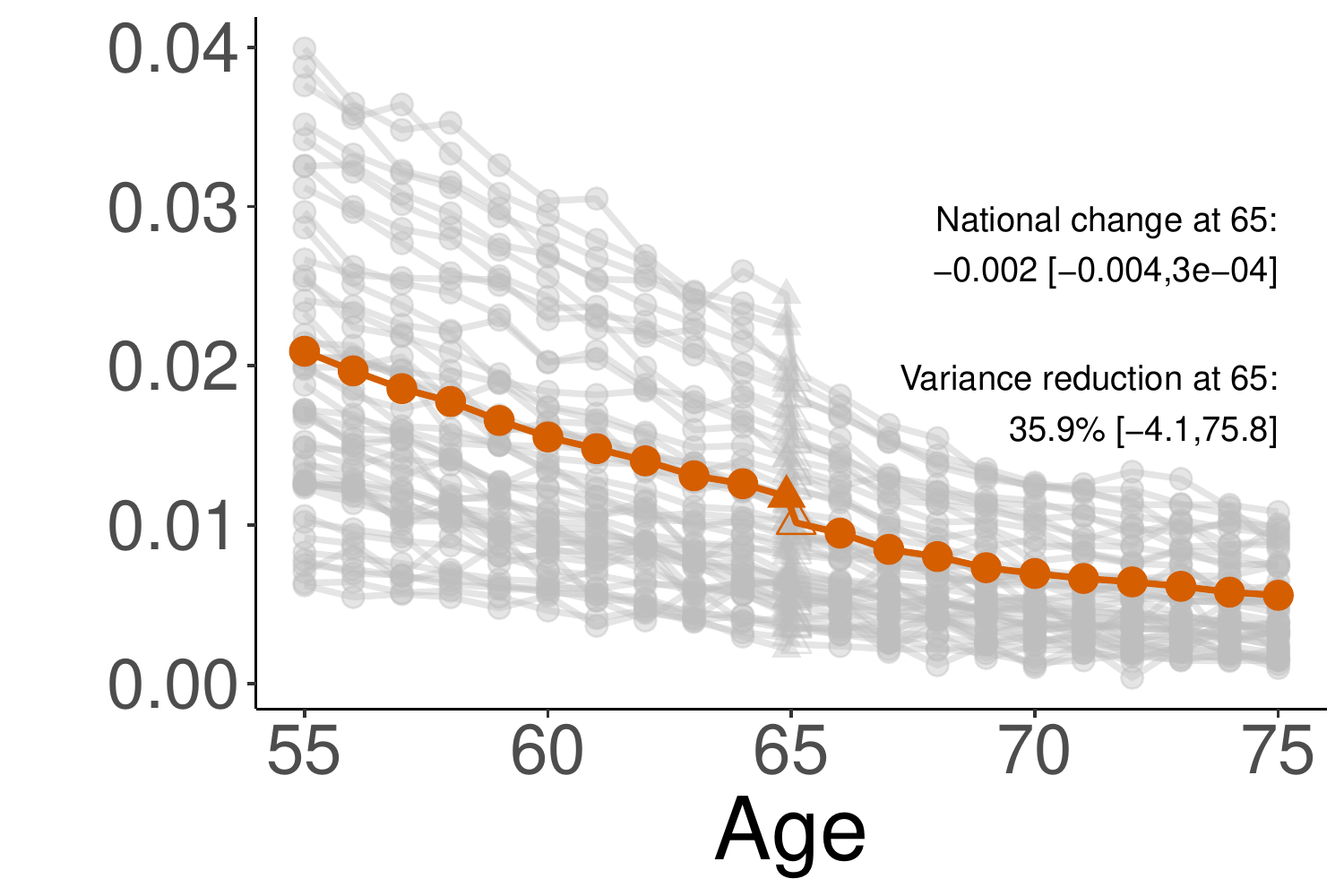} & \includegraphics[width=7.5cm]{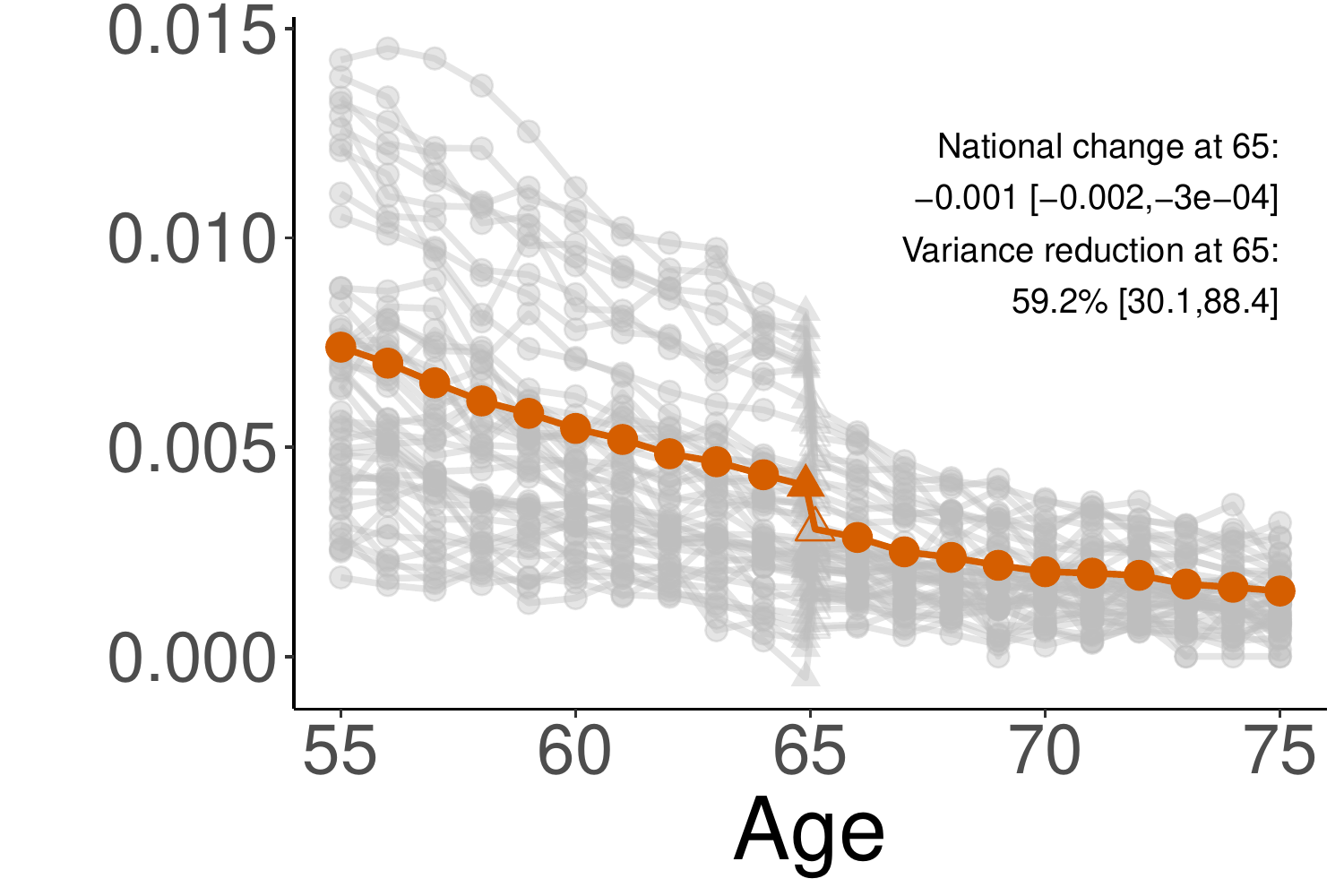}\\
  \textit{Panel E:} \$5,001-10,000 &  \textit{Panel F:} Share over \$10,000  \\
    \includegraphics[width=7.5cm]{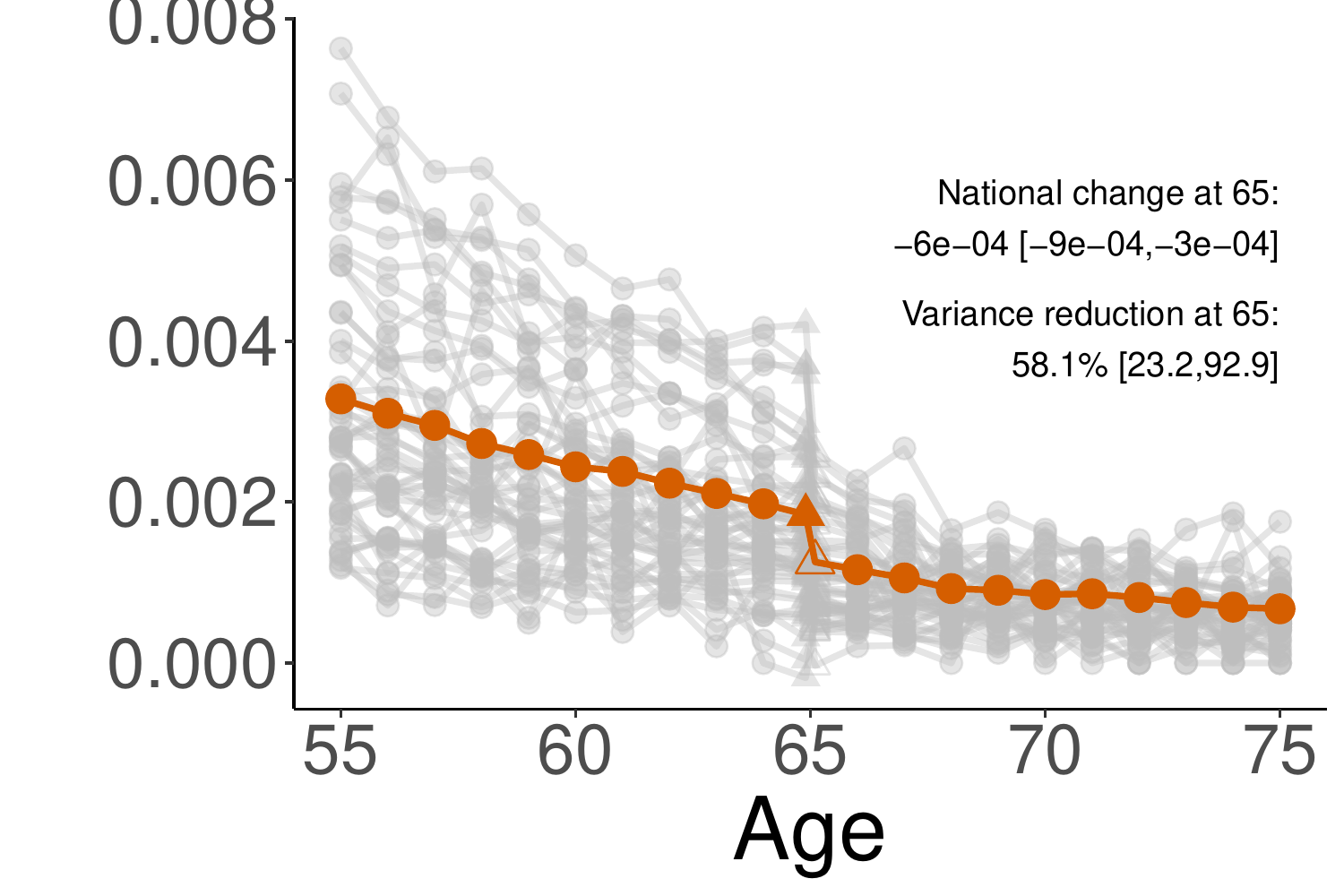} & \includegraphics[width=7.5cm]{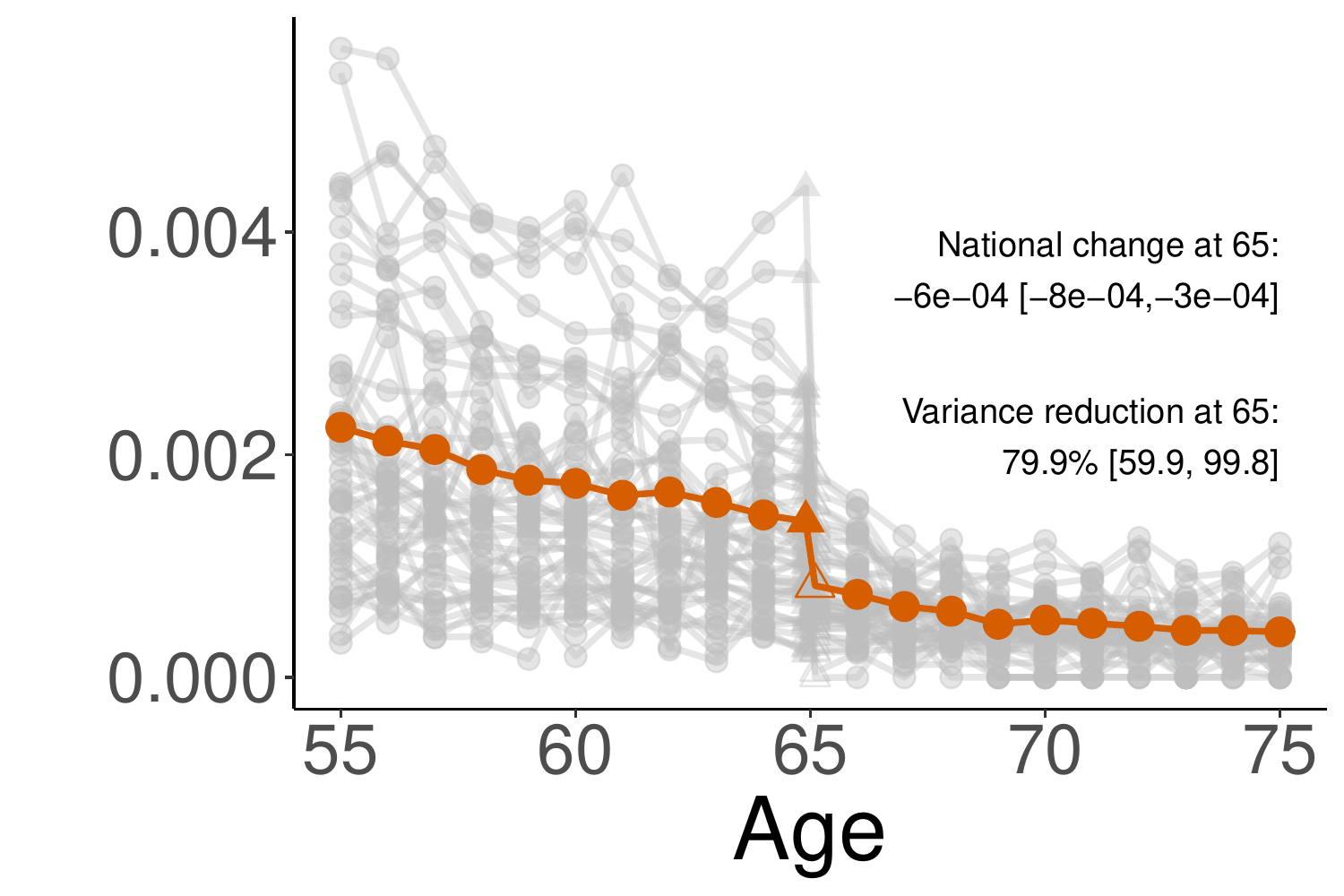}\\
\end{tabular}
\begin{minipage} {0.9\textwidth} \setstretch{.9} \medskip
  \footnotesize{\textbf{Note:} This figure plots the effect of Medicare eligibility at age 65 on the distribution of new collections debts within the past year. A local linear regression is fit on each side of the Medicare eligibility threshold using methods from \cite{kolesar2018inference}. We include hollow points that are the predicted counterfactual outcomes with and without Medicare at 65. The blue hollow dot is the predicted outcome without Medicare at age 65 and the red hollow dot is the predicted consumer credit outcome with Medicare at age 65. Panel A plots the share of individuals with collections debt between \$1-500 by age. Panel B plots the share of individuals with collections debt between \$500-1,000 by age. Panel C plots the share of individuals with collections debt between \$1,000-2,500 by age. Panel D plots the share of individuals with collections debt between \$2,501-5,000 by age. Panel E plots the share of individuals with collections debt between \$5,001-10,000 by age. Panel F plots the share of individuals with collections debt greater than \$10,000 by age. The sample includes individuals who were age 55-75 between 2008 and 2017. See Section \ref{background_data} for additional details on the outcomes and sample. Source: The financial health outcomes are based on 137,340,577 person-year observations from the New York Fed Consumer Credit Panel / Equifax, 2008-2017}
  \end{minipage}
\end{figure}

%%%%%%%%%%%%%%%%%%%%%%%%%%%%%%%%%%%%%%%%%%%%%
% Robustness of Main Age RD to Bound Scaling %
%%%%%%%%%%%%%%%%%%%%%%%%%%%%%%%%%%%%%%%%%%%%%%

\clearpage

\clearpage

%%%%%%%%%%%%%%%%%%%%%%%%%%%%%%%%%%%%%%%%%%%%%%%%%%%%%%%%%%%%%%%%%
% Robustness of Main Age RD Variance Estimates to Bound Scaling %
%%%%%%%%%%%%%%%%%%%%%%%%%%%%%%%%%%%%%%%%%%%%%%%%%%%%%%%%%%%%%%%%%
\clearpage

%%%%%%%%%%%%%%%%%%%%%%%%%%%%%%%%%%%%%%%%%%
% Robustness of Main Age RD to Bandwidth %
%%%%%%%%%%%%%%%%%%%%%%%%%%%%%%%%%%%%%%%%%%
\clearpage

%%%%%%%%%%%%%%%%%%%%%%%%%%%%
% Smoothness of covariates %
%%%%%%%%%%%%%%%%%%%%%%%%%%%%
\clearpage
\begin{figure}[htpb!]
  \centering
  \caption{Smoothness of covariates at age 65}
  \label{fig:covariates_ageRD_apx}
\begin{tabular}{cc}
  \textit{Panel A:}  Share homeowners by age in years &   \textit{Panel B:} Share married by age in years \\
\includegraphics[width=7.5cm]{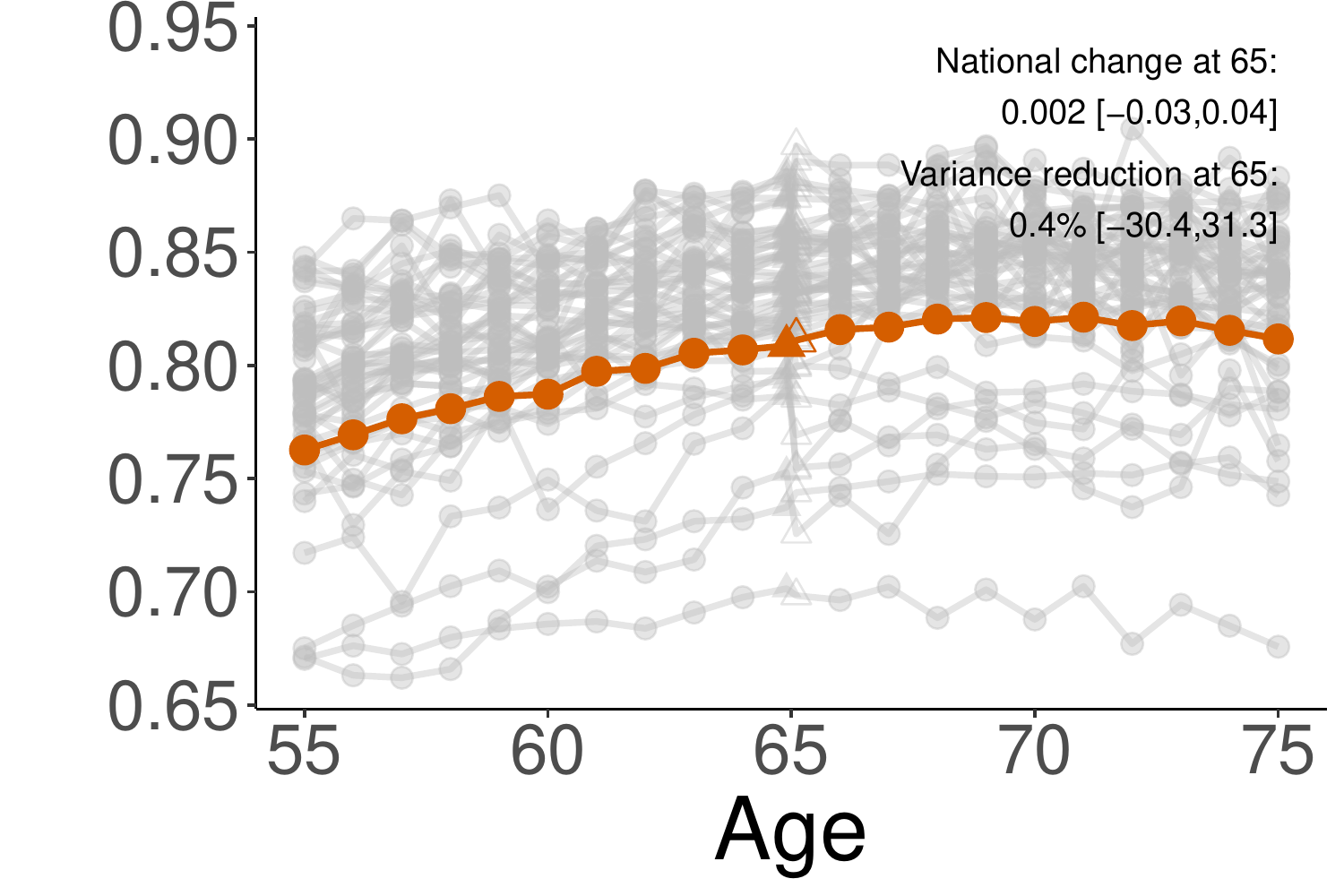} & \includegraphics[width=7.5cm]{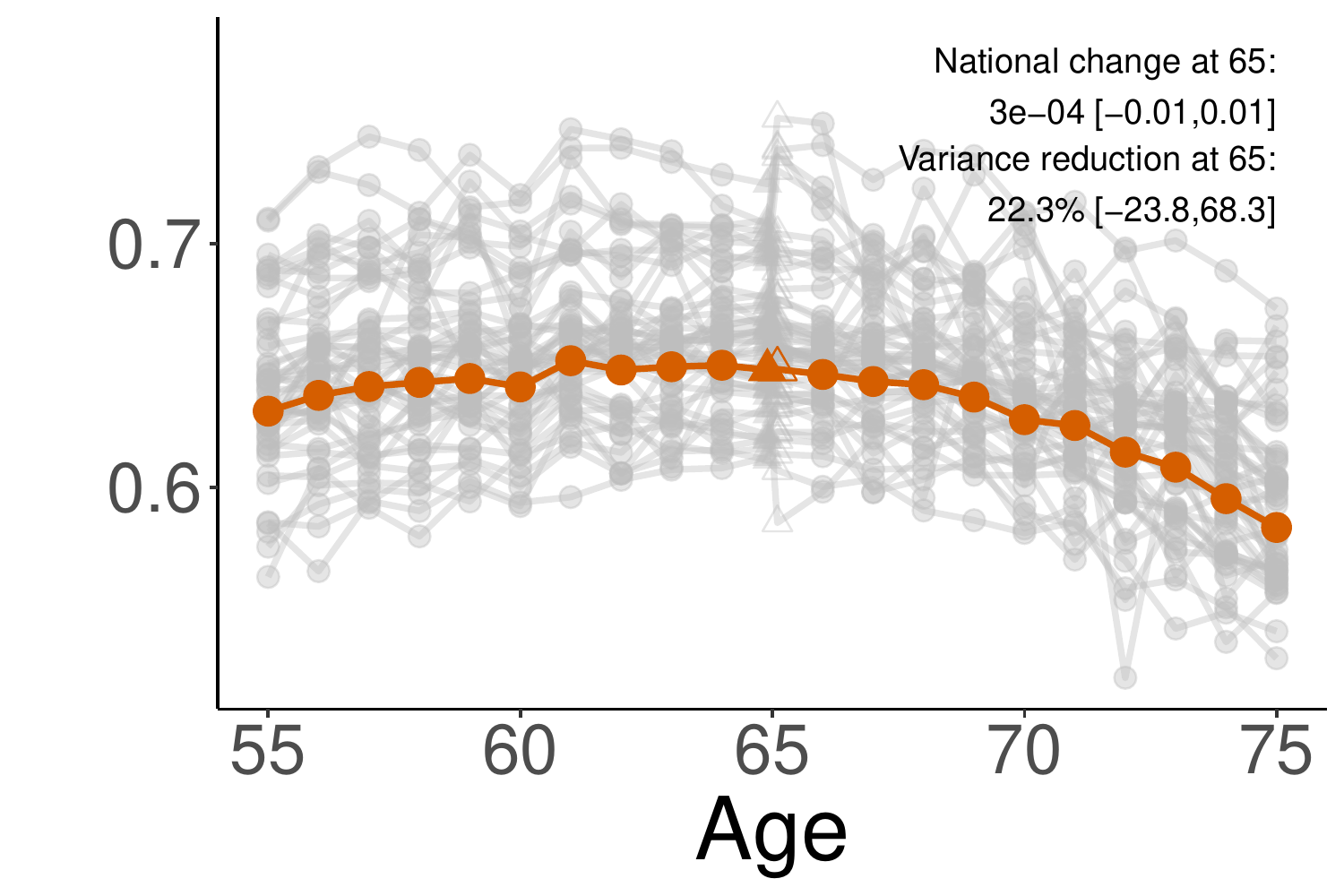}\\
  \textit{Panel C:} Usual hours worked per week &  \textit{Panel D:} Social Security income \\
 \includegraphics[width=7.5cm]{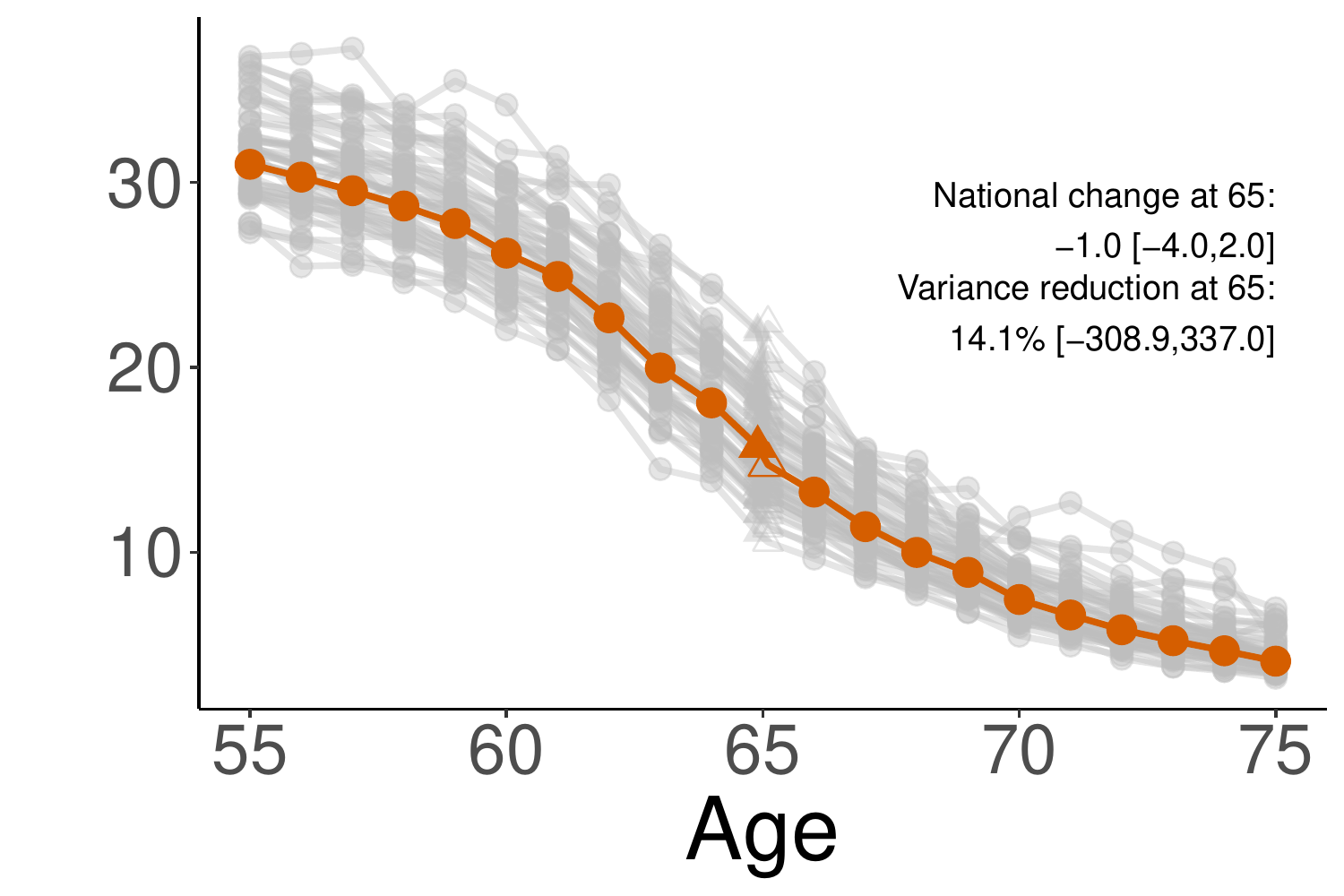} & \includegraphics[width=7.5cm]{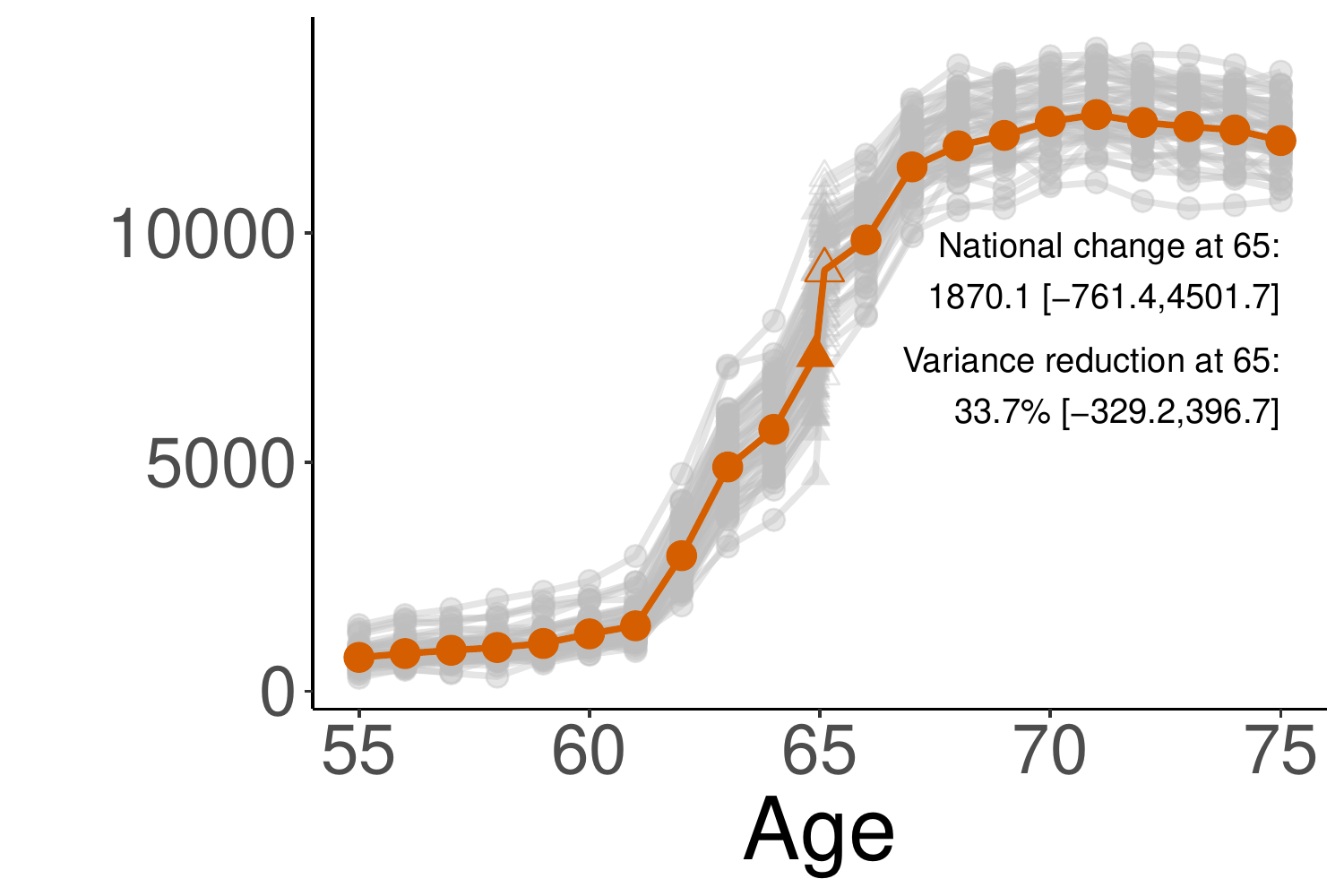} \\
\end{tabular}
\begin{minipage} {0.9\textwidth} \setstretch{.9} \medskip
  \footnotesize{\textbf{Note:} This figure plots a series of individual covariates by age. The horizontal axis denotes age in years. A local linear regression is fit on each side of the Medicare eligibility threshold using methods from \cite{kolesar2018inference}. We include hollow points that are the predicted counterfactual outcomes with and without Medicare at 65. The blue hollow dot is the predicted covariate without Medicare and the red hollow dot is the predicted covariate with Medicare. Panel A plots homeownership rates by age. Panel B reports the share married by age. Panel C plots weekly hours worked by age. Panel D plots social security income by age. The sample includes individuals who were age 55-75 between 2008 and 2017. See Section \ref{background_data} for additional details on the outcomes and sample. Source: American Community Survey, 2008-2017.}
  \end{minipage}
\end{figure}

%%%%%%%%%%%%%%%%%%%%%%%%%%%%%%%%%
% Covariate smoothness by state %
%%%%%%%%%%%%%%%%%%%%%%%%%%%%%%%%%
\clearpage
%\begin{landscape}
\begin{figure}[htpb!]
  \centering
  \caption{Covariate smoothness by state}
  \label{fig:cov_smoothness_state_apx}
\begin{tabular}{ccc}
  \textit{Panel A:} Homeowner &    \textit{Panel B:} Married &  \textit{Panel C:} Employed \\
 \includegraphics[width=4.5cm]{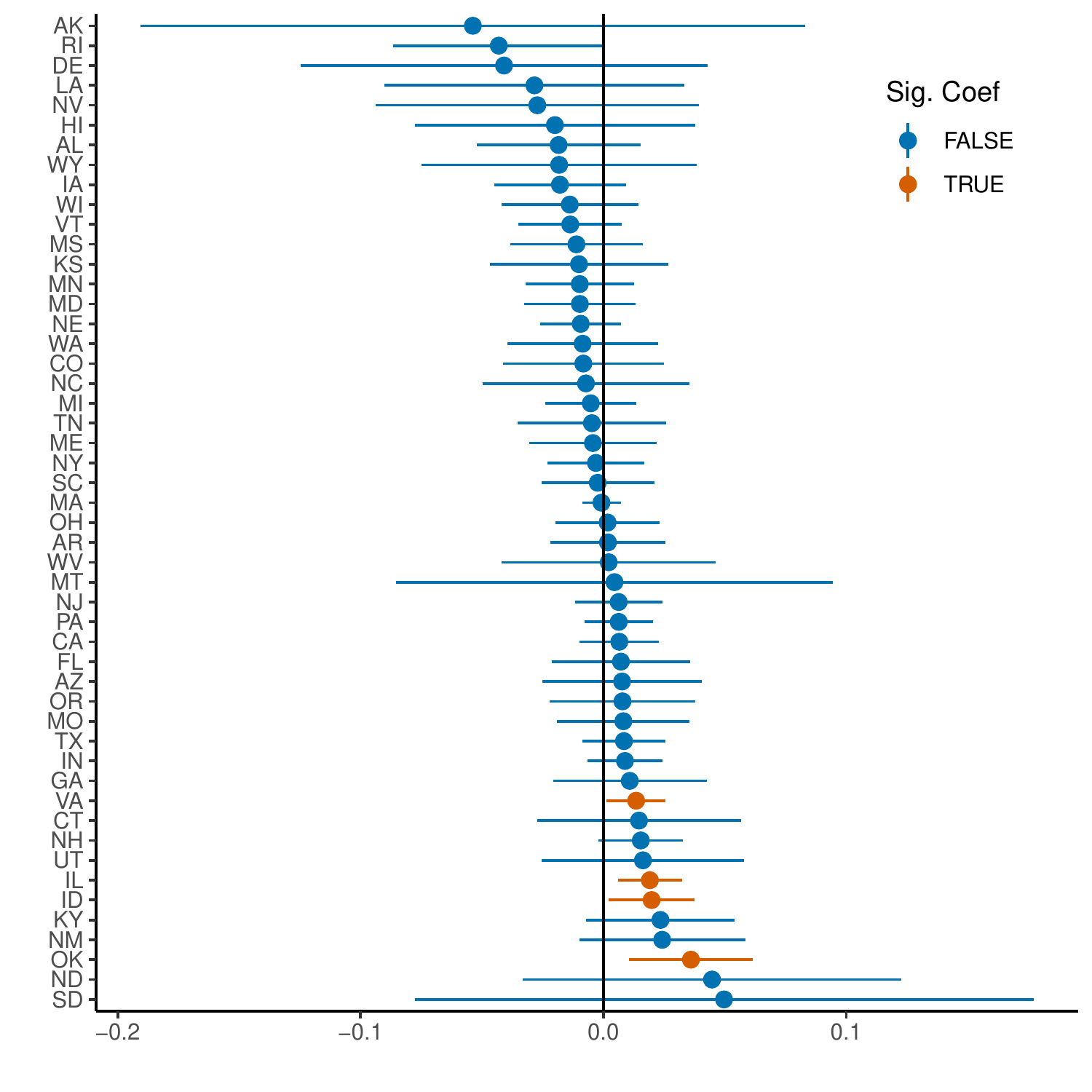} &   \includegraphics[width=4.5cm]{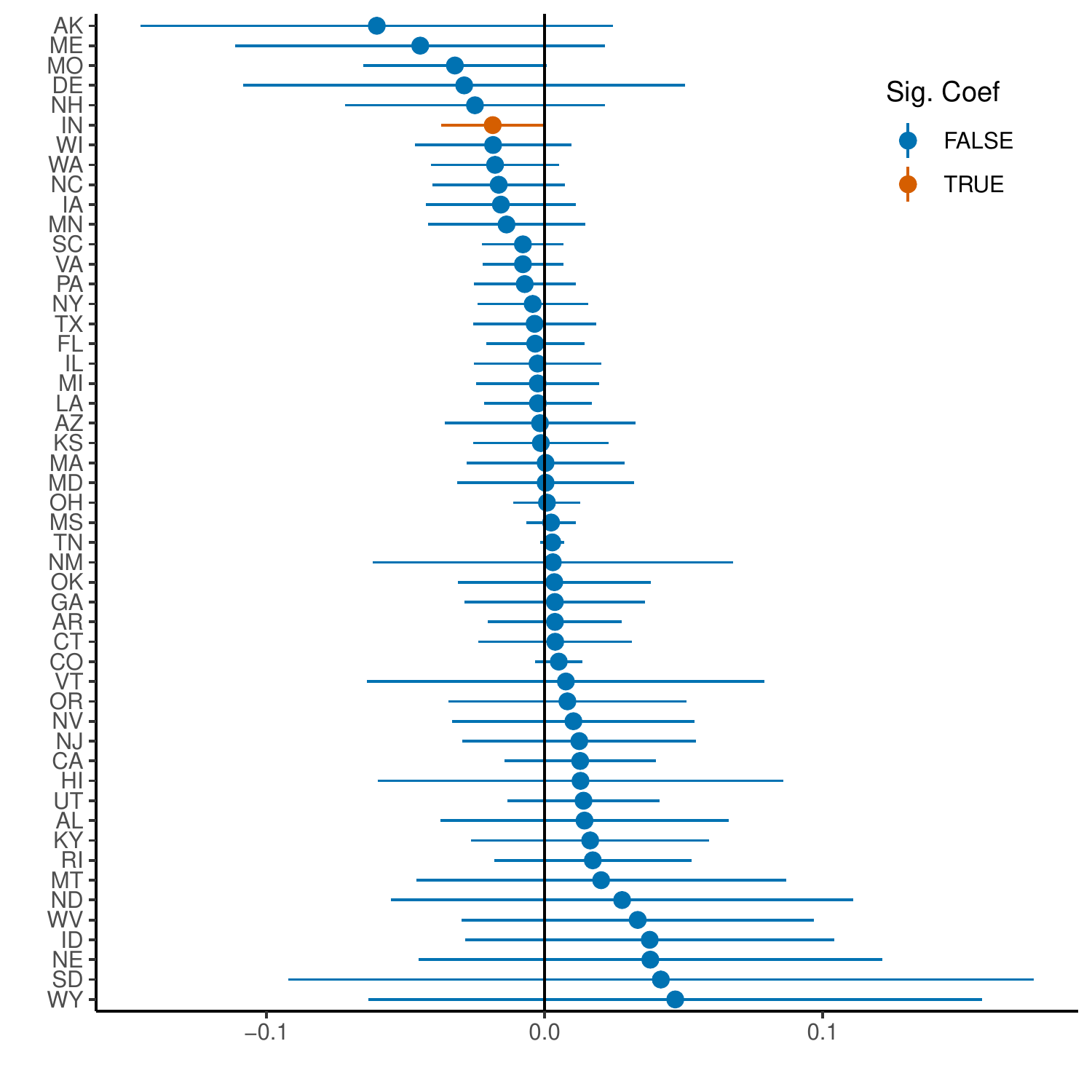} &   \includegraphics[width=4.5cm]{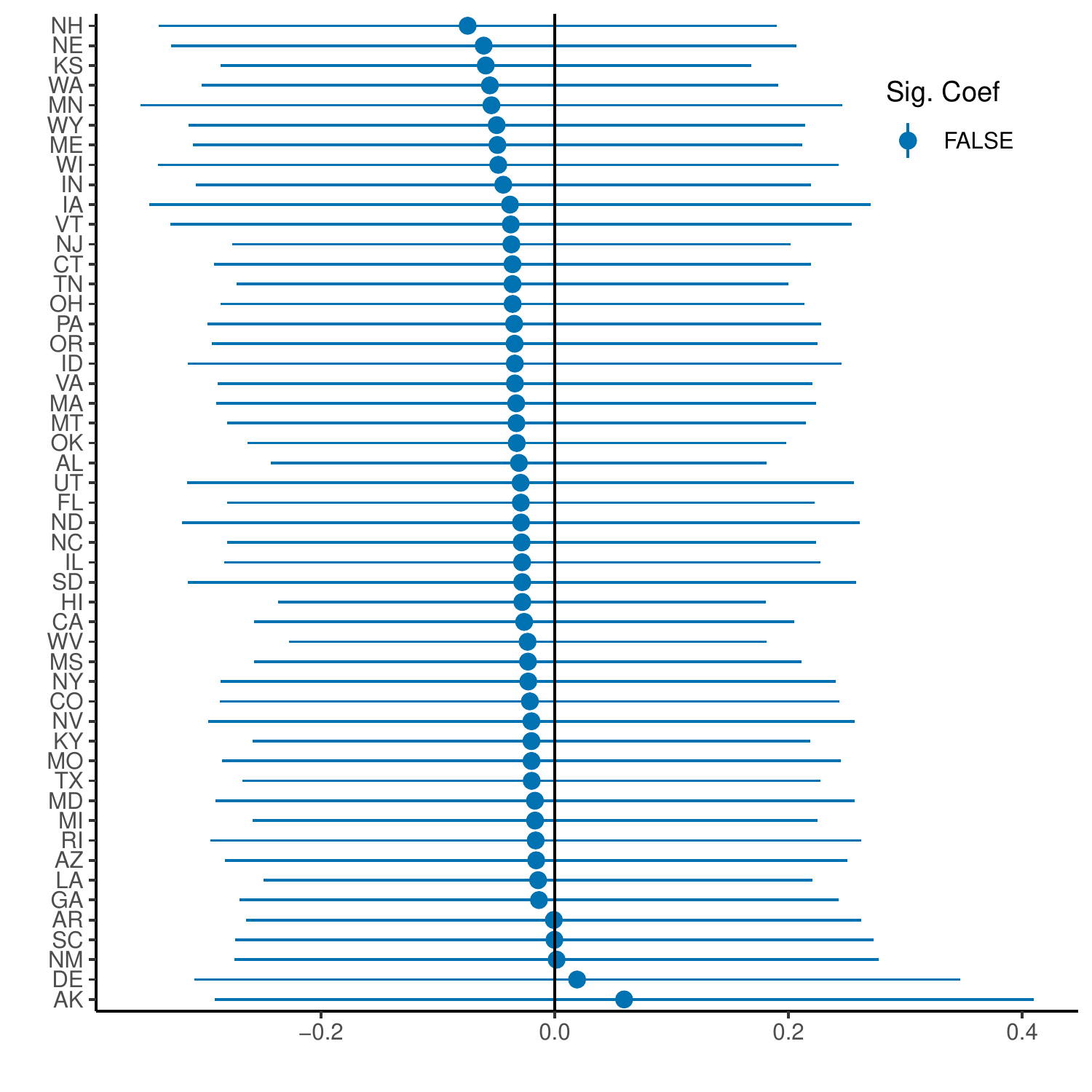} \\ 
\textit{Panel D:} Income &  \textit{Panel E:} Hours worked &  \textit{Panel F:} Social Security Income \\
 \includegraphics[width=4.5cm]{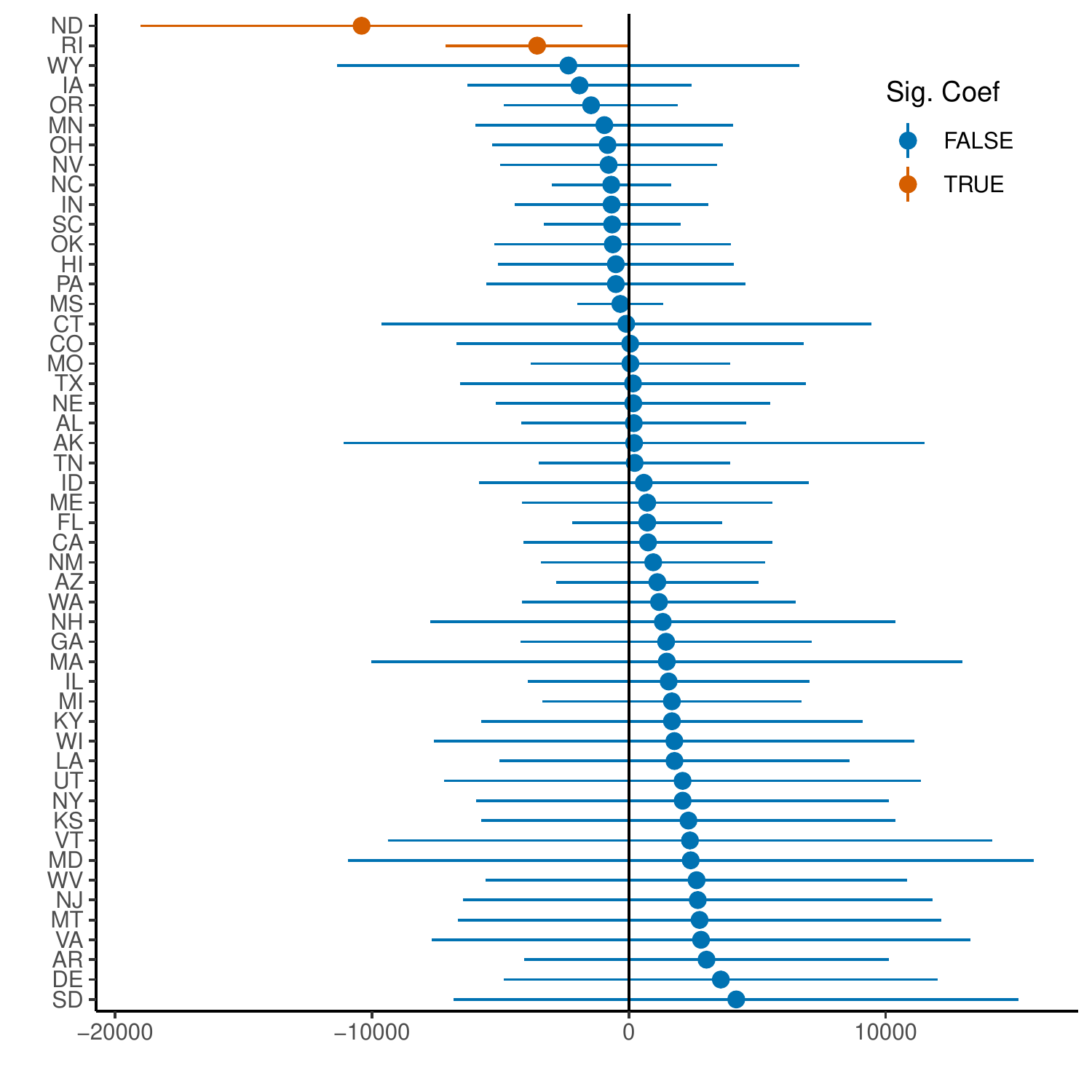} &   \includegraphics[width=4.5cm]{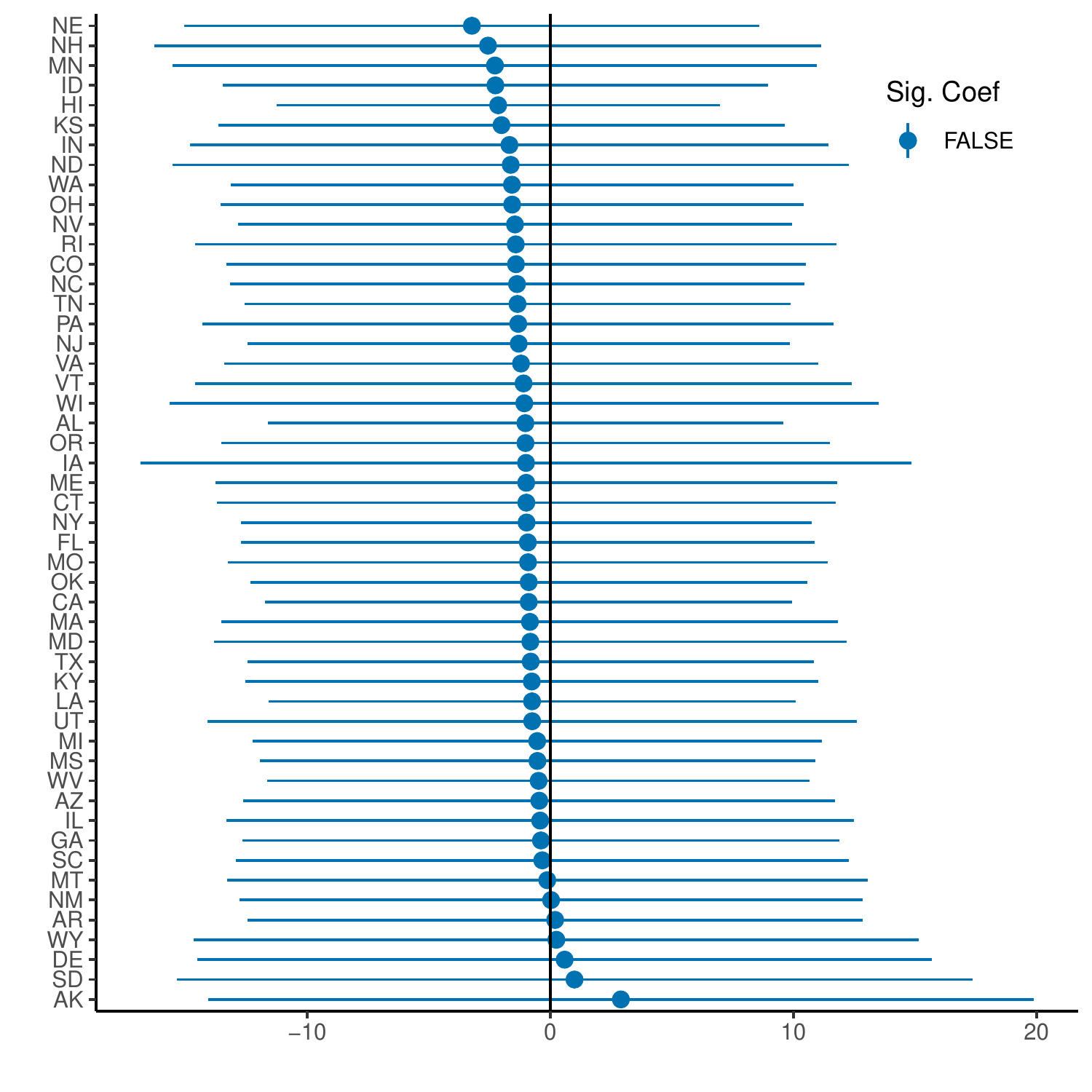} &   \includegraphics[width=4.5cm]{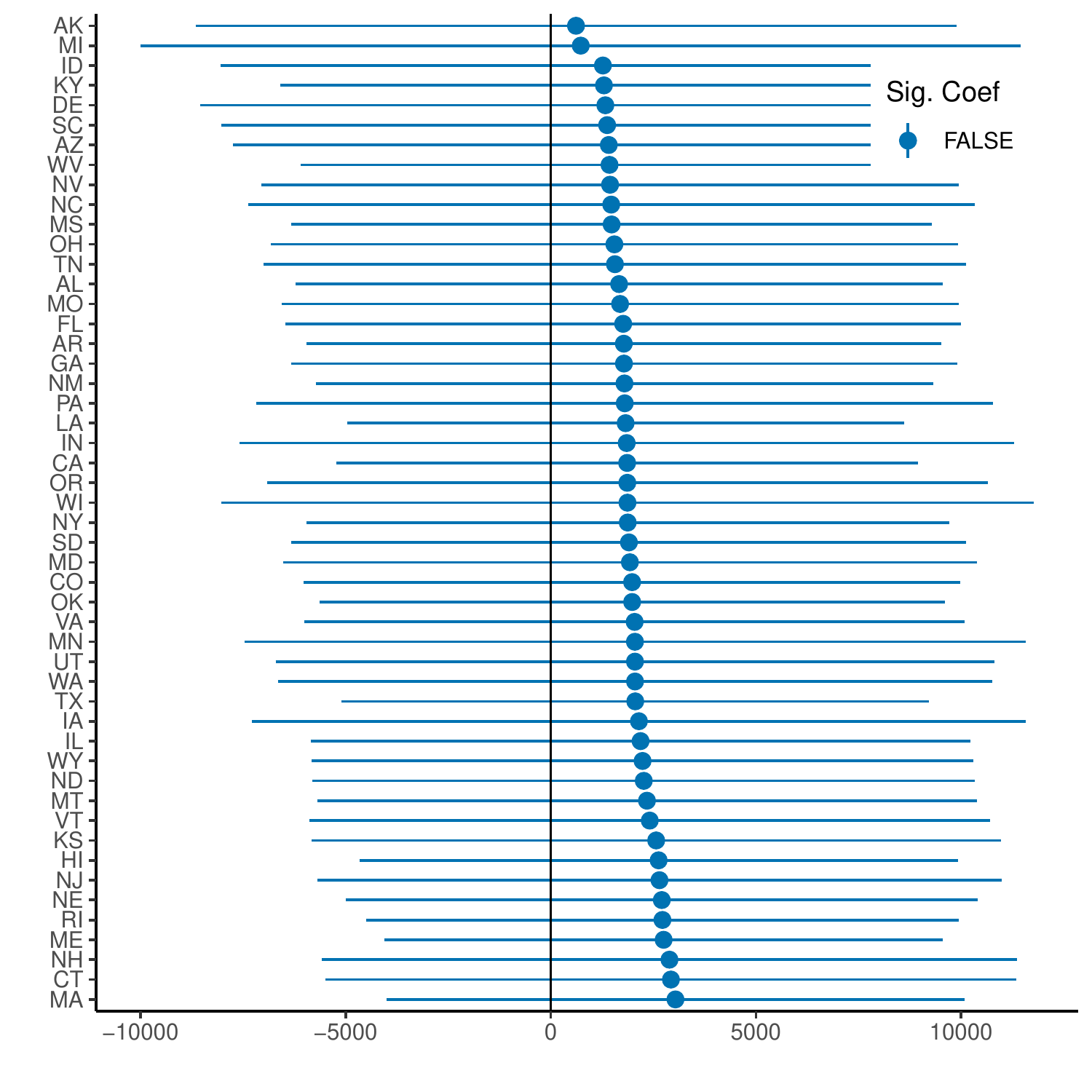} \\ [.5cm] 
\end{tabular}
	\begin{minipage} {0.9\textwidth} \setstretch{.9} \medskip
          \footnotesize{\textbf{Note:} This figure plots the estimated discontinuities at age 65 in individual covariates by state. A local linear regression is fit on each side of the Medicare eligibility threshold using methods from \cite{kolesar2018inference}. The red bars indicate statistically significant results at the 5\% level. Panel A plots the estimated discontinuities in homeownership rates by state. Panel B plots the estimated discontinuities in marriage rates by state. Panel C plots the estimated discontinuities in employment by state. Panel D plots the estimated discontinuities in income by state. Panel E plots the estimated discontinuities in hours worked by state. Panel F plots the estimate discontinuities in social security income by state. The sample includes individuals who were age 55-75 between 2008 and 2017. See Section \ref{background_data} for additional details on the outcomes and sample. Source: American Community Survey, 2008-2017.}
	\end{minipage}
\end{figure}
%\end{landscape}

%%%%%%%%%%%%%%%%%%%%%%%%%%%%%%%%%
% Covariate smoothness by CZ %
%%%%%%%%%%%%%%%%%%%%%%%%%%%%%%%%%
\clearpage
%\begin{landscape}
\begin{figure}[htpb!]
  \centering
  \caption{Covariate smoothness by commuting zone}
  \label{fig:cov_smoothness_cz_apx}
\begin{tabular}{cc}
  \textit{Panel A:} Homeowner &    \textit{Panel B:} Married  \\
 \includegraphics[width=7.5cm]{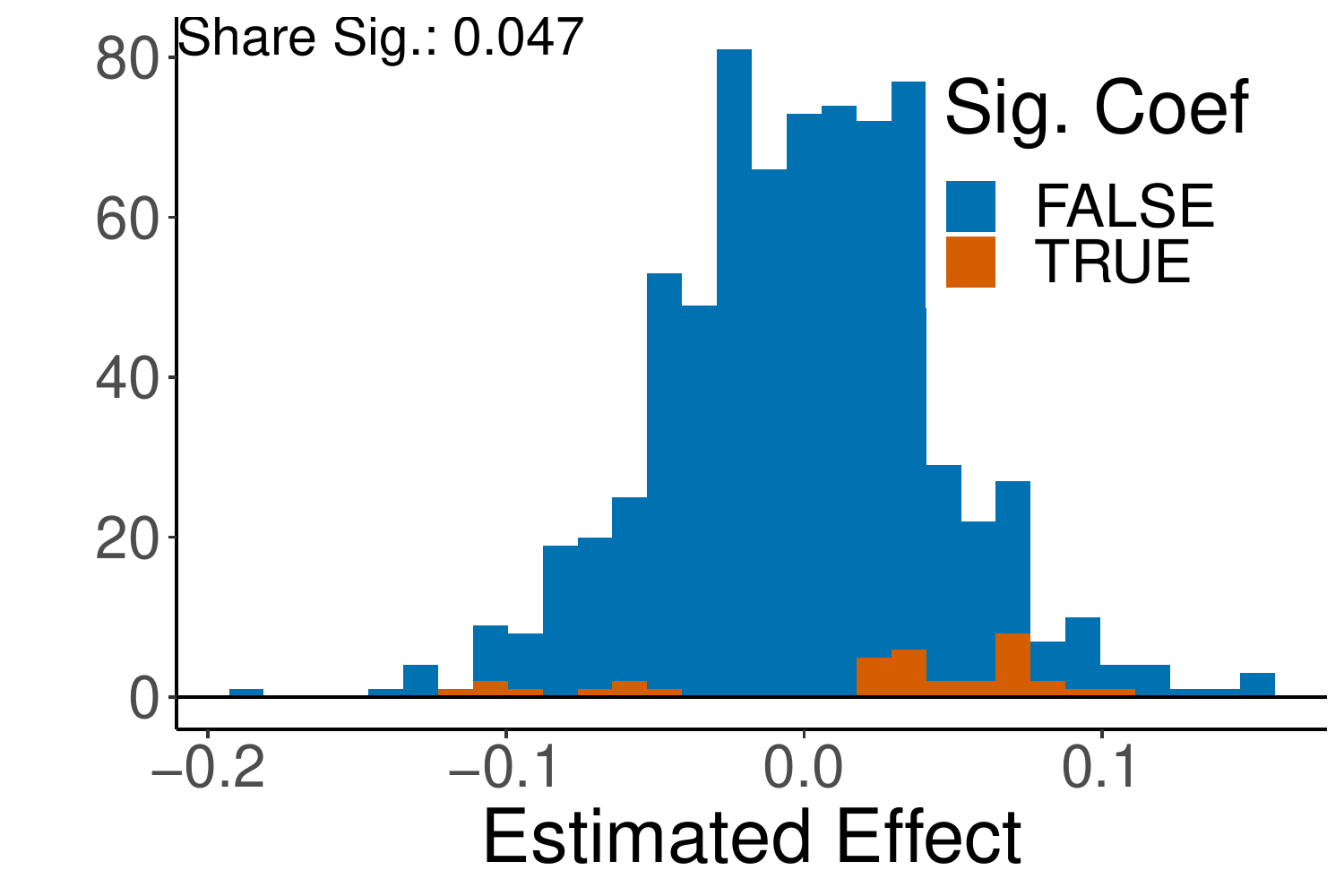} &   \includegraphics[width=7.5cm]{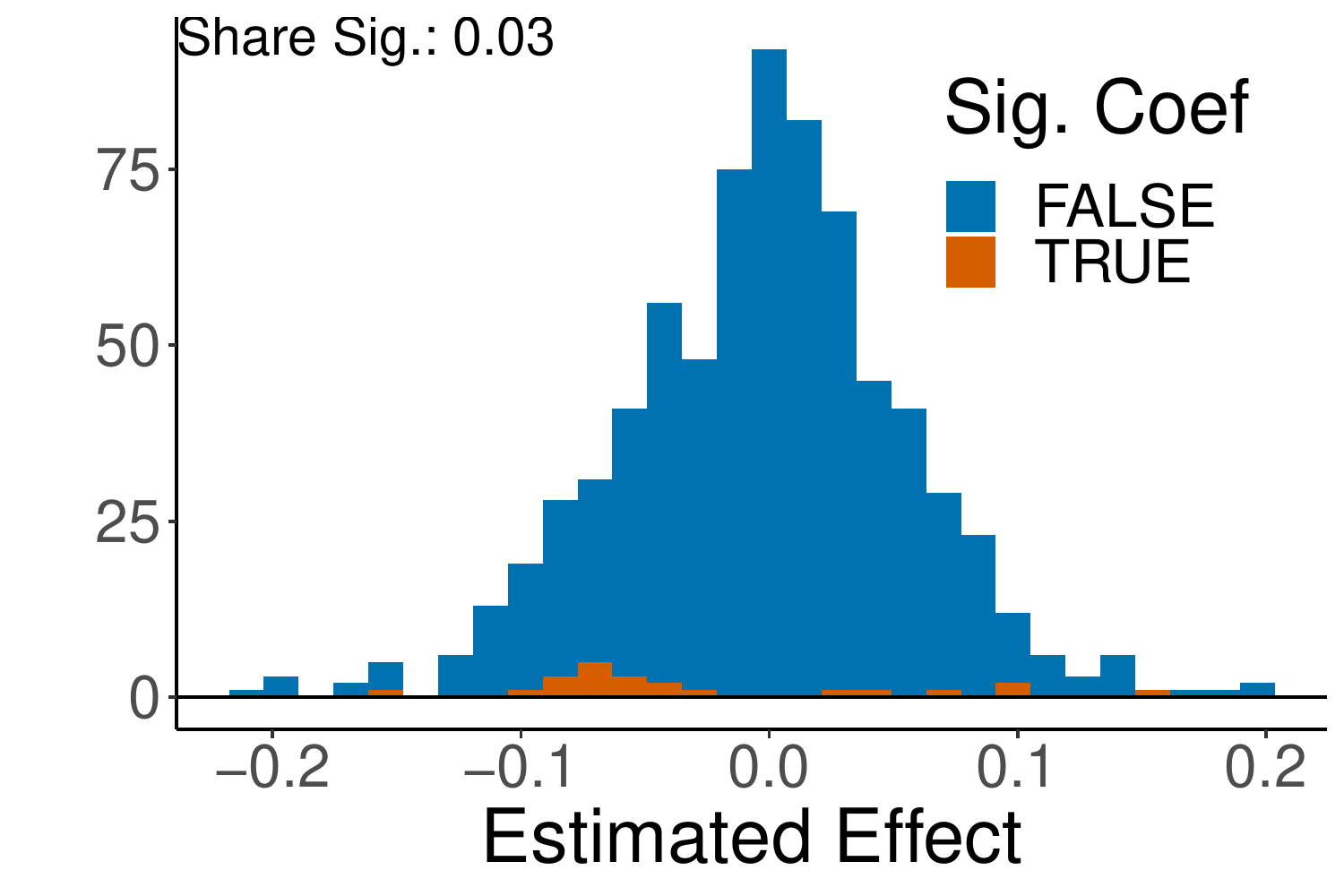} \\
 \textit{Panel C:} Employed & \textit{Panel D:} Income \\
 \includegraphics[width=7.5cm]{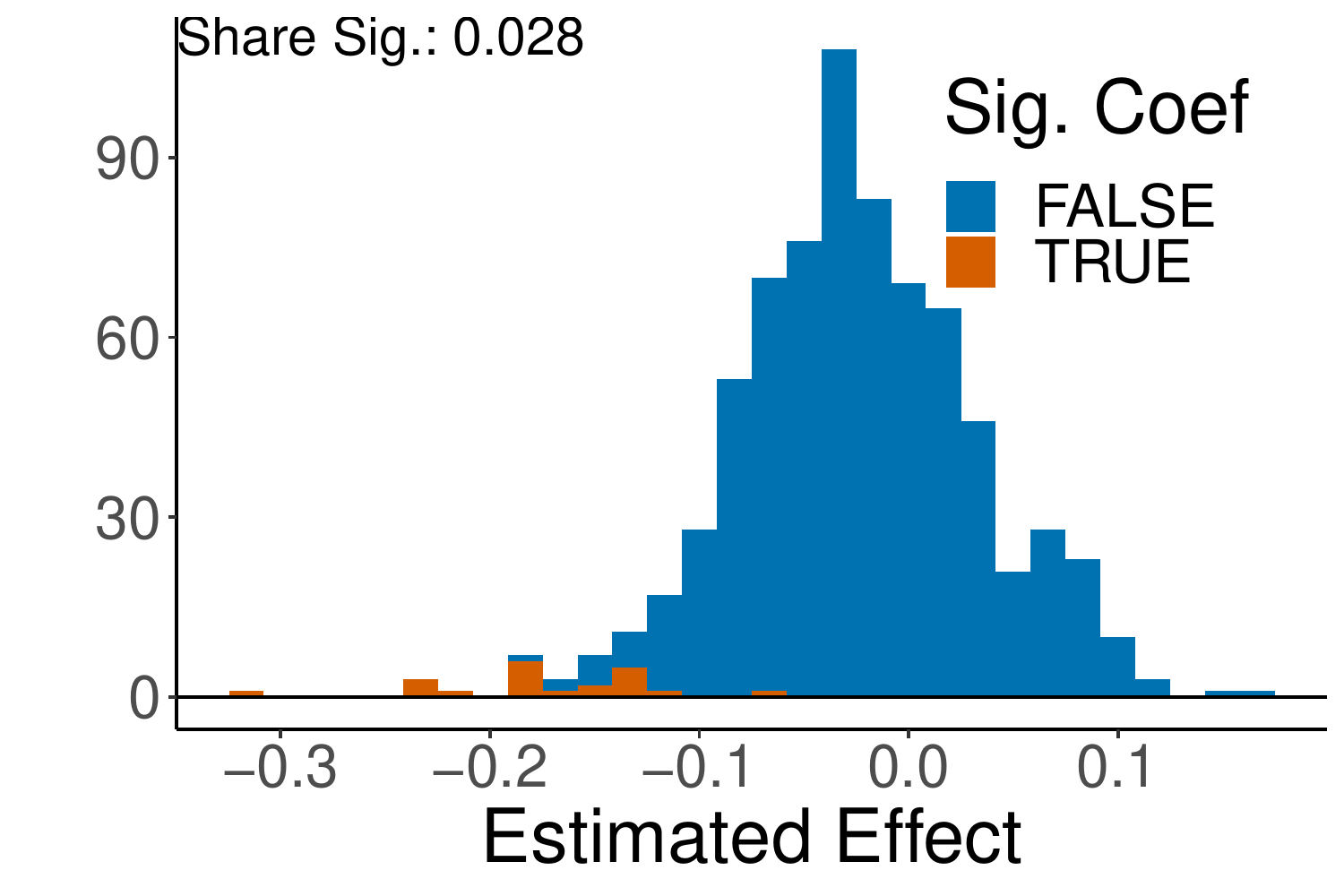} &
 \includegraphics[width=7.5cm]{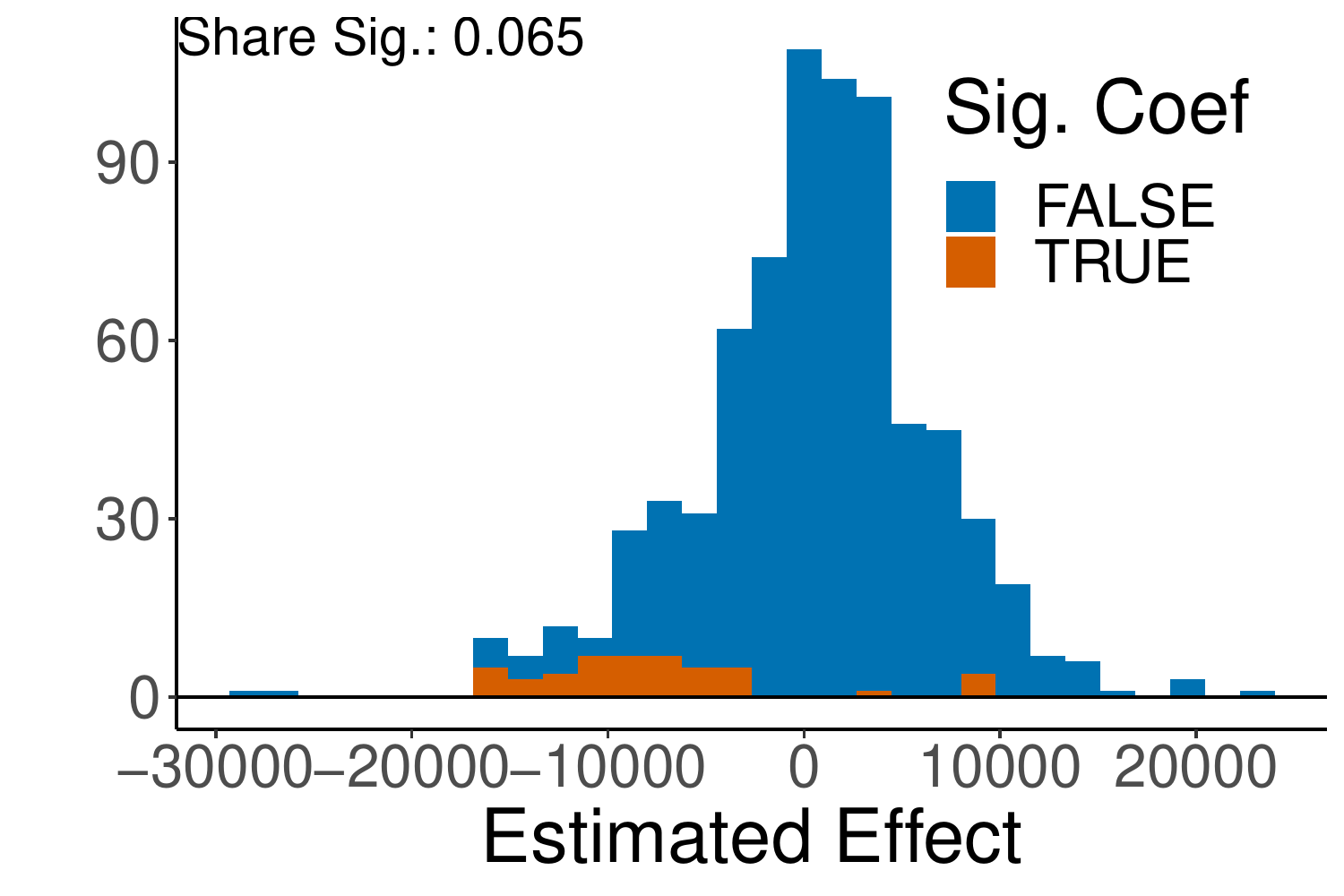} \\
 \textit{Panel E:} Hours worked & \textit{Panel F:} Social Security Income \\
 \includegraphics[width=7.5cm]{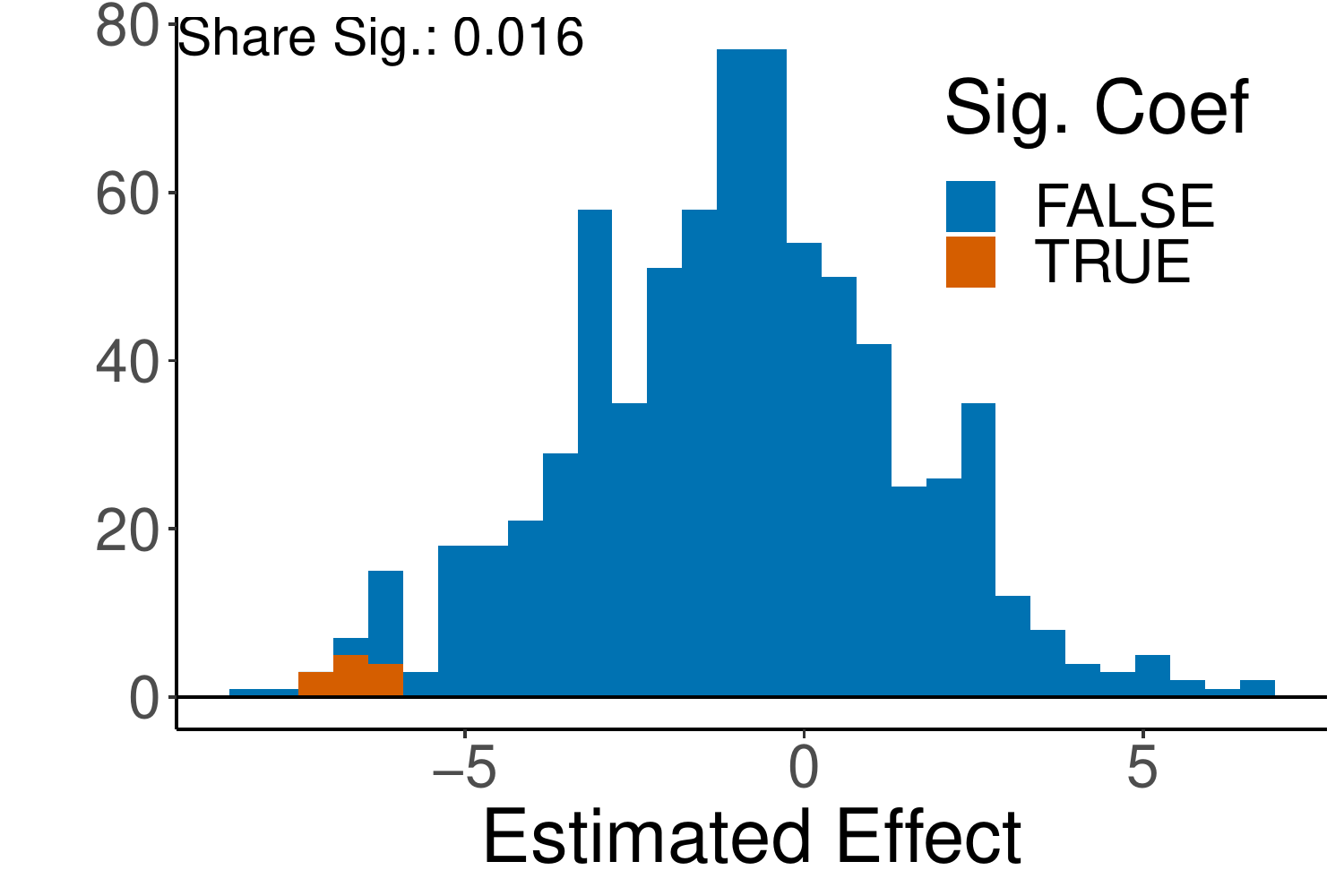} &   \includegraphics[width=7.5cm]{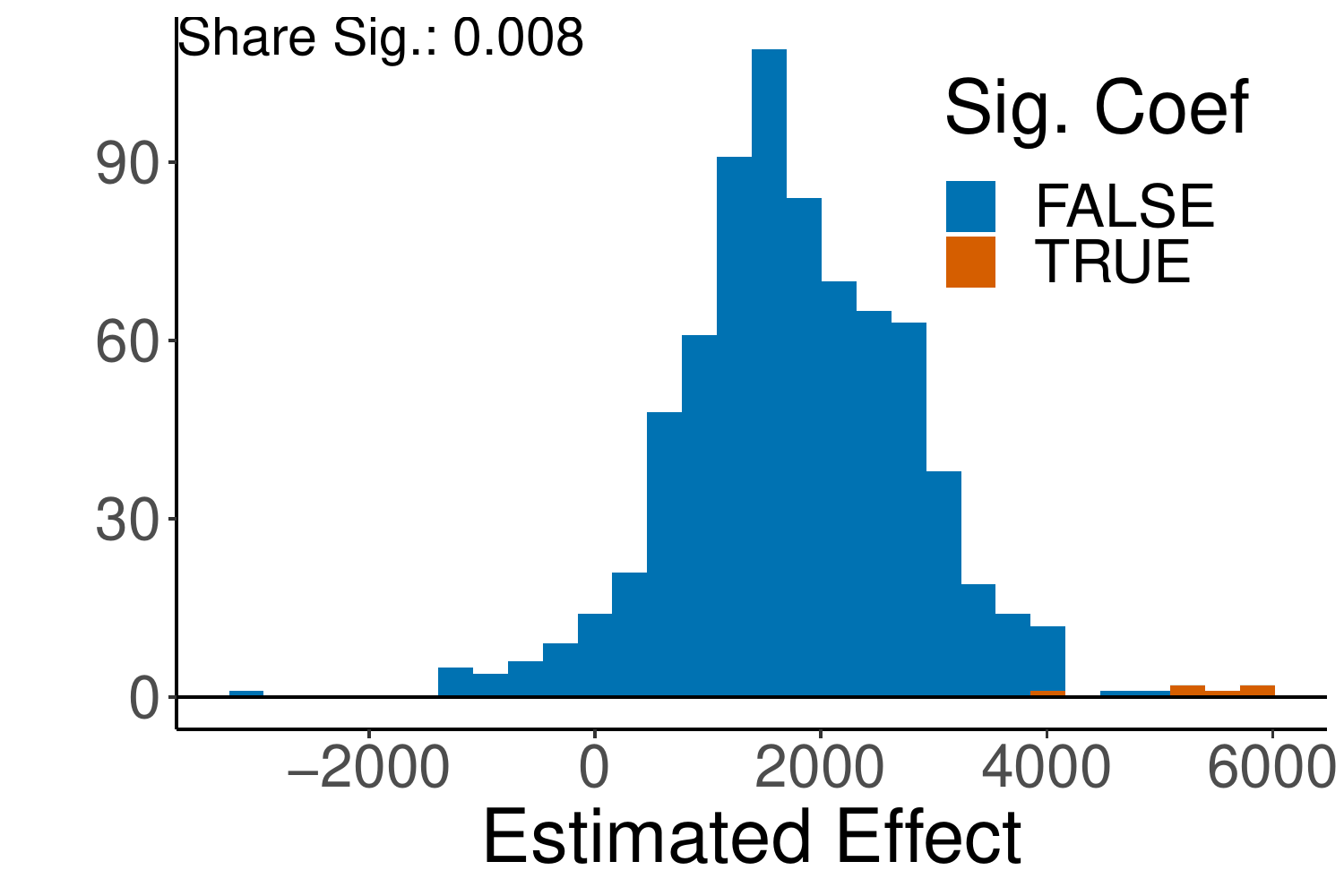} \\
\end{tabular}
	\begin{minipage} {0.95\textwidth} \setstretch{.9} \medskip
          \footnotesize{\textbf{Note:} This figure plots histograms of normalized estimated discontinuities at age 65 in individual covariates by commuting zone (CZ). A local linear regression is fit on each side of the Medicare eligibility threshold by CZ using methods from \cite{kolesar2018inference}. The red bars indicate statistically significant results at the 5\% level. The blue bars indicate statistically insignificant results at the 5\% level. Panel A plots the estimated discontinuities in homeownership rates by state. Panel B plots the estimated discontinuities in marriage rates by state. Panel C plots the estimated discontinuities in employment by state. Panel D plots the estimated discontinuities in income by state. Panel E plots the estimated discontinuities in hours worked by state. Panel F plots the estimate discontinuities in social security income by state. The sample includes individuals who were age 55-75 between 2008 and 2017. See Section \ref{background_data} for additional details on the outcomes and sample. Source: American Community Survey, 2008-2017.}
	\end{minipage}
\end{figure}
%\end{landscape}

%%%%%%%%%%%%%%%%%%%%%%%%%%%%%%%%%%%%%%%%%%%%%%%%%%%%%%%%%%%%%%%%%%%%%%%%%%%%%%%%%%%
%Reduced form relationship between consumer credit outcomes and uninsurance rates %
%%%%%%%%%%%%%%%%%%%%%%%%%%%%%%%%%%%%%%%%%%%%%%%%%%%%%%%%%%%%%%%%%%%%%%%%%%%%%%%%%%%
%% CURRENT VERSION IS FOR 55-64 YEAR-oLDS, WHAT DO WE ACTUALLY WANT FOR PITCH?
\clearpage
\begin{figure}[htpb!]
  \centering
  \caption{Consumer credit outcomes and uninsurance rates across states}
\label{fig:uninsurance_vs_finhealth_state_apx}
\begin{tabular}{cc}
  \textit{Panel A:} Debt collections &   \textit{Panel B:} Bankruptcy \\
\includegraphics[width=3in]{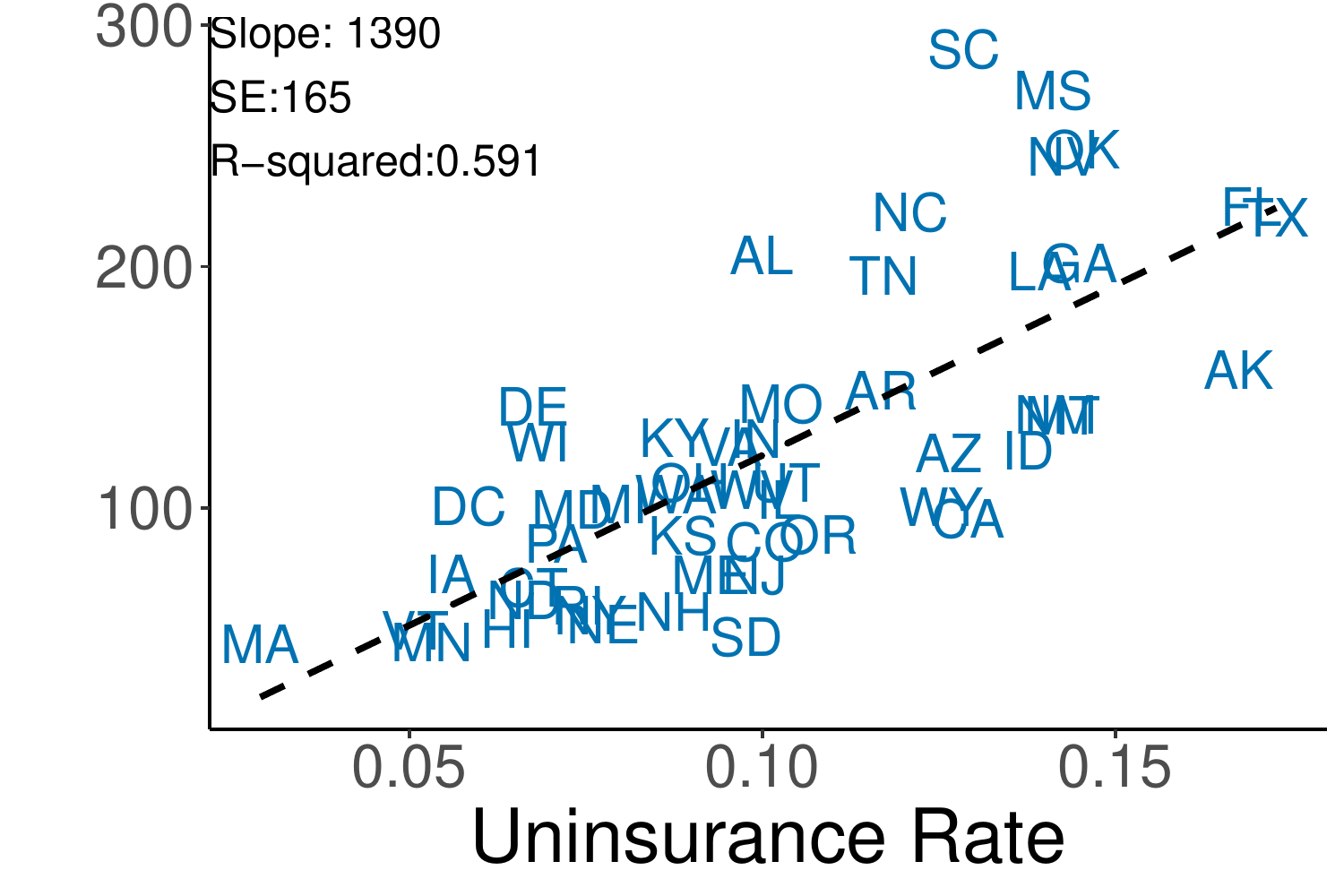} & \includegraphics[width=3in]{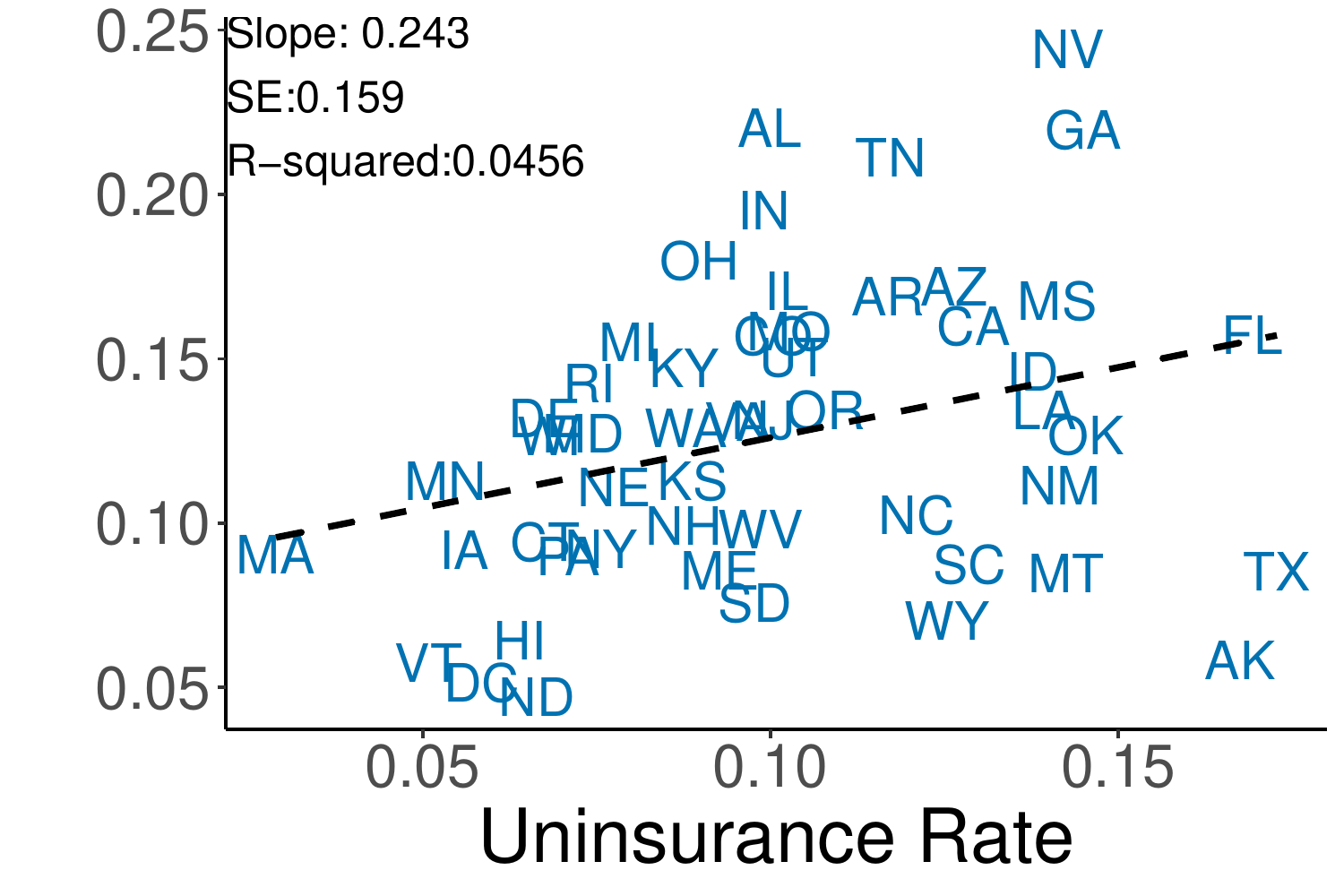}\\
  \textit{Panel C:} Share debt past due &  \textit{Panel D:} Credit Score \\
\includegraphics[width=3in]{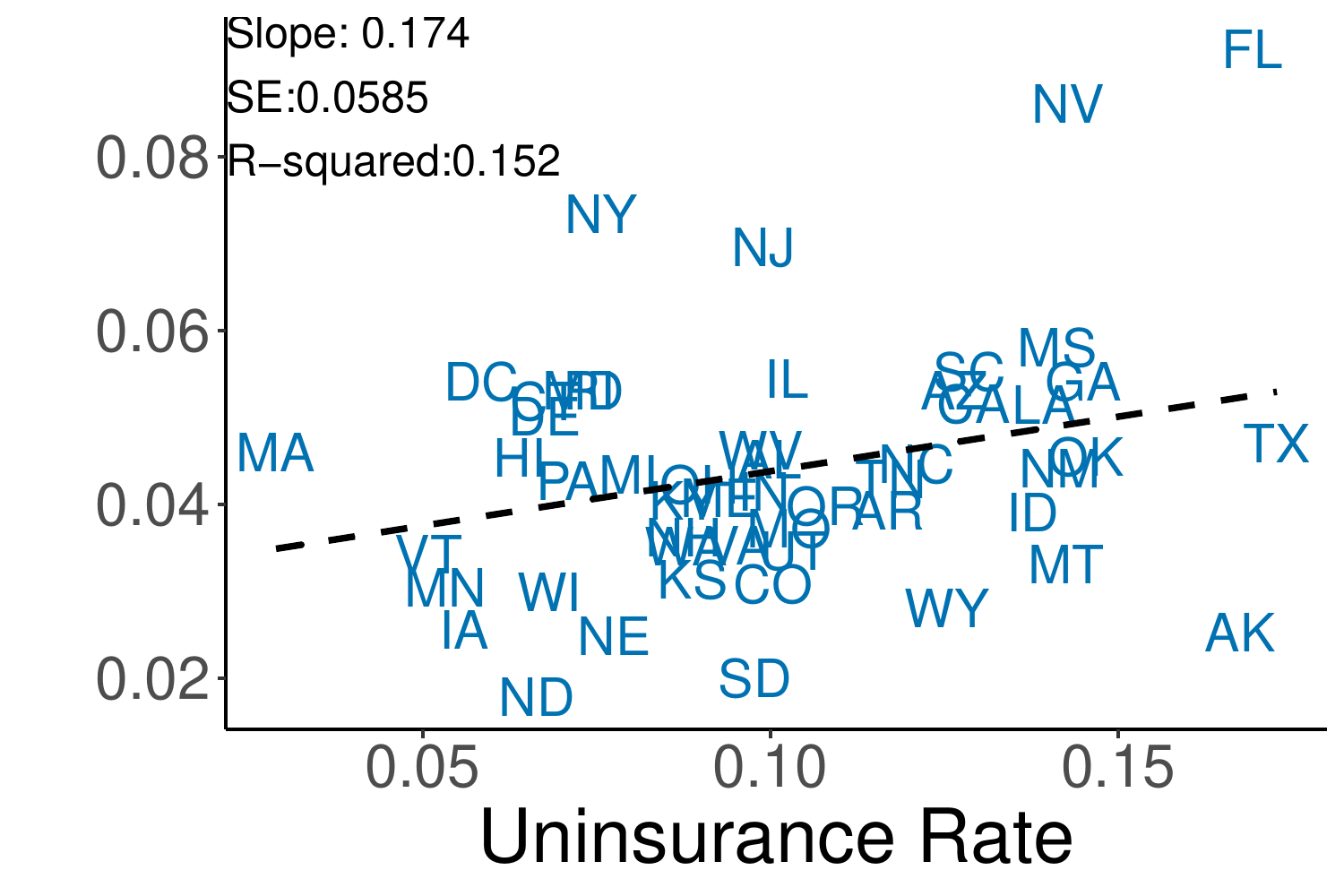} & \includegraphics[width=3in]{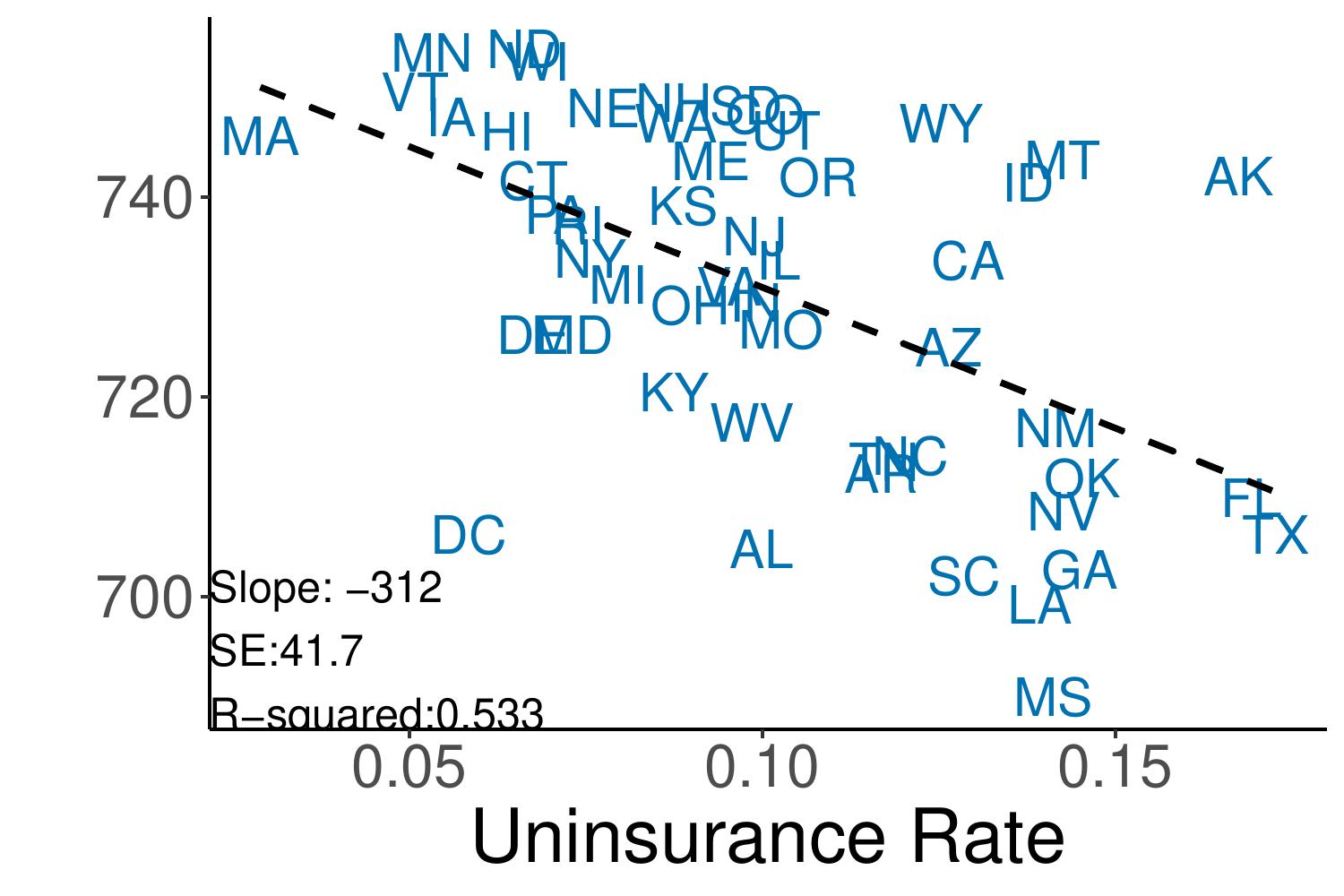} 
\end{tabular}
\begin{minipage} {0.9\textwidth} \setstretch{.9} \medskip
  \footnotesize{\textbf{Note:} This figure plots consumer credit outcomes against state uninsurance rates for individuals aged 55-64 years old. Panel A plots the dollar value of new collections debt reported on credit reports annually. Panel B plots the annual rate of new bankruptcies in percentage points. Panel C plots the share of debt that is more than 30 days past due. The share debt past due is calculated as the average individual's debt more than 30 days past due, divided by the average total debt of all individuals of the same age living in that state. %We divide by this average, rather than individuals' own debt levels, to avoid the divide-by-zero problem. 
  Panel D plots credit score data using the Equifax Risk Score 3.0.  See Section \ref{background_data} for additional details on the outcomes and sample. Source: Consumer credit outcomes are based on 137,340,577 person-year observations from the New York Fed Consumer Credit Panel / Equifax, 2008-2017. State-level uninsurance rates are from the American Community Survey, 2008-2017.
  }
\end{minipage}
\end{figure}

%%%%%%%%%%%%%%%%%%%%%%%%%%%%%%%%%%%%%%%%%%%%%%%%%%%%%%%%%%%%%%%%%%%%%%%
% State-level reductions in collections vs. change in insurance at 65 %
%%%%%%%%%%%%%%%%%%%%%%%%%%%%%%%%%%%%%%%%%%%%%%%%%%%%%%%%%%%%%%%%%%%%%%%
\clearpage
\begin{figure}[htpb!]
  \centering
  \caption{Effect of Medicare eligibility on the level of collections debt at age 65 vs. effects on insurance}
  \label{fig:drop_pctui_binscatter}
  \makebox[\textwidth][c]{
  \begin{tabular}{cc}
        \textit{Panel A:} State-level estimates &         \textit{Panel B:}  CZ-level estimates \\
        \includegraphics[width=3.5in]{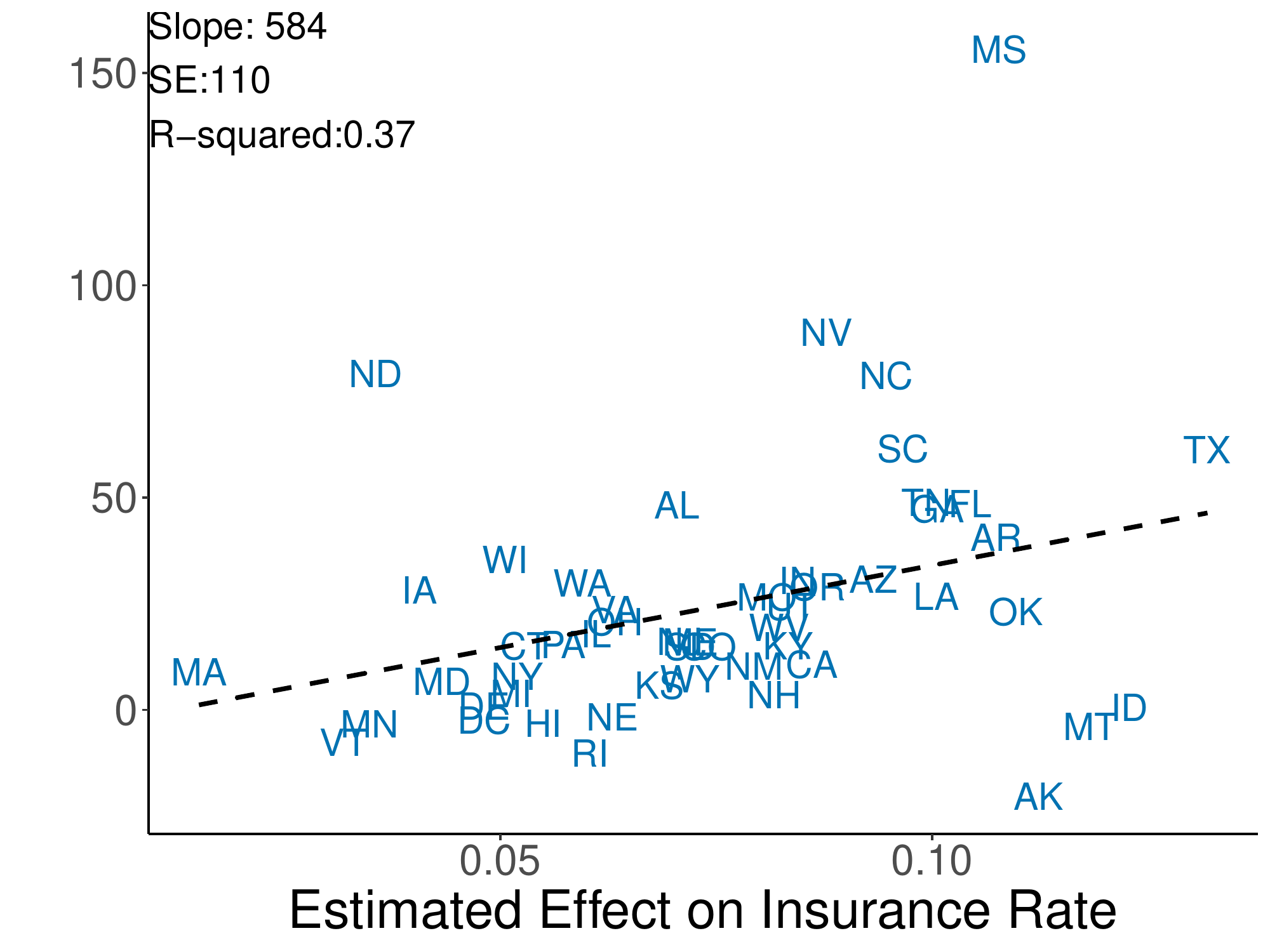} &
        \includegraphics[width=3.5in]{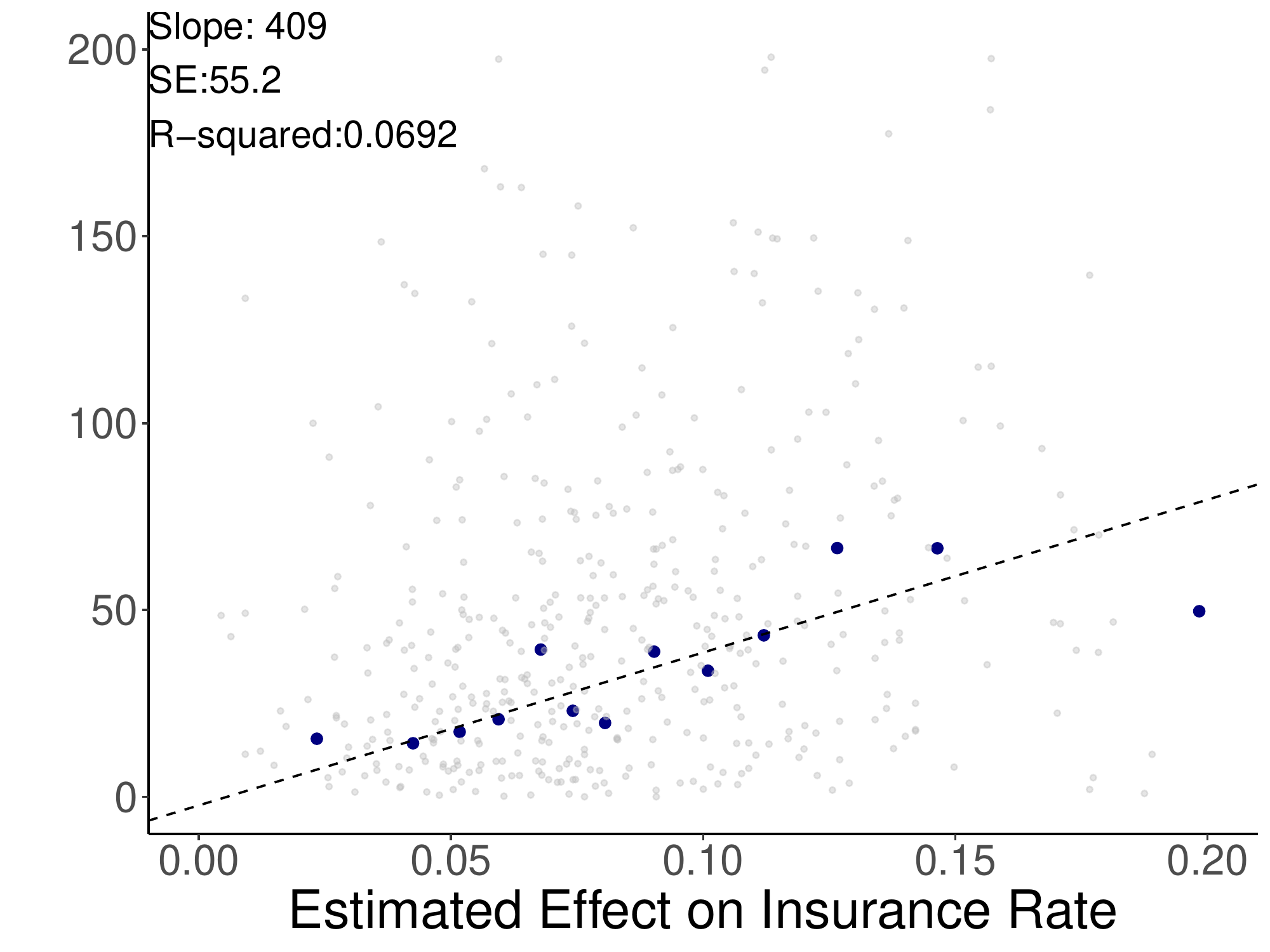}
\end{tabular}}
 
  \begin{minipage} {0.9\textwidth} \setstretch{.9} \medskip
    \footnotesize{\textbf{Note:} This figure plots point estimates of the reduction in the flow of newly-reported collections debt (within the past year) and the increase in the insurance rate at age 65 based on local linear regressions, done separately by state and commuting zone (CZ), using the methods from \cite{kolesar2018inference}. Panel A plots state-level estimates. Panel B plots CZ-level estimates, where the dark points are binned averages constructed using the \texttt{binsreg} command from \cite{cattaneo2019binscatter}. The horizontal axes are the estimated effect on the insurance rate at age 65 by locality. The vertical axes are the reduction in the flow of collections debt at age 65 by locality. Source: Consumer credit outcomes are based on 137,340,577 person-year observations from the New York Fed Consumer Credit Panel / Equifax, 2008-2017. State-level uninsurance rates are from the American Community Survey, 2008-2017.}
  \end{minipage}
\end{figure}

%%%%%%%%%%%%%%%%%%%%%%%%%%%%%%%%%%%%%%%%%%%%%%%%%%%%%%%%%%%%%%
% Correlates with area-level effects, robustness to area FEs %
%%%%%%%%%%%%%%%%%%%%%%%%%%%%%%%%%%%%%%%%%%%%%%%%%%%%%%%%%%%%%%
\clearpage
\clearpage
\begin{figure}[htpb!]
  \centering
  \caption{Correlates with reduction in collections debt at age 65, with Fixed Effects}
  \label{fig:correlates_drop_collections_cz_fe}
 \makebox[\textwidth][c]{ \begin{tabular}{c}
  \textit{Panel A:} Area-level demographic characteristics \\[.25cm]
\includegraphics[width=6in]{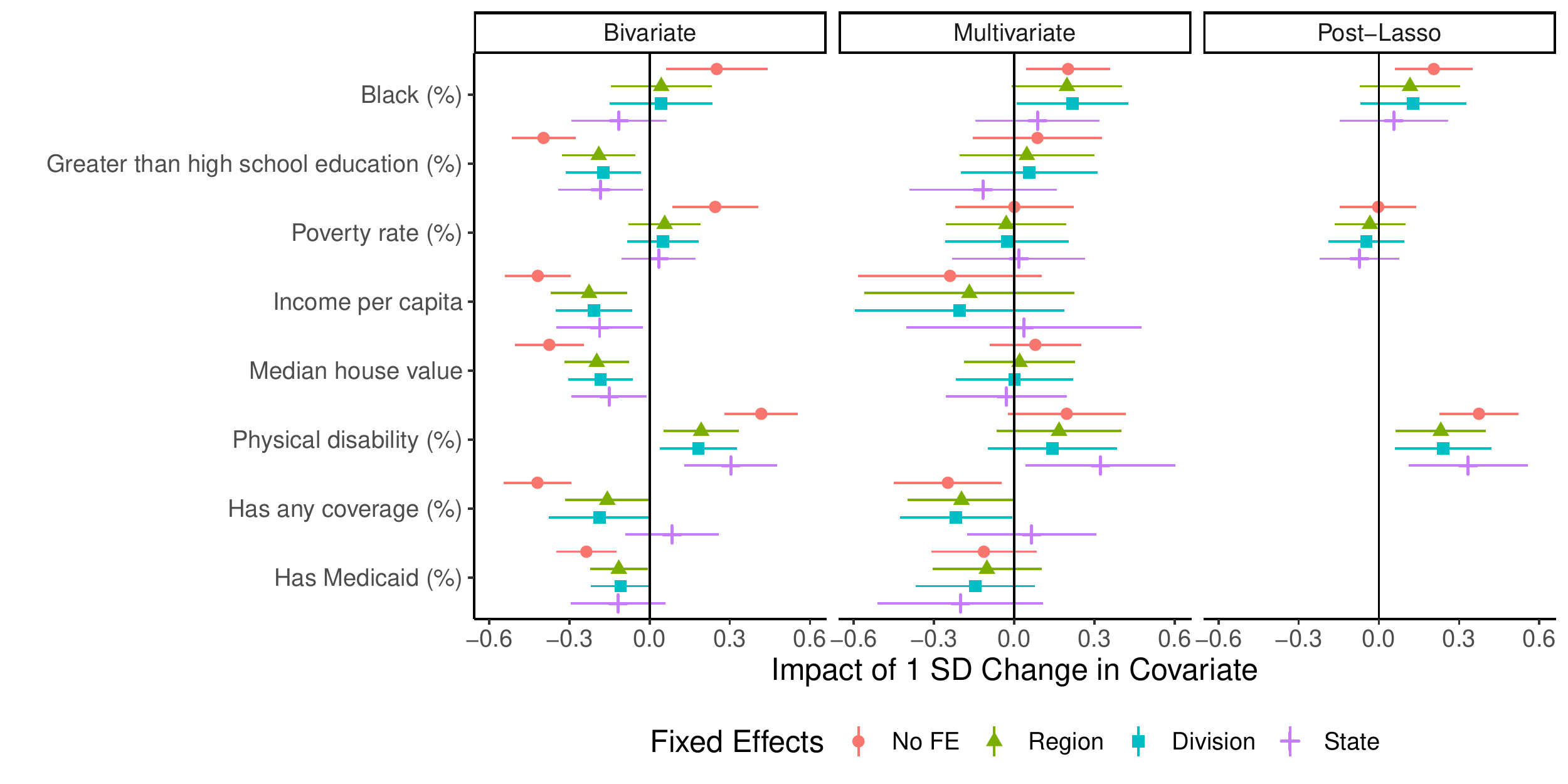} \\
\\
  \textit{Panel B:} Healthcare market characteristics \\[.25cm]
\includegraphics[width=6in]{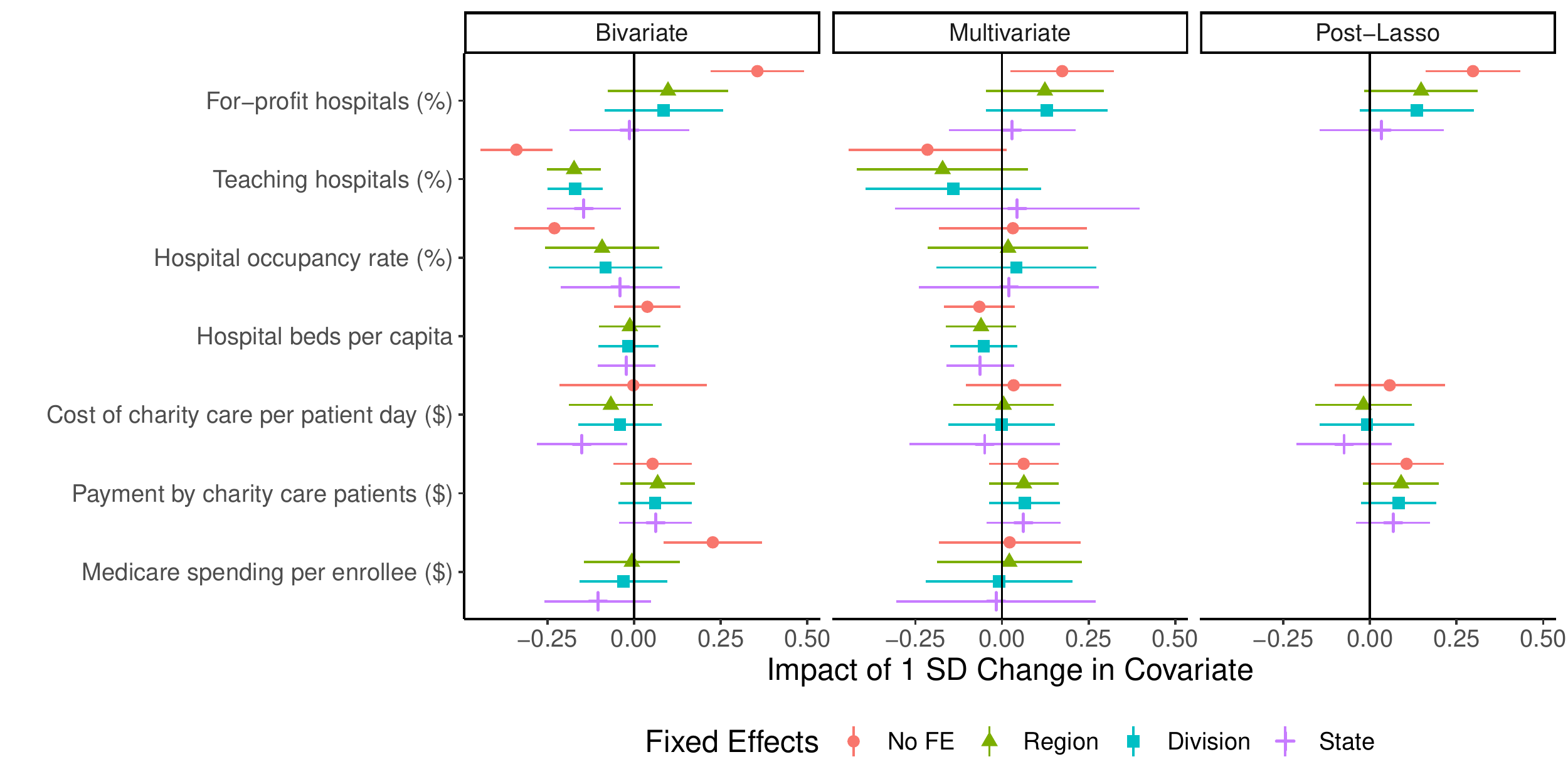} \\
\end{tabular}}
 
  \begin{minipage} {\textwidth} \setstretch{.9} \medskip
    \footnotesize{\textbf{Note:} This figure plots bivariate OLS regression results (left panel), multivariate OLS regression results (center panel), and post-Lasso multivariate regression results (right panel) of CZ-level estimated reductions in collections debt per capita on a set of CZ-level characteristics. We standardize all the variables so the coefficients reflect the strength of the association between a one standard deviation change in the covariate and the estimated reduction in collections debt at age 65. The horizontal bars are 95\% confidence intervals. The multivariate OLS regression results and post-Lasso multivariate regression results are both run on the full set of characteristics in Panels A and B. For post-Lasso, we first estimate a Lasso regression on the full set of characteristics and then report the results of multivariate OLS run on the characteristics chosen by the Lasso regression. For each correlate we report our primary results where there are area-level fixed effects (termed ``No FE'') and then results where we include fixed effects for Census Region, Census Division, and state, respectively. Tabular versions of these results are in Table \ref{tab:tau_beta_correlates_apx}. Source: Consumer credit outcomes are based on 137,340,577 person-year observations from the New York Fed Consumer Credit Panel / Equifax, 2008-2017. CZ-level uninsurance rates are from the American Community Survey, 2008-2017. Healthcare market characteristics are from the Healthcare Cost Report Information System (HCRIS) and the Dartmouth Atlas. For additional details on the data see Section \ref{background_data}.}
  \end{minipage}
\end{figure}

%%%%%%%%%%%%%%%%%%%%%%%%%%%%%%%%%%%%%%%%%%%%%%%%%%%%%%%%%%%%%%%
% CZ-level per-capita vs. per newly-insured plot %
%%%%%%%%%%%%%%%%%%%%%%%%%%%%%%%%%%%%%%%%%%%%%%%%%%%%%%%%%%%%%%%
\clearpage

\begin{figure}[htpb!]
  \centering
  \caption{Effect of Medicare eligibility on per-capita collections debt vs. per-newly insured collections debt at the CZ-level}
  \label{fig:tau_vs_beta_scatter}
    \begin{tabular}{c}
    \includegraphics[width=5.5in]{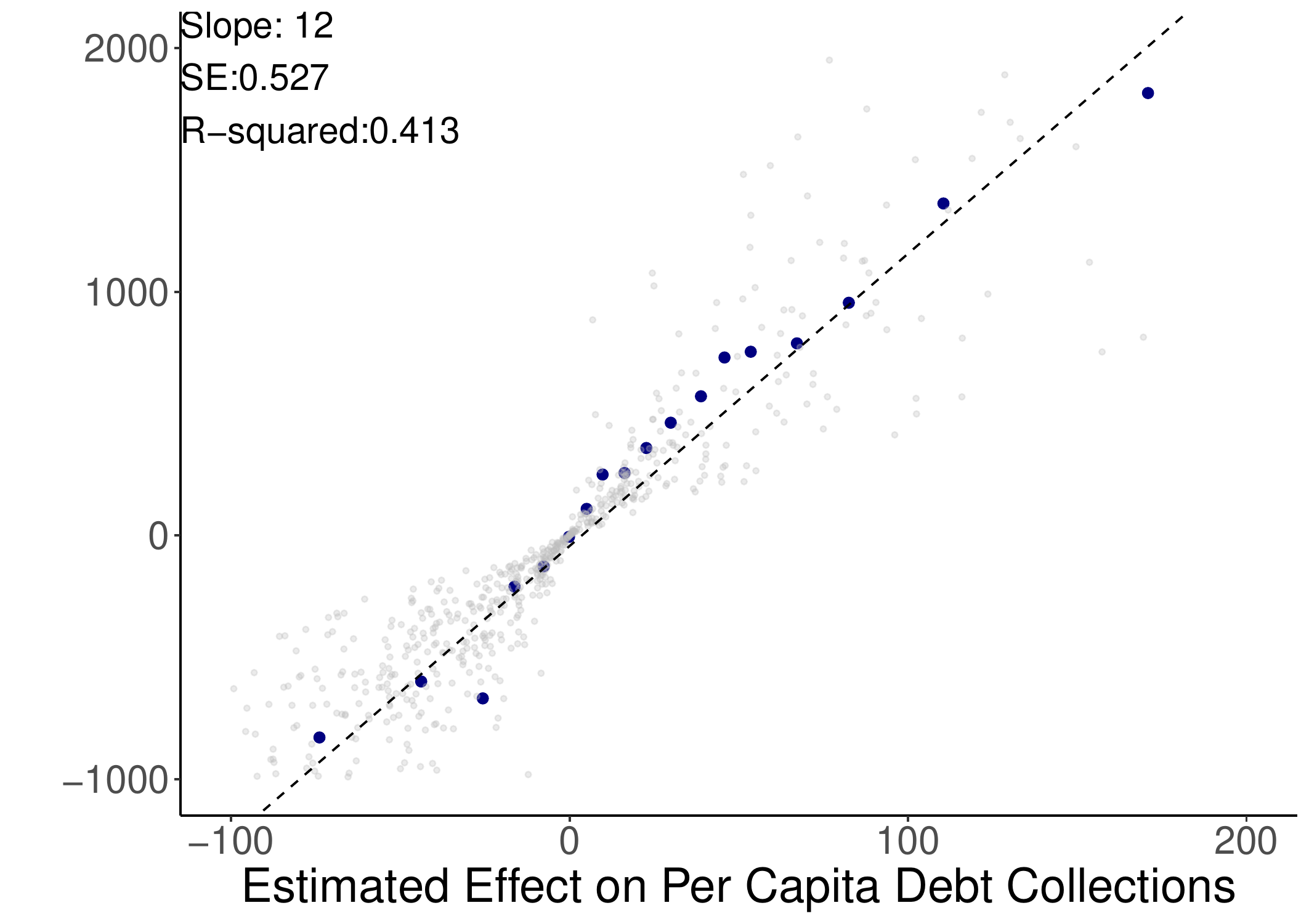} 
    \end{tabular}
  \begin{minipage} {0.9\textwidth} \setstretch{.9} \medskip
    \footnotesize{\textbf{Note:} This figure plots point estimates of the reduction in the flow of per-capita collections debt (within the past year) vs. point estimates of the reduction in the flow of per-capita collections debt (within the past year) per newly-insured. The estimates are based on local linear regressions, done separately by commuting zone (CZ), using the methods from \cite{kolesar2018inference}. The dark points are binned averages constructed using the \texttt{binsreg} command from \cite{cattaneo2019binscatter}. The horizontal axis is the estimated effect on per capita debt collections at age 65 by CZ. The vertical axis is the estimated reduction in the flow of collections debt per newly-insured at age 65 by CZ. Source: Consumer credit outcomes are based on 137,340,577 person-year observations from the New York Fed Consumer Credit Panel / Equifax, 2008-2017. State-level uninsurance rates are from the American Community Survey, 2008-2017.}
  \end{minipage}
\end{figure}

%%%%%%%%%%%%%%%%%%%%%%%%%%%%%%%%%%%%%%%%%%%%
% Forecasted taus before and after the ACA %
%%%%%%%%%%%%%%%%%%%%%%%%%%%%%%%%%%%%%%%%%%%%
\clearpage
\begin{figure}[htpb!]
  \centering
  \caption{Forecasts of causal reductions in collections debt from expanding health insurance to the near-elderly by commuting zone}
  \label{fig:ca_forecast_tau_aca}
  \begin{tabular}{c}
  \textit{Panel A:} Pre-ACA, 2008-2013 \\
\includegraphics[width=4.5in,trim=4 4 4 4,clip]{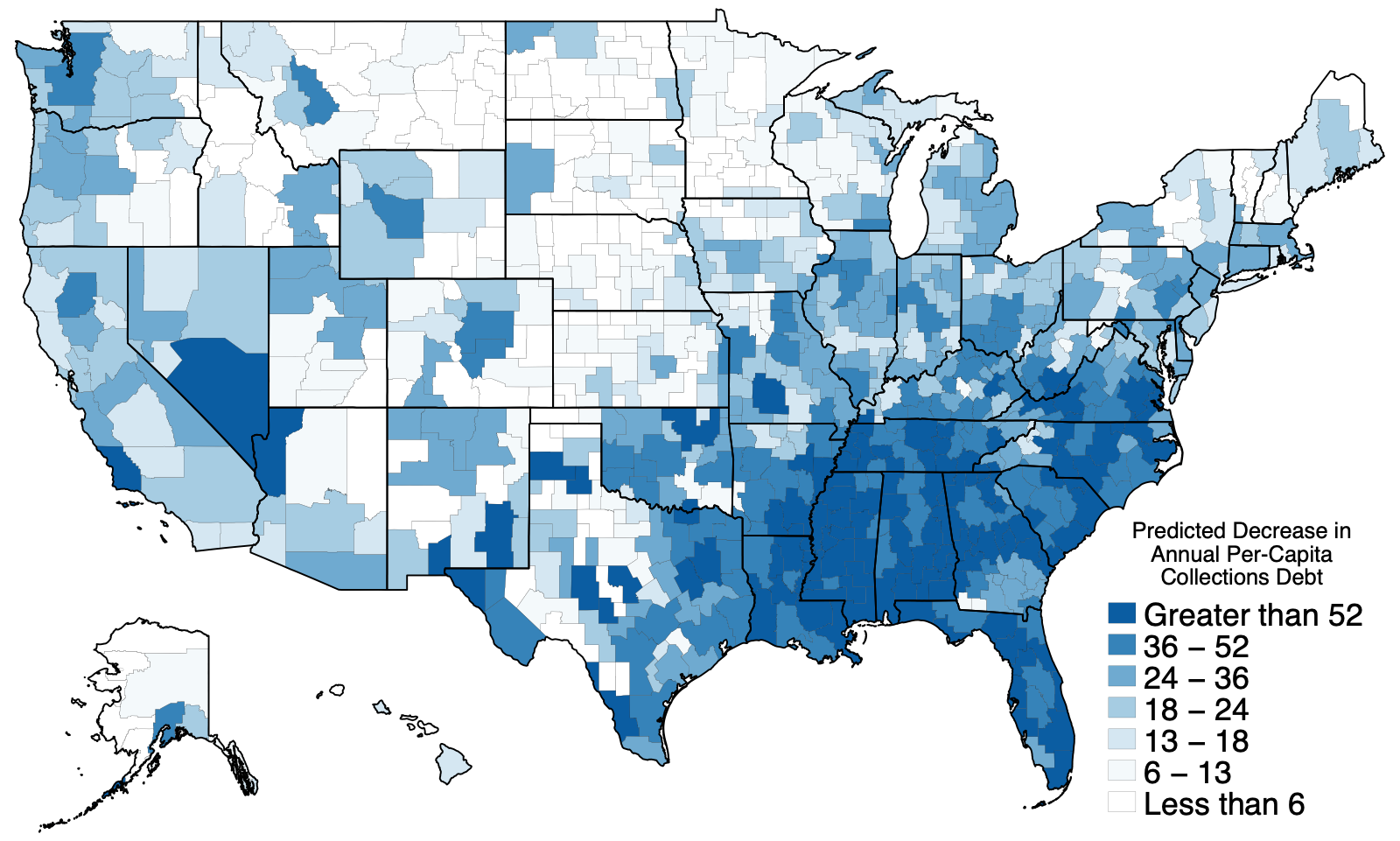} \\
\\
\textit{Panel B:} Post-ACA, 2014-2017 \\ 
\includegraphics[width=4.5in,trim=4 4 4 4,clip]{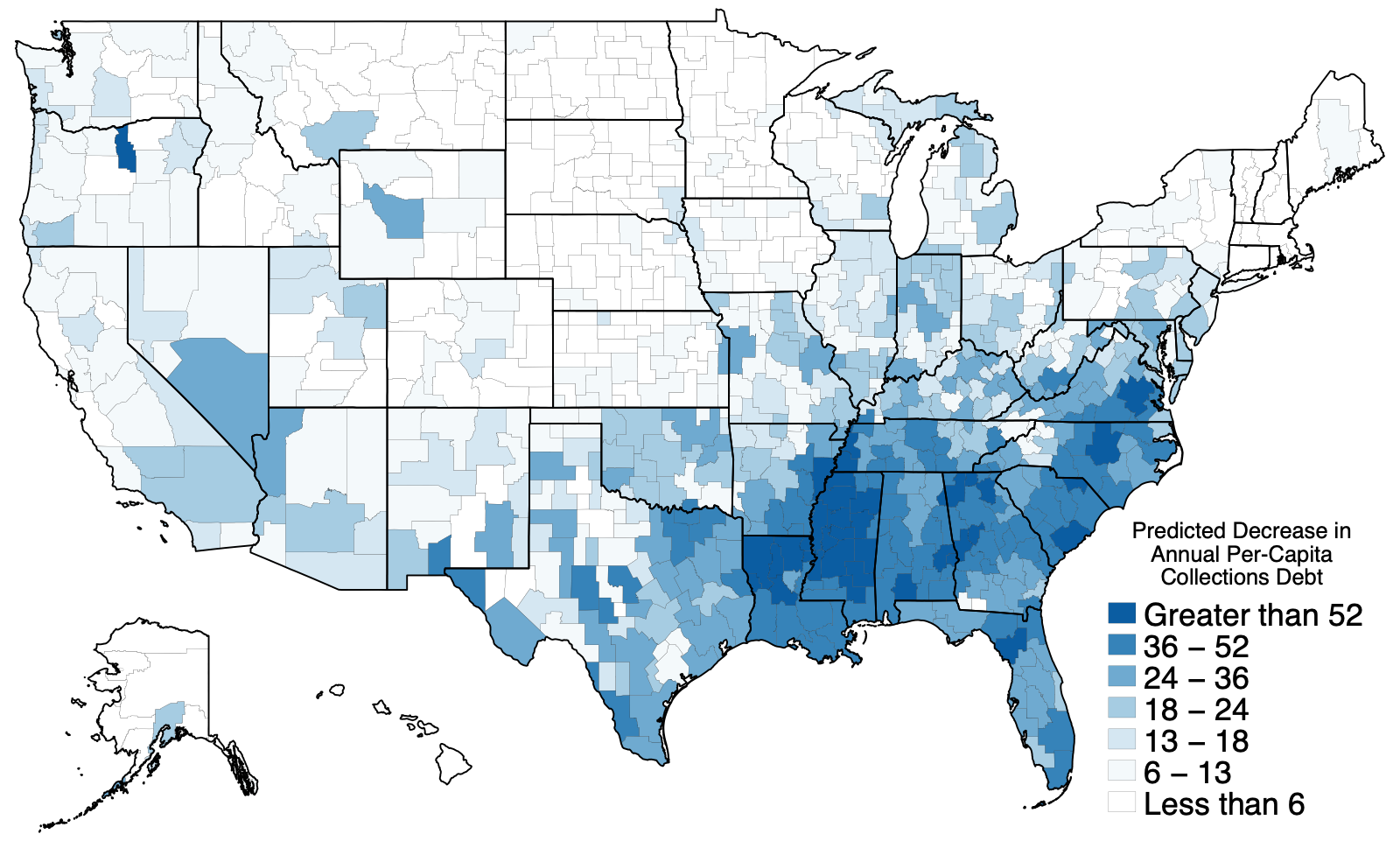}
\end{tabular}
 
  \begin{minipage} {0.9\textwidth} \setstretch{.9} \medskip
    \footnotesize{\textbf{Note:} This figure plots mean square error (MSE)-minimizing forecasts of the reductions in collections debt per capita in the pre-ACA (Panel A) and post-ACA period (Panel B). We construct the MSE-minimizing forecasting by first running a Lasso regression to predict the CZ-level reductions in collections debt per capita separate for each period. This generates a prediction for each CZ in each period, which we call $\hat{\gamma_{l}}$. Following \cite{chetty2018impactsb} we then combine the $\hat{\gamma_{l}}$ estimates with our estimates of $\gamma_{l}$ to construct the mean square error-minimizing forecast for each commuting zone in each period, $\gamma_{l}^{f}$. Source: Consumer credit outcomes are based on 137,340,577 person-year observations from the New York Fed Consumer Credit Panel / Equifax, 2008-2017. CZ-level uninsurance rates are from the American Community Survey, 2008-2017. Healthcare market characteristics are from the Healthcare Cost Report Information System (HCRIS) and the Dartmouth Atlas. For additional details on the data see Section \ref{background_data}..}
  \end{minipage}
\end{figure}

%%%%%%%%%%%%%%%%%%%%%%%%%%%%%%%%%%%%%%%%%%%%%%%%%%%%%%%%%%%%%%%%%%%%
% Counterfactual insurance levels pre-65, before and after the ACA %
%%%%%%%%%%%%%%%%%%%%%%%%%%%%%%%%%%%%%%%%%%%%%%%%%%%%%%%%%%%%%%%%%%%%
\clearpage
\begin{figure}[htpb!]
  \centering
  \caption{Counterfactual health insurance rates by commuting zone at age 65 without Medicare, pre- and post-ACA}
  \label{fig:insurance_pre_cz_map_shrink_aca}
  \begin{tabular}{c}
  \textit{Panel A:} Pre-ACA, 2008-2013 \\
\includegraphics[width=4.5in,trim=4 4 4 4,clip]{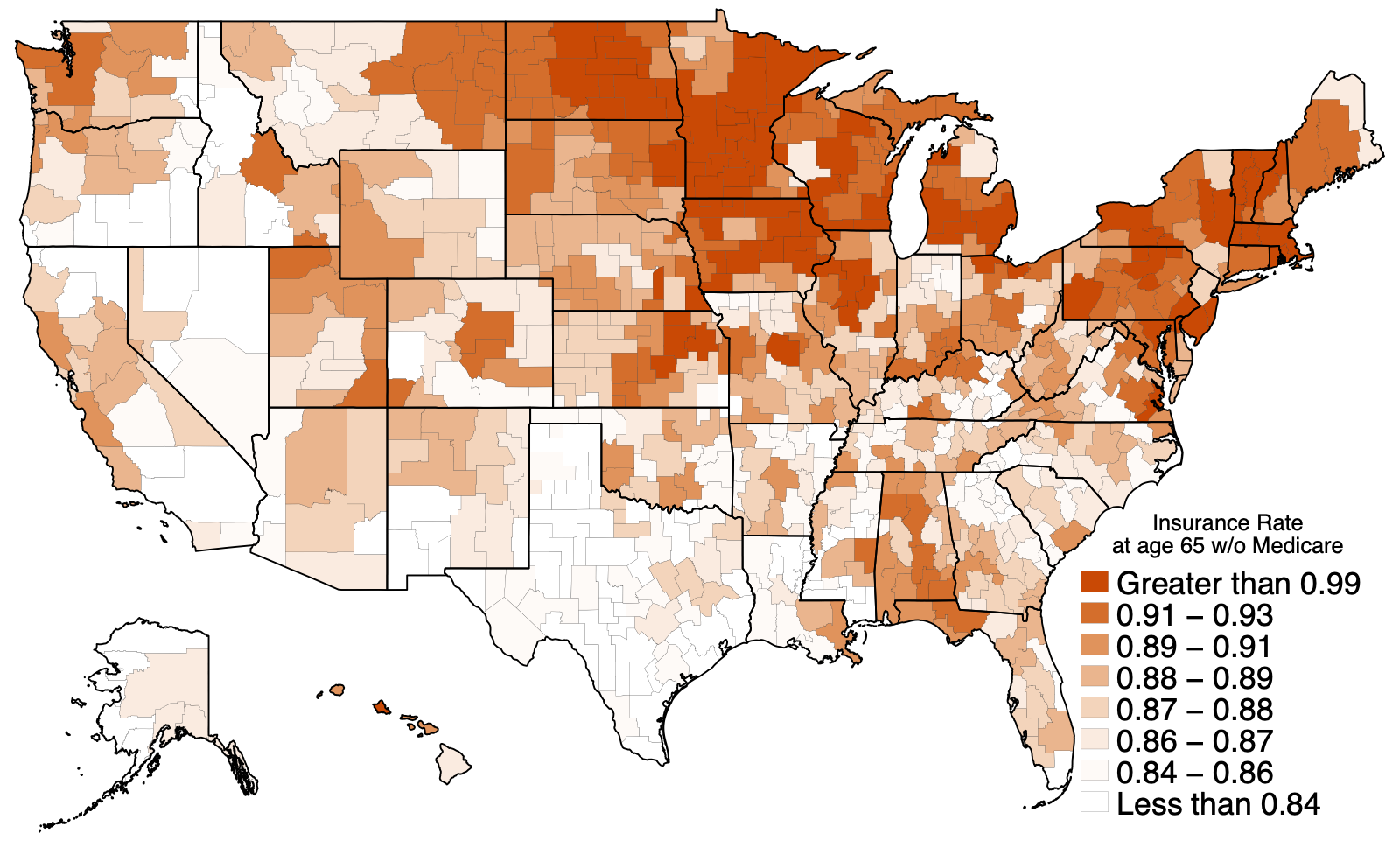} \\
\\
\textit{Panel B:} Post-ACA, 2014-2017 \\ 
\includegraphics[width=4.5in,trim=4 4 4 4,clip]{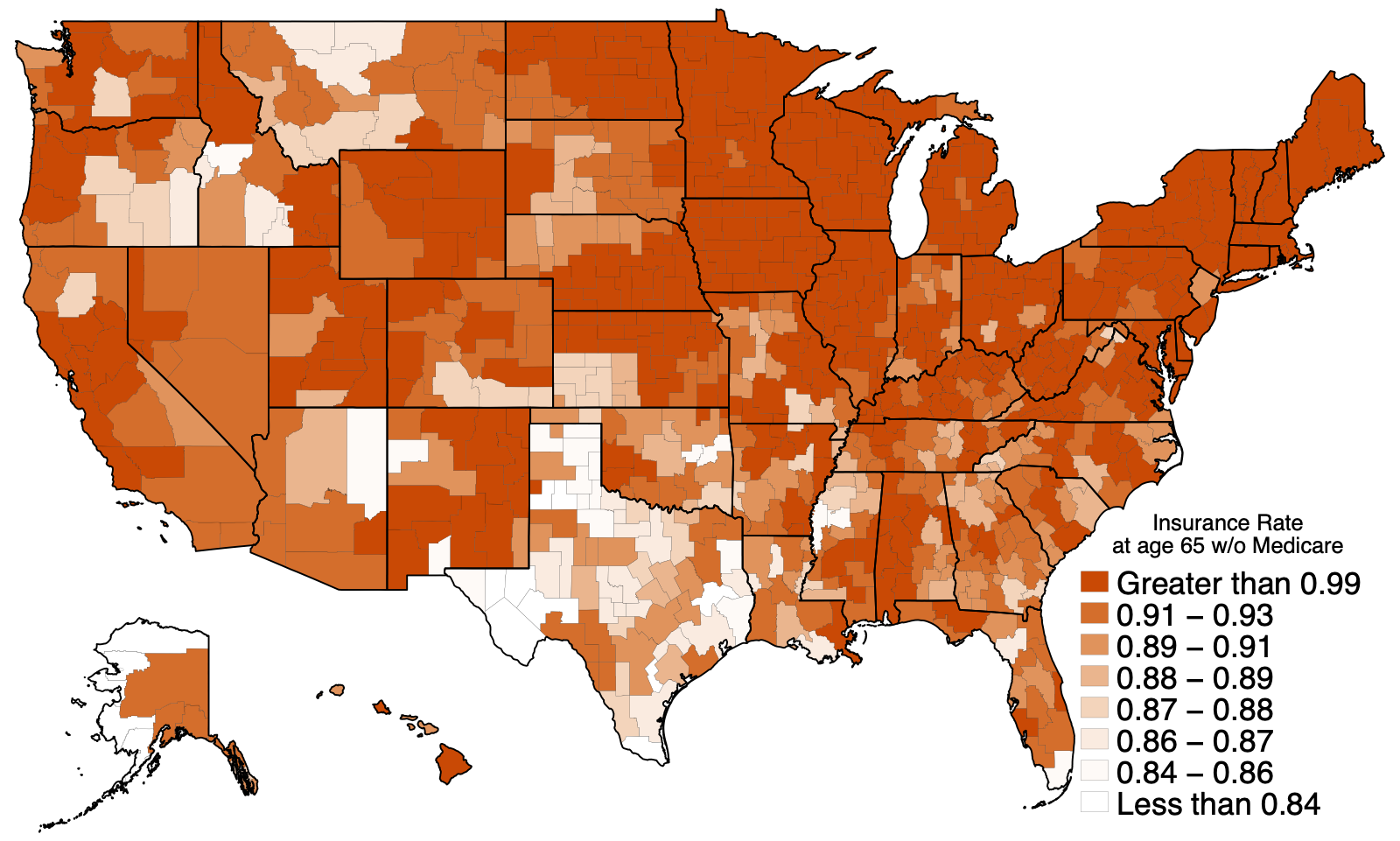}
\end{tabular}
 
  \begin{minipage} {0.9\textwidth} \setstretch{.9} \medskip
    \footnotesize{\textbf{Note:} This figure plots our counterfactual estimates of the share of the population with health insurance coverage at age 65, without Medicare, before and after the full implementation of the Affordable Care Act in 2014. The counterfactuals are based on local linear regressions, done separately by commuting zone, using the methods from \cite{kolesar2018inference}. These estimates are then shrunk using empirical Bayes, described in Section \ref{apx:methods}. Panel A. presents the counterfactuals from the pre-ACA period, 2008-2013. Panel B. presents the counterfactuals from the post-ACA period, 2014-2017. Darker shading corresponds to states with higher counterfactual health insurance rates. Source: CZ-level uninsurance rates are from the American Community Survey, 2008-2017.}
  \end{minipage}
\end{figure}

%%%%%%%%%%%%%%%%%%%%%%%%%%%%%%%%%%%%%%%%%%%%%%%%%%%%%%%%%%%%%
% CZ-level Diff-in-disc gains in coverage due to the ACA vs %
%%%%%%%%%%%%%%%%%%%%%%%%%%%%%%%%%%%%%%%%%%%%%%%%%%%%%%%%%%%%%
\clearpage
\begin{figure}[htpb!]
  \centering
  \caption{Difference-in-discontinuities estimates of increases in the near-elderly health insurance rate due to the ACA}
  \label{fig:diffindisc_insurance}
  \begin{tabular}{c}
    \includegraphics[width=5.5in,trim=4 4 4 4,clip]{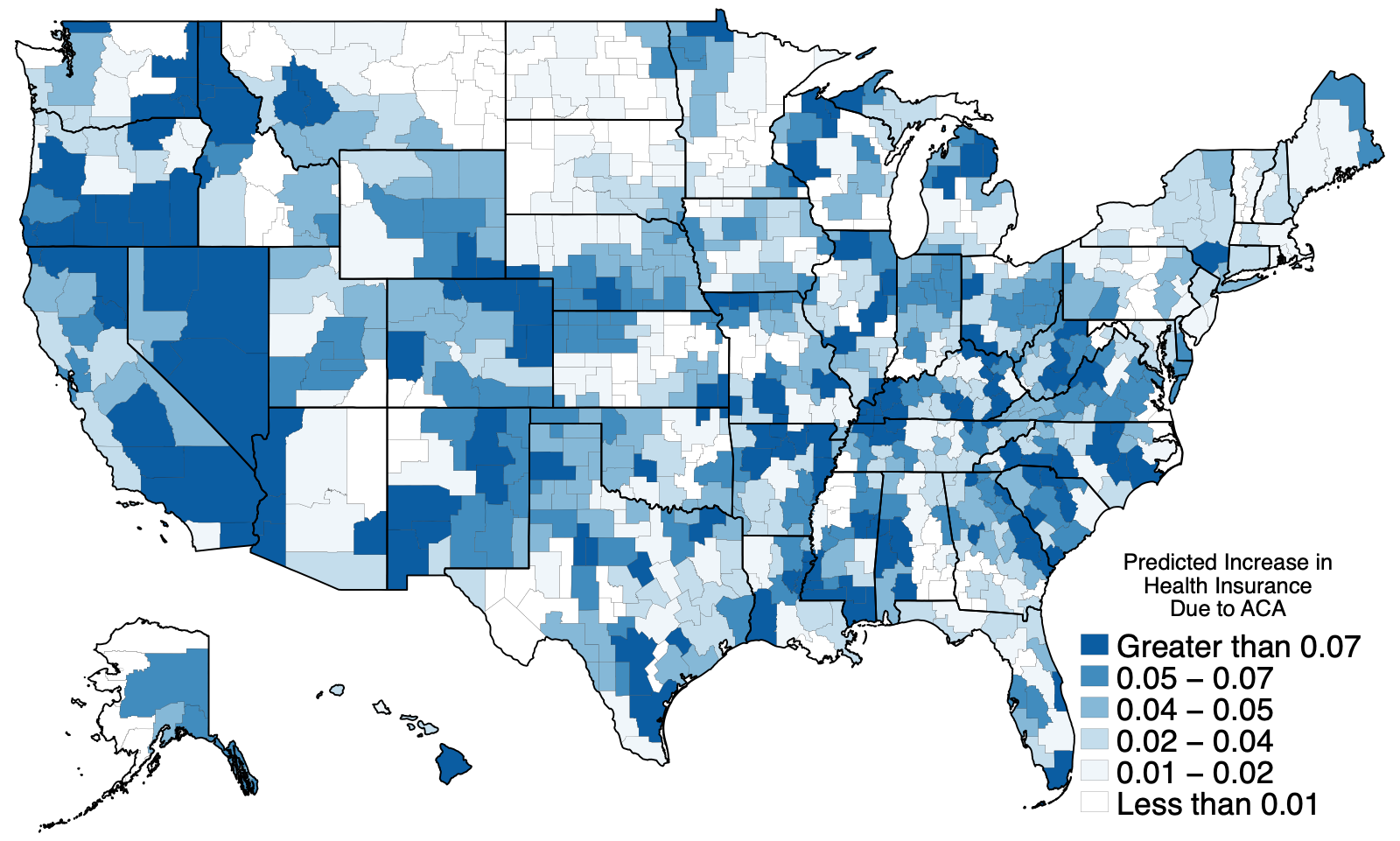} \\
  \end{tabular}
  \begin{minipage} {0.9\textwidth} \setstretch{.9} \medskip
    \footnotesize{\textbf{Note:} This figure plots CZ-level estimates of the increase in health insurance coverage  due to the Affordable Care Act using a difference-in-discontinuities design similar to \cite{duggan2019impact}. We compare the discontinuity in health insurance at age 65 in the post-ACA period (2014-2017) to the discontinuity in health insurance at age 65 prior to its full implementation (2008-2013). CZ-level uninsurance rates are from the American Community Survey, 2008-2017.}
  \end{minipage}
\end{figure}

% % %%%%%%%%%%%%%%%%%%%%%%%%%%%%%%%%%%%%%%%%%%%%%%%%%%%%%%%%%%%%%%%%%%%%
% % % CZ-level Diff-in-disc gains in coverage due to the ACA vs. betas %
% % %%%%%%%%%%%%%%%%%%%%%%%%%%%%%%%%%%%%%%%%%%%%%%%%%%%%%%%%%%%%%%%%%%%%
% \clearpage
% \begin{figure}[htpb!]
%   \centering
%   \caption{Difference-in-discontinuities estimates of increases in health insurance due to ACA vs. pre-ACA changes in coverage at 65}
%   \label{fig:diffindisc_insurance_vs_preACA_insdelta65}
%   \begin{tabular}{c}
%   \\
% \includegraphics[width=5.5in]{graphs/cz_diff_in_disc_has_ins_vs_preaca_has_ins.pdf} 
% \\
% \end{tabular}
 
%   \begin{minipage} {0.9\textwidth} \setstretch{.9} \medskip
%     \footnotesize{\textbf{Note:} This figure plots CZ-level estimates of the increase in health insurance coverage  due to the Affordable Care Act using a difference-in-discontinuities design similar to \cite{duggan2019impact} against the pre-ACA increase in the insurance rate at age 65 based on local linear regressions, done separately by state, using the methods from \cite{kolesar2018inference}. The horizontal axis is the estimated effect on the insurance rate at age 65 by state in the pre-ACA period. The vertical axis is the stata-level diff-in-disc estimate of the increase in the insurance rate due to the ACA. Source: State-level uninsurance rates are from the American Community Survey, 2008-2017.}
%   \end{minipage}
% \end{figure}

%%%%%%%%%%%%%%%%%%%%%%%%%%%%%%%%%%%%%%%%%%%%%
% Forecasted betas before and after the ACA %
%%%%%%%%%%%%%%%%%%%%%%%%%%%%%%%%%%%%%%%%%%%%%
\clearpage
\begin{figure}[htpb!]
  \centering
  \caption{Forecasts of causal reductions in collections debt per newly-insured near-elderly person by commuting zone, pre- and post-ACA}
  \label{fig:cz_forecast_beta_aca}
  \begin{tabular}{c}
  \textit{Panel A:} Pre-ACA, 2008-2013 \\
\includegraphics[width=4.5in,trim=4 4 4 4,clip]{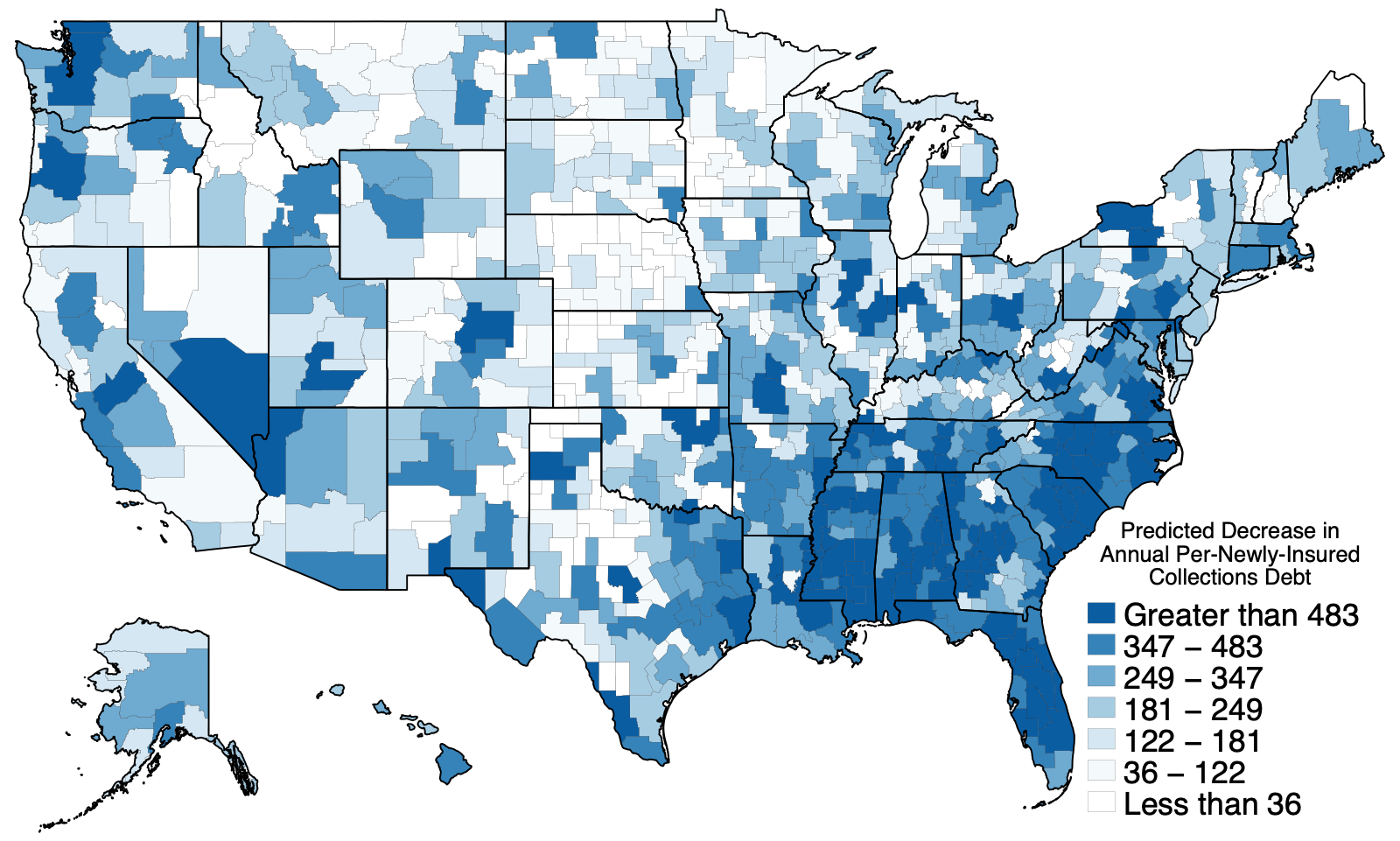} \\
\\
\textit{Panel B:} Post-ACA, 2014-2017 \\ 
\includegraphics[width=4.5in,trim=4 4 4 4,clip]{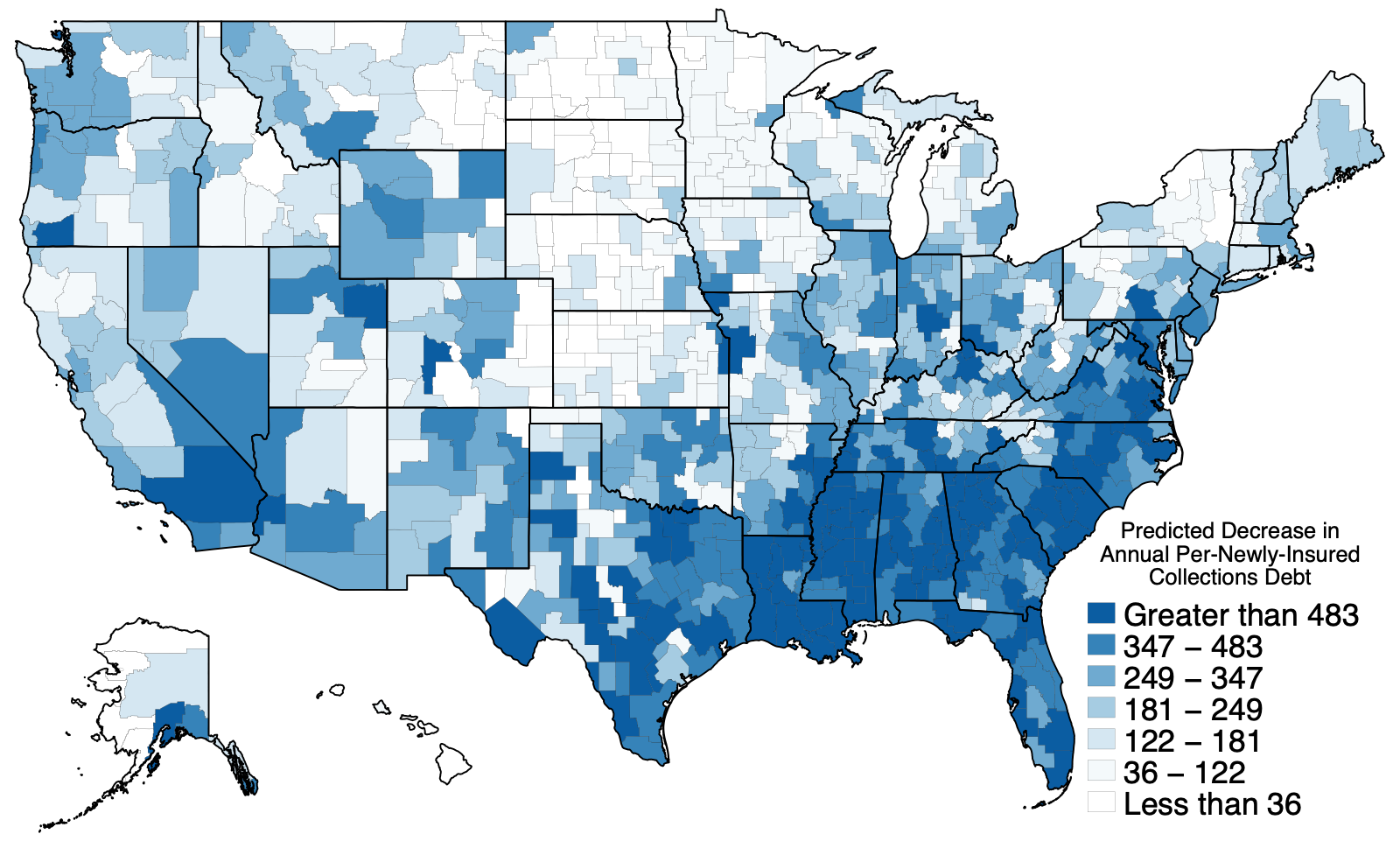}
\end{tabular}
 
  \begin{minipage} {0.9\textwidth} \setstretch{.9} \medskip
    \footnotesize{\textbf{Note:} This figure plots mean square error (MSE)-minimizing forecasts of the reductions in collections debt per newly-insured for the pre-ACA period (Panel A) and post-ACA period (Panel B). We construct the MSE-minimizing forecasting by first running a Lasso regression to predict the CZ-level reductions in collections debt per newly-insured. This generates a prediction for each CZ, which we call $\hat{\beta_{l}}$. Following \cite{chetty2018impactsb} we then combine the $\hat{\beta_{l}}$ estimates with our estimates of $\beta_{l}$ to construct the mean square error-minimizing forecast for each commuting zone, $\beta_{l}^{f}$. Source: Consumer credit outcomes are based on 137,340,577 person-year observations from the New York Fed Consumer Credit Panel / Equifax, 2008-2017. State-level uninsurance rates are from the American Community Survey, 2008-2017. Healthcare market characteristics are from the Healthcare Cost Report Information System (HCRIS) and the Dartmouth Atlas. For additional details on the data see Section \ref{background_data}.}
  \end{minipage}
\end{figure}

\makeatletter
\setlength{\@fptop}{5pt}
\makeatother

%%%%%%%%%%%%%%%%%%%%%%%%%%%%%%%%%%%%%%%%%%%%%%%%%%%%%%%%%%
% Robustness of main RD results to other specifications %%
%%%%%%%%%%%%%%%%%%%%%%%%%%%%%%%%%%%%%%%%%%%%%%%%%%%%%%%%%%
\clearpage
\begin{table}[hp]
\centering
\caption{Robustness of estimated changes in financial outcomes at age 65}
\label{tab:main_agerd_appendix_othermodels_apx}
\resizebox{\textwidth}{!}{
\begin{tabular}{ldddddddd}
\toprule
& \multicolumn{1}{c}{Main} & \multicolumn{1}{c}{Linear with} & \multicolumn{1}{c}{Quad. with} & \multicolumn{1}{c}{Cubic. with} & \multicolumn{1}{c}{Linear with} & \multicolumn{1}{c}{Quad. with} & \multicolumn{1}{c}{Cubic. with} & \multicolumn{1}{c}{Local linear}\\

& \multicolumn{1}{c}{Estimate} & \multicolumn{1}{c}{robust SEs} & \multicolumn{1}{c}{robust SEs} & \multicolumn{1}{c}{robust SEs} & \multicolumn{1}{c}{clustered SEs} & \multicolumn{1}{c}{clustered SEs} & \multicolumn{1}{c}{clustered SEs} & \multicolumn{1}{c}{with rdrobust}\\

\cmidrule{2-9}
& \multicolumn{1}{c}{(1)} & \multicolumn{1}{c}{(2)} & \multicolumn{1}{c}{(3)} & \multicolumn{1}{c}{(4)} & \multicolumn{1}{c}{(5)} & \multicolumn{1}{c}{(6)} & \multicolumn{1}{c}{(7)} & \multicolumn{1}{c}{(8)} \\
\midrule
Share with any coverage & 0.076* & 0.073* & 0.077* & 0.074* & 0.073* & 0.077* & 0.074* & 0.073*\\
 & (0.004) & (0.002) & (0.004) & (0.007) & (0.002) & (0.001) & (0.002) & (0.01)\\
 & [0.064, & [0.069, & [0.069, & [0.06, & [0.07, & [0.075, & [0.071, & [0.053,\\
 & 0.087] & 0.078] & 0.086] & 0.088] & 0.077] & 0.08] & 0.078] & 0.093]\\
Debt in collections & -28.546* & -32.981* & -32.121* & -23.58* & -32.981* & -32.121* & -23.58* & -21.558*\\
 & (6.813) & (2.578) & (4.407) & (7.467) & (2.212) & (2.498) & (2.549) & (10.83)\\
 & [-43.355, & [-38.034, & [-40.76, & [-38.216, & [-37.317, & [-37.018, & [-28.577, & [-42.783,\\
 & -13.737] & -27.928] & -23.483] & -8.945] & -28.645] & -27.224] & -18.584] & -0.332]\\
Credit score & 0.818 & 1.199 & 0.645 & 0.682 & 1.199* & 0.645* & 0.682* & 0.795\\
 & (1.943) & (0.87) & (1.515) & (2.57) & (0.437) & (0.135) & (0.108) & (3.663)\\
 & [-3.001, & [-0.506, & [-2.324, & [-4.357, & [0.342, & [0.381, & [0.471, & [-6.384,\\
 & 4.637] & 2.905] & 3.614] & 5.72] & 2.057] & 0.909] & 0.892] & 7.975]\\
Bankruptcy (pp) & -0.004 & -0.006 & -0.008 & 0.005 & -0.006* & -0.008 & 0.005 & 0.003\\
 & (0.007) & (0.003) & (0.006) & (0.01) & (0.003) & (0.005) & (0.006) & (0.013)\\
 & [-0.016, & [-0.012, & [-0.019, & [-0.014, & [-0.011, & [-0.017, & [-0.006, & [-0.023,\\
 & 0.008] & 0.001] & 0.003] & 0.024] & 0] & 0.001] & 0.016] & 0.029]\\
Share employed & -0.029 & -0.089* & -0.002 & -0.016 & -0.089* & -0.002 & -0.016 & -0.039\\
 & (0.008) & (0.004) & (0.008) & (0.014) & (0.017) & (0.01) & (0.014) & (0.022)\\
 & [-0.096, & [-0.097, & [-0.018, & [-0.043, & [-0.122, & [-0.022, & [-0.043, & [-0.083,\\
 & 0.038] & -0.08] & 0.014] & 0.01] & -0.055] & 0.018] & 0.01] & 0.004]\\
Income & 880.67 & 458.575 & 1572.808 & 343.741 & 458.575 & 1572.808* & 343.741 & 997.591\\
 & (1322.697) & (749.423) & (1315.303) & (2235.622) & (244.825) & (285.332) & (333.556) & (4308.174)\\
 & [-2002.838, & [-1010.295, & [-1005.185, & [-4038.078, & [-21.281, & [1013.557, & [-310.03, & [-7446.274,\\
 & 3764.179] & 1927.445] & 4150.801] & 4725.559] & 938.432] & 2132.06] & 997.511] & 9441.457]\\
Total debt past due & -220.973 & -213.784 & -230.164 & -200.976 & -213.784* & -230.164* & -200.976* & -233.868\\
 & (404.833) & (139.794) & (240.424) & (412.141) & (38.637) & (53.615) & (52.082) & (746.706)\\
 & [-935.78, & [-487.78, & [-701.396, & [-1008.772, & [-289.513, & [-335.25, & [-303.057, & [-1697.385,\\
 & 493.833] & 60.212] & 241.067] & 606.82] & -138.055] & -125.079] & -98.895] & 1229.649]\\
Mortgage debt past due & -191.289 & -176.85 & -207.937 & -177.245 & -176.85* & -207.937* & -177.245* & -195.622\\
 & (350.784) & (120.781) & (206.156) & (354.14) & (30.91) & (44.385) & (46.654) & (634.81)\\
 & [-810.149, & [-413.58, & [-612.002, & [-871.359, & [-237.433, & [-294.932, & [-268.687, & [-1439.827,\\
 & 427.571] & 59.881] & 196.128] & 516.87] & -116.266] & -120.942] & -85.803] & 1048.583]\\
Credit card debt past due & -13.922 & -27.734* & -11.793 & -6.079 & -27.734* & -11.793* & -6.079 & -4.517\\
 & (23.341) & (9.16) & (16.058) & (27.349) & (4.22) & (3.959) & (4.492) & (49.812)\\
 & [-60.867, & [-45.688, & [-43.265, & [-59.683, & [-36.005, & [-19.552, & [-14.883, & [-102.146,\\
 & 33.023] & -9.781] & 19.68] & 47.524] & -19.464] & -4.033] & 2.724] & 93.112]\\
Foreclosure & -0.005 & -0.003 & -0.001 & -0.01 & -0.003 & -0.001 & -0.01* & -0.015\\
 & (0.006) & (0.003) & (0.005) & (0.008) & (0.002) & (0.004) & (0.004) & (0.014)\\
 & [-0.015, & [-0.009, & [-0.011, & [-0.026, & [-0.007, & [-0.008, & [-0.017, & [-0.042,\\
 & 0.006] & 0.002] & 0.008] & 0.006] & 0] & 0.006] & -0.003] & 0.011]\\
Share of mortgage debt past due & -0.003 & -0.005* & -0.003 & -0.003 & -0.005* & -0.003* & -0.003* & -0.003\\
 & (0.005) & (0.002) & (0.003) & (0.005) & (0.001) & (0.001) & (0.001) & (0.008)\\
 & [-0.012, & [-0.008, & [-0.008, & [-0.012, & [-0.006, & [-0.005, & [-0.005, & [-0.019,\\
 & 0.005] & -0.002] & 0.002] & 0.006] & -0.003] & -0.001] & -0.001] & 0.013]\\
Share of cc debt past due & -0.003 & -0.008* & -0.001 & -0.001 & -0.008* & -0.001* & -0.001 & -0.001\\
 & (0.004) & (0.002) & (0.003) & (0.005) & (0.002) & (0.001) & (0.001) & (0.01)\\
 & [-0.011, & [-0.011, & [-0.007, & [-0.011, & [-0.011, & [-0.002, & [-0.002, & [-0.02,\\
 & 0.005] & -0.004] & 0.005] & 0.009] & -0.004] & 0] & 0.001] & 0.018]\\
\bottomrule
\end{tabular}

}
\begin{minipage} {0.95\textwidth} \setstretch{.9} \bigskip
\footnotesize{\textbf{Note:}
This table reports the sensitivity of our main regression discontinuity estimates to alternative specifications. Column 1 reports the point estimate, standard error, and bias-adjusted 95\% confidence interval from a local linear regression using techniques from \cite{kolesar2018inference}. Columns 2-4 report the results of estimating the discontinuity at 65 using three parametric models and robust standard errors with linear, quadratic, and cubic age trends, respectively. Columns 2-4 report the results of estimating the discontinuity at 65 using three parametric models and clustering standard errors by age (the running variable) as in \cite{lee2008regression} with linear, quadratic, and cubic age trends, respectively. Column 8 reports the results of estimating the discontinuity using the local linear regression model as in \cite{calonico2015rdrobust}. The sample includes individuals who were age 55-75 between 2008 and 2017. Credit score data used is from Equifax Riskscore 3.0. See Section \ref{background_data} for additional details on the outcomes and sample. Source: The financial health outcomes are based on 137,340,577 person-year observations from the New York Fed Consumer Credit Panel / Equifax, 2008-2017.
}
\end{minipage}
\end{table}

%%%%%%%%%%%%%%%%%%%%%%%%
% Top-50 CZ estimates  %
%%%%%%%%%%%%%%%%%%%%%%%%

\clearpage
\begin{table}[hp]
\centering
\caption{Location-specific estimates and forecasts for 50 largest CZs}
\label{tab:czone_estimates_table}
\resizebox{.5\textwidth}{!}{
\begin{tabular}{llcccc}
\toprule
 & & \multicolumn{2}{c}{Per capita} & \multicolumn{2}{c}{Per newly-insured} \\
\cmidrule(lr){3-4} \cmidrule(lr){5-6}
 &  & \multicolumn{1}{c}{$\gamma_{l}^{f}$} & \multicolumn{1}{c}{\mbox{RMSE}} & \multicolumn{1}{c}{$\beta_{l}^{f}$} & \multicolumn{1}{c}{\mbox{RMSE}}\\
\cmidrule{3-6}
State & CZ & \multicolumn{1}{c}{(1)} & \multicolumn{1}{c}{(2)} & \multicolumn{1}{c}{(3)} & \multicolumn{1}{c}{(4)} \\
\midrule
    
Arizona & Phoenix & -25 & 11 & -273 & 180\\
California & Los Angeles & -29 & 11 & -206 & 171\\
California & Sacramento & -17 & 8 & -259 & 112\\
California & San Diego & -27 & 7 & -443 & 158\\
California & San Francisco & -8 & 10 & -108 & 176\\
California & San Jose & -14 & 12 & -243 & 247\\
Colorado & Denver & -25 & 6 & -409 & 87\\
Connecticut & Bridgeport & -15 & 3 & -272 & 61\\
District of Columbia & Washington DC & -22 & 10 & -468 & 208\\
Florida & Jacksonville & -42 & 13 & -614 & 248\\
Florida & Miami & -62 & 12 & -419 & 139\\
Florida & Orlando & -44 & 13 & -472 & 267\\
Florida & Port St. Lucie & -58 & 13 & -719 & 243\\
Florida & Sarasota & -53 & 13 & -485 & 256\\
Florida & Tampa & -53 & 13 & -484 & 222\\
Georgia & Atlanta & -43 & 12 & -548 & 206\\
Illinois & Chicago & -17 & 9 & -277 & 159\\
Indiana & Indianapolis & -41 & 11 & -614 & 194\\
Maryland & Baltimore & -23 & 12 & -599 & 265\\
Massachusetts & Boston & -14 & 8 & -403 & 263\\
Michigan & Detroit & -21 & 11 & -417 & 242\\
Michigan & Grand Rapids & -8 & 12 & -106 & 233\\
Minnesota & Minneapolis & -3 & 9 & -152 & 225\\
Missouri & Kansas City & -40 & 13 & -412 & 240\\
Missouri & St. Louis & -30 & 10 & -436 & 180\\
Nevada & Las Vegas & -53 & 12 & -591 & 196\\
New Hampshire & Manchester & -11 & 11 & -148 & 194\\
New Jersey & Newark & -18 & 7 & -221 & 98\\
New Jersey & Toms River & -14 & 10 & -280 & 191\\
New York & Buffalo & -14 & 3 & -449 & 62\\
New York & New York City & -12 & 6 & -197 & 114\\
North Carolina & Charlotte & -57 & 13 & -729 & 268\\
North Carolina & Raleigh & -53 & 11 & -956 & 167\\
Ohio & Cincinnati & -26 & 9 & -456 & 140\\
Ohio & Cleveland & -16 & 10 & -281 & 199\\
Ohio & Columbus & -37 & 6 & -759 & 61\\
Ohio & Dayton & -24 & 11 & -381 & 208\\
Oregon & Portland & -10 & 5 & -139 & 82\\
Pennsylvania & Philadelphia & -18 & 6 & -344 & 84\\
Pennsylvania & Pittsburgh & -21 & 11 & -289 & 243\\
Rhode Island & Providence & -2 & 7 & -46 & 171\\
Tennessee & Nashville & -52 & 8 & -611 & 78\\
Texas & Austin & -43 & 13 & -436 & 248\\
Texas & Dallas & -54 & 13 & -502 & 227\\
Texas & Fort Worth & -50 & 12 & -563 & 125\\
Texas & Houston & -49 & 12 & -435 & 153\\
Texas & San Antonio & -43 & 13 & -305 & 254\\
Utah & Salt Lake City & -31 & 13 & -276 & 245\\
Washington & Seattle & -25 & 10 & -467 & 213\\
Wisconsin & Milwaukee & -28 & 10 & -532 & 215\\
\bottomrule
\end{tabular}
 }
\begin{minipage} {0.95\textwidth} \setstretch{.9} \bigskip
\footnotesize{\textbf{Note:} This table reports the mean square error (MSE)-minimizing forecasts of the reductions in collections debt per capita and the reduction in collections debt per newly-insured for the 50 most populous CZs based on their near-elderly population. We construct the MSE-minimizing forecasting by first running a Lasso regression to predict the CZ-level reductions in collections debt per capita (or per newly-insured). This generates a prediction for each CZ, which we call $\hat{\gamma_{l}}$. Following \cite{chetty2018impactsb} we then combine the $\hat{\gamma_{l}}$ estimates with our estimates of $\gamma_{l}$ to construct the mean square error-minimizing forecast for each commuting zone, $\gamma_{l}^{f}$, which we present in Column 1. Column 2 presents the root-mean-square error (RMSE) which is calculated using methods from \cite{chetty2018impactsb}. Column 3 reports the mean square error-minimizing forecast of the reduction in collections debt per newly-insured associated with a (nearly) universal health insurance expansion, $\beta_{l}^{f}$. Column 4 presents the RMSE for $\beta_{l}^{f}$. Source: Consumer credit outcomes are based on 137,340,577 person-year observations from the New York Fed Consumer Credit Panel / Equifax, 2008-2017. CZ-level uninsurance rates are from the American Community Survey, 2008-2017. Healthcare market characteristics are from the Healthcare Cost Report Information System (HCRIS) and the Dartmouth Atlas. For details on the data see Section \ref{background_data}.}
\end{minipage}
\end{table}

%%%%%%%%%%%%%%%%%%%%%%%%%%%%%%%%
% Description of CCP variables %
%%%%%%%%%%%%%%%%%%%%%%%%%%%%%%%%
\clearpage
\begin{table}[hp]
\centering
\caption{Description of the Federal Reserve Bank of New York's Equifax Consumer Credit Panel (CCP)}
\label{tab:ccp_descriptions}
\resizebox{\textwidth}{!}{
\begin{tabular}{ll}
\toprule
\mbox{Variable} & \mbox{Description} \\
\midrule
%Number of accounts sent to collection agencies & the number of 3rd party collections within the past 12 months.\\
Amount in collections & the total collection amount of these 3rd party collection accounts. \\
Number of delinquent accounts & the count of all non-current loans.\\
Amount delinquent & the sum of all non-current loan balances.\\
Total credit card balance past due & the difference of total bankcard balance and current bankcard balance.\\
Total mortgage account balance past due & the difference of total mortgage account balance (incl. home equity installment) and current mortgage balance.\\
Foreclosure & flag for if an individual recorded a foreclosure in the past 24 months.\\
New Foreclosure & Number of people that recorded a foreclosure in the current quarter, but not the two previous quarters.\\
Bankruptcy & flag for if an individual recorded a bankruptcy in the past 24 months.\\
New Bankruptcy & Number of people that recorded a bankruptcy in the current quarter, but not the two previous quarters.\\
Equifax Risk Score & always refers to Equifax Risk Score 3.0.\\
\bottomrule
\end{tabular}
}
\begin{minipage} {\textwidth} \setstretch{.9} \bigskip
\scriptsize{\textbf{Note:} 
  This table reports definitions for the financial variables used from the New York Fed Consumer Credit Panel. The dataset consists of 137,340,577 person-year observations from the New York Fed Consumer Credit Panel / Equifax, 2008-2017. 
}
\end{minipage}
\end{table}

%%%%%%%%%%%%%%%%%%%%%%%%%%%%%%%%%%%%%%%%%%%%%
% PLACEHOLDER: Correlates of tau/beta table %
%%%%%%%%%%%%%%%%%%%%%%%%%%%%%%%%%%%%%%%%%%%%%
\clearpage
\begin{table}[hp]
\centering
\caption{Correlates with reduction in collections debt at age 65}
\label{tab:tau_beta_correlates_apx}
\resizebox{\textwidth}{!}{
\begin{tabular}{llrrrrrr}
\toprule
& & \multicolumn{2}{c}{Bivariate} &\multicolumn{2}{c}{Multivariate} & \multicolumn{2}{c}{Post-Lasso} \\
\cmidrule{3-8}
Covariate & Estimate Type & Estimate & S.E.  & Estimate & S.E. & Estimate & S.E. \\
\midrule
Black (\%) & Per Capita & -7.17 & (2.77) & -5.74 & (2.28) & -6.23 & (2.08)\\
Greater than high school education (\%) & Per Capita & 11.30 & (1.74) & -2.47 & (3.52) & 4.86 & (2.46)\\
Has any coverage (\%) & Per Capita & 12.00 & (1.86) & 7.09 & (2.94) &  & \\
Has Medicaid (\%) & Per Capita & 6.75 & (1.65) & 3.24 & (2.87) &  & \\
Hospital beds per capita & Per Capita & -1.09 & (1.4) & 1.86 & (1.48) &  & \\
\addlinespace
Income per capita & Per Capita & 11.90 & (1.79) & 6.86 & (5.01) &  & \\
Median house value & Per Capita & 10.70 & (1.88) & -2.25 & (2.49) &  & \\
Hospital occupancy rate (\%) & Per Capita & 6.56 & (1.68) & -0.90 & (3.12) &  & \\
Physical disability (\%) & Per Capita & -11.90 & (2) & -5.60 & (3.21) & -7.41 & (2.56)\\
Poverty rate (\%) & Per Capita & -7.01 & (2.34) & -0.01 & (3.24) & 1.02 & (2.16)\\
\addlinespace
Payment by charity care patients (\$) & Per Capita & -1.52 & (1.65) & -1.78 & (1.46) & -2.70 & (1.53)\\
Medicare spending per enrollee (\$) & Per Capita & -6.48 & (2.08) & -0.63 & (2.98) &  & \\
For-profit hospitals (\%) & Per Capita & -10.20 & (1.96) & -4.96 & (2.17) & -8.29 & (1.97)\\
Teaching hospitals (\%) & Per Capita & 9.69 & (1.51) & 6.14 & (3.32) &  & \\
Cost of charity care per patient day (\$) & Per Capita & 0.07 & (3.1) & -0.96 & (2) & -1.26 & (2.21)\\
\midrule
Black (\%) & Per Newly Insured & -62.20 & (37.3) & -54.20 & (33.7) & -53.50 & (31.8)\\
Greater than high school education (\%) & Per Newly Insured & 76.10 & (25) & -47.20 & (49.8) & -5.16 & (39.9)\\
Has any coverage (\%) & Per Newly Insured & -3.87 & (29.3) & -127.00 & (52.6) &  & \\
Has Medicaid (\%) & Per Newly Insured & 91.80 & (28.4) & 101.00 & (52.8) &  & \\
Hospital beds per capita & Per Newly Insured & 21.00 & (38.2) & 46.80 & (38.7) &  & \\
\addlinespace
Income per capita & Per Newly Insured & 95.50 & (24.9) & 125.00 & (64.3) &  & \\
Median house value & Per Newly Insured & 97.20 & (30.9) & -58.50 & (44.9) &  & \\
Hospital occupancy rate (\%) & Per Newly Insured & 45.20 & (34.5) & -6.50 & (47.8) &  & \\
Physical disability (\%) & Per Newly Insured & -113.00 & (24.6) & -39.50 & (50.8) & -123.00 & (38.6)\\
Poverty rate (\%) & Per Newly Insured & -19.30 & (28.6) & -66.60 & (44.4) & 48.80 & (26.3)\\
\addlinespace
Payment by charity care patients (\$) & Per Newly Insured & -39.30 & (33.5) & -38.00 & (32.4) & -49.20 & (31.5)\\
Medicare spending per enrollee (\$) & Per Newly Insured & -29.50 & (32) & -40.20 & (45.1) &  & \\
For-profit hospitals (\%) & Per Newly Insured & -70.00 & (28.6) & -65.30 & (30.3) & -69.40 & (26.6)\\
Teaching hospitals (\%) & Per Newly Insured & 110.00 & (18.6) & 92.70 & (55.1) &  & \\
Cost of charity care per patient day (\$) & Per Newly Insured & 29.50 & (37.2) & -59.30 & (39.1) & 0.70 & (31.4)\\
\bottomrule
\end{tabular}}
\begin{minipage} {\textwidth} \setstretch{.9} \bigskip
\scriptsize{\textbf{Note:} 
  This table reports the CZ-level correlates with our RD estimated reductions in collections debt at age 65 plotted in Figure \ref{fig:correlates_drop_collections_cz}. The ``Estimate Type'' column indicates whether the row presents correlates with our ``per capita'' or ``per newly-insured'' estimates. For each row, we we present the estimates and standard errors for bivariate, multivariate, and post-Lasso models. We standardize all the variables so the coefficients reflect the strength of the association between a one standard deviation change in the covariate and the estimated reduction in collections debt at age 65.  The multivariate OLS regression results and post-Lasso multivariate regression results are both run on the full set of characteristics. For post-Lasso, we first estimate a Lasso regression on the full set of characteristics and then report the results of multivariate OLS run on the characteristics chosen by the Lasso regression. Source: Consumer credit outcomes are based on 137,340,577 person-year observations from the New York Fed Consumer Credit Panel / Equifax, 2008-2017. CZ-level uninsurance rates are from the American Community Survey, 2008-2017. Healthcare market characteristics are from the Healthcare Cost Report Information System (HCRIS) and the Dartmouth Atlas. For additional details on the data see Section \ref{background_data}.
}
\end{minipage}
\end{table}

%%% TABLE FOR DECOMPOSITION
\clearpage
\begin{table}[hp]
    \centering    
    \caption{Changes in the forecast reductions in collections debt at age 65, pre and post-ACA, used for decomposition in Figure \ref{fig:collections_prepost_aca}}
    \label{tab:decomp_components_apx}
\resizebox{\textwidth}{!}{
 \begin{tabular}{ccccccccccccccccc}
    \toprule
          &  \multicolumn{3}{c}{Per Capita}  & \multicolumn{3}{c}{Insurance Effect} & \multicolumn{3}{c}{Per Newly Insured}& \multicolumn{3}{c}{Covariance}&\multicolumn{4}{c}{Decomposition} \\
          \cmidrule(lr){2-4}\cmidrule(lr){5-7}\cmidrule(lr){8-10}\cmidrule(lr){11-13}\cmidrule(lr){14-17}
        Location & Pre & Post & Diff & Pre & Post & Diff & Pre & Post & Diff &  Pre & Post & Diff & $\eta$ & $\eta_{1}$ & $\eta_{2}$ & $\eta_{3}$\\
        \midrule
South & 43.61 & 30.66 & -12.95 & 0.13 & 0.08 & -0.05 & 365.95 & 408.95 & 43.0 & -0.95 & -0.95 & 0.00 & -0.3 & -0.38 & 0.08 & 0.00\\
All Others & 14.36 & 7.23 & -7.13 & 0.09 & 0.06 & -0.04 & 168.76 & 167.66 & -1.1 & 0.19 & -0.30 & -0.49 & -0.5 & -0.43 & 0.00 & -0.06\\
Difference & -29.25 & -23.43 & 5.82 & -0.03 & -0.02 & 0.01 & -197.20 & -241.30 & -44.1 & 1.15 & 0.66 & -0.49 & -0.2 & -0.05 & -0.08 & -0.06\\
\bottomrule
    \end{tabular}}
    
\begin{minipage} {\textwidth} \setstretch{.9} \bigskip
\footnotesize{\textbf{Note:} 
  This table reports the components involved in the decomposition presented in Panel B of Figure \ref{fig:collections_prepost_aca} and discussed in Appendix \ref{sec:decomp_apx}. Averages are constructed using unweighted means across commuting zones. South is defined using Census regions, and includes Alabama, Arkansas, Delaware, Florida, Georgia, Kentucky, Louisiana, Maryland, Mississippi, North Carolina, Oklahoma, South Carolina, Tennessee, Texas, Virginia, and West Virginia.
}
\end{minipage}
\end{table}

\end{document}